\documentclass[twocolumn,tighten]{aastex631}

\usepackage{graphicx}
\usepackage{bm}
\usepackage{url}
\usepackage{threeparttable}
\usepackage{amsmath,amssymb}
\usepackage{xspace}
\usepackage{ulem}
\usepackage{subfigure}
\usepackage{longtable}
\usepackage{afterpage}

\newcommand{\ergs}{erg s$^{-1}$}

\shorttitle{Nuclear Mm-wave Emission of Hard-X-ray Selected Nearby AGNs}
\shortauthors{Kawamuro et al.}

\begin{document}
\title{
    BASS XXXII: 
    Studying the Nuclear Mm-wave Continuum Emission of AGNs with ALMA at 
    Scales $\lesssim$ 100--200\, pc
} 

\correspondingauthor{Taiki Kawamuro}
\email{taiki.kawamuro@mail.udp.cl}

\author[0000-0002-6808-2052]{Taiki Kawamuro}
\altaffiliation{FONDECYT postdoctoral fellow}
\affil{Nu\'{c}leo de Astronom\'{i}a de la Facultad de Ingenier\'{i}a, Universidad Diego Portales, Av. Ej\'{e}ercito Libertador 441, Santiago, Chile}
\affil{National Astronomical Observatory of Japan, Osawa, Mitaka, Tokyo 181-8588, Japan}

\author[0000-0001-5231-2645]{Claudio Ricci} 
\affil{Nu\'{c}leo de Astronom\'{i}a de la Facultad de Ingenier\'{i}a, Universidad Diego Portales, Av. Ej\'{e}ercito Libertador 441, Santiago, Chile}
\affil{Kavli Institute for Astronomy and Astrophysics, Peking University, Beijing 100871, People’s Republic of China}

\author{Masatoshi Imanishi} 
\affil{National Astronomical Observatory of Japan, Osawa, Mitaka, Tokyo 181-8588, Japan}
\affil{Department of Astronomy, School of Science, Graduate University for Advanced Studies (SOKENDAI), 2-21-1 Osawa, Mitaka, Tokyo 181-8588, Japan}

\author{Richard F. Mushotzky}
\affil{Department of Astronomy, University of Maryland, College Park, MD 20742, USA}

\author{Takuma Izumi}
\altaffiliation{NAOJ fellow} 
\affil{National Astronomical Observatory of Japan, Osawa, Mitaka, Tokyo 181-8588, Japan}
\affil{Department of Astronomy, School of Science, Graduate University for Advanced Studies (SOKENDAI), 2-21-1 Osawa, Mitaka, Tokyo 181-8588, Japan}

\author[0000-0001-5742-5980]{Federica Ricci}
\affil{Dipartimento di Fisica e Astronomia, Università di Bologna, via Gobetti 93/2, 40129 Bologna, Italy}
\affil{INAF Osservatorio Astronomico di Bologna, via Gobetti 93/3, 40129 Bologna, Italy}

\author[0000-0002-8686-8737]{Franz E. Bauer}
\affiliation{Instituto de Astrof\'{\i}sica  and Centro de Astroingenier{\'{\i}}a, Facultad de F\'{i}sica, Pontificia Universidad Cat\'{o}lica de Chile, Casilla 306, Santiago 22, Chile}
\affiliation{Millennium Institute of Astrophysics (MAS), Nuncio Monse{\~{n}}or S{\'{o}}tero Sanz 100, Providencia, Santiago, Chile}
\affiliation{Space Science Institute, 4750 Walnut Street, Suite 205, Boulder, Colorado 80301, USA}

\author{Michael J. Koss} 
\affil{Eureka Scientific, 2452 Delmer Street Suite 100, Oakland, CA 94602-3017, USA}

\author[0000-0002-3683-7297]{Benny Trakhtenbrot}
\affil{School of Physics and Astronomy, Tel Aviv University, Tel Aviv 69978, Israel}

\author{Kohei Ichikawa}
\affil{Frontier Research Institute for Interdisciplinary Sciences, Tohoku University, Sendai 980-8578, Japan}
\affil{Astronomical Institute, Tohoku University, Aramaki, Aoba-ku, Sendai, Miyagi 980-8578, Japan}
\affil{Max-Planck-Institut f\"{u}r extraterrestrische Physik (MPE), Giessenbachstrasse 1, D-85748 Garching bei M{\" u}nchen, Germanyn}

\author{Alejandra F. Rojas}
\altaffiliation{FONDECYT postdoctoral fellow}
\affil{Centro de Astronom\'{i}a (CITEVA), Universidad de Antofagasta, Avenida Angamos 601, Antofagasta, Chile}

\author{Krista Lynne Smith}
\altaffiliation{Einstein fellow}
\affil{KIPAC at SLAC, Stanford University, Menlo Park, CA 94025, USA}
\affil{Southern Methodist University, Department of Physics, Dallas, TX 75205, USA}

\author{Taro Shimizu}
\affil{Department of Physics and Astronomy, University College London, Gower Street, London WC1E 6BT, UK}

\author[0000-0002-5037-951X]{Kyuseok Oh}
\affil{Korea Astronomy and Space Science Institute, Daedeokdae-ro 776, Yuseong-gu, Daejeon 34055, Republic of Korea}

\author{Jakob S. den Brok}
\affil{Institute for Particle Physics and Astrophysics, ETH Z\"{u}rich, Wolfgang-Pauli-Strasse 27, CH-8093 Z\"{u}rich, Switzerland}
\affil{Argelander Institute for Astronomy, Auf dem H\"{u}gel 71, 53231, Bonn, Germany}

\author[0000-0002-9850-6290]{Shunsuke Baba}
\affil{Graduate School of Science and Engineering, Kagoshima University, Korimoto, Kagoshima, Kagoshima 890-0065, Japan}

\author{Mislav Balokovi\'{c}}
\affil{Yale Center for Astronomy \& Astrophysics, 52 Hillhouse Avenue, New Haven, CT 06511, USA}
\affil{Department of Physics, Yale University, P.O. Box 208120, New Haven, CT 06520, USA}

\author[0000-0001-9910-3234]{Chin-Shin Chang}
\affil{Joint ALMA Observatory, Alonso de Cordova 3107, Vitacura, Santiago, Chile}

\author{Darshan Kakkad}
\affil{European Southern Observatory, Alonso de Cordova 3107, Vitacura, Casilla 19001, Santiago de Chile, Chile}
\affil{European Southern Observatory, Karl-Schwarzschild-Strasse 2, Garching bei M{\" u}nchen, Germany}

\author{Ryan W. Pfeifle}
\affil{Department of Physics \& Astronomy, George Mason University, 4400 University Drive, MSN 3F3, Fairfax, VA 22030, USA}

\author{George C. Privon}
\affil{Department of Astronomy, University of Florida, P.O. Box 112055, Gainesville, FL 32611, USA}

\author[0000-0001-8433-550X]{Matthew J. Temple}
\altaffiliation{FONDECYT postdoctoral fellow}
\affil{Nu\'{c}leo de Astronom\'{i}a de la Facultad de Ingenier\'{i}a, Universidad Diego Portales, Av. Ej\'{e}ercito Libertador 441, Santiago, Chile}

\author{Yoshihiro Ueda}
\affil{Department of Astronomy, Kyoto University, Kyoto 606-8502, Japan}

\author{Fiona Harrison}
\affil{Cahill Center for Astronomy and Astrophysics, California Institute of Technology, Pasadena, CA 91125, USA}

\author{Meredith C. Powell}
\affil{Institute of Particle Astrophysics and Cosmology, Stan- ford University, 452 Lomita Mall, Stanford, CA 94305, USA}

\author{Daniel Stern}
\affil{Jet Propulsion Laboratory, California Institute of Technology, 4800 Oak Grove Drive, MS 169-224, Pasadena, CA 91109, USA}

\author{Meg Urry}
\affil{Yale Center for Astronomy \& Astrophysics, Physics Department, New Haven, CT 06520, USA}

\author{David B. Sanders}
\affil{Institute for Astronomy, 2680 Woodlawn Drive, University of Hawaii, Honolulu, HI 96822, USA}

\begin{abstract}

To understand the origin of nuclear ($\lesssim$ 100\,pc) millimeter-wave (mm-wave) continuum emission in active galactic nuclei (AGNs), we systematically analyzed sub-arcsec resolution Band-6 (211--275\,GHz) ALMA data of 98 nearby AGNs ($z <$ 0.05) from the 70-month Swift/BAT catalog.
The sample, almost unbiased for obscured systems, provides the largest number of AGNs to date with high mm-wave spatial resolution sampling ($\sim$ 1--200 pc), and spans broad ranges of 14--150\,keV luminosity \{$40 < \log[L_{\rm 14-150}/({\rm erg\,s^{-1}})] < 45$\}, black hole mass [$5 < \log(M_{\rm BH}/M_\odot) < 10$], and Eddington ratio ($-4 < \log \lambda_{\rm Edd} < 2$). We find a significant correlation between 1.3\,mm (230\,GHz) and 14--150\,keV luminosities. Its scatter is $\approx$ 0.36\,dex, and the mm-wave emission may serve as a good proxy of the AGN luminosity, free of dust extinction up to $N_{\rm H} \sim 10^{26}$ cm$^{-2}$. While the mm-wave emission could be self-absorbed synchrotron radiation around the X-ray corona according to past works, we also discuss different possible origins of the mm-wave emission; AGN-related dust emission, outflow-driven shocks, and a small-scale ($<$ 200\,pc) jet. 
The dust emission is unlikely to be dominant, 
as the mm-wave slope is generally flatter than expected. 
Also, due to no increase in the mm-wave luminosity with the Eddington ratio, a radiation-driven outflow model is possibly not the common mechanism. 
Furthermore, we find independence of the mm-wave luminosity on 
indicators of the inclination angle from the polar axis of the nuclear structure, which is inconsistent with a jet model whose luminosity depends only on the angle.

\end{abstract}

\keywords{galaxies: active -- X-rays: galaxies -- submm/mm: galaxies} 

\section{Introduction}\label{sec:int}

Active galactic nuclei (AGNs) emit radiation over a wide range of wavelengths \cite[e.g., ][]{Elv94,Ho08,Mul11,Ber21}. 
By decomposing their spectral energy distributions (SEDs),  
it has been known that the most prominent spectral components originate from an accretion disk (optical and UV: ultra-violet), a corona of hot electrons (X-ray), and surrounding heated dust on scales of $\sim 0.01$--1\,pc  \citep[IR: infrared; e.g.,][]{Kos14rev,Alm17}. 
However, millimeter-wave (mm-wave hereafter) emission has not been often considered in these studies \cite[e.g., see Figure 1 of][]{Hic18}. 
One of the main reasons is that star-formation (SF) processes in the host galaxy could significantly contribute to the mm-wave emission. 
Emission of dust heated by stellar radiation, often represented as a power law with $\alpha_{\rm mm} \sim -3.5$\footnote{We define $\alpha_{\rm mm}$ as $S_\nu \propto \nu^{-\alpha_{\rm mm}}$ in flux density.}, can be significant \citep[e.g.,][]{Con98,Mul11}. Also, free-free emission from H{\sc{ii}} regions appears as an almost flat spectrum with $\alpha_{\rm mm} = 0.1$, and the synchrotron emission component from supernova remnants and other stellar processes \cite[e.g.,][]{Con92,Pan19} extends from the centimeter-wave (cm-wave hereafter) band ($<$ 10\,GHz) with $\alpha_{\rm mm} \sim 0.8$ \cite[][]{Tab17}.

To separate the mm-wave emission due to an AGN from that of SF, it is crucial to use high-resolution observations. 
As an extreme case, \cite{Eht19I} revealed emission at 230 GHz at the very center of M\,87 on a scale of $\sim 40$ $\mu$arcsec ($\approx 3\times10^{-3}$\,pc), by coordinating mm-wave telescopes distributed across the globe to form an Earth-sized virtual telescope. 
The scale is comparable to the expected horizon-scale ($\approx$ 5 Schwarzschild radii) structure of a supermassive black hole (SMBH) with $\sim 7\times 10^9$ solar masses \citep[$M_\odot$; e.g.,][]{Geb11}. 
This radiation was interpreted as synchrotron emission. 
The Atacama Large Millimeter/submillimeter Array (ALMA), which observed many more objects than the Event Horizon Telescope Collaboration, has provided supportive results for the presence of AGN-related mm-wave emission. \cite{Ino18} identified mm-wave (in particular at $\sim$ 100--300 GHz) emission that exceeds the extrapolation of 
a component in the lower frequency band ($\sim$ 1--10\,GHz) in two nearby AGNs (IC~4329A and NGC~985). The authors then interpreted their excesses as due to self-absorbed synchrotron radiation from compact regions on scales of $\sim$ 40--50 Schwarzschild radii \cite[see also][]{Lao08,Beh15,Beh18,Ino14,doi16,Wu18,Ino20}. 
Although direct imaging at extreme resolutions better than $\sim$ 40 $\mu$arcsec \citep{Eht19I} and spectral decomposition are powerful tools to isolate AGN emission, one can also identify AGN mm-wave emission at high-spatial resolution by using the time variability. 
For example, by observing the nearby bright AGN NGC\,7469, \cite{Beh20} reported that there could be a 14-day delay in X-ray emission behind mm-wave emission \cite[see also][]{Bal15,Izu17}. While the 
corresponding light travel time of $\sim$ 0.01\,pc is consistent with the scale of a broad line region \citep{Pet14}, the authors discussed that the mm-wave and X-ray emission was produced on a scale of a few gravitational radii. Their idea is based on a similar phenomenon in stellar coronae, where mm-wave emitting electrons diffuse and lose energy slowly in magnetic fields, producing X-rays.

Although various mechanisms, such as dust heated by an AGN, outflow-driven shocks, and a jet, have also been discussed as the origin \citep[e.g.,][]{Jia10,Nim15}, the above observational results suggest the presence of the mm-wave emission on the scale of an X-ray-emitting hot corona \citep[$\sim$ 10 Schwarzschild radii; e.g.,][]{Mor08,Mor12}. The coronal scenario is interestingly consistent with the idea that magnetic reconnection contributes to the formation of the X-ray corona  \cite[e.g.,][]{Liu02,Liu03,Liu16,Che20}, predicting a link between the mm-wave and X-ray emission. 

Understanding the origin of the mm-wave emission is crucial to providing a complete picture of the AGN phenomenon and also could have several important applications. 
If it is confirmed that the mm-wave emission can serve as a good measure of the AGN luminosity, that could play a valuable role in constraining AGN activity, particularly for buried systems with thick gas layers (e.g., $N_{\rm H} > 10^{25}$ cm$^{-2}$). Such objects may be associated with rapidly growing SMBHs in merging galaxies  \citep[e.g.,][]{San88,Hop10,Ric17b,Ric21b,Yam21} and therefore may be crucial for understanding the growth of SMBHs \citep[e.g.,][]{Hop10,Tre12,Ble18}. 
In fact, mm-wave emission is almost unaffected by dust extinction up to a hydrogen column density of $N_{\rm H} \sim 10^{26}$ cm$^{-2}$, where the optical depth becomes $\sim$ 1, considering a Galactic dust-to-gas ratio \citep{Hil83}.
This column density is much larger than $N_{\rm H} \sim 10^{24}$ cm$^{-2}$ corresponding to optical depth of $\sim$ 1 for hard X-rays at 10\,keV \citep[e.g.,][]{Mor83,Bur11,Ric15}, which currently provide the least-biased samples for AGN studies in the nearby Universe \cite[$z < 0.1$; e.g., ][]{Geo15,Air15,Kaw13,Kaw16a,Kaw16c,Kaw21,Kos16,ric17c,Oh17,Kam18,Gar19}.
However, previous observational studies, using the Combined Array for Research in Millimetre-wave Astronomy (CARMA) and the Australia Telescope Compact Array (ATCA) telescopes at resolutions of $\gtrsim$ 1\arcsec\ \citep{Beh15,Beh18}, found only tentative relations between mm-wave and AGN X-ray luminosities. 

In this paper, we assess correlations of nuclear ($\lesssim$ 100\,pc)\footnote{Throughout this paper, we refer to a region within a radius of $\lesssim 100$\,pc as ``nuclear".} mm-wave luminosity 
with AGN luminosities, using high-resolution ($\lesssim$ 0\farcs6) ALMA Band-6 (211--275\,GHz) data for a large sample of nearby ($z < 0.05$) AGNs, selected from the 70-month Swift/BAT catalog \citep{Bau13}. Then, we investigate the origin of the nuclear mm-wave emission. 
This study is part of the BAT AGN Spectroscopic Survey (BASS) project \cite[e.g., ][]{Kos17,ric17c,Kos22_overview}, 
providing one of the best-studied samples of nearby AGNs by 
collecting a large set of multi-wavelength data for Swift/BAT-detected AGNs, from radio to the $\gamma$-rays \citep[e.g., ][]{Oh17,ric17c,Kos17,lamperti17,Shi17,Ich19,pal19,bae19,roj20,smi20,Kos22_dis,Kos22_catalog}. 
Thus, our sample is not only almost unbiased for obscured systems thanks to the hard X-ray ($>$ 10\,keV) selection but also allows us to explore potential relations between the mm-wave emission and various physical properties of the AGNs, such as X-ray and bolometric luminosities ($L_{\rm bol}$), black hole mass ($M_{\rm BH}$), and Eddington ratio ($\lambda_{\rm Edd}$).

This paper is structured as follows. In Sections~\ref{sec:sample}, 
\ref{sec:anc}, and \ref{sec:alma_data}, we introduce our sample, ancillary data, and 
our analysis of the archival ALMA data, respectively.
With these data, empirical relations of the mm-wave luminosity with AGN luminosities are presented and discussed in Section~\ref{sec:cor}. 
In Sections~\ref{sec:sf_cont} and \ref{sec:cor4sigAGN}, 
we discuss whether the AGN is the dominant contributor to the mm-wave emission compared to the SF. After this, in Section~\ref{sec:reg_sum}, we summarize the mm-wave relations so that one can use them to estimate the AGN luminosity. As a final discussion, we discuss four possible AGN mechanisms as the origin of the mm-wave emission in Section~\ref{sec:agn_mec}. 
Finally, we demonstrate the potential of mm-wave observations through ALMA and ngVLA to identify obscured AGNs in Section~\ref{sec:future}, and our findings are summarized in Section~\ref{sec:sum}. All data in the mm-wave band produced through this work (e.g., ALMA images, fluxes, spectral indices, and luminosities) are summarized in an accompanying paper (Kawamuro et al., submitted, hereafter Paper~II).

Throughout the paper, we adopt standard cosmological parameters ($H_0$ = 70 km\,s$^{-1}$\,Mpc$^{-1}$, $\Omega_{\rm m} = 0.3$, $\Omega_\Lambda = 0.7$). 
Particularly for objects at distances below 50\,Mpc, redshift-independent distances are adopted by referring to the Extragalactic Distance Database \citep{Tul09}, a catalog of the Cosmicflows project \citep{Cou17}, and NASA/IPAC Extragalactic Database in this order \citep[see][for detailes]{Kos22_catalog}. 
Also, we define a correlation as ``significant" if the two-sided Spearman correlation test returns a $p$-value smaller than 1\%. 
Lastly, uncertainties are quoted at the 1$\sigma$ equivalent values unless otherwise stated.

\section{Sample}\label{sec:sample}

We assembled a nearby AGN sample ($z < 0.05$) by selecting, from an X-ray spectral catalog of AGNs detected in the Swift/BAT 70-month catalog \citep{Bau13,ric17c}, all the nearby ($<$ 200\,Mpc) objects for which archival ALMA Band-6 (211--275 GHz) data with angular resolutions $< 1$\arcsec\ are available as of 2021 April. 
By performing a systematic broad-band ($\sim$ 0.5--200\,keV) spectral analysis, \cite{ric17c} provide accurately estimated intrinsic AGN X-ray luminosities, which are crucial for our study. 
ALMA Band 6 was selected because AGN emission is expected to be prominent around the frequency band \citep[i.e., $\gtrsim$ 100\,GHz; e.g.,][]{Ino18}, and the band provides the largest sample of sources with $\gtrsim$ 100\,GHz data between Band 3 and Band 10. 
We searched for the ALMA data of BAT AGNs using a radius of 5\arcsec, corresponding to half the radius of the typical ALMA primary beam size in Band 6 ($\sim$ 20\arcsec). 
As a result, 98 AGNs are selected for our study and are listed in Table~\ref{tab_app:sample} in the appendix. 
We note that \cite{Kos22_catalog} listed some BAT AGNs as those having Blazar-like properties by confirming that their SEDs, 
consisting of at least radio and X-ray data, are dominated by non-thermal emission from radio to $\gamma$-rays and their radio properties are consistent with relativistic beaming. 
For identification, they specifically referred to the Roma Blazar Catalog \cite[BZCAT;][]{Mas09} and the follow-up work by \cite{pal19}. 
Although there are four Blazar-like BAT-detected AGNs that are at $z < 0.05$ and are observable with ALMA (Decl. $<$ 40\arcdeg),
none of them are included in our sample because they do not have publicly available Band-6 data. 
In Section~\ref{sec:ll_cor}, we however mention that they seem much more mm-wave luminous than our targets.

\begin{figure}
    \centering
    \includegraphics[width=8.1cm]{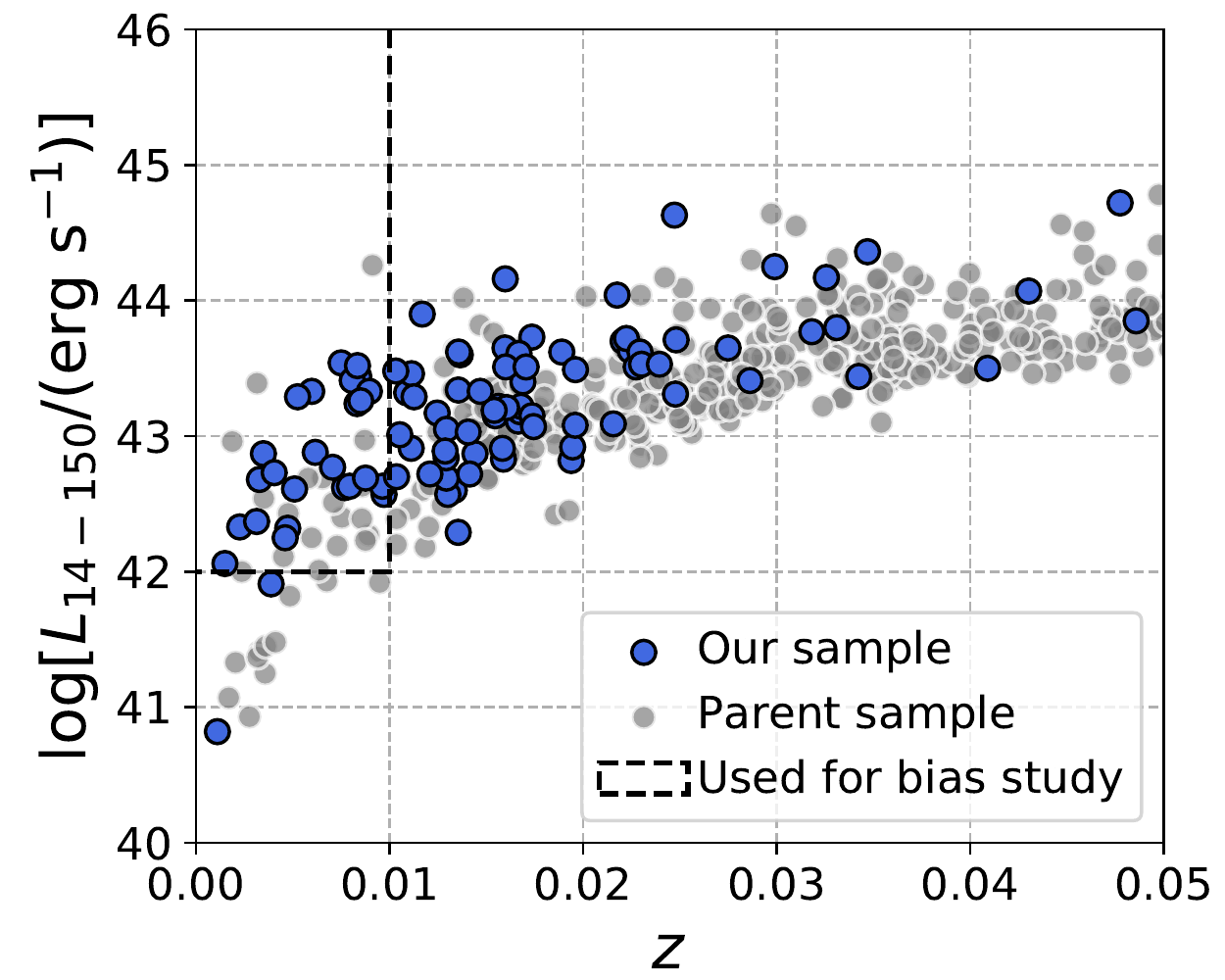} 
    \caption{Plot of 14--150\,keV luminosity versus redshift for our sample (blue) and the parent sample of \cite{ric17c} (gray).
    The area covered by $\log [L_{\rm 14-150}/({\rm erg}\,{\rm s}^{-1})] > 42$ and $z < 0.01$ is adopted to select AGNs to investigate the Malmquist bias (Section~\ref{sec:bias}).
    }
    \label{fig:z_vs_lx}
\end{figure}

Figure~\ref{fig:z_vs_lx} plots our AGNs in the 14--150\,keV luminosity versus redshift plane.
Compared with the most up‐to‐date BASS DR2 catalog of \cite{Kos22_catalog}, our sample comprises $\approx$ 34\% of the non-Blazar AGNs in $z < 0.05$ and Decl. $< 40^\circ$.
Also, among the DR2 AGNs in these $z$ and Decl. ranges, 22\% of the type-1 AGNs, 48\% of the type-1.9 AGNs, and 25\% of the type-2 AGNs are included in our sample.

\begin{figure*}
    \centering 
    \includegraphics[width=5.8cm]{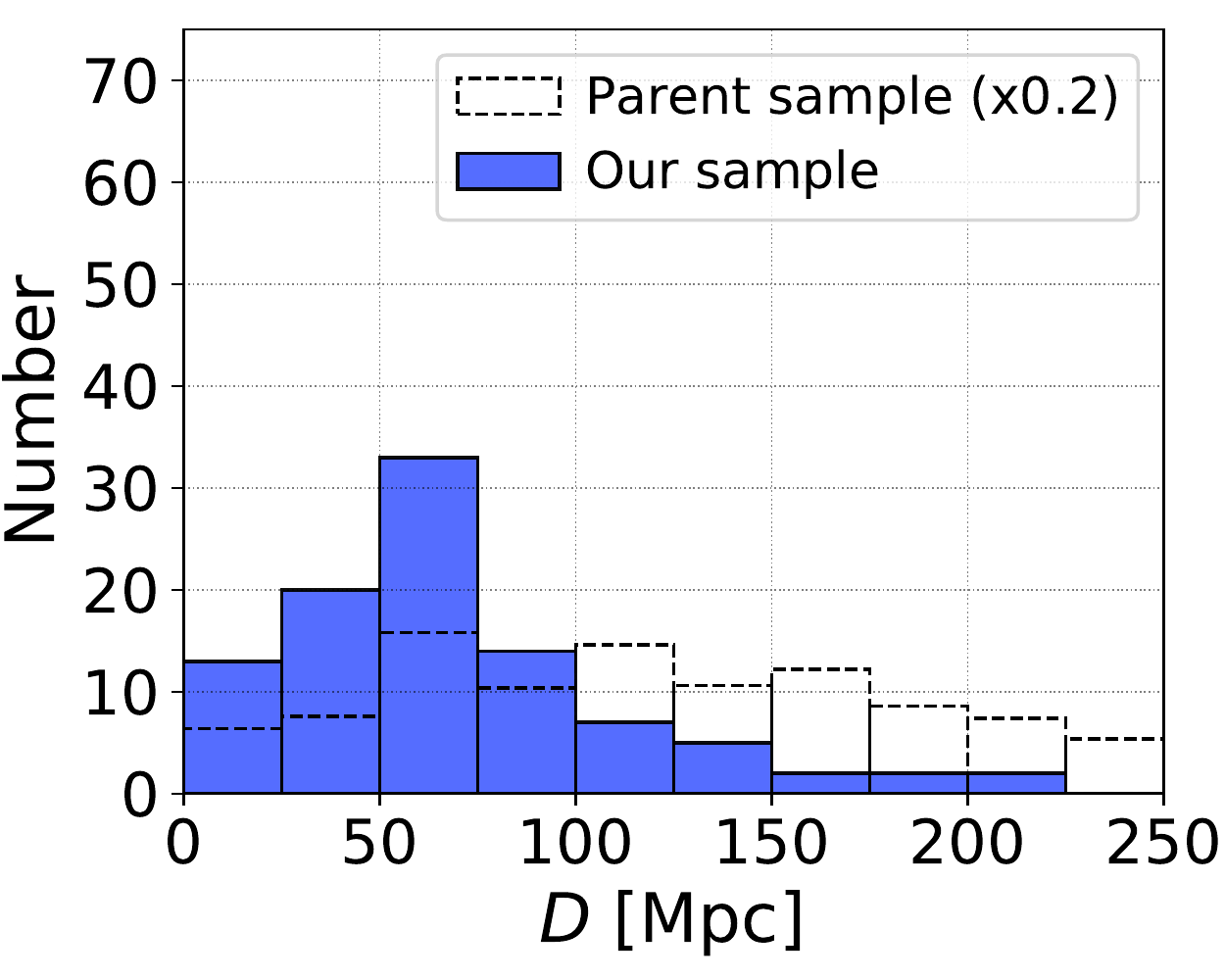}
    \includegraphics[width=5.8cm]{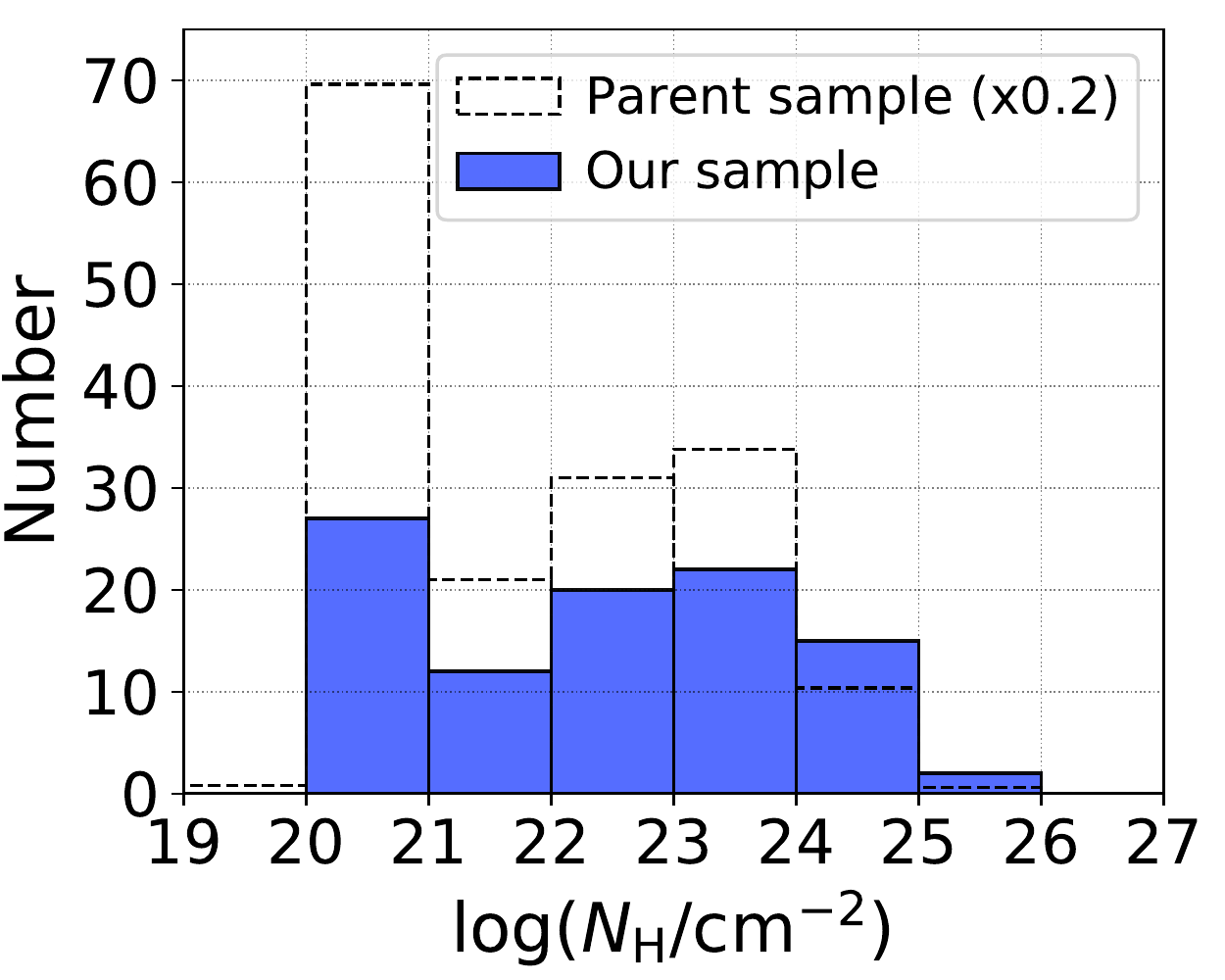}
    \includegraphics[width=5.8cm]{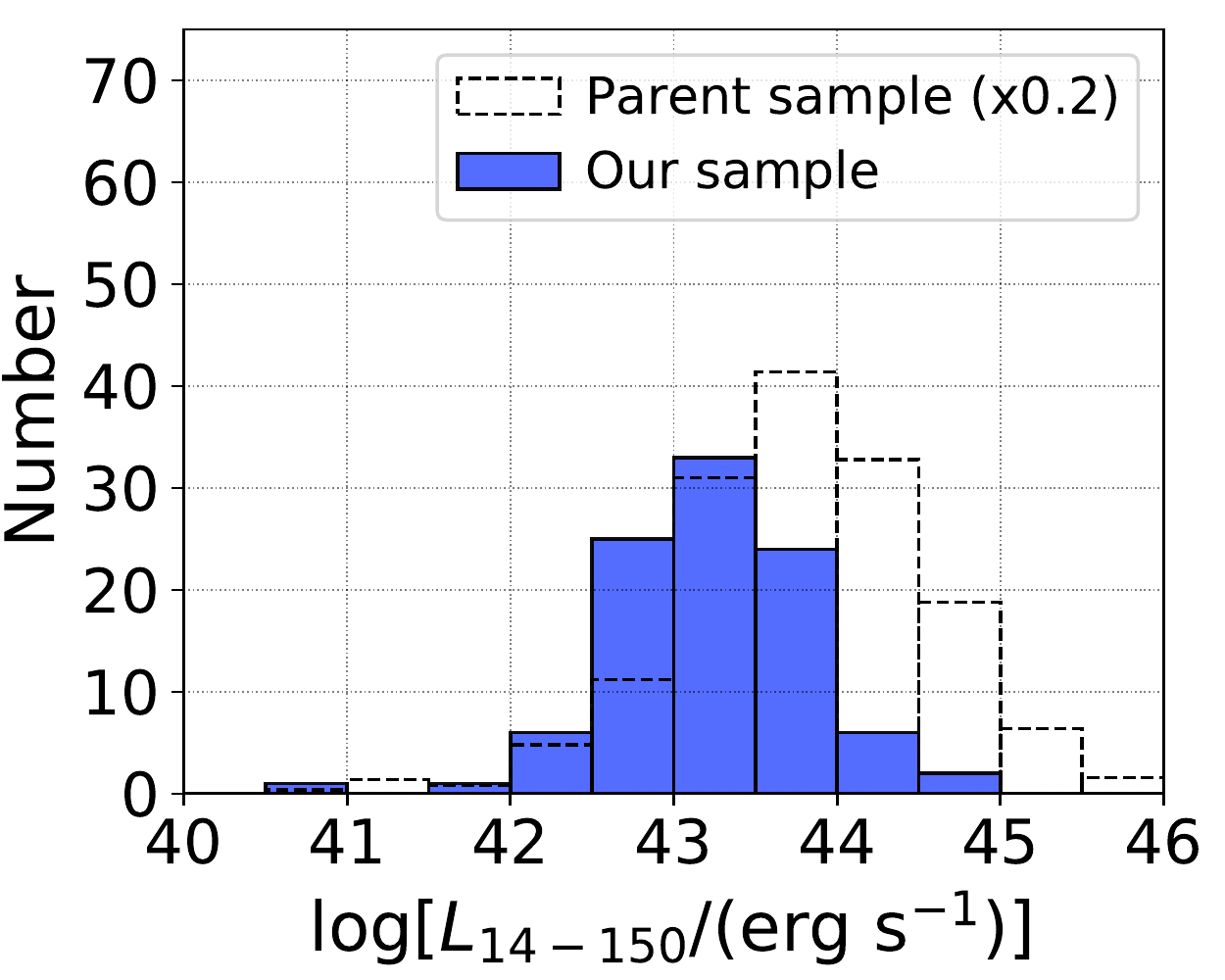}
    \includegraphics[width=5.8cm]{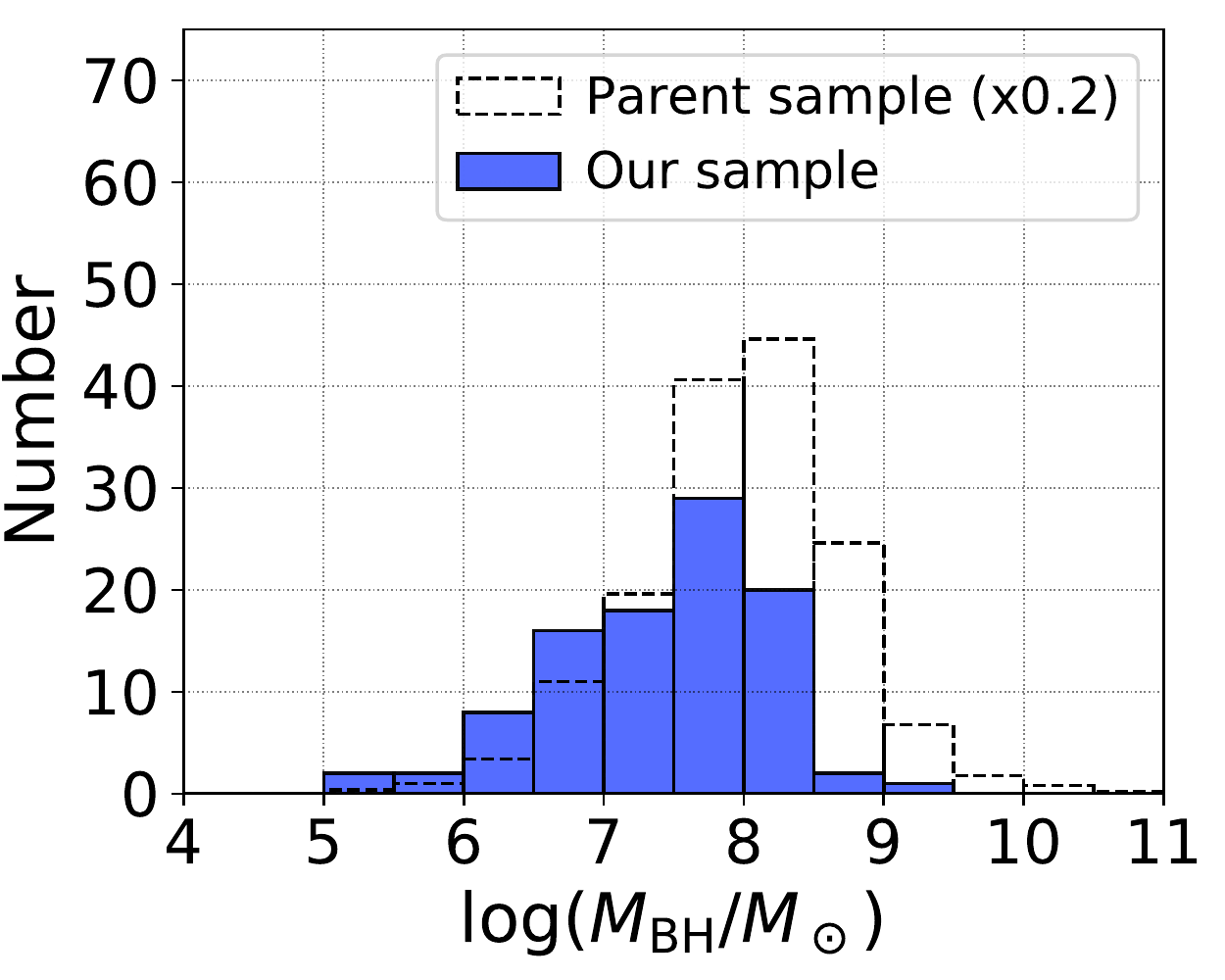}
    \includegraphics[width=5.8cm]{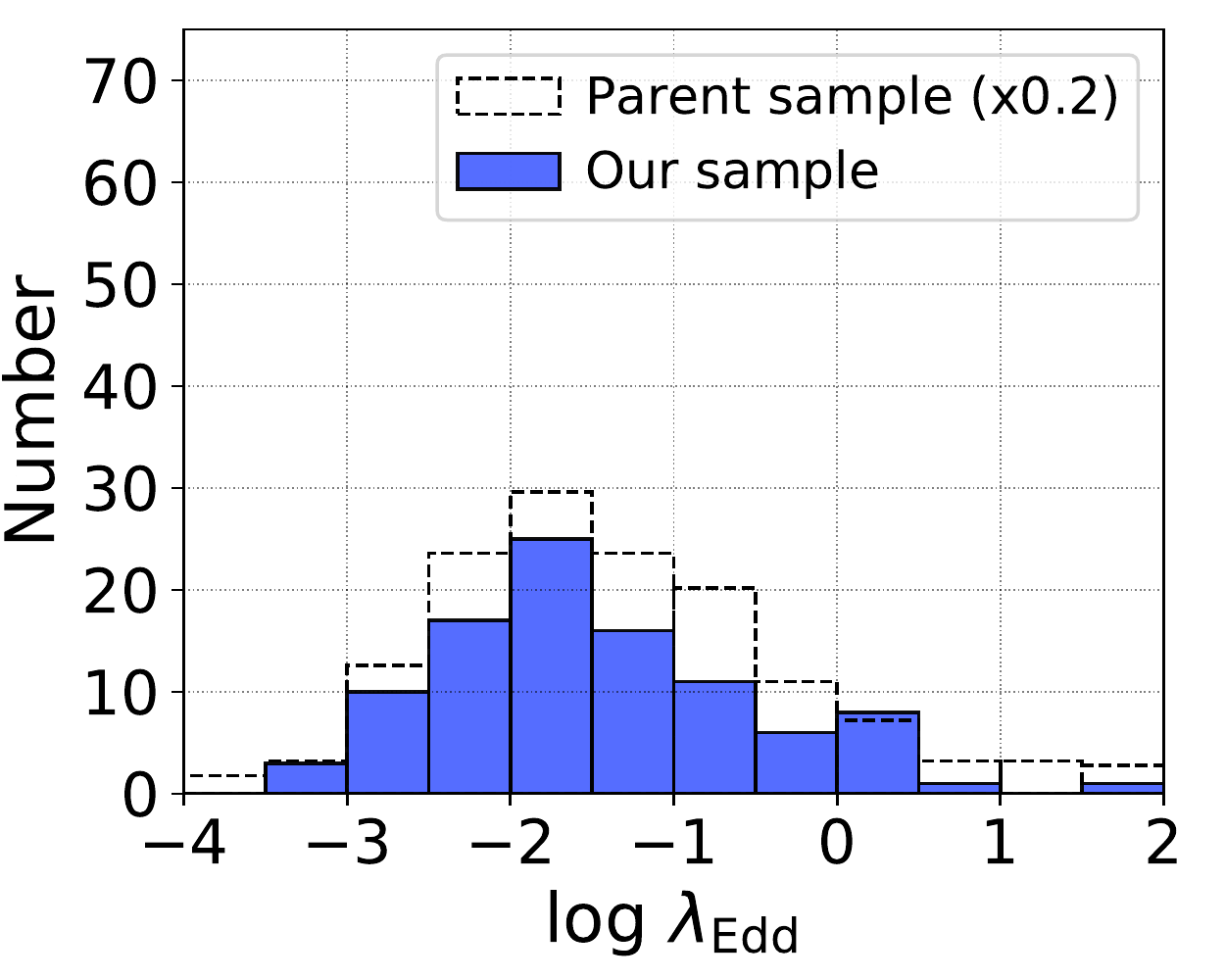}
    \includegraphics[width=5.8cm]{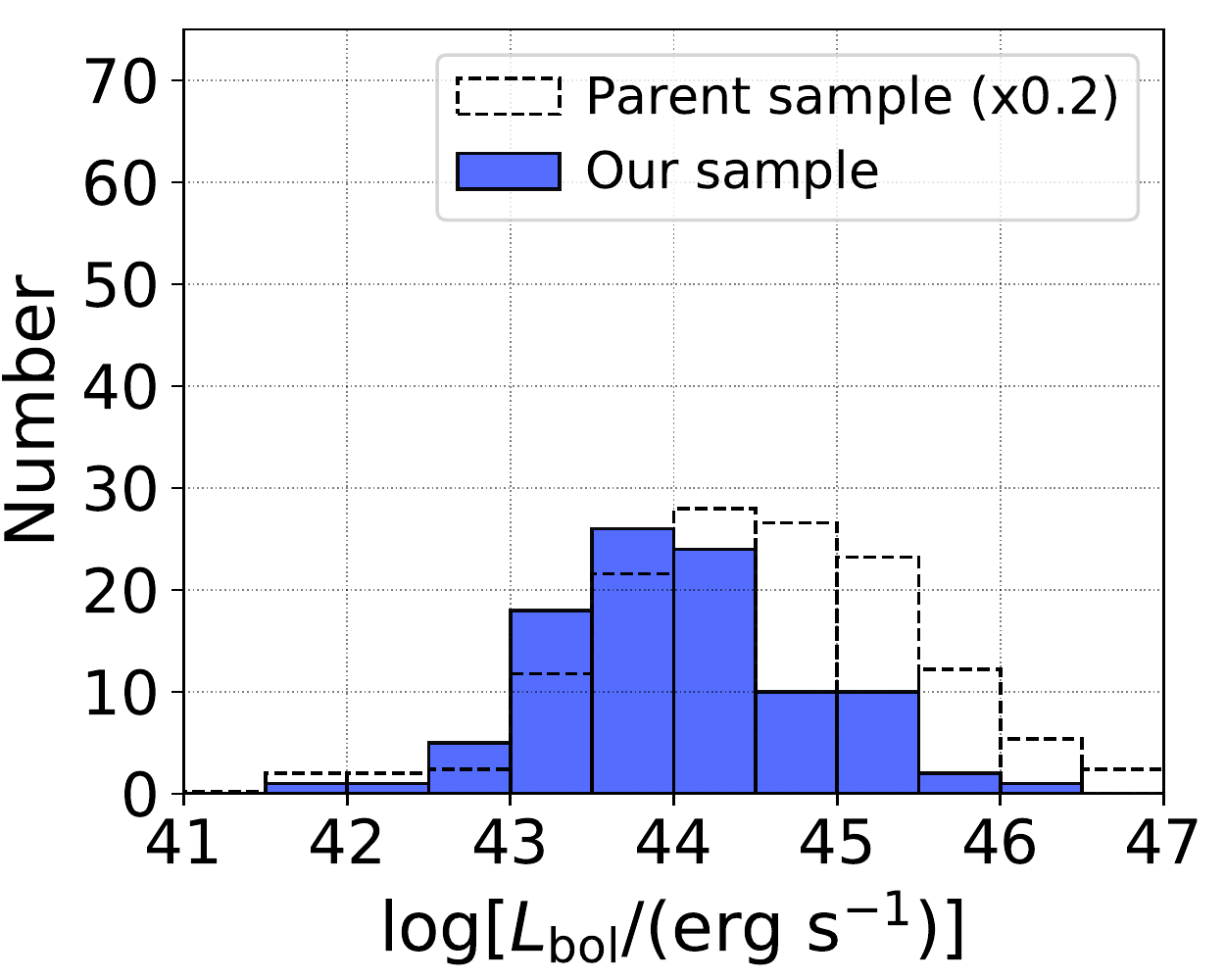}
    \caption{Histograms of essential parameters for our sample and the parent sample of \cite{ric17c} (from the left top to the right bottom: distance, absorbing hydrogen column density measured in the X-ray band, 14--150\,keV luminosity, black hole mass, Eddington ratio, and bolometric luminosity). 
    The histograms of $D$, $M_{\rm BH}$, $\lambda_{\rm Edd}$, and $L_{\rm bol}$ are composed of all 98 AGNs, whose black hole masses 
    were measured either by \cite{Kos17} or by \cite{Kos22_catalog}. 
    Those of $N_{\rm H}$, and $L_{\rm 14-150}$ also include all 98 AGNs, and the values were taken from \cite{ric17c}. 
    The bolometric luminosities, necessary to derive the Eddington ratios, were calculated based on a 2--10\,keV-to-bolometric correction factor of \cite{Dur20}. 
    The histograms for the parent sample are rescaled by a factor of 0.2 to be easily compared with those of our sample. 
    }
    \label{fig:hist}
\end{figure*}

Our parent BAT sample is a flux-limited one that is almost unbiased for obscured systems, up to the Compton-thick ($N_{\rm H} \sim 10^{24}$ cm$^{-2}$) level \citep{Ric15}. 
However, the characteristics of our sample should be different from that, given that 
the archived ALMA data are the results of accepted proposals, which selected appropriate objects to achieve each objective and thus possibly produce selection biases. For example, there were a proposal that included preferentially the nearest AGNs ($D <$ 40\,Mpc; 2019.1.01742.S) and another that focused on nearby luminous AGNs ($z < 0.05$ and $L_{\rm bol} > 10^{43}$ erg\,s$^{-1}$; 2017.1.01439.S). 
To assess possible biases in our sample, in 
Figure~\ref{fig:hist}, we show the histograms of some basic properties of our AGNs, including distance ($D$), line-of-sight absorbing hydrogen column density ($N_{\rm H}$; \citealp{ric17c}), 14--150\,keV luminosity ($L_{\rm 14-150}$; \citealp{ric17c}), black hole mass ($M_{\rm BH}$; \citealp{Kos17,Kos22_catalog}), Eddington ratio ($\lambda_{\rm Edd}$), and bolometric luminosity ($L_{\rm bol}$). 
The bolometric luminosities are estimated by considering a 2--10 keV bolometric correction function that depends on the Eddington ratio, with a scatter of 0.31\,dex, derived by \cite{Dur20}. Here, we use 2--10\,keV luminosities estimated from 14--150\,keV ones via X-ray photon indices. The choice is because the 14--150\,keV intrinsic luminosities were measured based on BAT 70-month averaged spectra and can be readily used without considering the possible effects of short time variability.
As indicated in Figure~\ref{fig:hist}, our sample covers a wide range in luminosity \{$40 \lesssim \log[L_{\rm 14-150}/({\rm erg\,s^{-1}})] \lesssim 45$\}, black hole mass [$5 \lesssim \log(M_{\rm BH}/M_\odot) \lesssim 10$] and Eddington ratio ($-4 \lesssim \log \lambda_{\rm Edd} \lesssim 2$).
Compared with the parent sample \citep{ric17c},
our sample is strongly biased in favor of objects with distances below 100\,Mpc.
Accordingly, extremely luminous and massive objects, which are typically rarer and hence preferentially found within larger volumes, 
do not seem to be well covered by our sample.
Also, our sample preferentially covers the highest end in column density, $\log(N_{\rm H}/{\rm cm}^{-2}) > 24$. This bias may be because we preferentially sample less luminous objects, which are often obscured with column densities greater than $10^{22}$ cm$^{-2}$, as shown in Figure~14 of \cite{ric17c}. 
To quantitatively discuss whether the biases of $L_{\rm 14-150}$, $M_{\rm BH}$, and $N_{\rm H}$ can be attributed to the distance bias, we define two subsamples of nearby AGNs ($D <$ 100\,Mpc) and distant AGNs ($D >$ 100\,Mpc), and compare their 
distributions for $L_{\rm 14-150}$, $M_{\rm BH}$ and $N_{\rm H}$ using the Kolmogorov-Smirnov test (KS-test). 
As a result, the $p$-values are found to be $\ll$ 0.01, supporting that the distributions are significantly different for the investigated parameters.
Additionally, the Eddington-ratio distributions are compared and are found to be statistically indistinguishable according to a $p$-value larger than 0.01, consistent with the trend seen in Figure~\ref{fig:hist}.

\begin{figure}
    \centering 
    \includegraphics[width=8.5cm]{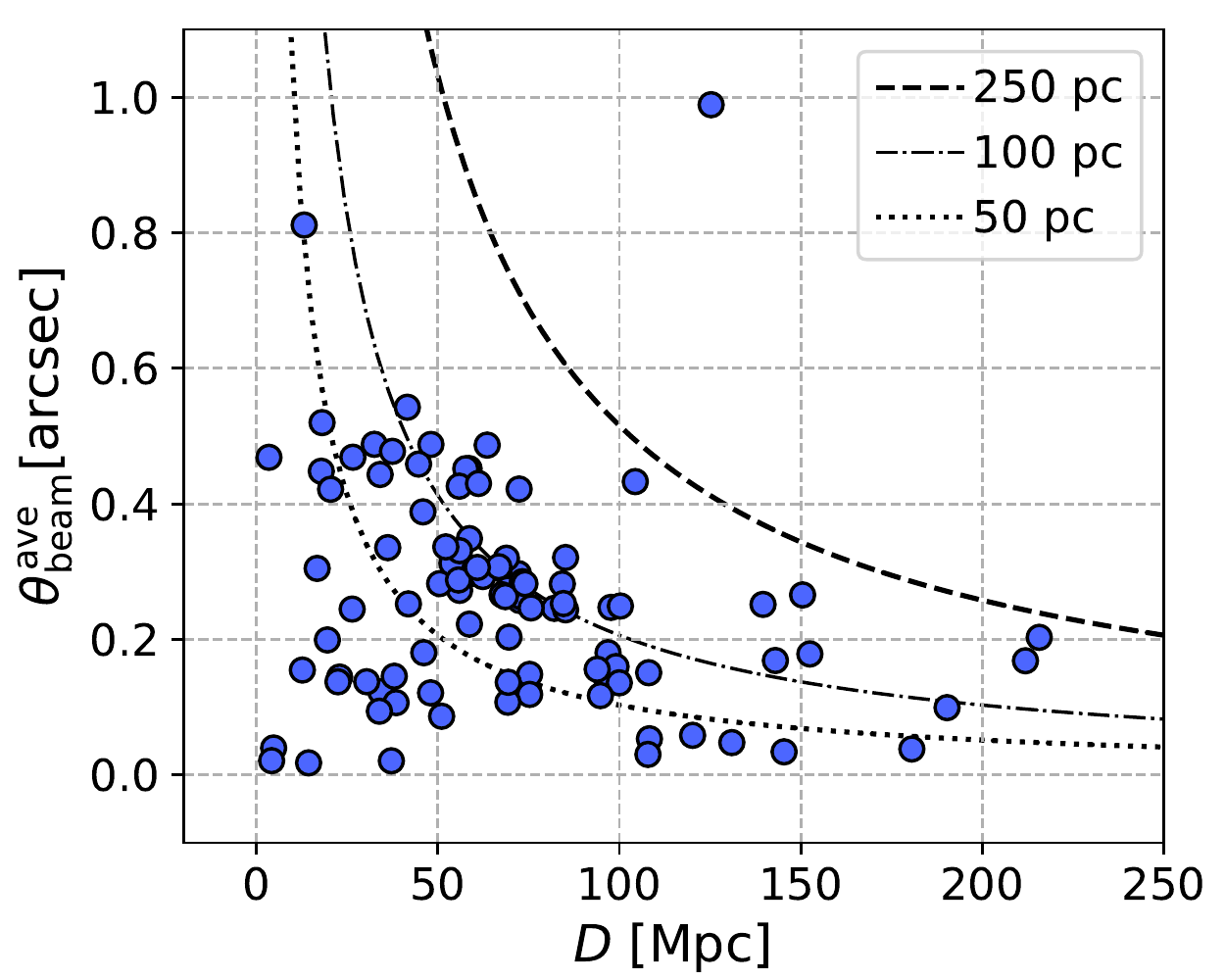}
    \caption{
    Scatter plot of the average beam size $(\theta^{\rm min}_{\rm beam}\times\theta^{\rm maj}_{\rm beam})^{1/2}$  versus the object distance ($D$).
    Dotted, dot-dashed, and dashed lines indicate required beam sizes to achieve the physical resolutions of 50 pc, 100 pc, and 250 pc, respectively, as a function of distance. 
    }
    \label{fig:dist_vs_res}
\end{figure}

Figure~\ref{fig:dist_vs_res} shows the average ALMA beam size $\theta^{\rm ave}_{\rm beam}$, which we define as $(\theta^{\rm maj}_{\rm beam}\times\theta^{\rm min}_{\rm beam})^{1/2}$ ($\theta^{\rm maj}_{\rm beam}$ and $\theta^{\rm min}_{\rm beam}$ are the full width at the half maximum (FWHM) of a beam along the major and minor axes, respectively) following the custom in ALMA operations, versus the distance to the source ($D$). The distances were taken from \cite{Kos22_catalog}, the BASS DR2 catalog paper. 
The figure shows that spatial resolutions better than 250\,pc are achieved for all objects, except Mrk 705 for which $\theta^{\rm ave}_{\rm beam} \approx$ 1\arcsec, $\approx$ 630\,pc at $D \sim$ 130 Mpc. 
Throughout this paper, we include Mrk 705, and confirm that any of our conclusions do not change even if that is excluded.
The median value of the average beam sizes is $\approx$ 80\,pc, which cannot resolve  warm dust emission around the AGN that is traced in the mid-IR (MIR) band \citep[$\sim$ 1\,pc;][]{Kis13}.

The sample is superior to those of previous mm-wave studies \citep{Beh15,Beh18} in three aspects. (1)
Our current sample size is more than three times larger than the past ones. (2) The sub-arcsec resolutions of our ALMA data are significantly better than achieved previously \cite[i.e., $\sim$ 1\arcsec--2\arcsec;][]{Beh18}. 
The corresponding physical sizes, $\lesssim 100$ pc for almost all our targets, can reduce the contaminating light even from circumnuclear disks on scales of $\gtrsim$ 100\,pc \cite[e.g.,][]{Gar14,Gar16,Com19}. 
(3) The choice of the Band-6 (211--275\,GHz) observation could be more advantageous than the $\sim100$\,GHz frequency adopted in the previous studies. 
In the higher frequency band, a smaller contribution of the synchrotron emission component extending from the cm-wave band is expected, given its possible negative spectral slope \citep[e.g.,][]{Chi20}, and thermal dust emission would still be insignificant \cite[e.g.,][]{Gar16,Ino20}. Thus, the band has the potential to better probe self-absorbed synchrotron components from AGNs.

\section{Data at Different Wavelengths}\label{sec:anc}

X-ray data we use are taken from \cite{ric17c}, who analyzed XMM-Newton, Swift/XRT, ASCA, Chandra, and Suzaku data together with Swift/BAT data and tabulated various physical quantities (e.g., intrinsic luminosities and fluxes in the 14--150\,keV and 2--10\,keV bands, and the hydrogen column density). 
As errors in X-ray luminosities and fluxes, we consider 0.1\,dex and 0.4\,dex for less obscured ($\log [N_{\rm H}/({\rm cm}^{-2})] < 23.5$) and heavily obscured ($\log [N_{\rm H}/({\rm cm}^{-2})] \geq 23.5$) sources, respectively.

\begin{figure*}
    \centering
    \includegraphics[width=8.5cm]{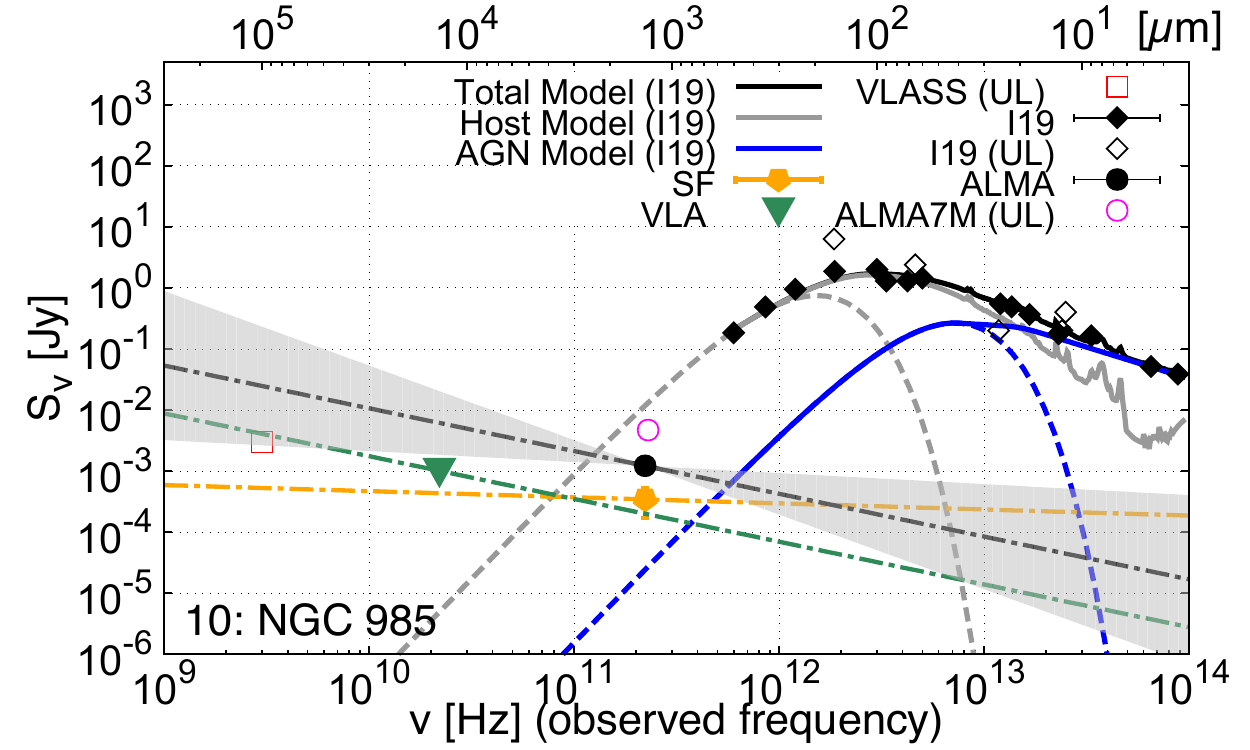}
    \includegraphics[width=8.5cm]{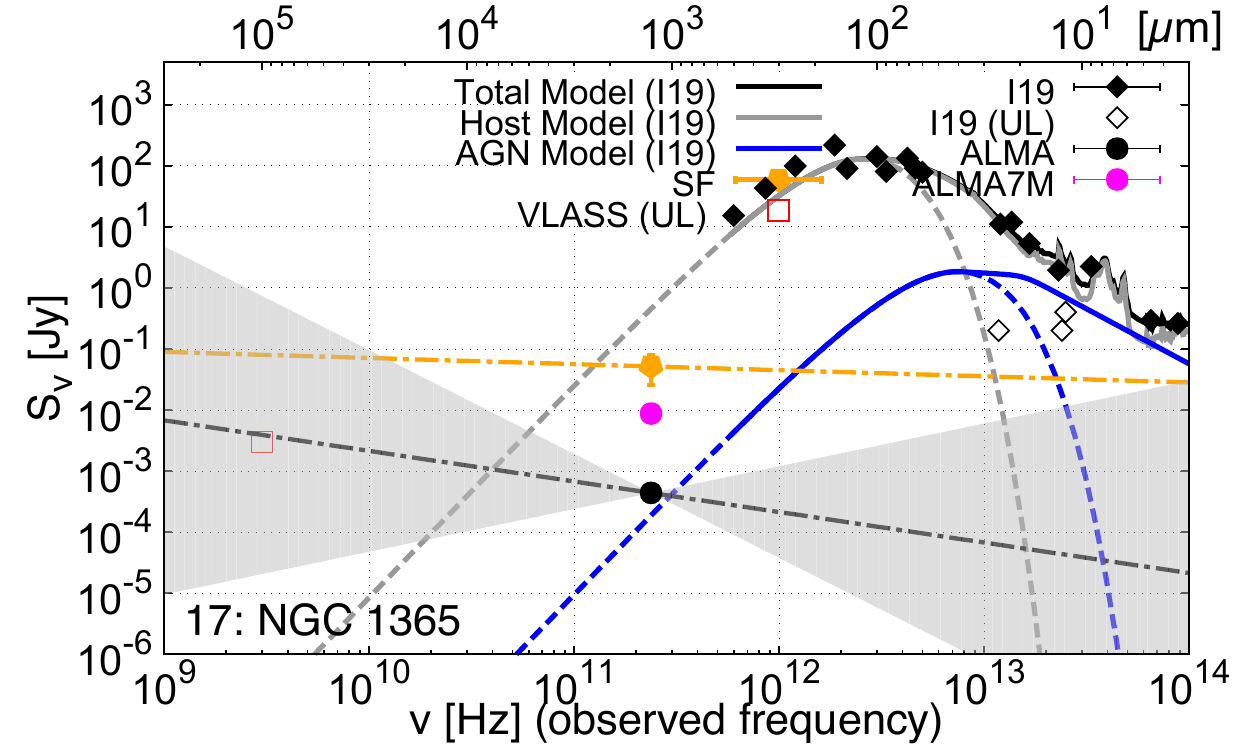}
    \includegraphics[width=8.5cm]{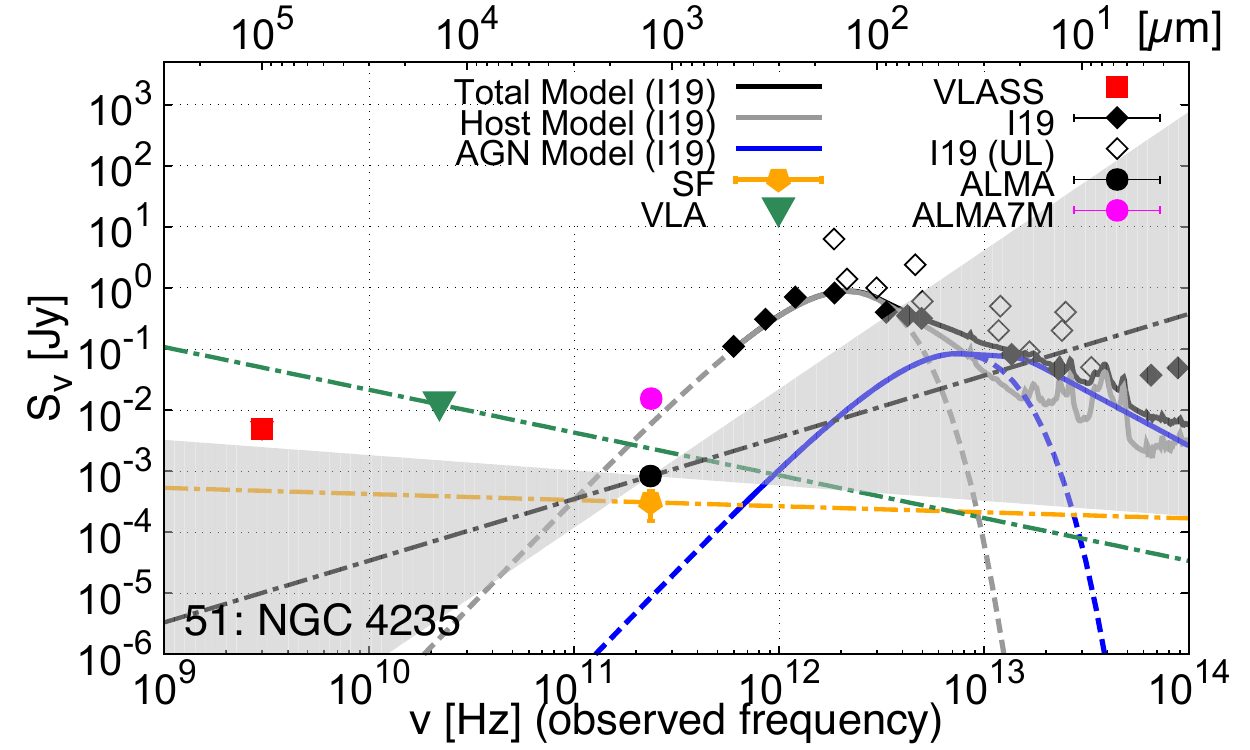}
    \includegraphics[width=8.5cm]{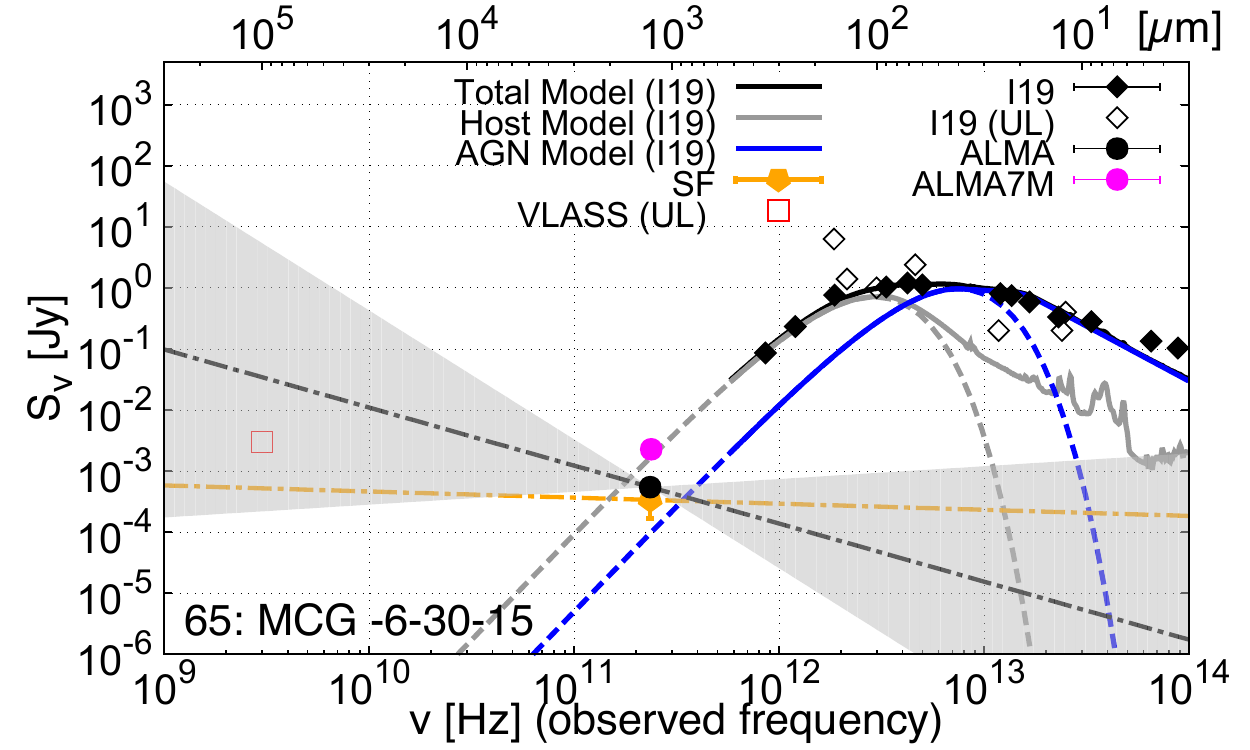}
    \caption{
    Radio-to-IR SEDs of NGC~985, NGC 1365, NGC 4235, and MCG $-$6$-$30$-$15. 
    In each panel, the data points in the IR band ($\lesssim$ 500\,$\mu$m), indicated by diamonds, are taken from \cite{Ich19}. 
    Filled and unfilled marks denote detected values and upper limits, respectively, and in the same way, the other data are plotted. Errors are shown for all data, except for the 22\,GHz radio band as the error is unavailable from the original paper of \cite{smi20}. 
    The IR flux densities were measured with apertures of FWHMs $\gtrsim$ 6\arcsec, much larger than that for the mm-wave flux density (i.e., $\lesssim$ 0\farcs6). 
    Best-fit AGN and host-galaxy templates are presented by solid blue and gray lines, respectively. 
    Modified black-body components, used to extrapolate the best-fit models to the lower-frequency regime, are presented by dashed lines with the original colors. 
    A black-filled circle shows the peak flux density measured by the ALMA 12-m array. Its accompanying dot-dashed black line represents a power law obtained by fitting the ALMA 12-m data in different spectral windows. The shaded area represents the uncertainty in the spectral slope. 
    Additionally, the peak flux density measured by using an ALMA 7-m array data with a beam size of $\sim$ 4\farcs5--6\farcs5\ is shown as a magenta circle. We note that only 22 objects have 7-m data. 
    An orange pentagon indicates the flux density due to synchrotron plus free-free emission, expected by the SFR from the IR data obtained at resolutions coarser than 6\arcsec. Thus, this should be regarded as upper contribution limit to the mm-wave flux density. 
    The accompanying orange dot-dashed line is a power law with an index of 0.1, which is expected for the free-free emission, likely dominant in the mm-wave band (see Equation~\ref{eqn:sfr2lmm}).  
    Lower frequency data at 3\,GHz and 22\,GHz obtained by VLA are shown as a red square and a green triangle, respectively. Particularly for the 22\,GHz data, a power law with an index of 0.7 is shown by a green dot-dashed line. 
    }
    \label{fig:sed}
\end{figure*}

In addition, we utilize the IR data of \cite{Ich19}, who compiled IR photometry data from four observatories (WISE, IRAS, AKARI, and Herschel) \cite[see also][]{Mus14,Mel14,Shi16,Shi17} and decomposed the IR (3--500\,$\mu$m) SEDs into AGN and host-galaxy components. 
To obtain physical quantities in many objects via the SED analysis, they reduced the number of free parameters as much as possible so that the fitting could be performed even with a few data points. Specifically, they fitted five host-galaxy templates plus one AGN template, and each template had normalization as the only free parameter, except for the case of high-luminosity AGNs ($L_{\rm 14-150} > 10^{44}$ \ergs; i.e., a slope of the AGN template at short wavelengths as an additional free parameter). 
As a result, the SED fitting was performed for many objects (606 AGNs). However, due to this simplified procedure, there are some caveats in this analysis. First, while the IR AGN emission should depend on various physical parameters, this fact was not considered. However, we stress that the IR luminosities derived from the AGN templates are consistent with those derived from high-resolution IR photometry \citep{Ich19}, suggesting the accuracy of the luminosity measurement. Second, since the resolution is coarser at longer wavelengths, the result may overestimate the flux at long wavelengths to that expected from the data at shorter wavelengths. 
Of our 98 objects, the SED fit was performed for 88, of which 64 (i.e., 73\%) have good quality SEDs consisting of ten or more detected points covering long-wavelength bands above 140\,$\mu$m. Some examples of the SEDs are shown in Figure~\ref{fig:sed}. 
\cite{Ich19} provide the AGN-related 12\,$\mu$m luminosities for our sources, and following them, we adopt 0.22\,dex as an uncertainty of the MIR luminosities. We note that three AGNs for which \cite{Ich19} provided only the lower limits for their MIR AGN luminosities are excluded from our assessment of relations involving MIR emission.
In addition, \cite{Ich19} derived integrated host-galaxy 
8--1000\,$\mu$m far-IR (FIR) luminosities, and the corresponding star formation rates (SFRs) can be derived with a conversion factor of \cite{Ken98}. As with the 12\,$\mu$m luminosities, we adopt the uncertainty of 0.22\,dex for the two quantities.

Furthermore, 3\,GHz radio data were compiled from a catalog produced by the Very Large Array Sky Survey \cite[VLASS;][]{Gor21}. The project plan is to scan the entire sky north of $-40^\circ$ around 3\,GHz three times, and a catalog created using data obtained in the first epoch between 2017 and 2019 is publicly available. 
The achieved resolution is 2\farcs5, and is the highest among the other cm-wave large sky surveys
\citep[FIRST, NVSS;][]{Whi97,Con98,hel15}. 
Thus, the catalog allows us to infer radio loudness in a region as close to the nucleus as possible for a large number of objects. 
We define radio loudness as the ratio of 3\,GHz and 14--150\,keV luminosities in log scale ($R_{\rm 14-150}$). If the 3\,GHz flux density of an object is below 3\,mJy beam$^{-1}$, including non-detection, we consider an upper limit of 3\,mJy beam$^{-1}$. 
This treatment is motivated by the fact that there is greater uncertainty in the flux below 3\,mJy beam$^{-1}$ \citep{Gor21}. The radio loudnesses of the objects with 3\,GHz data are distributed from $-6.5$ to $-3.9$. 
A canonical radio loudness value that distinguishes between radio loud objects (RL) and radio quiet objects (RQ) is $\log[\nu L_{\nu} (5\,{\rm GHz})/L_{2-10}] = -4.5$, as proposed by \cite{ter03}, and the corresponding $R_{\rm 14-150}$ is $\approx -4.9$. Here, we extrapolate the 5\,GHz and 2--10\,keV luminosities with spectral indices of 0.7 and 0.8, respectively \citep[e.g.,][]{Ued14,ric17c,Chi20}. 
However, in this study, we adopt $R_{\rm 14-150} = -5.3$ as a threshold so that we can have subsamples of RL and RQ objects with similar sizes of 34 and 35.


Lastly, we also use fluxes of [O {\sc{iii}}]$\lambda$5007, [Si {\sc{vi}}]$\lambda$1.96, and [Si {\sc{x}}]$\lambda$1.43, to examine their correlations with mm-wave emission. 
The fluxes of the ionized lines were measured within the BASS framework
\citep{Oh22,den22}. 
Among our 98 objects, we find extinction-corrected oxygen fluxes, calculated by \cite{Oh22}, for 90 objects and observed silicate fluxes for 17 objects from 
\cite{den22}.

\begin{figure}
    \centering
    \includegraphics[width=7cm]{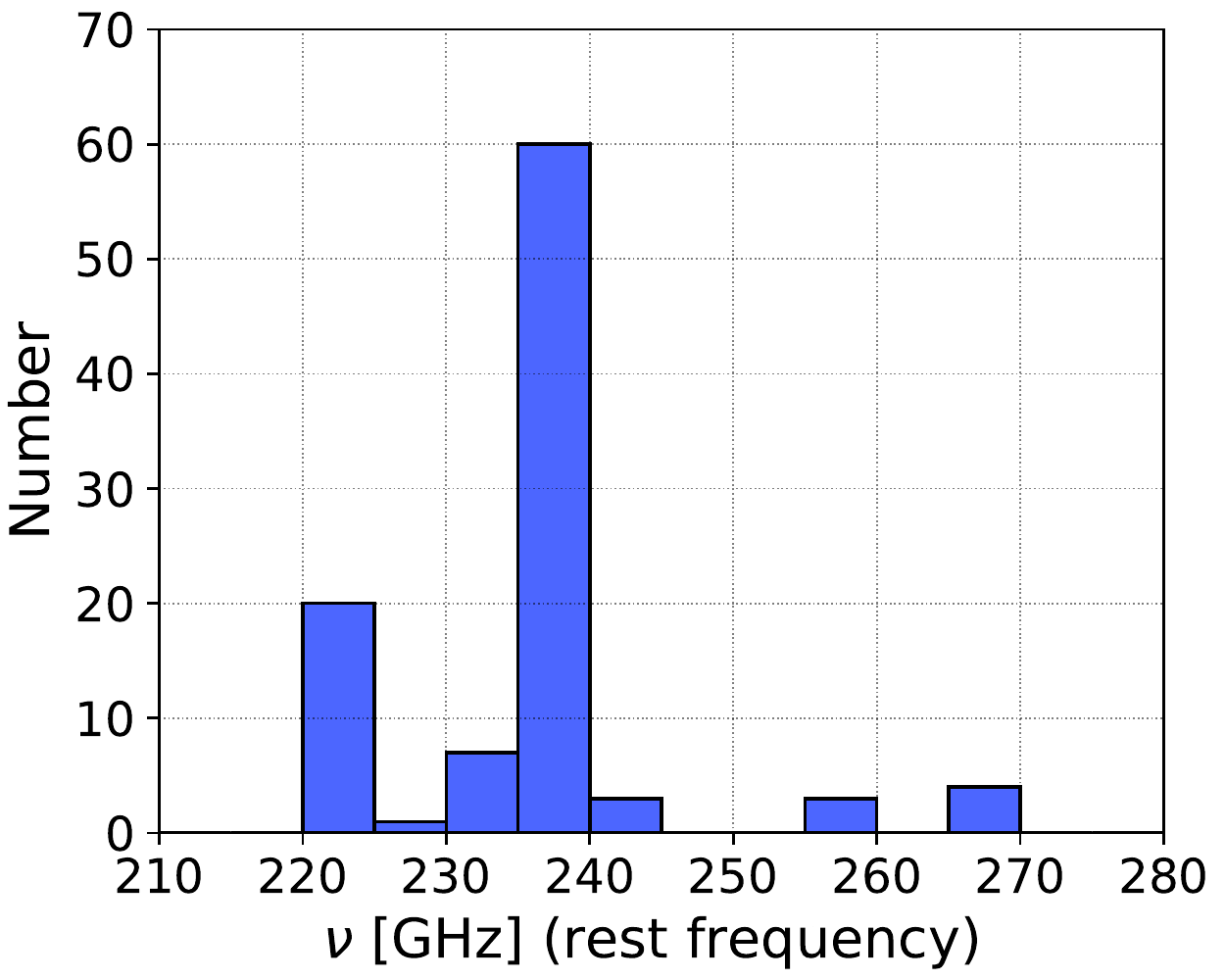} 
    \caption{
    Histogram of the rest frequencies at which our targets were observed by ALMA.
    }
    \label{fig:freq}
\end{figure}

\section{ALMA Data Analysis}\label{sec:alma_data}

For each target, we measured the peak mm-wave flux density ($S^{\rm peak}_{\nu,\rm mm}$) from an ALMA data as follows. 
The Common Astronomy Software Applications package \citep[CASA; ][]{McM07} was used for our analysis. Following the standard procedure, we first reduced and calibrated the raw data using the scripts used for quality verification by the ALMA Regional Center. 
In the above two processes, we adopted the version of CASA that was suitable to run the scripts, but for the rest of the analysis, we used CASA v.6.1.0.118.
Then, from the reprocessed visibility data, we created dirty images (i.e., an observed image, corresponding to a true image convolved with a point spread function produced by the sampled visibilities) and carefully identified spectral channels free of any strong emission lines, based on nuclear spectra within 1\,kpc. In these channels, continuum emission was imaged using \textsc{tclean}
with the deconvolver \textsc{clark} in the multi-frequency synthesis mode in the same 1\,kpc region. 
We used the Briggs weighting method with \textsc{robust} = 0.5. For data obtained in the mosaic mode, we set \textsc{gridder} to mosaic. 
We set \textsc{cell} to be small enough to divide the beam size into at least three pixels (i.e., $\approx$ 0\farcs004--0\farcs2). 
The parameter \textsc{threshold} was set to be within 3--4$\sigma_{\rm mm}$, where $\sigma_{\rm mm}$ is the noise level derived from regions devoid of emission.
If an AGN was observed with multiple spectral windows in its observation, we individually reconstructed an image for each window, in addition to the one considering all available windows. Finally, for the cleaned images, primary-beam correction was applied.

\begin{figure*}
    \centering 
    \includegraphics[width=5.8cm]{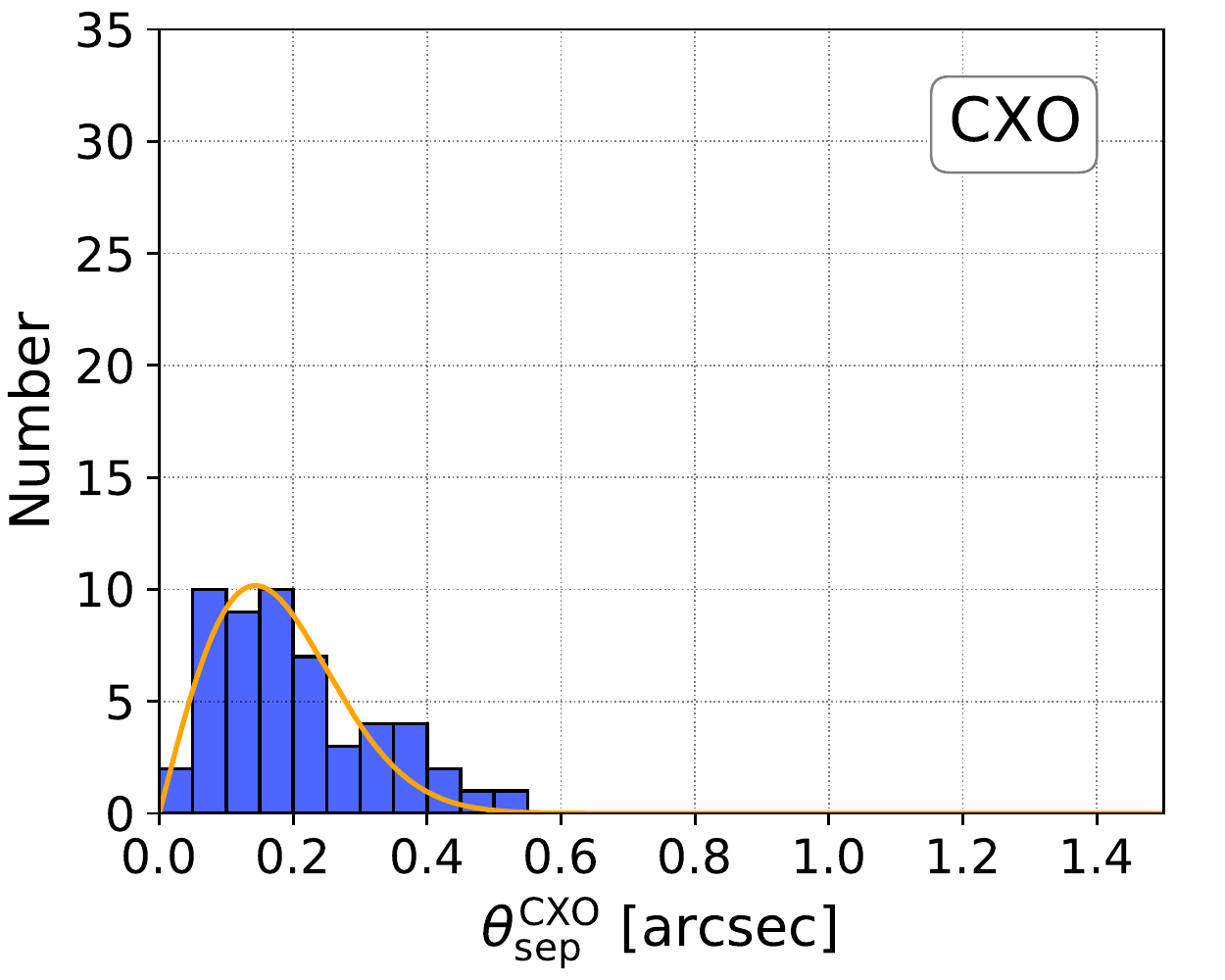}
    \includegraphics[width=5.8cm]{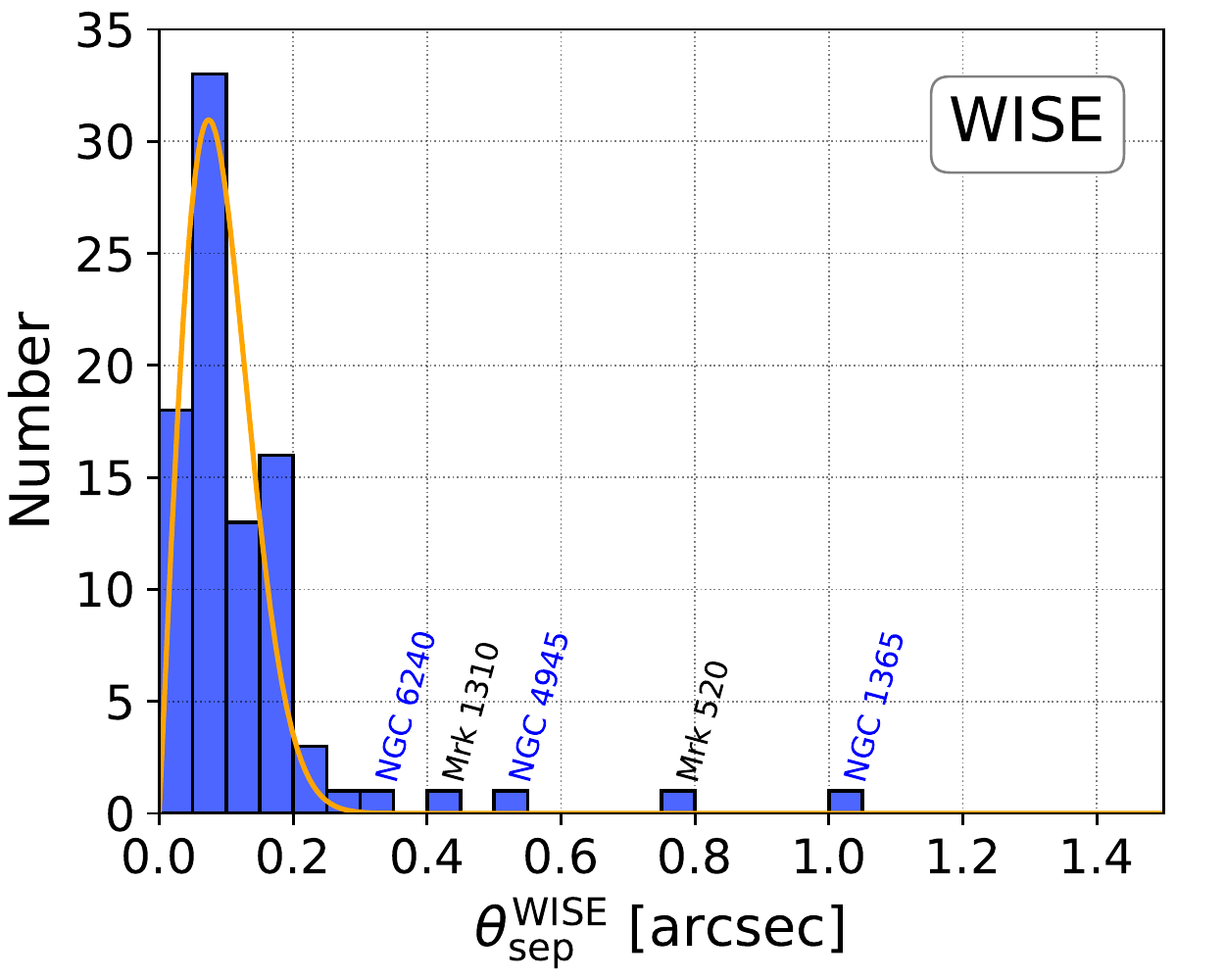}
    \includegraphics[width=5.8cm]{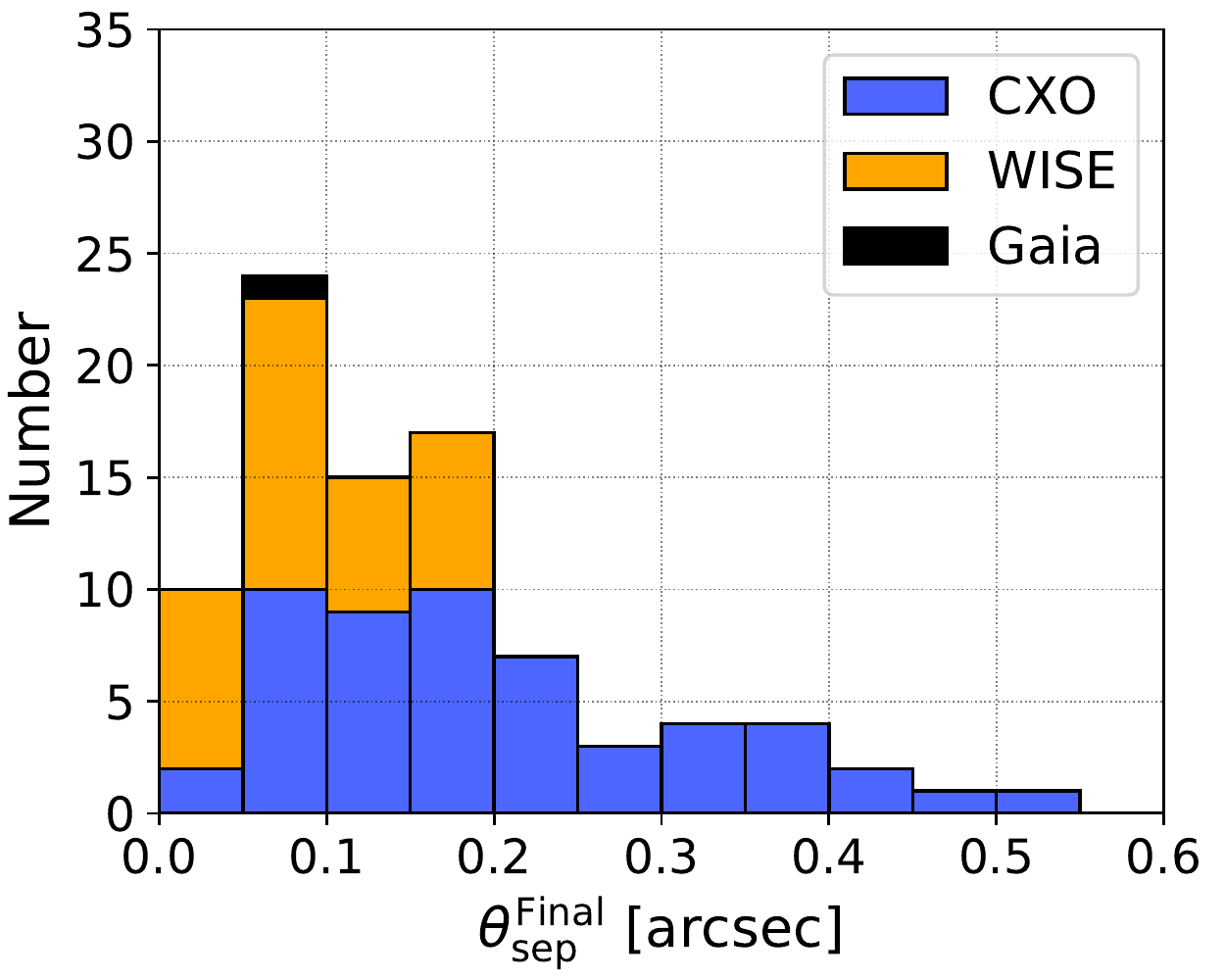}
    \caption{
    Left: Histogram of the separation angle between the 
    Chandra X-ray position and the identified mm-wave peak for 53 AGNs. 
    The orange line shows a fitted Rayleigh distribution. 
    Middle: Same as the left panel, but for the WISE positions. The fitted Rayleigh distribution (orange line) has a smaller width than that fitted to the Chandra result, but several outliers appear. 
    The objects named in blue have Chandra data, and such data are not found for those in black (Mrk 1310 and Mrk 520). 
    Right: Histogram on the separation angles for the AGN positions that we finally adopted in searching for mm-wave peaks. 
    Chandra, WISE, and Gaia only for Mrk 1310 are adopted in that order for the AGN positions. 
    }
    \label{fig:sep_angle}
\end{figure*}

Figure~\ref{fig:freq} shows the histogram of the rest frequencies at which our AGNs were observed. For those observed with multiple spectral windows, we adopt the central frequency of the collapsed spectral window, after removing any emission line flux. The peak around $\sim 230-240$ GHz would be due to the frequent observations of nearby AGNs for CO($J$=2--1) at the rest frequency of 230.538 GHz.

To identify nuclear emission in each ALMA image, we first defined AGN positions by using Chandra X-ray data. 
Chandra data were preferentially used because X-rays are an excellent probe of AGNs, and Chandra has the best available angular resolution in the X-ray band. 
For 56 targets, we found Chandra data where 
the offsets between targets and the focal planes are less than 1\arcmin. According to the Chandra X-Ray Center\footnote{https://cxc.harvard.edu/cal/ASPECT/celmon/},
in such on-axis observations, a target should be located within $\sim$ 1\farcs4 at the 99\% level, and thus 
we searched for nuclear emission in the ALMA images
within a radius of 1\farcs4 from the X-ray AGN positions. 
As the astrometric accuracy of ALMA is 
$\lesssim$ 0\farcs1\footnote{https://help.almascience.org/kb/articles/what-is-the-astrometric-accuracy-of-alma}, we ignored its positional error. 
We calculated the signal-to-noise ratio for a peak flux on the resulting images within the search radius. 
If the ratio is above 5$\sigma_{\rm mm}$, 
we regarded the peak emission as a detection considering 
the thresholds of 3--4$\sigma_{\rm mm}$ adopted for the clean process.  
We note that because NGC 3393 and NGC 7582 have their brightest mm-wave peaks around radio ($\approx$ 1--8\,GHz) lobes   \citep[e.g.,][]{Coo00,Ric18}, we ignored the mm-wave components in the search for nuclear mm-wave emission. 
The left panel of Figure~\ref{fig:sep_angle} shows a histogram of the separation angle between the Chandra position and the mm-wave peak identified. 
The histogram appears well-fitted by a Rayleigh distribution (orange line in the figure), which is expected if the positional error follows a Gaussian distribution. 
This result supports the hypothesis that we have successfully identified AGN-related mm-wave emission.

For those without Chandra data, we relied on the ALLWISE catalog in the near-to-mid infrared band \citep{Wri00,Mai11}. In this band, emission from dust heated by an AGN is expected. 
The search radius for WISE positions was set to 1\farcs5. 
Although the positional accuracy of the catalog was estimated to be $\approx$ 0\farcs3 by cross-matching with the 2MASS catalog\footnote{https://wise2.ipac.caltech.edu/docs/release/allsky/expsup/sec6\_4.html}, the larger radius was adopted because the 2MASS accuracy for a peak, or the nucleus, of a spatially resolved galaxy at the 99\% limit is $\sim$ 1\farcs5 \citep{She17}. 
For all 98 targets, we searched for mm-wave peaks around their WISE positions, and the middle panel of Figure~\ref{fig:sep_angle} shows a histogram on the obtained separation angles for the identified mm-wave peaks. 
Unlike the result based on the Chandra observations, several objects (NGC 6240, Mrk 1310, NGC 4945, Mrk 520, and NGC 1365) appear to be outliers against a Rayleigh distribution fitted to the entire histogram. 
The larger separation angles for NGC 6240, NGC 4945, and NGC 1365 are likely because their WISE positions deviate from the nuclei due to active star formation bright in the MIR band across their host galaxies \cite[e.g.,][]{Kra01,Gal05,Ega06}. 
Although the reason for this discrepancy for Mrk 1310 is unclear, the mm-wave emission cross-matched with the WISE position possibly originates around the nucleus. This argument is based on the fact that Mrk 1310 is a type-1 Seyfert galaxy, and its optical Gaia position \citep[][]{Gai18}, which would locate the nucleus, is close to the mm-wave emission with a separation angle of 0\farcs06. 
Lastly, Mrk 520 is a type-2 AGN; therefore, its Gaia position would be unreliable, unlike Mrk 1310. Thus, whether the mm-wave peak found for Mrk 520 is located around the galaxy center is ambiguous. Still, we assume that the mm-wave peak identified by WISE originates from the nucleus based on the above mm-wave searches using WISE, where the mm-wave peak seems to be found around an AGN generally.

The comparison of the Chandra and WISE results infers some important points. The width of the Rayleigh distribution obtained for the WISE positions is narrower than that for the Chandra positions, indicating a more accurate positioning by WISE.
However, WISE may misidentify the positions of AGNs due to surrounding active SF, as suggested for NGC 1365, NGC 4945, and NGC 6240. 
On the other hand, Chandra can locate AGN positions without being affected by star formation more than WISE, while its accuracy is slightly worse. 
Thus, WISE and Chandra have a trade-off relation (accuracy vs. precision).

\begin{figure}
    \centering 
    \includegraphics[width=8.5cm]{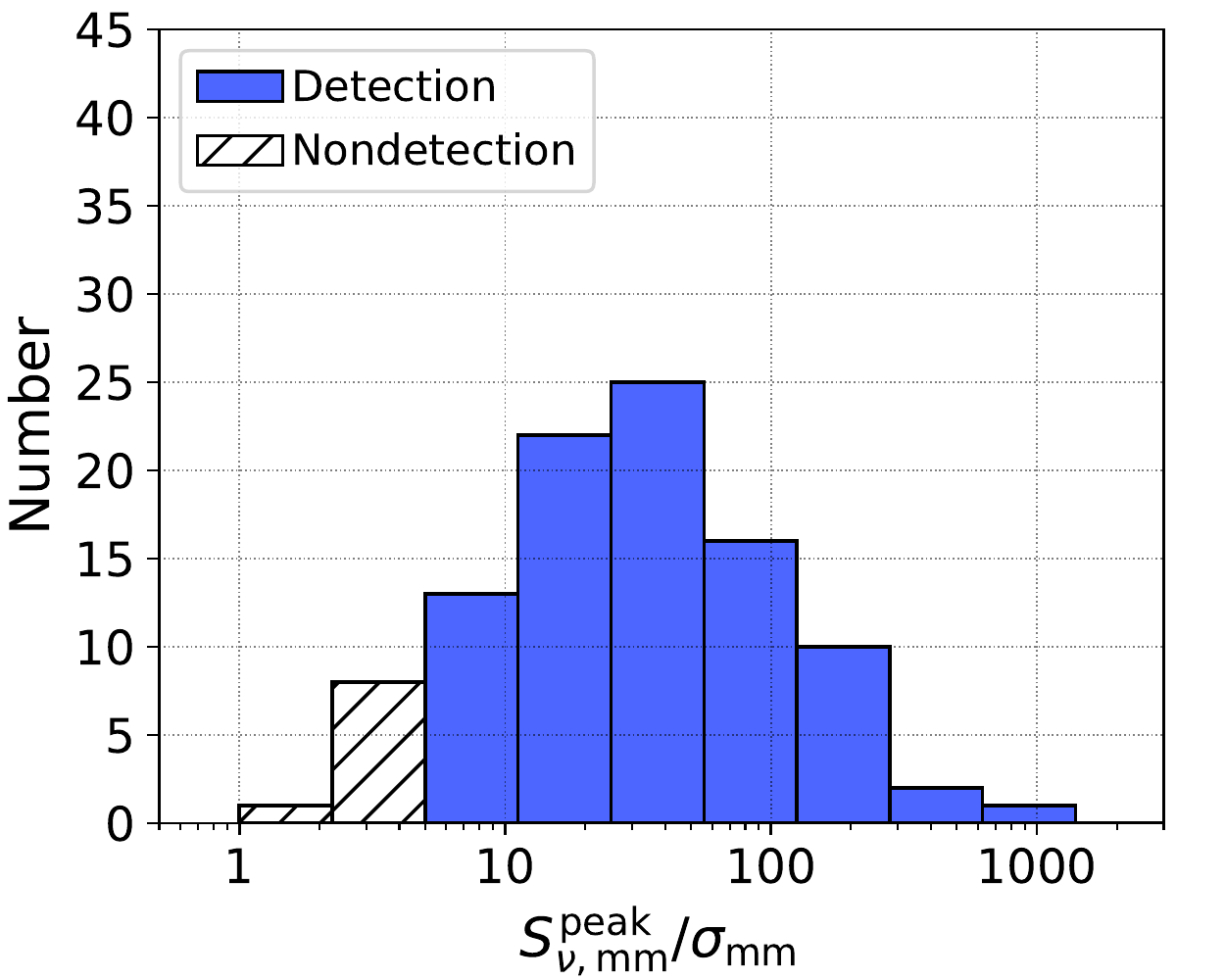}
    \caption{Histogram for the significance of peak emission. We regard emission with peak flux density less than $5\sigma_{\rm mm}$ as a nondetection. 
    Among our 98 objects, significant 
    emission was detected for 89 objects. 
    }
    \label{fig:sig}
\end{figure}

Taking into account the above results, we eventually adopted the mm-wave peaks identified by Chandra and WISE in this order, and then Gaia, particularly for Mrk 1310. The resultant histogram for our final mm-wave search is shown in the right panel of Figure~\ref{fig:sep_angle}. 
Eventually, for each of 75 AGNs (i.e., $\approx$ 77\%), 
significant ($\geq 5\sigma_{\rm mm}$) nuclear emission was identified in all available spectral windows. 
For other 14 AGNs, nuclear emission was detected by merging all available spectral windows. 
Thus, significant nuclear emission was detected above 5$\sigma_{\rm mm}$ for 89 AGNs, corresponding to a high detection rate of $\approx$ 91\%. For those without significant mm-wave emission, we assign an upper limit of a peak flux within a search radius (i.e., 1\farcs4 or 1\farcs5) plus its 1$\sigma_{\rm mm}$ error times 5.
Figure~\ref{fig:sig} shows the distribution of the significances. The median of the significances for the detected sources is $\approx$ 31, and the Circinus galaxy was detected with the highest significance of 724.

\begin{figure*}
    \centering 
    \includegraphics[width=5.6cm]{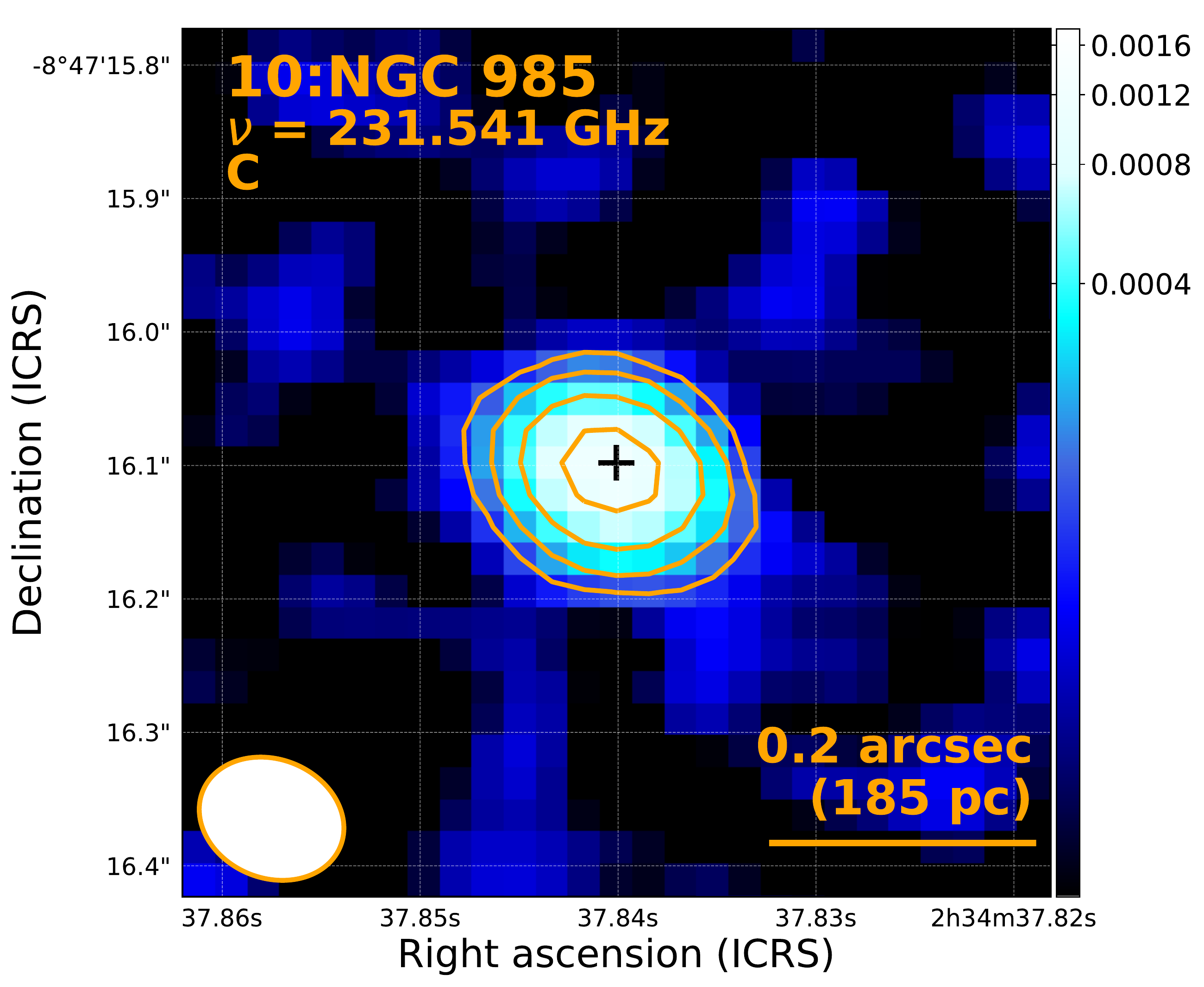}
    \includegraphics[width=5.6cm]{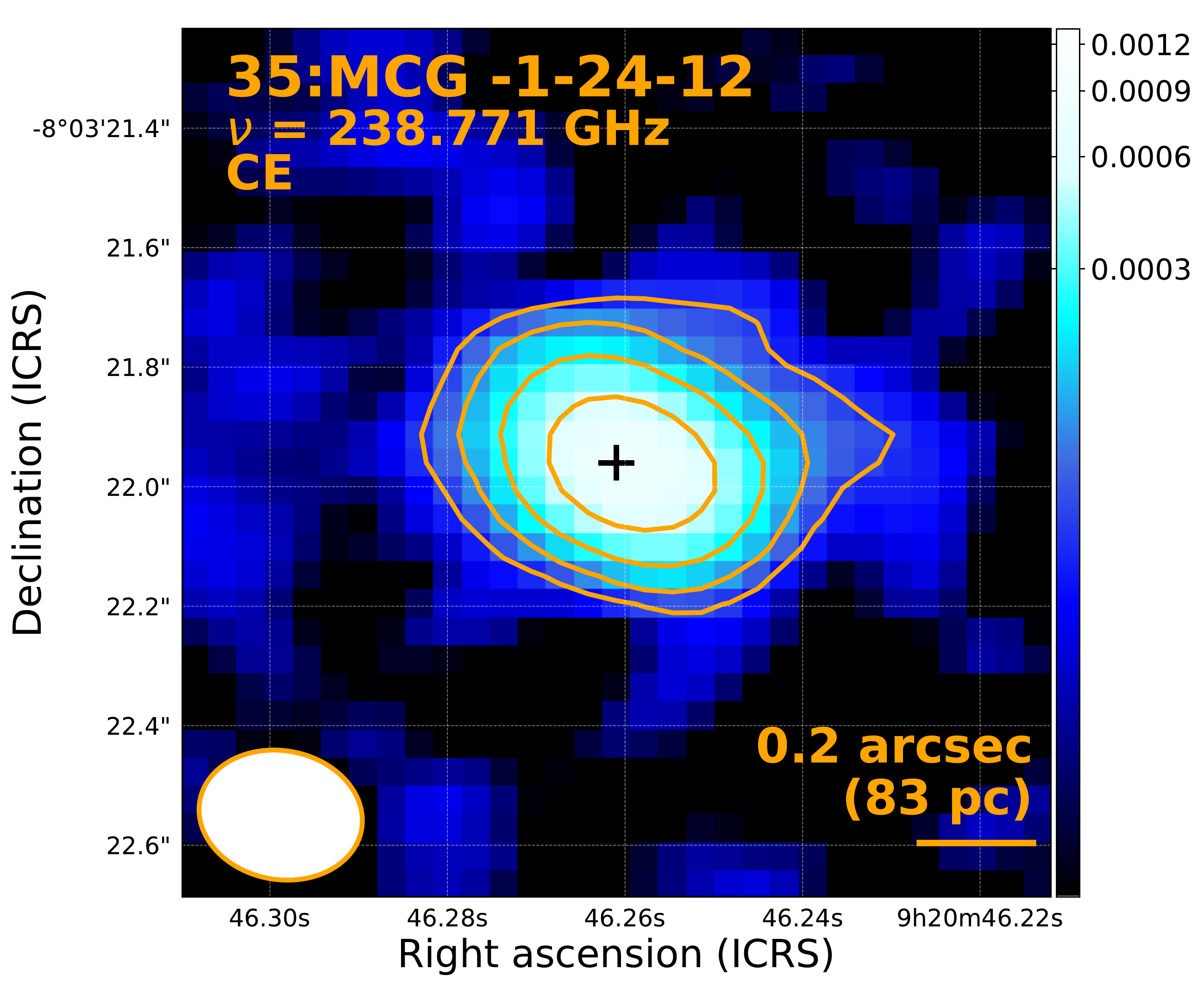}
    \includegraphics[width=5.6cm]{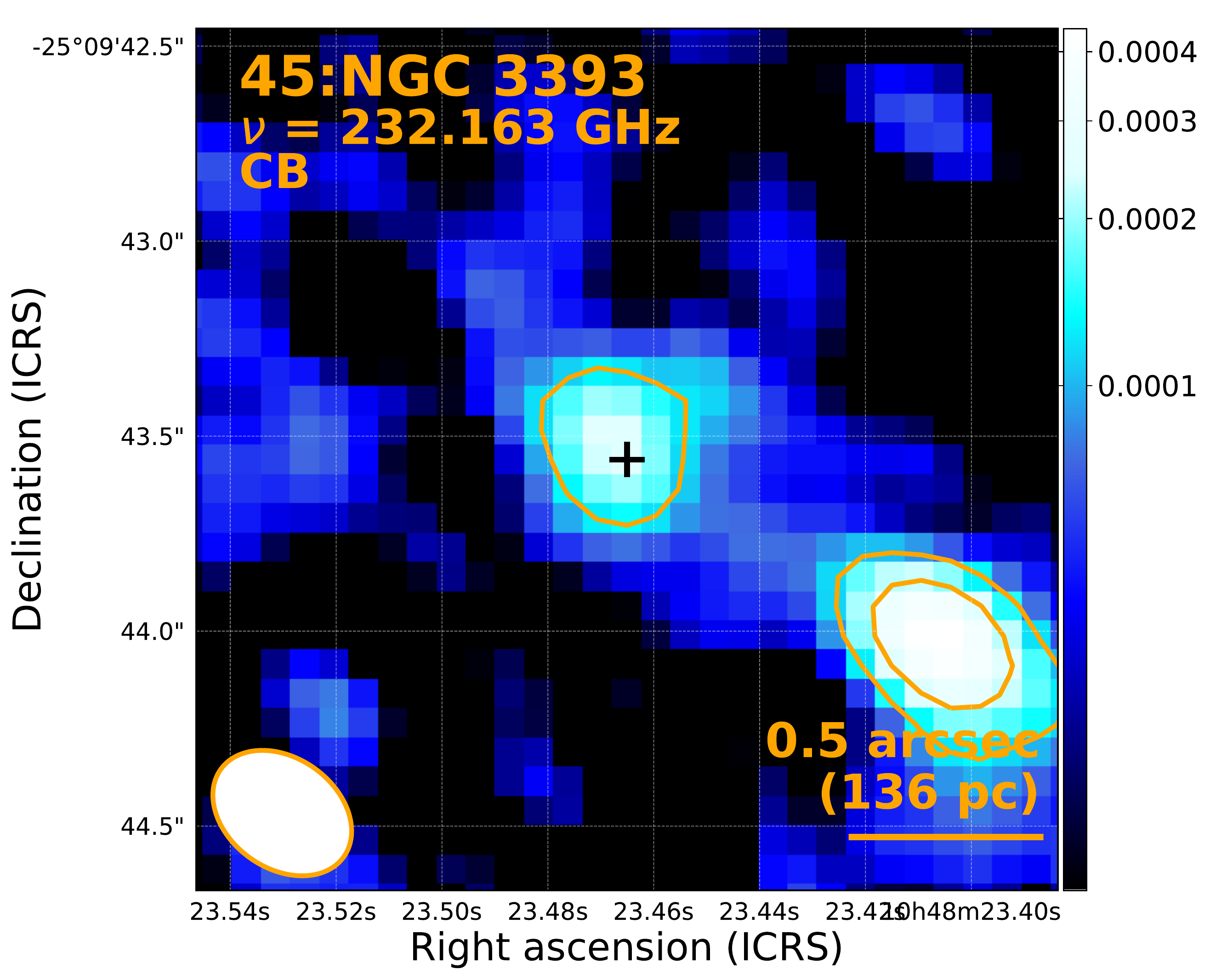} 
    \caption{
    ALMA 600\,pc$\times$600\,pc Band-6 images for NGC\,985, MCG\,$-$1$-$24$-$12, and NGC\,3393 obtained with beams (left bottom corners) of $\approx$ 100\,pc, a typical resolution achieved in our sample (Figure~\ref{fig:dist_vs_res}).
    The numbers to the left of the object names are the numbers assigned in this study (Table~\ref{tab_app:sample}) and Paper~II. 
    The indicated frequency in the figure is the rest-frame central frequency of the collapsed spectral window adopted, after removing any strong emission line flux. Each central black cross indicates the peak of the nuclear mm-wave emission, expected to be the AGN position.  
    C, CE, and CB are morphological parameters assigned. 
    C, E, and B indicate nuclear core emission, extended emission visually connected to the core component, and a blob (or blobs) separated from the core, respectively.
    Colors are assigned according to flux density in units of Jy beam$^{-1}$ following the color bar on the right side. The orange contours indicate where flux densities are $5\sigma_{\rm mm}$, $10\sigma_{\rm mm}$, $20\sigma_{\rm mm}$, $40\sigma_{\rm mm}$, $80\sigma_{\rm mm}$ and $160\sigma_{\rm mm}$. 
    For NGC\,985, MCG\,$-$1$-$24$-$12, and NGC\,3393, $\sigma_{\rm mm}$ = 0.02 mJy beam$^{-1}$, $\sigma_{\rm mm}$ = 0.01 mJy beam$^{-1}$, and $\sigma_{\rm mm}$ = 0.02 mJy beam$^{-1}$, respectively. 
    }
    \label{fig:image}
\end{figure*}

\begin{figure}
    \centering 
    \hspace{-1.0cm}  
    \includegraphics[width=8.2cm]{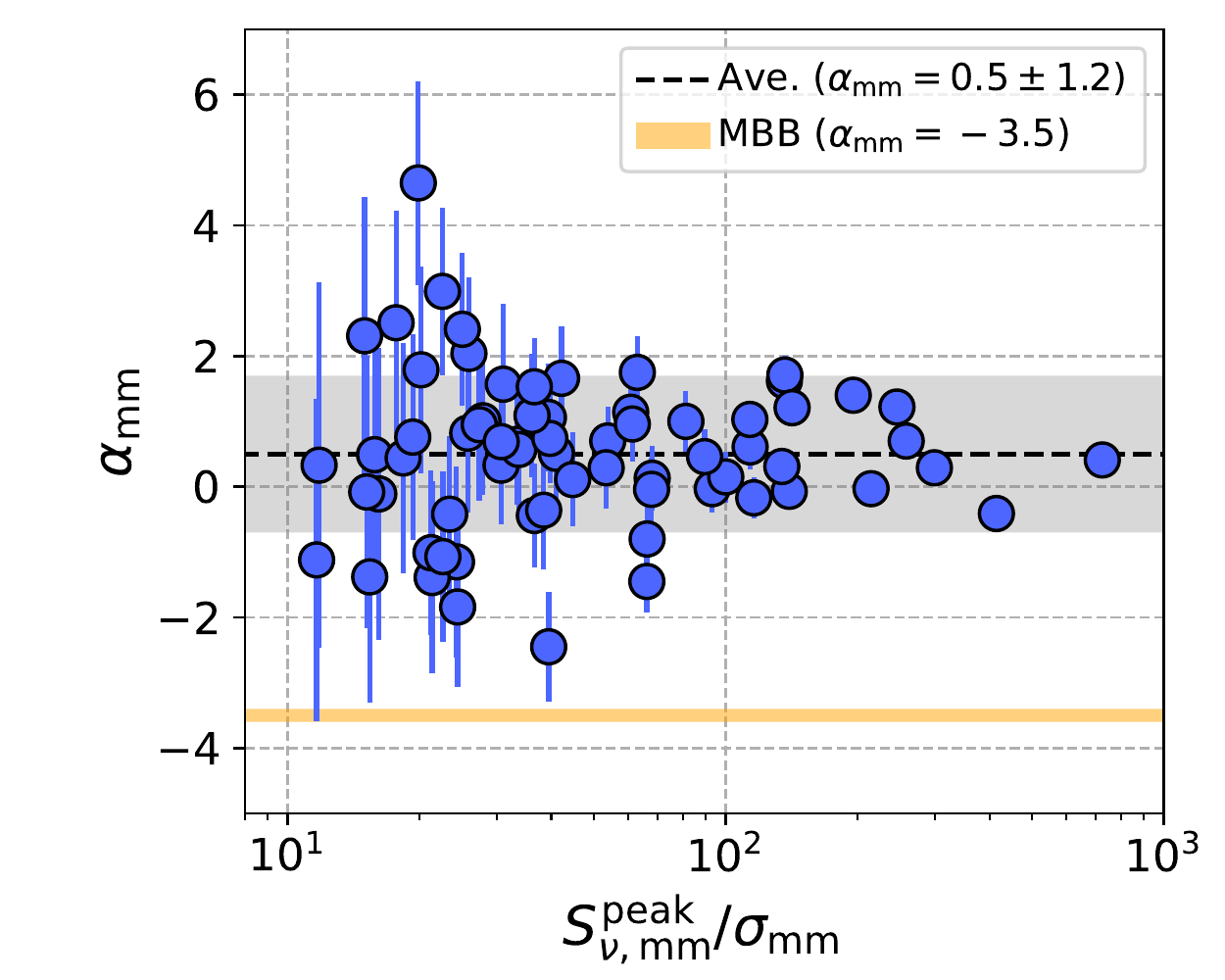} \\
    \hspace{-1.0cm}  
    \includegraphics[width=8.2cm]{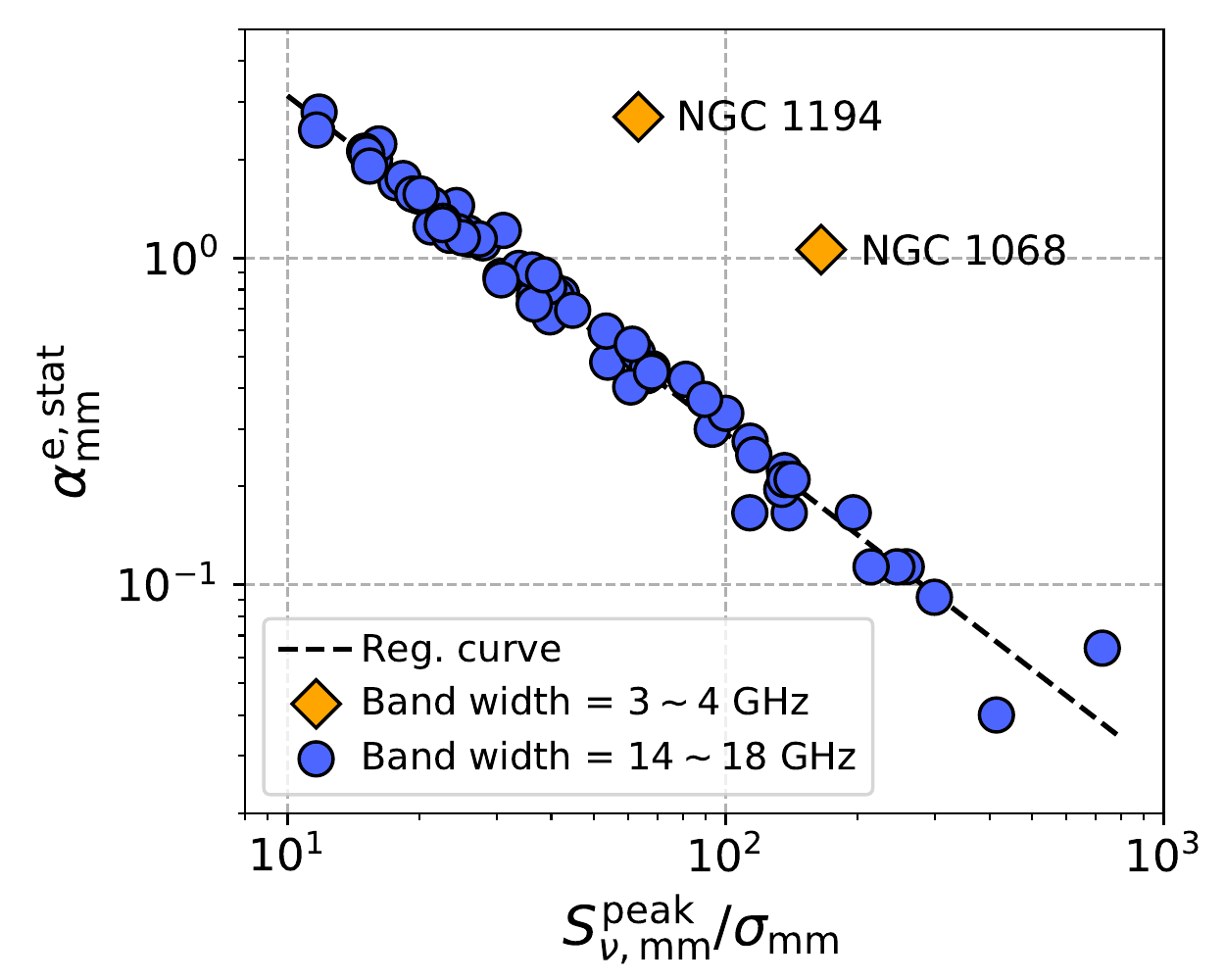}
    \caption{
    Top: Spectral index ($\alpha_{\rm mm}$ for $S^{\rm peak}_{\nu,\rm mm} \propto \nu^{-\alpha_{\rm mm}}$) against detection significance for the AGNs which were observed with more than two spectral windows and whose nuclear emission was significantly detected in all windows.
    The orange line indicates $\alpha_{\rm mm} = -3.5$, expected for thermal emission (modified black body emission) from dust. 
    Bottom: The statistical error of the mm-wave spectral index as a function of detection significance. 
    Blue data points indicate objects observed with spectral windows covering 14\,GHz $\sim$ 18\,GHz, while NGC 1068 and NGC 1194 (orange) were observed with narrower band widths (3\,GHz $\sim$ 4\,GHz). 
    A fitted line to all objects, except NGC 1068 and NGC 1194, is indicated by the dashed line and 
    is expressed as $\log \alpha^{\rm e,stat}_{\rm mm} =
    1.53 - 1.03\times \log(S^{\rm peak}_{\nu,\rm mm}/\sigma_{\rm mm})$. 
    }
    \label{fig:freq1}
\end{figure}

As an example, Figure~\ref{fig:image} shows high-spatial-resolution ALMA Band-6 images of NGC\,985, MCG\,$-$1$-$24$-$12, and NGC\,3393. These images were obtained with beams (left bottom corners) of $\approx$ 100\,pc, a typical resolution achieved in our sample (Figure~\ref{fig:dist_vs_res}).  
The figure demonstrates the ability to identify a nuclear component in high-spatial-resolution ALMA images ($\sim$ 100 pc) and isolate it from others, if any. 
For each object, we visually classified its mm-wave emission based on the image created by considering all available image(s). We considered three morphological features: (i) nuclear core (C), (ii) extended emission visually connected with the core component (E), and (iii) blob separated from the core (B). This information is tabulated in Paper~II. NGC\,985 was classified as C, while we considered MCG $-$1$-$24$-$12 as a CE object due to the presence of a faint extended component in addition to the core. Regarding NGC 3993, there is a blob-like structure, which is likely to be associated with a radio ($\approx$ 1--8\,GHz) lobe \cite[][]{Coo00}, and also a fainter core appears. Thus, we adopted CB. 
A more detailed discussion of extended emission is presented in Paper~II so that this paper can focus on nuclear emission. Of the 89 AGNs with significant nuclear emission, $\sim$ 46\% and $\sim$ 54\% were classified as C and the others, respectively. 


For 69 objects that were observed with more than two spectral windows and for which nuclear emission was detected in all windows, we derived spectral indices ($\alpha_{\rm mm}$), defined as $S^{\rm peak}_{\nu,\rm mm} \propto \nu^{-\alpha_{\rm mm}}$ in flux density (e.g., in units of Jy). In the fits, we adopted the chi-square method. 
In addition to the statistical error of the index ($\alpha^{\rm e,stat}_{\rm mm}$) obtained by the fits, the systematic error of 0.2 is considered due to the possible flux calibration uncertainty between  spectral windows at $\sim$ 230\,GHz by following \cite{Fra20}.
The top panel of Figure~\ref{fig:freq1} shows the spectral index against the significance of the detection. 
The average of the derived indices and their standard deviation are 0.5 and 1.2, respectively. 
These values are adopted as our representative value and error for the AGNs for which we could not determine the indices due to either a non-detection or an insufficient number of spectral windows.  
Although detailed in Section~\ref{sec:posindex}, the figure suggests that almost all constrained indices are inconsistent with that expected from thermal dust emission (e.g., $\alpha_{\rm mm} \sim -3.5$). 
As shown in the bottom panel of Figure~\ref{fig:freq1}, objects observed in band widths of $\sim$ 14--18\,GHz form a relation between $\log \alpha^{\rm e,stat}_{\rm mm}$ and $\log S^{\rm peak}_{\nu,\rm mm}/\sigma_{\rm mm}$. This can be represented as $\log \alpha^{\rm e,stat}_{\rm mm} =
1.53 - 1.03\times \log(S^{\rm peak}_{\nu,\rm mm}/\sigma_{\rm mm})$. However, NGC 1194 and NGC 1068, observed with a narrower frequency coverage ($<$ 4\,GHz), deviate from the relation.

\begin{figure}
    \centering 
    \includegraphics[width=7.3cm]{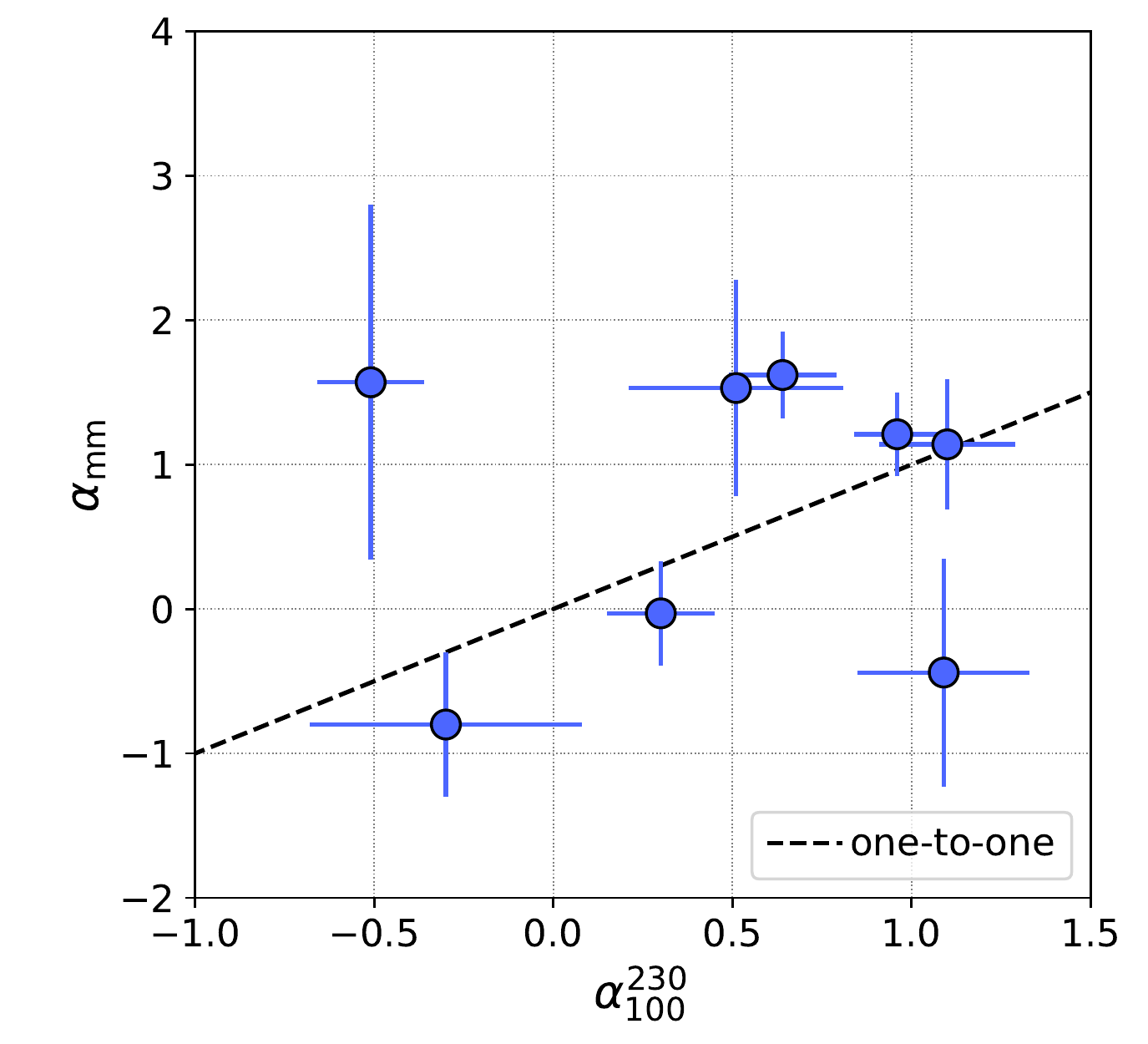}
    \vspace{-.5cm} 
    \caption{
    Scatter plot of the mm-wave index and the index derived from $\sim$ 230\,GHz and 100\,GHz data.
    The dashed line indicates the one-to-one relationship. 
    }
    \label{fig:indices}
\end{figure}

To roughly check whether the derived indices are reasonable, we compare some of them with indices derived by combining our peak flux densities at $\sim$ 230\,GHz and peak ones at 100\,GHz of \cite{Beh18} ($\alpha^{230}_{100}$). For eight objects, both values are obtained.
Figure~\ref{fig:indices} shows a scatter plot of the two indices ($\alpha_{\rm mm}$ and $\alpha^{\rm 230}_{100}$). The 100\,GHz flux densities were measured with larger beams ($\approx$ 1\arcsec--2\arcsec), but if a compact component ($\lesssim$ 0\farcs6) is dominant, the larger beams would not be a serious issue in interpreting $\alpha^{230}_{100}$. We can see that most of the data points are distributed around a one-to-one relationship. Quantitatively, except for one object, the two indices are consistent within $\approx$ 2$\sigma$. 
Given that the observations at 100\,GHz and $\sim$ 230\,GHz are not simultaneous, some deviations could be explained by time variability. 
Although a larger sample is preferred to conclude this, the result could support the conclusion that the indices constrained even in narrow bands (14\,GHz $\sim$ 18\,GHz) are reasonable. 
In contrast, as previously mentioned, the measurements for NGC 1194 and NGC 1068 would not be so reliable. The index derived for NGC 1068 is $2.7\pm1.1$, and this negative slope is inconsistent with a positive slope found from a SED analysis of \cite{Ino20}. Thus, for NGC 1068, we adopt $\alpha_{\rm mm}$ = $-1.3$, inferred from the modeling of \cite{Ino20}, and $\alpha^{\rm e}_{\rm mm} = 0.3$ calculated by combining $\alpha^{\rm e,stat}_{\rm mm}$ predicted from the relation with $S^{\rm peak}_{\nu,\rm mm}/\sigma_{\rm mm}$ and the systematic error. 
For NGC 1194, as no meaningful data are available, we adopt the representative values of $\alpha_{\rm mm} = 0.5\pm1.2$.

\begin{figure}
    \centering 
    \includegraphics[width=8cm]{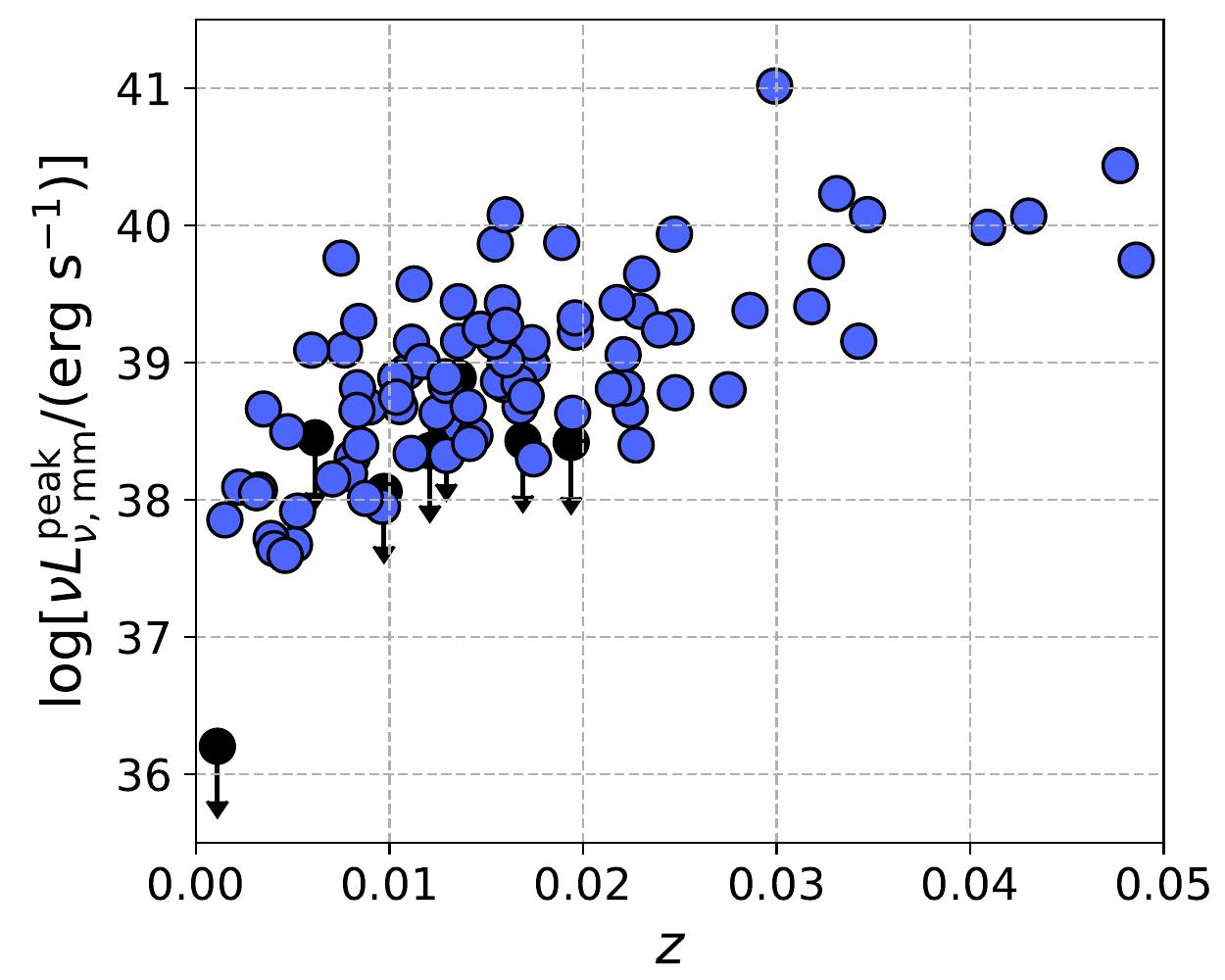}
    \caption{
    Mm-wave luminosity versus redshift  for our sample. 
    }
    \label{fig:z_vs_lmm}
\end{figure}

The determined indices were then used to calculate mm-wave luminosities, or 230\,GHz luminosities, using the equation \citep[e.g.,][]{Nov17}: 
\begin{equation}
    \nu L^{\rm peak}_{\nu,{\rm mm}} = \nu \frac{4\pi D^2}{(1+z)^{1-\alpha_{\rm mm}}}  \left(\frac{\nu}{\nu'}\right)^{-\alpha_{\rm mm}} S^{\rm peak}_{\nu',\rm mm},
\end{equation}
where $\nu$ is set to 230\,GHz as our representative frequency, and $S^{\rm peak}_{\nu',\rm mm}$ is the peak flux density at the observed frequency of $\nu'$. Each flux density is derived from an image consisting of all available spectral window(s). 
By following the definition of the luminosity, the flux is defined as follows:
\begin{equation}
    \nu F^{\rm peak}_{\nu,{\rm mm}} = \nu \frac{1}{(1+z)^{1-\alpha_{\rm mm}}}  \left(\frac{\nu}{\nu'}\right)^{-\alpha_{\rm mm}} S^{\rm peak}_{\nu',\rm mm}.
\end{equation}
In addition to statistical uncertainty, the systematic uncertainty of 10\% is included for luminosities and fluxes by following the suggestions in the ALMA technical handbook\footnote{http://almascience.org/documents-and-tools/cycle8/alma-technical-handbook}.
In the case of non-detection, we consider the peak flux plus $5\sigma_{\rm mm}$ as the upper limit.  
Figure~\ref{fig:z_vs_lmm} shows a plot of mm-wave luminosity versus redshift. Our AGNs cover a mm-wave luminosity range of $\log[\nu L^{\rm peak}_{\nu,{\rm mm}}/({\rm erg}\,{\rm s}^{-1}) ] \sim$ 37.5--41.0, except for the faintest AGN NGC 4395. 

We mention that in Section~\ref{sec_app:dep_inedx} of the appendix, we briefly introduce a study of whether there are relations of the spectral index with some AGN and host-galaxy parameters. The result is that  no correlations are found, and as the results are not closely related to the discussion in this main text, we omit the description here. 

\begin{figure*}
    \centering 
    \hspace{-.6cm} 
    \includegraphics[width=15cm]{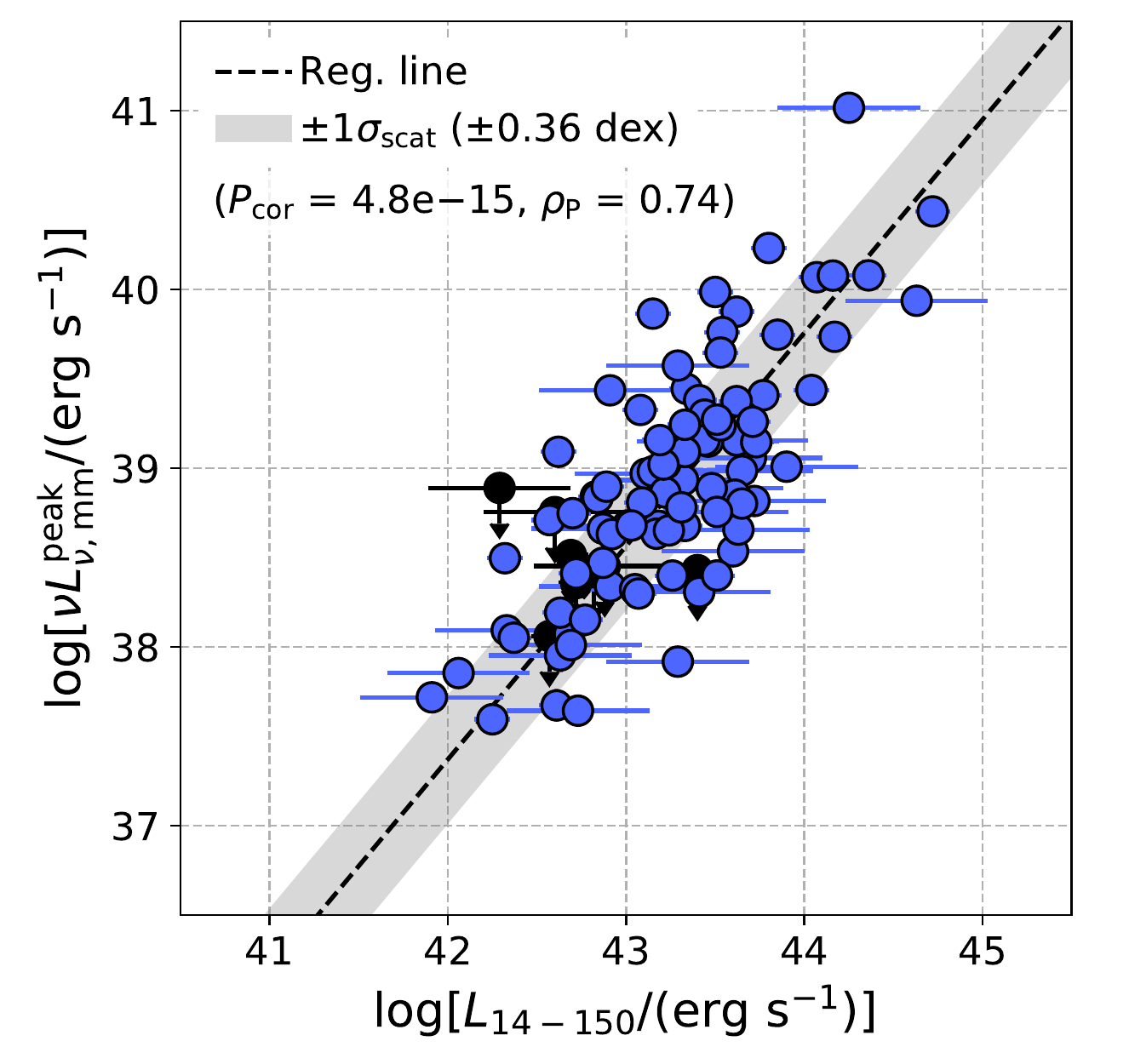} \hspace{-.5cm}
    \caption{
    Correlation of mm-wave and 14--150\,keV luminosities, derived by using ALMA and BAT, respectively. 
    AGNs with upper limits are shown as black circles. 
    The black dashed line indicates 
    the best-fit linear regression line, while the gray region denotes the $\pm 1\sigma_{\rm scat}$ range. 
    }
    \label{fig:var_lum1}
\end{figure*}

\begin{figure*}
    \centering 
    \includegraphics[width=5.9cm]{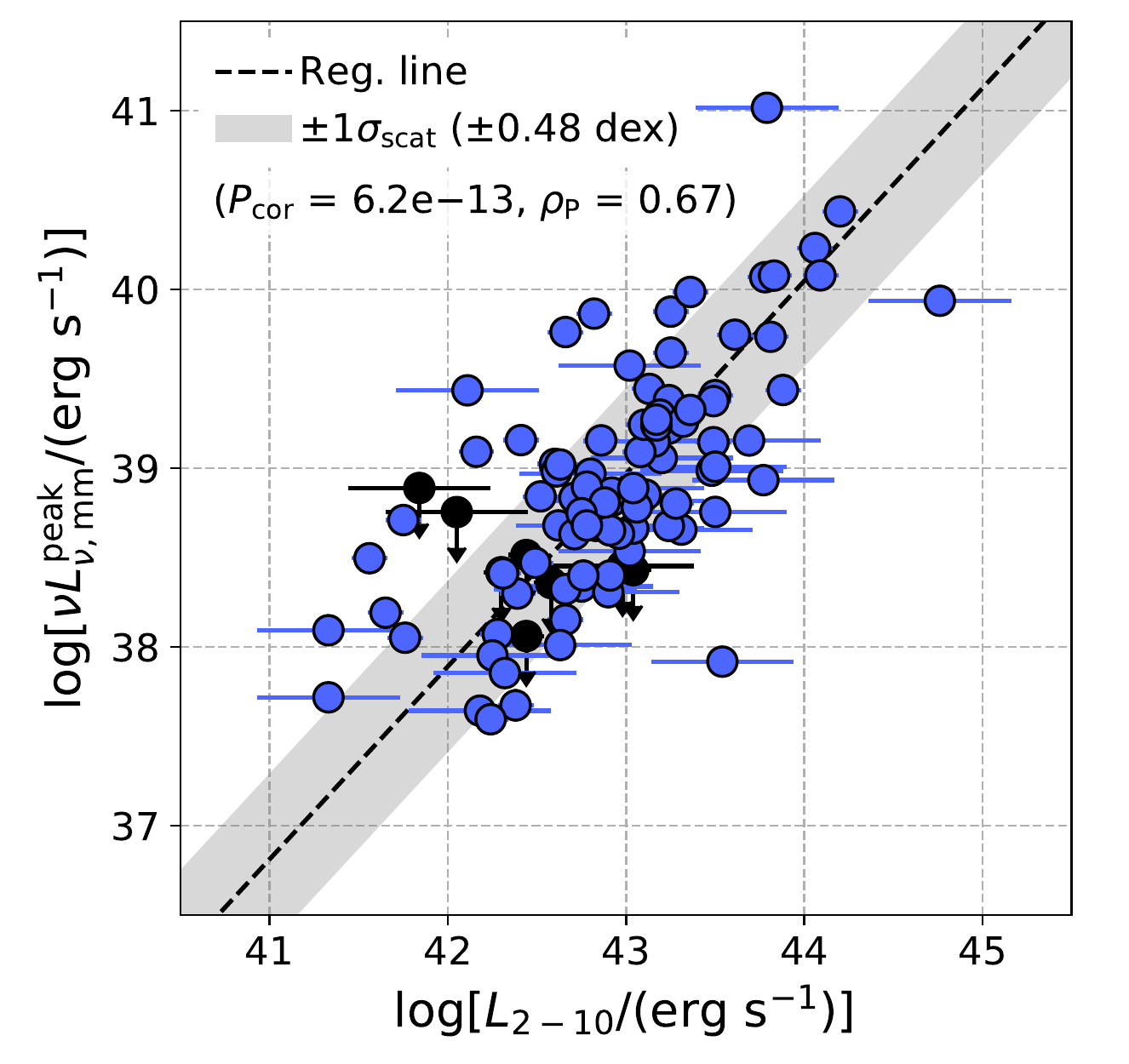}  \hspace{-.4cm}
    \includegraphics[width=5.9cm]{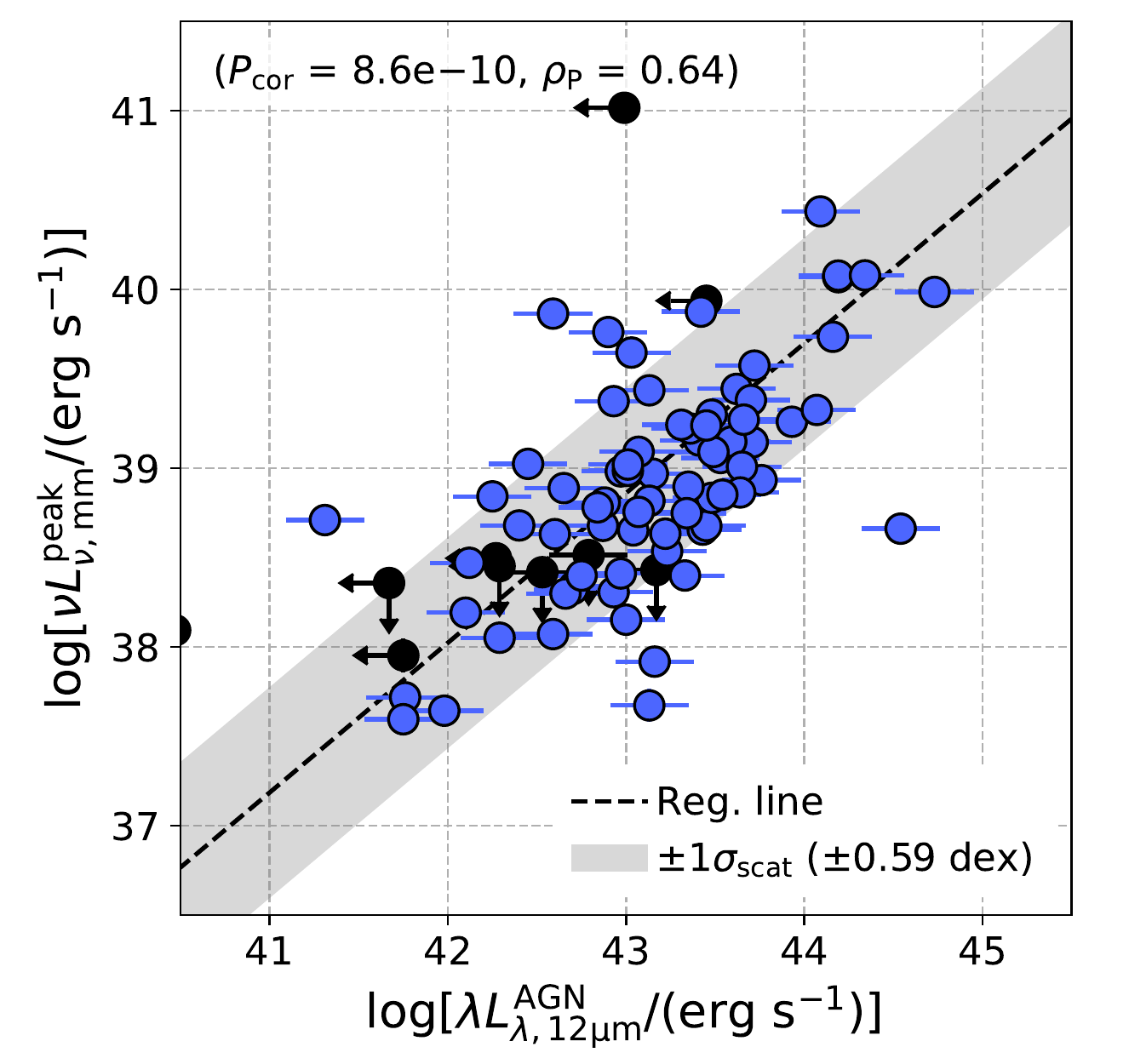} \hspace{-.4cm}
    \includegraphics[width=5.9cm]{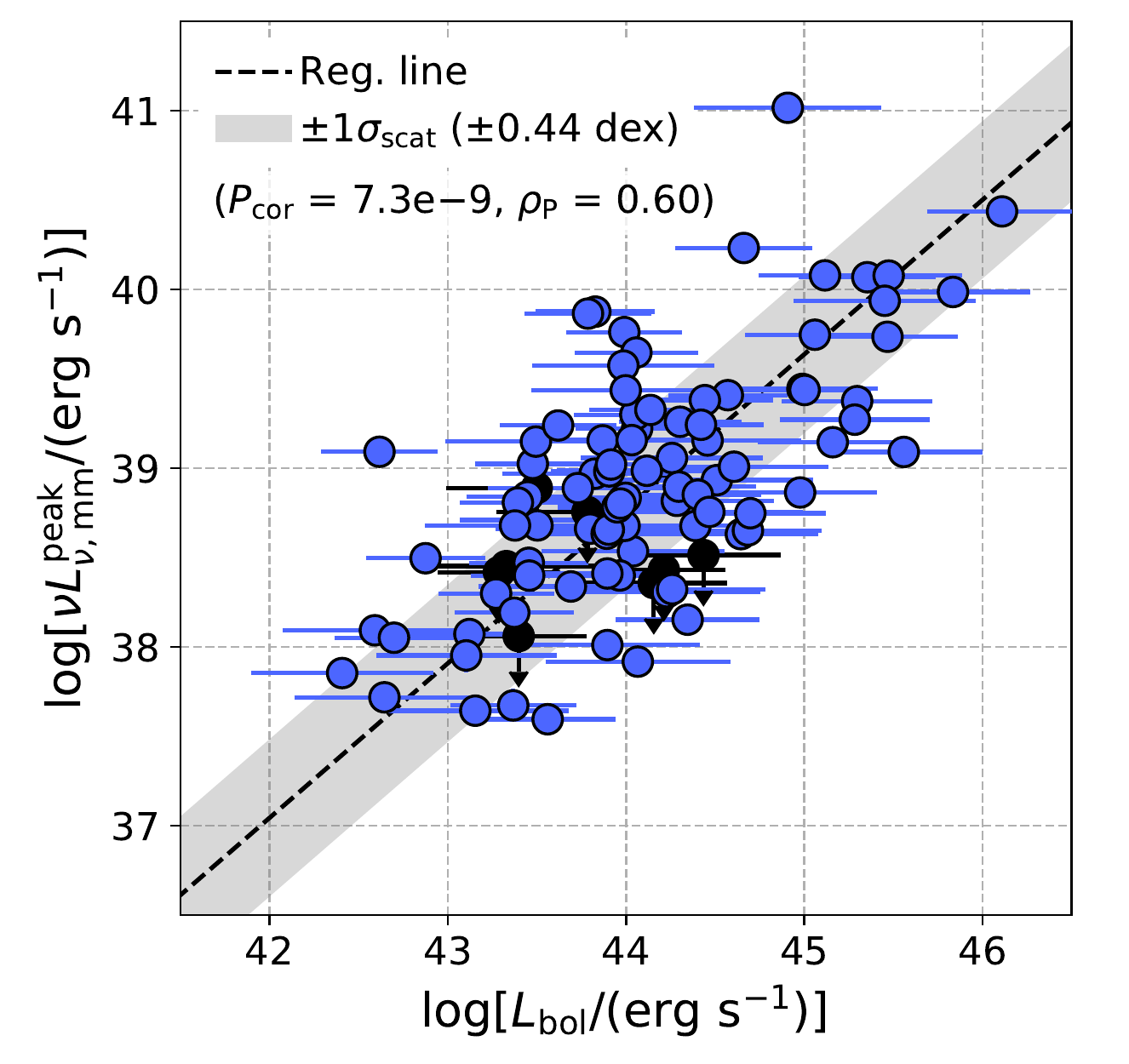} 
    \caption{
    Mm-wave luminosity versus 2--10\,keV, 12\,$\mu$m, and bolometric luminosities. 
    The 12\,$\mu$m luminosities take into account only an AGN component determined by the SED analysis of \cite{Ich19}.
    AGNs with upper limits are shown as black circles. 
    The black dashed lines indicate the best-fit linear regression lines, and the gray regions denote the $\pm1\sigma_{\rm scat}$ ranges.
    }
    \label{fig:var_lum2}
\end{figure*}

\section{Correlations of Nuclear Mm-wave and AGN Emission}\label{sec:cor}

We evaluate correlations of the mm-wave emission with different AGN components by using the results obtained from the Band-6 data (i.e., flux and luminosity) and also the other ancillary data (Section~\ref{sec:anc}). 
We derive many statistical values during the assessment but present only important ones in discussion. A list of obtained statistical values is provided in Table~\ref{tab_app:corr} of the appendix.

\subsection{The Tight Correlations between Mm-wave and AGN Luminosities}\label{sec:ll_cor}

We study the relations of the nuclear peak mm-wave luminosity with representative AGN ones: 14--150\,keV, 2--10\,keV, 12\,$\mu$m, and bolometric luminosities, and also their flux relations. 
For quantitative discussions, we calculate the $p$-value ($P_{\rm cor}$) and the Pearson correlation coefficient ($\rho_{\rm P}$) by using a bootstrap method \cite[e.g.,][]{Ric14b,Gup21,Kaw21}. 
This method draws many datasets from actual data, considering their uncertainties, and we derive the statistical values for each drawn dataset. For actual data with upper and lower errors, we randomly draw values from a Gaussian distribution where the mean and standard deviation are the best value and the 1$\sigma$ error, respectively.
For data with only an upper limit, we use a uniform distribution between zero and the upper limit. 
For each draw, we also derive a regression line of the form $\log {\rm Y} = \alpha \times \log {\rm X} + \beta$, based on the ordinary least-squares bisector regression fitting algorithm \citep{Iso90}. Moreover, an intrinsic scatter ($\sigma_{\rm scat}$), considering the uncertainties in actual data, is derived. 
By drawing 1000 datasets, we adopt the median value of the distribution for a parameter (i.e., $P_{\rm cor}$, $\rho_{\rm P}$, $\alpha$, $\beta$, or $\sigma_{\rm scat}$) 
as the best and their 16th and 84th percentiles as its lower and upper errors, respectively.

Figures~\ref{fig:var_lum1} and \ref{fig:var_lum2} show the correlations of the peak mm-wave luminosity for $L_{14-150}$, $L_{2-10}$, $\lambda L^{\rm AGN}_{\lambda, 12\,\mu{\rm m}}$, and $L_{\rm bol}$. All are found to be significant as quantified by very low $p$-values ($P_{\rm cor} \ll$  0.01; Table~\ref{tab_app:corr}). 
Also, for the fluxes, significant correlations are confirmed.
These are supplementarily shown in Section~\ref{sec_app:flux} of the appendix. 

Among the intrinsic scatters of the four luminosity correlations, that for the 14--150\,keV luminosity (0.36\,dex) is the smallest compared to the others (i.e., 0.48\,dex for $L_{2-10}$, 0.59\,dex for $\lambda L^{\rm AGN}_{\lambda, 12\,\mu{\rm m}}$, and 0.44\,dex for $L_{\rm bol}$). 
The smaller scatter compared with that for $L_{2-10}$ would be because the 2--10\,keV X-ray fluxes were measured based on single-epoch data, unlike the 14--150\,keV luminosities, which were derived from the 70-month averaged BAT spectra. 
Such single-epoch data may have observed short-time variability, and such variability can add scatter to the correlation.
Regarding the 12\,$\mu$m luminosity, the larger scatter would be in part because the emission strongly depends on the properties of the surrounding dust \citep[e.g., covering factor and optical depth; e.g.,][]{Sta16}. 
Lastly, the bolometric luminosity result gives an interesting insight. 
The luminosity can be used as an indicator of optical/UV luminosity, and thus its larger scatter than found for the 14--150\,keV could suggest that the mm-wave emission is more strongly coupled with the X-ray emission than with the optical/UV emission.
Furthermore, to examine whether the 14--150\,keV emission is most correlated with the mm-wave emission, we compare Pearson correlation coefficients based on the Fisher $r$-to-$z$ transformation\footnote{For given correlation strengths of $r_1$ and $r_2$ for two samples with sizes of $n_1$ and $n_2$ (in this paper,  $\rho_{\rm P}$ corresponds to $r$), the function ($z_1-z_2$)/(1/($n_1-3$)+1/($n_2-3$))$^{1/2}$, where $z_{[1,2]} = 1/2\ln (1+r_{[1,2]})/(1-r_{[1,2]})$, follows the standard normal distribution. Consequently, the difference between $r_1$ and $r_2$ can be statistically examined.}. Here, we define the $p$-value returned by this test as $P_{\rm rz}$ to distinguish this from the $p$-value for correlations ($P_{\rm cor}$). 
We then find that while the correlation strength for the 14--150\,keV X-ray luminosity ($\rho_{\rm P} = 0.74$) is the highest, it is not statistically stronger than those for the other luminosities  ($\rho_{\rm P} = $ 0.67 for $L_{2-10}$, $\rho_{\rm P} = $ 0.64 for $\lambda L^{\rm AGN}_{\lambda, 12\,\mu{\rm m}}$, $\rho_{\rm P} = $ 0.60 for $L_{\rm bol}$). 
In discussions on the origin of the mm-wave emission hereafter, we adopt the 14--150\,keV luminosity ($L_{\rm 14-150}$) as an indicator of the AGN activity given the smallest scatter and the highest correlation strength.

\begin{figure}
    \centering
    \includegraphics[width=8.3cm]{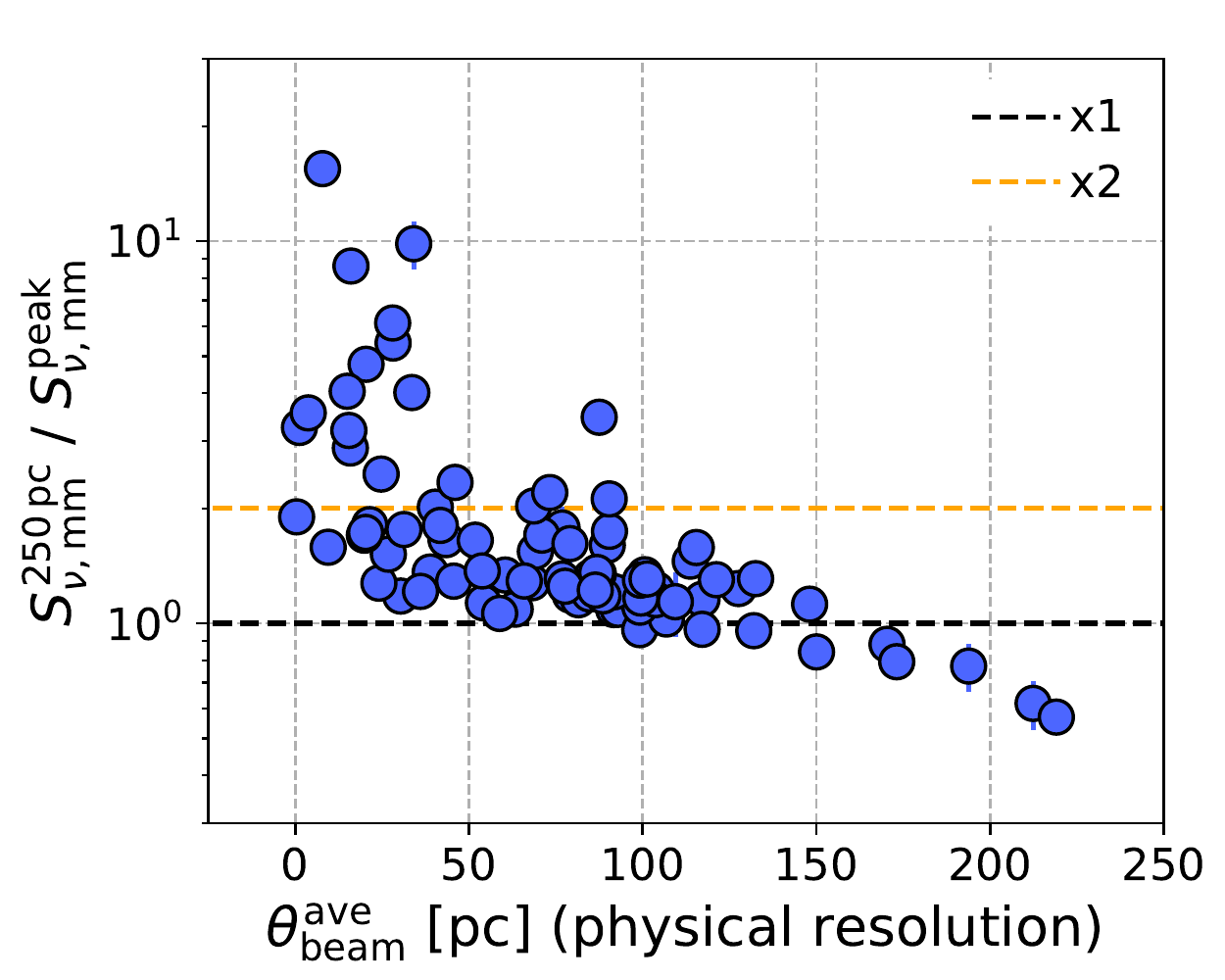}
    \caption{
    Ratio of the flux density ($S^{\rm 250\,pc}_{\nu, \rm mm}$) measured by an aperture with a diameter of 250\,pc and the peak one as a function of the spatial resolution achieved. 
    Values below $\sim 1$ particularly in a range $\theta^{\rm ave}_{\rm beam} > 150$\,pc are just due to the prescription to calculate the aperture flux density (see text). 
    }
    \label{fig:ratio2res}
\end{figure}

\begin{figure*}
    \centering 
    \includegraphics[width=18.cm]{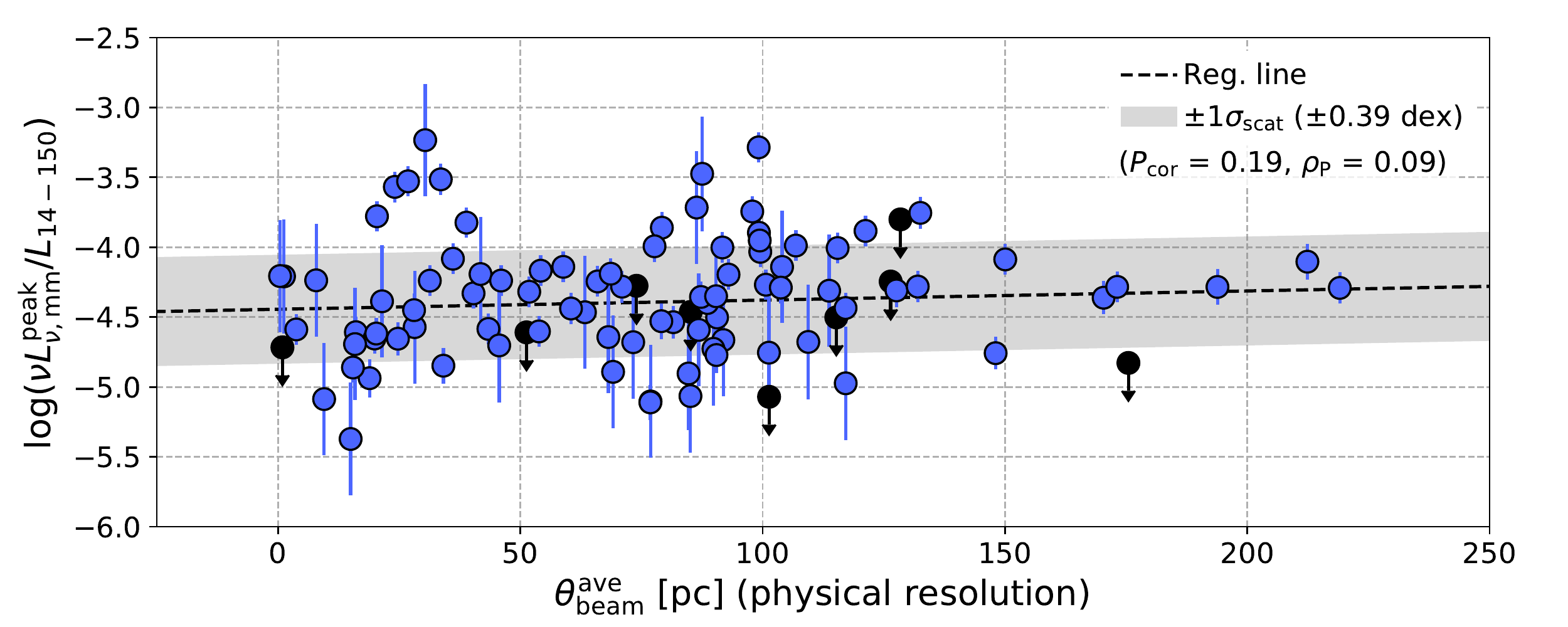}
    \caption{
    Mm-wave-to-X-ray luminosity ratio as a function of the physical resolution achieved in our ALMA data. 
    AGNs with upper limits are shown as black circles. 
    No significant correlation is found with $P_{\rm cor} \sim$ 0.19. 
    An important point of this plot is that the slope of the regression line (dashed line) is $\approx 0.0007$. Thus, the difference in the ratio is $\approx$ 0.1\,dex between 200\,pc and 10\,pc. 
    The flat slope can be interpreted as the ubiquitous presence of 
    a compact component on a scale of $\lesssim$ 10\,pc that is related to AGN X-ray emission. 
    }
    \label{fig:ratio_vs_res}
\end{figure*}

As described in Section~\ref{sec:sample}, where we have introduced our sample selection, no Blazar-like AGNs are included in our sample due to the absence of publicly available Band-6 ALMA data. However, we here briefly discuss mm-wave-to-X-ray ratios of Blazar-like AGNs. 
As representative Blazar-like AGNs, we refer to well-studied nearby objects of Mrk 421 and Mrk 501 ($z \approx$ 0.03). The SEDs of Mrk 421 and Mrk 501 are presented in \cite{Abd11_421} and \cite{Abd11_501}, respectively, and appear to have $\nu\,L_{\nu, \rm mm}/L_{\rm X} \sim 0.1$. For both objects, the size of a mm-wave emitting region was constrained to be $< $ 0.1\,pc. Therefore, the ratios may be observed with our achieved resolutions ($\lesssim$ 200\,pc) and may be fairly compared with those found for our targets. 
As a result, 
the comparison suggests that Blazar-like AGNs are more mm-wave luminous by approximately three orders of magnitude. 

\subsection{Insignificant Impact of Malmquist Bias}\label{sec:bias}

As our sample was originally based on the flux-limited Swift/BAT catalog \citep{Bau13}, we examine whether the Malmquist bias produces the correlation of the mm-wave and 14--150\,keV X-ray luminosities. For this purpose, we create a subsample by selecting 25 objects with $\log[L_{\rm 14-150}/({\rm erg}\,{\rm s}^{-1})] > 42$ and $z < 0.01$, corresponding to the area covered by the dashed lines in Figure~\ref{fig:z_vs_lx}. 
For the subsample, we assess the correlation between the mm-wave and 14--150\,keV luminosities, and then find a significant one with $P_{\rm cor} \approx 4\times10^{-3}$.
The Pearson correlation coefficient is $\rho_{\rm P}$ = 0.56.
Although that is less than what was obtained for the entire sample ($\rho_{\rm P}$ = 0.74), no statistical difference is found between the two values, quantified as $P_{\rm rz} = 0.18$. 
Thus, the Malmquist bias would not strongly affect the correlation.

\subsection{Mm-wave Luminosity vs. Physical Resolution}\label{sec:res}

It is important to discuss whether or not the origin of the nuclear mm-wave emission that is significantly correlated with the 14--150\,keV emission is diffuse emission that can be resolved with our spatial resolutions ($\sim$ 1--200\,pc). Here, we assess how much diffuse emission is resolved depending on the spatial resolution. Figure~\ref{fig:ratio2res} shows the ratio of a flux measured within an aperture of 250\,pc to a peak flux as a function of the beam size achieved. The aperture diameter is determined to be larger than any of our average beam sizes. In the figure, objects in a range above $\theta^{\rm ave}_{\rm beam} \sim$ 150\,pc tend to show flux ratios below $\sim$ 1, but this would be just due to the method adopted to calculate the flux density within the aperture. The flux density is calculated as $\Sigma S_i / N \times N/n$, where $\Sigma S_i/N$ is the average of the flux densities $S_i$ (Jy/beam) in each pixel within the aperture composed of $N$ pixels, and $N/n$ indicates the number of beams, each having $n$ pixels. Thus, for example, if an aperture and a beam are comparable in size (i.e., $N \sim n$), the flux density can be smaller than the peak one as $\Sigma S_i / N$ may be less than $S^{\rm peak}_i$. 
In addition to this, a clear trend seen in Figure~\ref{fig:ratio2res} is that half of the 30 objects below $\theta^{\rm ave}_{\rm beam} = 50$\,pc show ratios greater than 2. 
One might suggest that there appears to be a significant contribution of diffuse emission, such as that observed at $\theta^{\rm ave}_{\rm beam} < 50$\,pc, to the flux measured with a larger beam. 
However, this result can be explained if the sensitivities achieved are similar among different sources and a high spatial resolution is achieved preferentially for closer objects (i.e., brighter objects). 
To confirm this, we make high-resolution and low-resolution subsamples consisting of targets observed at $\theta^{\rm ave}_{\rm beam} < 50$\,pc and those observed at $\theta^{\rm ave}_{\rm beam} > 100$\,pc, respectively. 
Using the KS-test, we find that the subsamples have 
similar sensitivity distributions, and the high-resolution sample is significantly biased for closer objects. 
The original expectation (more diffuse contribution for larger beams) thus is not necessarily true. 
A consistent result can be obtained by deriving regression lines between $\nu L^{\rm peak}_{\nu,\rm mm}$ and $L_{\rm 14-150}$ for the two subsamples. By fixing their slopes to 1.17, the average of the slopes of two independently determined regression lines, we find that the intercepts obtained are almost the same. 
This result is consistent with the idea that even for the low-resolution subsample, a compact component related to X-ray emission, like that detected in the high-resolution subsample, contributes significantly to the observed emission.
 
Furthermore, we examine a relation between $\log (\nu L^{\rm  peak}_{\nu,{\rm mm}}/L_{\rm 14-150})$ and the physical resolution, as shown in Figure~\ref{fig:ratio_vs_res}. Applying the bootstrap method, we obtain $P_{\rm cor} = 0.19$, suggesting no significant correlation. 
The most important quantity is the slope of the regression line. 
By adopting the chi-square method with $\log (\nu L^{\rm peak}_{\nu,{\rm mm}}/L_{\rm 14-150})$ being set as the dependent variable, we obtain $\log (\nu L^{\rm peak}_{\nu,{\rm mm}}/L_{\rm 14-150}) \propto 0.0007(\pm0.0003) \times \theta^{\rm ave}_{\rm beam}$. The slope indicates that the difference in $\log (\nu L^{\rm peak}_{\nu,{\rm mm}}/L_{\rm 14-150})$ is at most 0.1\,dex for the highest and lowest resolutions (i.e., $\sim $ 1\,pc and $\sim$ 200\,pc). 
This is smaller than the intrinsic scatter of 0.36\,dex for 
the correlation of $\nu L^{\rm peak}_{\nu,{\rm mm}}$ and $L_{\rm 14-150}$ (Section~\ref{sec:ll_cor}). 

We additionally examine whether the conclusion depends on the achieved sensitivity in the mm-wave data because a higher sensitivity may detect extended emission and a steeper relation is expected if that is significant and is resolved more with increasing resolution. 
We perform a bootstrap analysis for the AGNs that were observed with a higher sensitivity than 0.027 mJy beam$^{-1}$, which is the median sensitivity of our ALMA data. The slope obtained is $\sim$ 0. Thus, the conclusion does not depend on the sensitivity. 

Lastly, we mention the beam shape, which can be elongated in some ALMA observations, for example, those toward objects outside the recommended declination range $-70^\circ \sim +20^\circ$\footnote{Figure 7.8 of the ALMA Technical Handbook available from http://almascience.org/documents-and-tools/cycle9/alma-technical-handbook}. The beams achieved for $\sim$ 60 objects, in fact, have aspect ratios of $> 1.2$, and in such cases the average value of $(\theta^{\rm min}_{\rm beam}\times\theta^{\rm maj}_{\rm beam})^{1/2}$ may not represent a linear resolution. Thus, for a clearer discussion, we make a sample of 37 AGNs observed with nearly circular beams of 
$\theta^{\rm maj}_{\rm beam}/\theta^{\rm min}_{\rm beam} \leq 1.2$, and derive a regression line in a resolution range 4--220\,pc. 
The resultant slope is $\approx 0.001$, supporting the conclusion drawn for the whole sample (i.e., no strong dependence on the achieved resolution).

\begin{figure}
    \centering 
    \includegraphics[width=8cm]{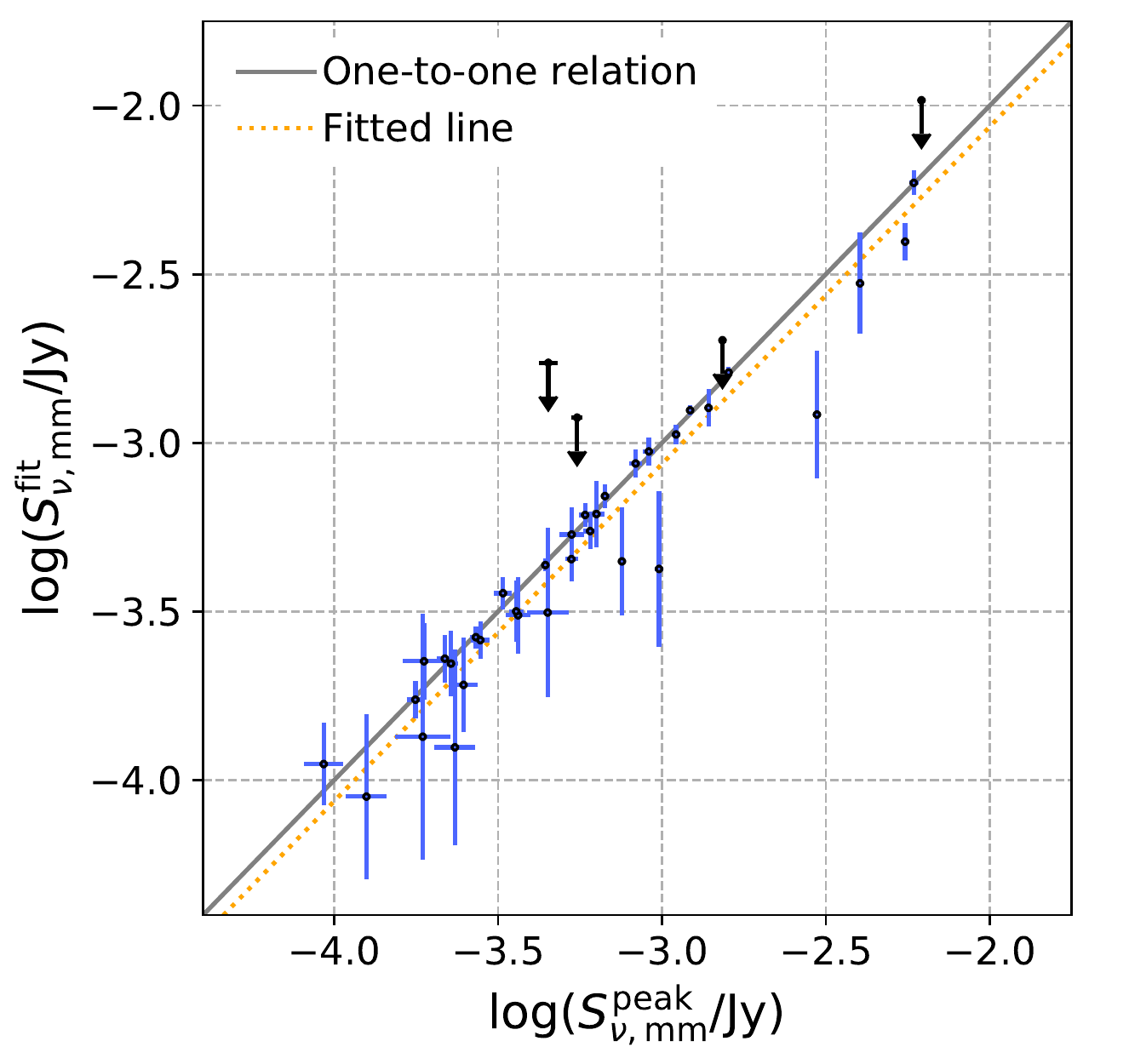}
    \caption{
    Comparison of the flux density of an un-resolved component constrained by the visibility fitting and the observed peak one. 
    The gray solid and yellow dotted lines 
    indicate the one-to-one relation and the fitted line with $S^{\rm fit}_{\nu,\rm mm}$ being the dependent parameter, respectively. 
    }
    \label{fig:visfit2}
\end{figure}  

\begin{figure}
    \centering 
    \includegraphics[width=7.6cm]{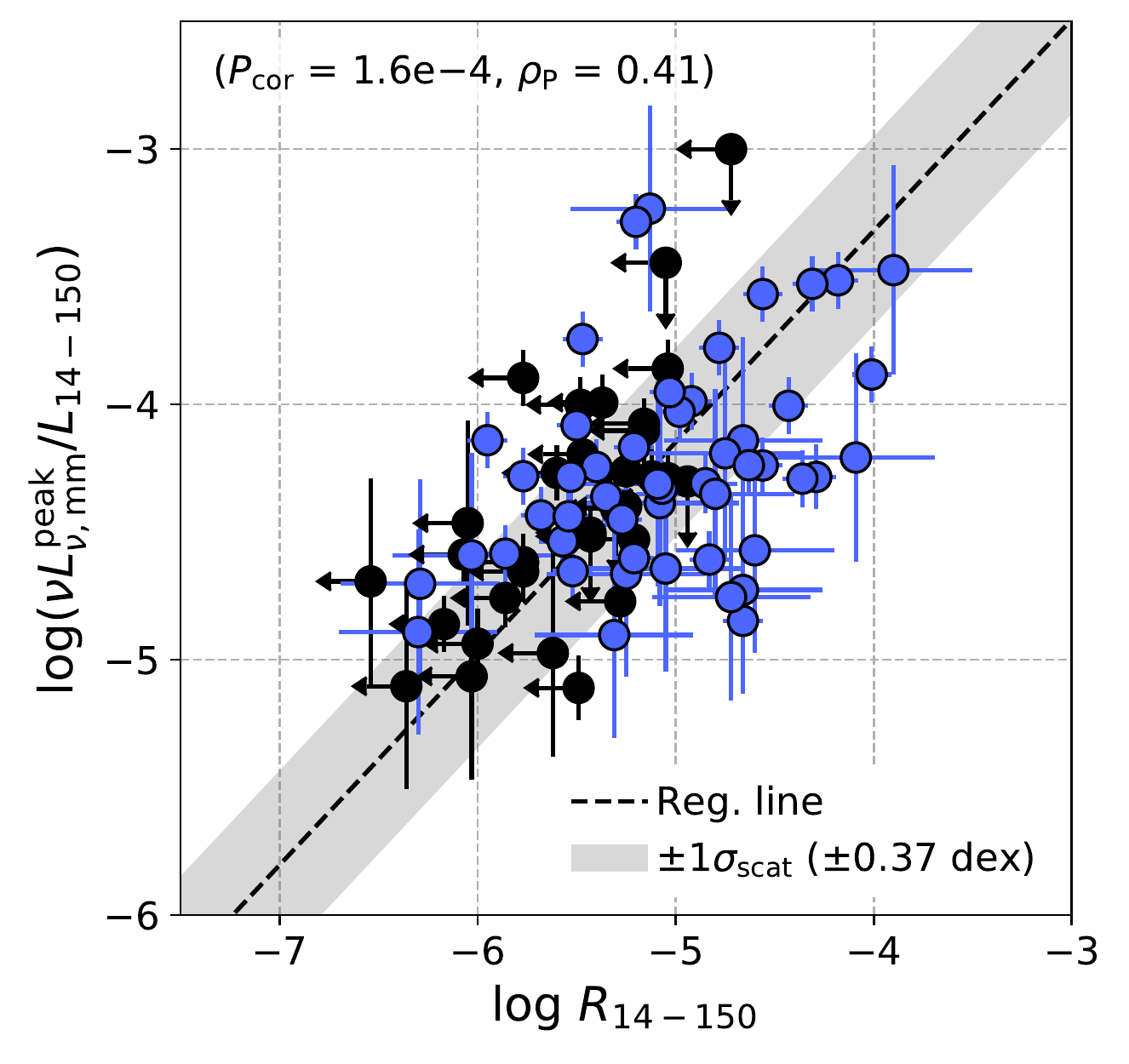}
    \includegraphics[width=7.6cm]{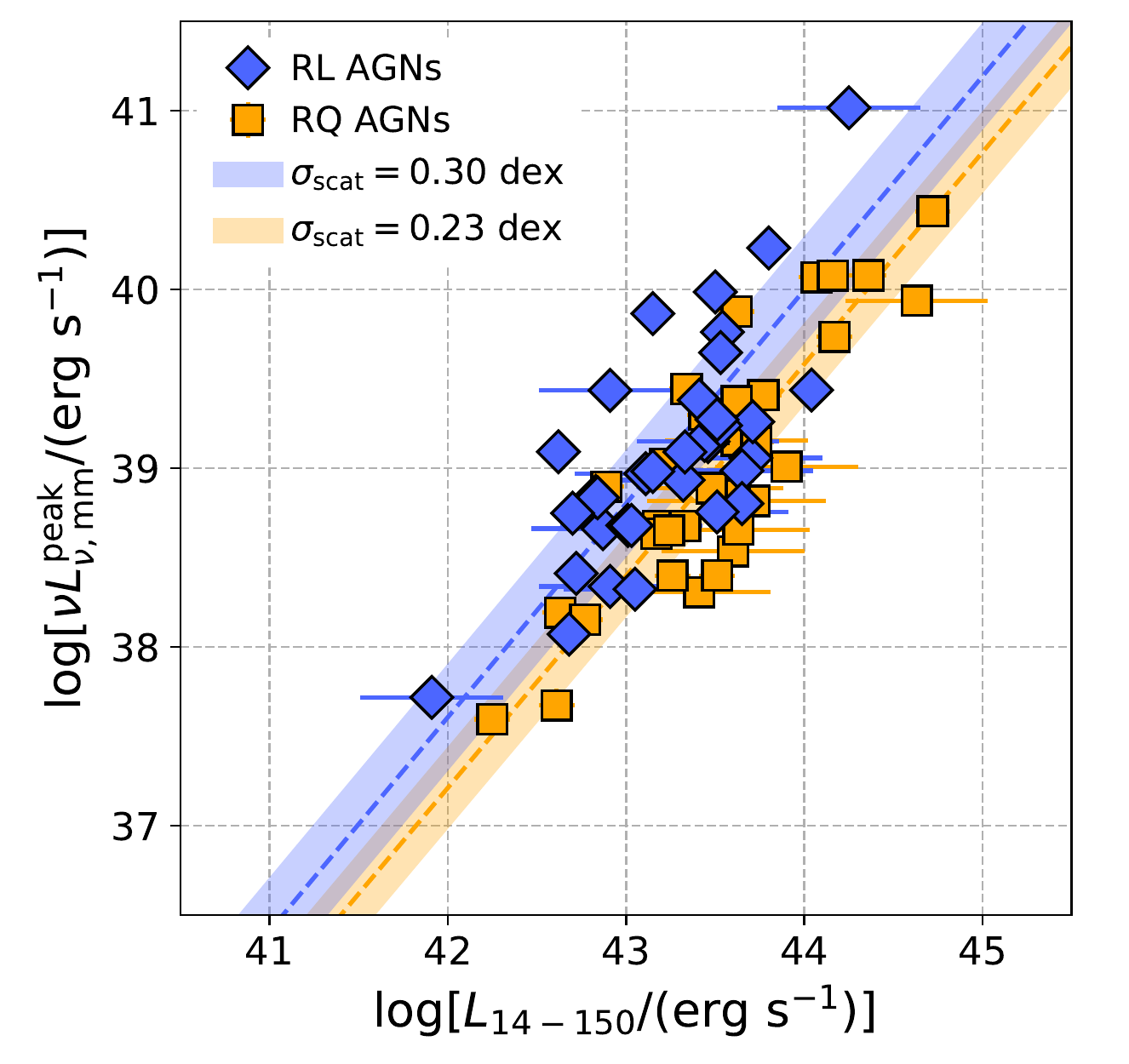}
    \caption{
    Top: Plot of the $\nu L^{\rm peak}_{\nu,{\rm mm}}/L_{\rm 14-150}$ ratio versus the radio loudness based on the 3\,GHz and 14--150\,keV luminosities. 
    AGNs with upper limits are shown in black. 
    Bottom: Correlations between the mm-wave and 14--150\,keV luminosities for RQ AGNs (orange) and RL AGNs (blue). 
    In both panels, fitted regression lines and $\pm1\sigma_{\rm scat}$ ranges are indicated by dashed lines and shaded areas, respectively. 
    }
    \label{fig:ratio_vs_rl}
\end{figure}

\subsection{Mm-wave Luminosity vs. Morphology}\label{sec:mor}

The high spatial resolutions of the ALMA data can resolve mm-wave emitting regions for some of our objects (Figure~\ref{fig:image}), and it is important to investigate whether such extended components affect the correlation, as they may not necessarily be related to AGN activity. 
Thus, we compare the correlations of a subsample of AGNs that show extended emission and that of the other AGNs. 
As the visual inspection performed in Section~\ref{sec:alma_data} should depend on the sensitivity and spatial resolution, 
we only consider AGNs detected above 20$\sigma_{\rm mm}$ and observed with beam sizes of $> 10$\,pc. 
Consequently, based on the KS-test, we confirm that the detection significance and resolution distributions of the two subsamples are statistically indistinguishable ($P_{\rm KS} \gtrsim 0.03$), and this is not true if a more relaxed criterion is considered. 
For the two subsamples of 26 AGNs with C or CB and 28 AGNs with E plus some of the others, we find significant correlations between $\nu L^{\rm peak}_{\nu,{\rm mm}}$ and $L_{\rm 14-150}$ with $P_{\rm cor} \ll 0.01$. 
Furthermore, their correlation strengths are found to be statistically the same. 
Thus, the visually identifiable extended components would not be significant in the observed correlations. 

We discuss the above result in more depth. 
A significant fraction of AGNs ($\sim$ 50\%) have been classified as having non-nuclear emission by eyes. Thus, even the other objects (i.e., C objects) may also have a similar contribution from a component that has not been visually identified. If true, this can lead to the same correlations between the two AGN types, as previously reported. 
Hence, to investigate in more detail whether the non-nuclear emission is insignificant for AGNs classified as C, we fit the observed visibility data using \textsc{UVMULTIFIT} \citep{Mar14} and constrain the flux density solely from an unresolved component. For this analysis, we select 38 AGNs that are classified as purely C and were not observed in the mosaic observation mode.
The former is considered so that we can avoid complex distributions for which a simple Gaussian function is insufficient. 
The latter is because \textsc{UVMULTIFIT} is experimental for the observing mode. 
Our fitting is detailed in Section~\ref{sec_app:fit}. 
Figure~\ref{fig:visfit2} shows the flux density of an unresolved component ($S^{\rm fit}_{\nu,\rm mm}$) versus the peak flux density. 
It can be seen that the flux of an unresolved component generally dominates the peak flux. 
To quantify the contribution, we perform a linear fit using the chi-square method with $S^{\rm fit}_{\nu,\rm mm}$ being the dependent parameter. 
The intercept obtained is $-$0.06\,dex, indicating an $\approx 13$\% contribution from the resolved emission on average. 
As inferred from this result, a significant correlation between the 
luminosity of the unresolved component and 14--150\,keV luminosity is found ($P_{\rm cor} \sim 10^{-3}$). Also, its constrained scatter is 0.27\,dex, close to that obtained with the peak flux density (i.e., 0.36\,dex).
After all, this result, as well as the similar correlations found for morphologically different AGN types, are consistent with the hypothesis that a significant fraction of the observed mm-wave emission originates from a compact ($\lesssim$ 10\,pc) region.

\subsection{Dependence on Radio Loudness}\label{sec:rad}

In the last of this section, we examine the dependence of $\nu L^{\rm peak}_{\nu,{\rm mm}}/L_{\rm 14-150}$ on the radio loudness ($R_{\rm 14-150}$), introduced in Section~\ref{sec:anc}.
A synchrotron component extending from the cm-wave band would contribute to the mm-wave emission, as inferred from the SED data presented by \cite{Ino18}. 
Figure~\ref{fig:ratio_vs_rl} shows $\nu L^{\rm peak}_{\nu,{\rm mm}}/L_{\rm 14-150}$ versus $R_{\rm 14-150}$, and there exists a significant correlation with $P_{\rm cor} \approx 2\times10^{-4}$. 
To detail this correlation, we derive the statistical parameters for individual RQ-AGN and RL-AGN subsamples and find low $p$-values of $P_{\rm cor} \sim 10^{-7}$ and $P_{\rm cor} \sim 10^{-4}$, respectively (Figure~\ref{fig:ratio_vs_rl}). 
There is no significant difference between their correlation coefficients of $\rho_{\rm P} =$ 0.87 and $\rho_{\rm P} =$ 0.64 as $P_{\rm rz} \approx 0.02$, and also the slopes obtained are $1.19^{+0.08}_{-0.06}$ for the RQ AGNs and $1.20^{+0.13}_{-0.10}$ for the RL AGNs,
consistent with each other within 1$\sigma$. 
Furthermore, to focus on the difference in the intercept, 
we derive them by fixing the slopes at 1.19, the average of the independently determined values for the subsamples. The difference of $\approx$ 0.4\,dex is then found to be significant at 5.6 sigma level (i.e., $p$-value is less than 0.01). Intrinsic scatters for the RL-AGN and RQ-AGN samples are $\approx$ 0.30\,dex and 0.23\,dex, and, indeed, the sum of two Gaussian distributions that have these scatters and are separated by 0.4\,dex can be approximated by a single Gaussian function with a scatter of 0.34\,dex. This is almost the same as the scatter of 0.36\,dex obtained for the entire sample. Thus, a spectral component extending from the cm-wave to mm-wave bands would contribute to the scatter of the entire sample. In the following discussion, however, we do not separate the RQ and RL AGNs to retain a larger sample while accepting their possible difference by $\approx$ 0.3--0.4\,dex.

\section{Mm-wave Emission FROM SF}\label{sec:sf_cont}

We have found a possible relation between the nuclear mm-wave emission and the AGN activity traced by the hard X-ray (14--150\,keV) emission. 
To understand the origin of the mm-wave emission, we first discuss three possible SF mechanisms for the mm-wave emission rather than AGN mechanisms: (1) thermal emission from heated dust in SF regions, 
(2) free-free emission from H{\sc{ii}} regions created by massive stars above $\sim$ 8 $M_\odot$, (3) synchrotron emission by cosmic-rays accelerated by supernova remnants and other galactic sources \citep[e.g., ][]{Con92,Gor90,Gre19,Tab17,Dom21}. Then, we identify the strongest contributor among the three SF processes. 

\subsection{Expected Fluxes and Spectral Indices of the SF Components}\label{sec:sf}

To estimate the contribution of the SF components or their expected flux densities, the radio-to-IR SEDs constructed by combining the VLA, ALMA, and IR data (Section~\ref{sec:anc}) are helpful. 
As an example, four SEDs are shown in Figure~\ref{fig:sed}.
The FIR data around 60--100~$\mu$m ($\sim$ 80\%) were taken at resolutions coarser than $\sim$ 6\arcsec\ using Herschel/PACS. However, according to an imaging analysis of Herschel data for BAT-selected nearby AGNs ($z < 0.05$) by \cite{Mus14}, thus resembling our sample, 
a significant fraction ($>$ 50\%) of their 70\,$\mu$m emission originates from an aperture with 6\arcsec.
The contribution of SF-related dust emission can be estimated by using the host-galaxy SED model, constrained in the IR band by \cite{Ich19}. 
Because these models are not available in the mm-wave band, we extrapolated them by adopting a modified black body model, expressed as $S_\nu \propto \nu^{\beta_{\rm BB}} F_{\rm BB}$, where $F_{\rm BB}$ is the Planck function, and $\beta_{\rm BB}$ is fixed to 1.5.  
The modified black-body model with $\beta_{\rm BB} = 1.5$ was adopted to create the host-galaxy models above $40~\mu$m used in \cite{Ich19} \citep[see ][for more details]{Mul11}.
Also, the $\beta_{\rm BB}$ value is well within the range of indices found for star-forming galaxies \cite[e.g., $1.60\pm0.38$;][]{Cas12}. This fact makes $\beta_{\rm BB} = 1.5$ a reasonable option for the host-galaxy model.

\begin{figure}
    \centering 
    \includegraphics[width=8.5cm]{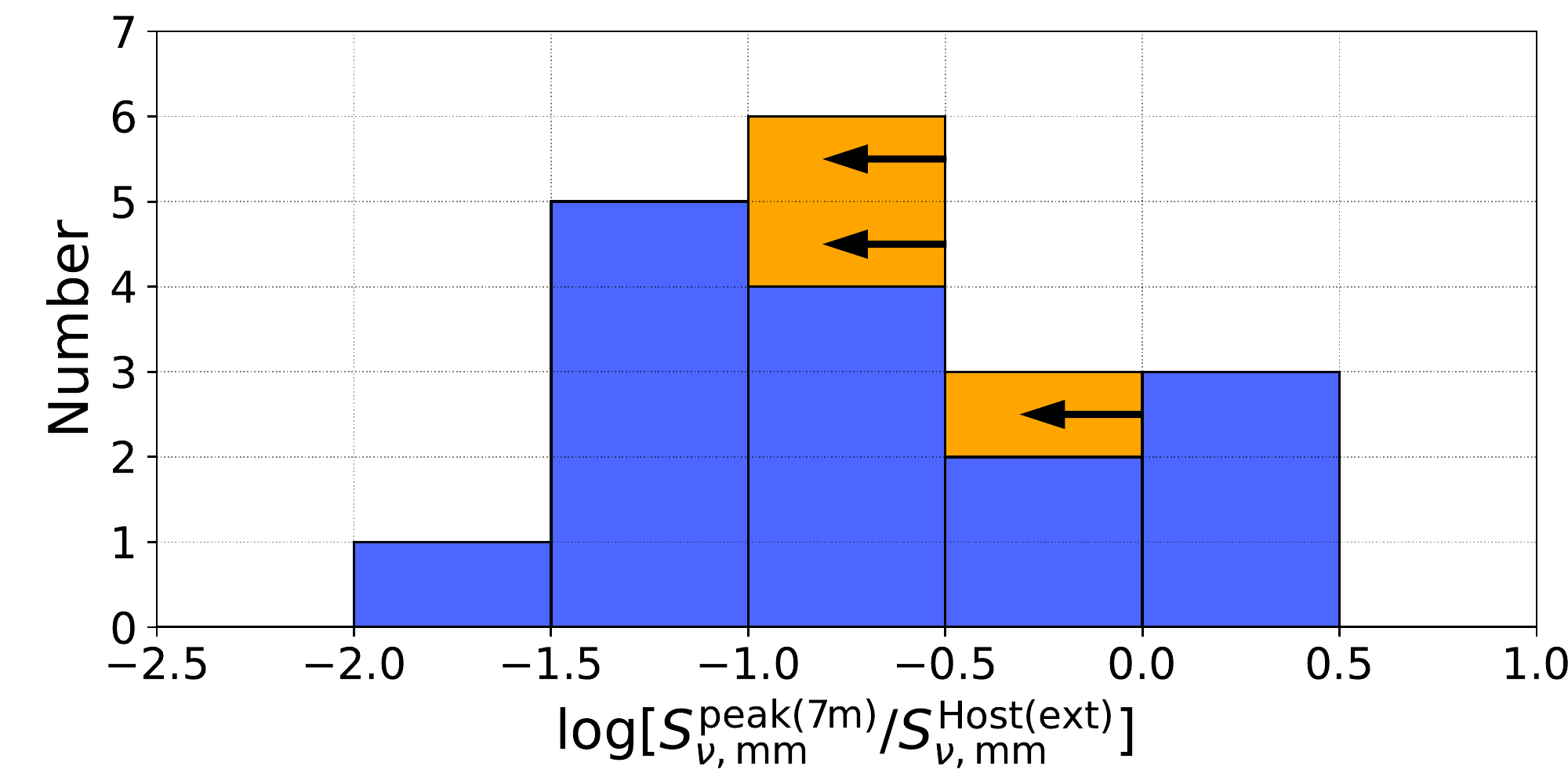}     
    \includegraphics[width=8.5cm]{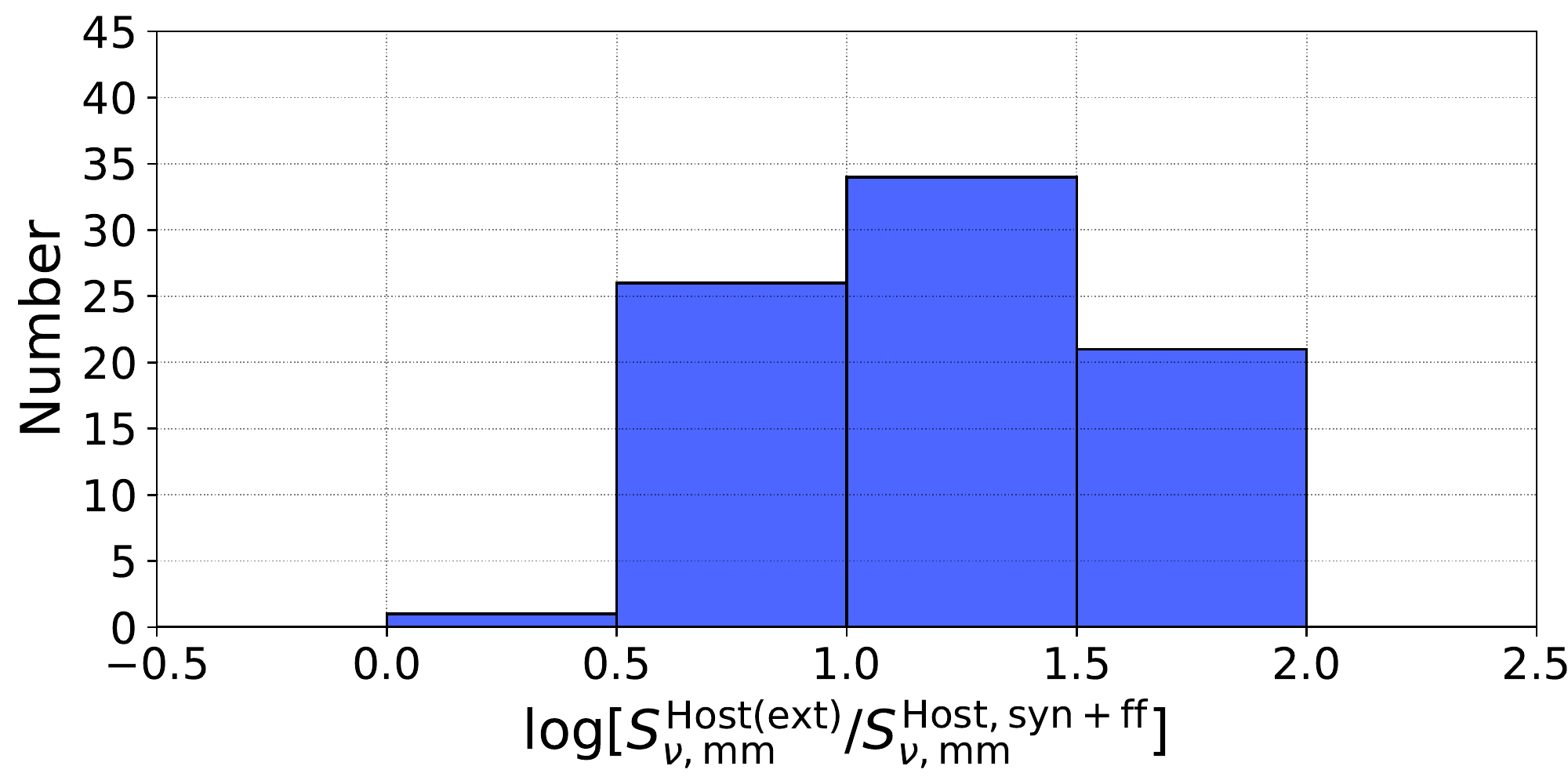}
    \includegraphics[width=8.5cm]{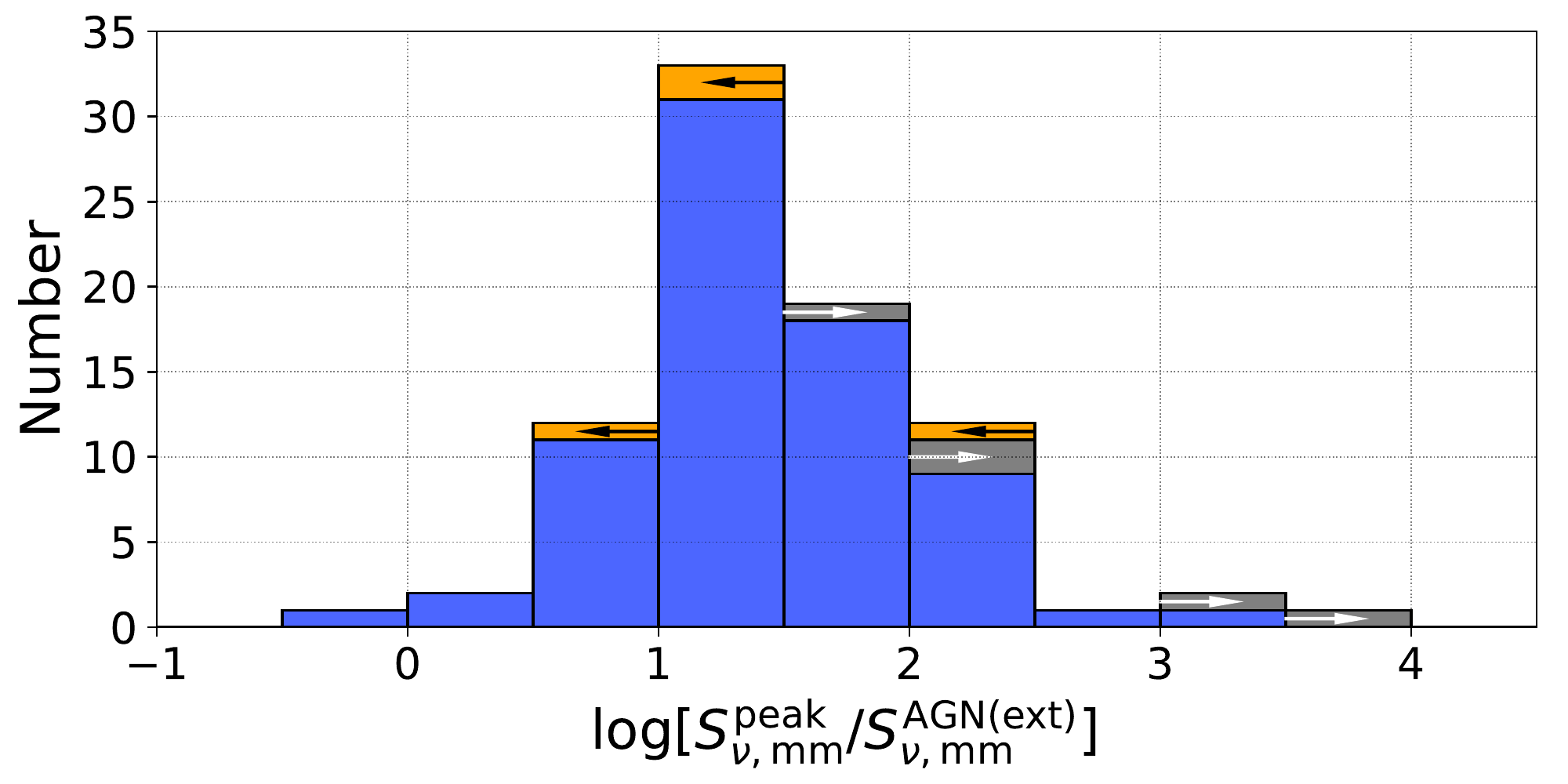}
    \caption{
    Top: 
    Histogram of the logarithmic ratio of the mm-wave peak flux density measured by using an ALMA 7-m array data and that expected from the host-galaxy SED model. 
    Sources with upper limits are indicated in orange. 
    Middle: 
    Histogram of the logarithmic ratio of the mm-wave 
    flux density expected from the host-galaxy SED model (i.e., thermal dust emission) and that predicted from synchrotron plus free-free emission. 
    In each SED of Figure~\ref{fig:sed}, these two components are shown as a gray dashed line and an orange pentagon, respectively. 
    The thermal emission is generally stronger than the sum of the synchrotron and free-free components. 
    Bottom: 
    Histogram of the logarithmic ratio of 
    the mm-wave peak flux density and 
    that predicted from the AGN SED model (blue dashed line in Figure~\ref{fig:sed}). 
    Sources with upper and lower limits are indicated in orange and gray, respectively. 
    }
    \label{fig:ratios} 
\end{figure}

To examine whether the above extrapolation ($\beta_{\rm BB} = 1.5$) is appropriate, we use Band-6 ALMA 7-m array data. Their attainable angular resolutions are $\gtrsim$ 4\arcsec, closer to the typical FIR emitting size of BAT-selected AGNs 
\cite[e.g., 6\arcsec;][]{Mus14}. 
Of our 98 objects, 22 had ALMA 7-m data, and among them, 18 also have their SED analysis results. 
The 7-m data were analyzed in the same way as the 12-m data to obtain peak fluxes ($S^{\rm peak(7m)}_{\nu,\rm mm}$). The actual average resolutions of $(\theta^{\rm min}_{\rm beam}\times\theta^{\rm maj}_{\rm beam})^{1/2}$ are in a range 4\farcs5--6\farcs5. The top panel of Figure~\ref{fig:ratios} shows a histogram of the ratio of the obtained peak flux density to the extrapolated one ($S^{\rm peak(7m)}_{\nu,\rm mm}/S^{\rm Host(ext)}_{\nu,\rm mm}$). According to \cite{Mus14}, a ratio of $\sim$ 1 is expected, but the ratio is less than half for more than half of the objects. We suspect that this is partly because the size of a FIR emitting region of these sources is exceptionally larger than the typical angular size seen for the BAT AGNs of \cite{Mus14}. 
Originally, FIR PACS data at 70\,$\mu$m and 160\,$\mu$m for BAT-selected AGNs were analyzed by \cite{Mel14}, and they measured the entire 70\,$\mu$m fluxes by either adopting a fiducial aperture of 12\arcsec or manually defining an emitting area. Then, the measured fluxes were used in \cite{Mus14}. 
By checking the original paper of \cite{Mel14}, 
we find that for sources with $S^{\rm peak(7m)}_{\nu,\rm mm}/S^{\rm Host(ext)}_{\nu,\rm mm} < 0.5$, a much larger aperture ($\sim 12$\arcsec--160\arcsec) was generally adopted. 
On the other hand, the fiducial aperture (12\arcsec) was adopted for objects 
with $S^{\rm peak(7m)}_{\nu,\rm mm}/S^{\rm Host(ext)}_{\nu,\rm mm} > 0.5$. 
Thus, the larger ratio may be due to the 7-m data still missing diffuse emission, and  there seems to be no strong evidence for a discrepancy between the extrapolation and the ALMA data. Rather, the consistent sources ($S^{\rm peak(7m)}_{\nu,\rm mm}/S^{\rm Host(ext)}_{\nu,\rm mm} > 0.5$) seem to support the validity of the extrapolation.

Regarding the potentially existing free-free and synchrotron components related to SF, we estimate the sum of their expected luminosities in units of 
${\rm erg}\,{\rm s}^{-1}\,{\rm GHz}^{-1}$ for a given SFR using the equation:  
\begin{eqnarray}\label{eqn:sfr2lmm}
    L_\nu & = & 
    {\rm SFR\,(}M_\ast>0.1 M_\odot{\rm )}/(M_\odot {\rm yr}^{-1})     \nonumber \\
    & & \times  [9.5\times10^{36} (\nu/{\rm GHz})^{-\alpha_{\rm syn}} \nonumber \\ 
    & & + 1.6\times10^{36} (T_{\rm e}/10^4 {\rm K})^{-0.59}  (\nu/{\rm GHz})^{-0.1}].
\end{eqnarray}
This is derived by following \cite{Con92}.
The first and second terms within the parentheses 
consider synchrotron emission and free-free emission, respectively. 
The electron temperature ($T_{\rm e}$) expected in H{\sc{ii}} regions, or for ionized gas around massive stars, is set to $10^4$\,K \citep[e.g.,][]{Gor90}. 
The spectral index of $\alpha_{\rm syn}$ is set to 0.8 by 
following \cite{Tab17}, who constrained radio (1--10\,GHz) slopes for nearby star-forming galaxies, finding a typical spectral slope of 0.8 with a standard deviation of 0.2. 
Similar indices were observed for synchrotron emission from supernova remnants \cite[e.g.,][]{Gre19,Dom21}. We confirm that our discussion and conclusion are not affected by the choice of $\alpha_{\rm syn}$ within the possible range of $\sim$ 0.6--1.0. 
To calculate the expected luminosities for our AGNs, we use the SFRs and errors obtained from the SED fittings in the IR band \citep{Ich19}. 
In Figure~\ref{fig:sed}, showing four SEDs, an expected mm-wave flux is plotted 
for each object as an orange pentagon, together with a power law with an expected index of 0.1 around $\sim$ 200--300\,GHz (Equation~\ref{eqn:sfr2lmm}). The calculated flux should be considered as an upper contribution limit to the observed mm-wave emission since the SFRs were measured at apertures larger than $\gtrsim$ 6\arcsec.

Based on the data obtained so far, we calculate the ratio between the flux density of the thermal emission and the sum of the synchrotron and free-free emission to identify the strongest SF component. 
The middle panel of Figure~\ref{fig:ratios} shows a histogram of the ratios, and the thermal emission is seen to be the strongest in all objects. 
A crucial indication of this result is that the spectral index of the sum of the SF components is expected to be $\alpha_{\rm mm} \approx -3.5$ for most objects. This result is used in the next section to identify AGNs for which the SF emission is expected to be negligible. 

We note that to identify AGNs with little SF emission, one might compare the fluxes of the observed mm-wave emission and the expected SF emission, but it is difficult to draw a robust conclusion in that way. 
For example, the correction for the large difference in aperture between the ALMA and IR data (i.e., $\lesssim$ 0\farcs6 and $\gtrsim$ 6\arcsec) needs to be considered under the assumption of a radial distribution of SF emission. 
Moreover, it is needed to estimate how much extended emission is resolved out in the ALMA data. 
Therefore, we do not adopt this method as the main approach to examine the relative strength of the SF and AGN components but present a brief discussion in
Section~\ref{sec_app:sed} of the appendix.

\begin{figure}
    \centering 
    \includegraphics[width=8.8cm]{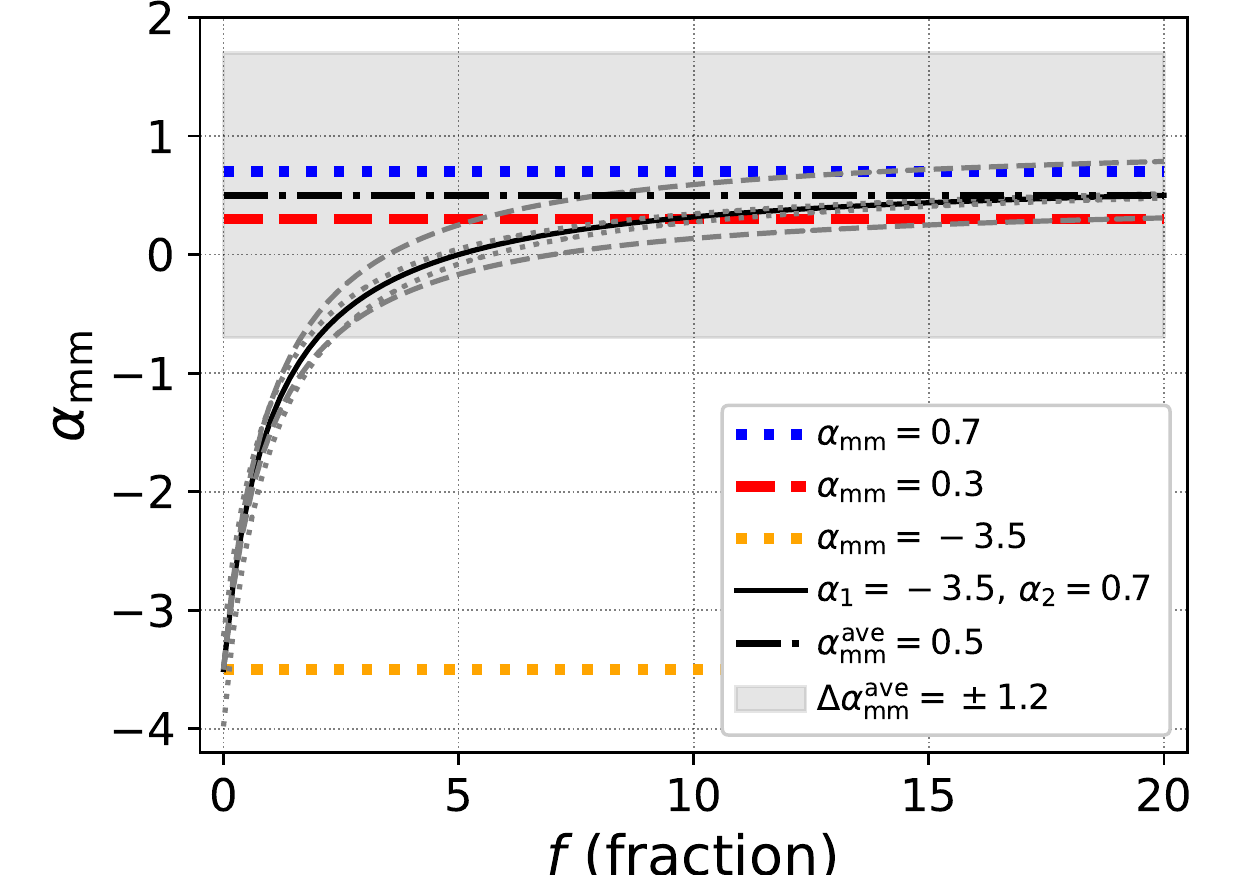}
    \caption{
    Spectral index expected from the sum of two power-law components as a function of the fraction between them ($f$).
    Specifically, the sum is represented as 
    $S_\nu \propto (\nu/\nu_0)^{-\alpha_1} + 
    f \times (\nu/\nu_0)^{-\alpha_2}
    $, where $\alpha_1 = -3.5$ (orange) and $\alpha_2 = 0.7$ (blue) are considered the fiducial values for SF and an AGN, respectively.
    The result with these parameters is shown by a black solid curve. 
    The red dashed line indicates $\alpha_{\rm mm} = 0.3$, suggesting that the emission from objects with indices above $\approx$ 0.3 would be dominated by the AGN emission with $\alpha_2 \sim$ 0.7 rather than the SF emission with $\alpha_1 = -3.5$. 
    Considering a possible range of $\alpha_1 = -3.60\pm0.38$ from $\beta_{\rm BB} = 1.60\pm0.38$ \citep{Cas12}, we also plot the corresponding lines by dotted gray lines. 
    In addition, the cases where the power-law indices of the AGN component are 0.5 and 1.0, expected from the SED models of IC 4329A and NGC 985 \citep{Ino18}, are also indicated by dashed gray lines. 
    For comparison, the black dot-dashed line and 
    gray shaded area represent the average of the observed values for our AGNs and their standard deviation, respectively. 
    }
    \label{fig:slope_vs_f}
\end{figure}

\section{Observational Evidence Supporting the Relation between Nuclear Mm-wave Emission and AGN Activity}\label{sec:cor4sigAGN}

\begin{figure}
    \centering 
    \includegraphics[width=7.2cm]{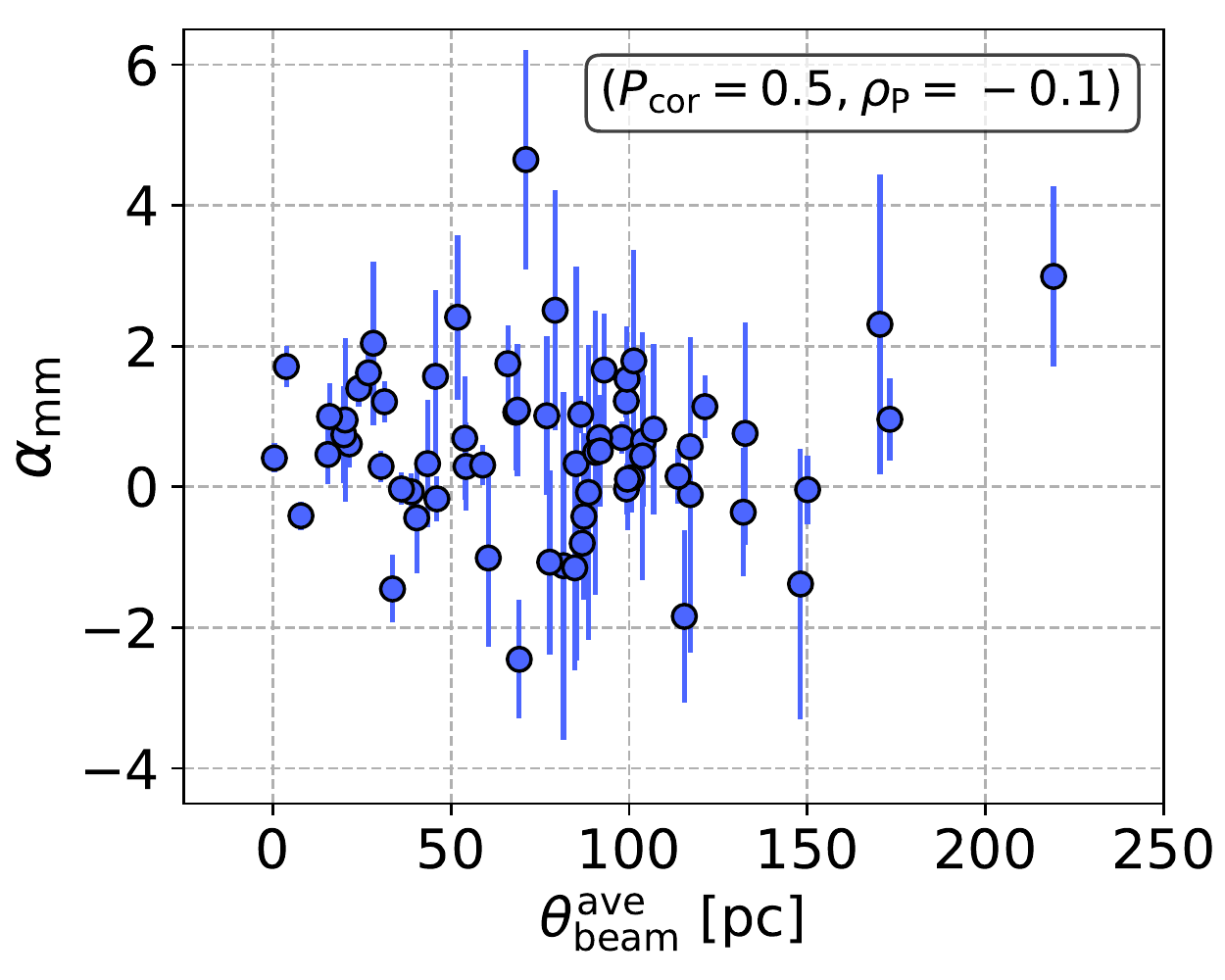}
    \caption{
    Scatter plot of the spectral index versus the 
    achieved beam size in units of pc. 
    The spectral index is not correlated with the beam size. 
    }
    \label{fig:freq2}
\end{figure}

Throughout this section, while considering the discussion on the SF contribution, we discuss whether the AGN emission dominates the observed mm-wave flux. Three approaches are adopted and are separately discussed in the following subsections. In the first approach, we assume that the 
host-galaxy component important in discussing its contribution is dust emission represented by $\alpha_{\rm mm} \sim -3.5$, which we have discussed in the previous section. 
Then, in the subsequent approaches, we assume that synchrotron and free-free emission is important. This assumption is complementary to the first assumption and would be important at the current stage where it cannot be completely ruled out that the dust emission could be weaker in the mm-wave band than the other synchrotron and free-free emission.

\subsection{Positive Spectral Index as an Indicator for AGN-dominant Objects}\label{sec:posindex}

Based on the first assumption that the dust, or modified black-body, emission from the host galaxy is the strongest SF component, 
we restrict a sample to AGNs whose mm-wave emission would have little contamination from SF. 
The modified black-body emission can be expressed approximately by a power law with an index of $\sim$ $-$3.5 in the mm-wave band. Such indices are quite different from those expected for synchrotron components of AGNs. For example, \cite{Ino18} found that synchrotron emission from an AGN can be characterized by a spectral index of $\sim 0.5$--$0.9$ (see their Figure~4). 
Due to the expected large difference, we can use the observed spectral index to  select objects whose mm-wave fluxes are dominated by synchrotron emission from AGNs. 
This kind of study was carried out in \cite{Eve20}, who
classified extragalactic objects with 95\,GHz, 150\,GHz, and 220GHz data from the South Pole Telescope. 
We note that our particular focus on the synchrotron emission is because thermal emission due to AGN-related dust is unlikely to be the origin of the mm-wave emission, as discussed later in Section~\ref{sec:agn_dust}.


Specifically, as the sum of the SF and AGN components, we consider $S_\nu \propto (\nu/\nu_0)^{-\alpha_1} + f \times (\nu/\nu_0)^{-\alpha_2}$ by introducing $f$ to represent their relative strength and assume $\alpha_1 = -3.5$ (SF) and $\alpha_2 = 0.7$ (AGN). 
Figure~\ref{fig:slope_vs_f} shows that the observed spectral index increases with the fraction of $f$, and that the emission with an index above $\approx$ 0.3 would be dominated by the AGN synchrotron emission (i.e., $f \sim 10$). This result does not strongly depend on the choice of $\alpha_1$ (SF) in a possible range between $-3$ and $-4$ (gray lines of Figure~\ref{fig:slope_vs_f}). This considers $\beta_{\rm BB} \sim 1.60\pm0.38$ (the index for modified black-body emission included as $S_\nu \propto \nu^{\beta_{\rm BB}} F_{\rm BB}$; Section~\ref{sec:sf}), found for nearby SF galaxies \citep{Cas12}. The choice of $\alpha_2$ (AGN) is motivated by the results of \cite{Ino18}. 
In contrast to $\alpha_1$, the assumption of $\alpha_2$ has a non-negligible impact. 
The figure shows two cases where $\alpha_2$ = 0.5 and $\alpha_2$ = 1.0, which we derive as the minimum and maximum values by simulating synchrotron-emission spectra in a range of the power-law index for an electron distribution, constrained for IC 4329A and NGC 985 \citep{Ino18}.
The result shows that the spectral index increases with the fraction more rapidly,
particularly for $\alpha_2 = 1.0$.

We note that a negative index does not always suggest modified black-body emission, as the optically thick part of synchrotron emission should have a spectral index of $\approx -5/2$. Indeed, such values were suggested for some nearby AGNs \citep{Ino18,Ino20}. Thus, by selecting sources based on their spectral index, we miss some AGNs whose emission is characterized by a negative index at the observed frequency but is dominated by an AGN component. 
However, to conservatively select only sources whose mm-wave emission does not have significant SF contamination, we consider the assumption of $\alpha_2 = 0.7$ reasonable. 

Based on Figure~\ref{fig:slope_vs_f}, we create a sample by selecting 19 AGNs with $\alpha_{\rm mm} - \alpha^{\rm e}_{\rm mm} > 0.3$ whose mm-wave emission is thus expected to be dominated by the AGN ($f > 10$) and examine the correlation between mm-wave and X-ray luminosities for the sample. 
Even if $\alpha_2 = 1.0$ is the case, only a slightly larger contribution of SF, indicated with $f \approx 5$, is expected. 
For the sample, we find a significant correlation between $\nu L^{\rm peak}_{\nu,{\rm mm}}$ and $L_{\rm 14-150}$, with $P_{\rm cor} \approx  3\times10^{-4}$. 
Interestingly, no significant difference in $\rho_{\rm P}$ is found between the restricted AGN sample ($\rho_{\rm P} = 0.76$) and the entire sample 
($\rho_{\rm P} = 0.74$; Figure~\ref{fig:var_lum1}). 
Therefore, the results obtained from the entire sample may also reflect a strong coupling between the AGN-related mm-wave and X-ray emission. 

Consistent results with the above argument can be obtained from a scatter plot of the spectral index and the spatial resolution achieved, as shown in Figure~\ref{fig:freq2}. 
If the contaminating light from the SF component gets stronger with increasing beam size, a negative trend may be seen but is not found.

\begin{figure}
    \centering
    \includegraphics[width=7cm]{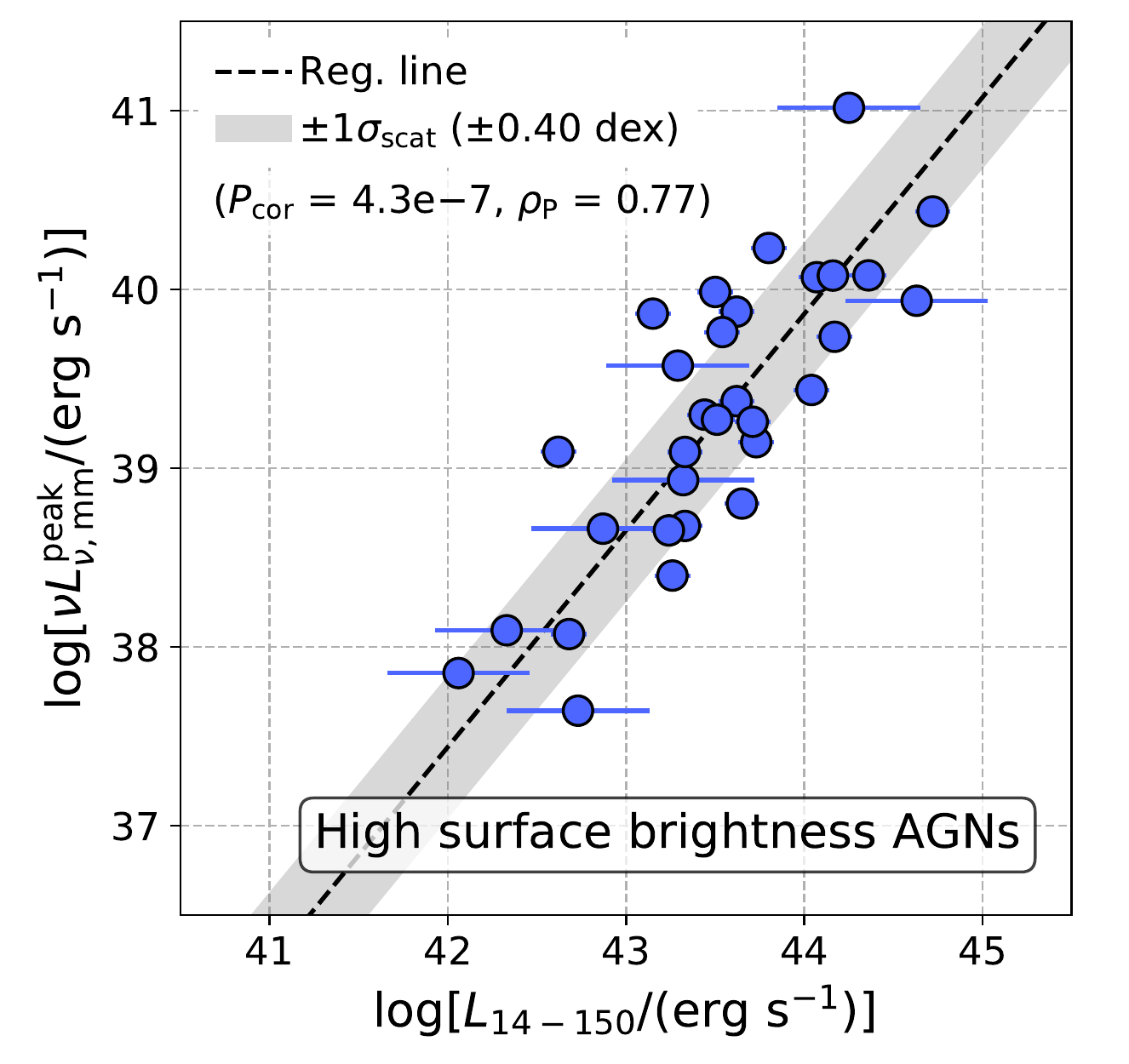}
    \caption{
    Luminosity correlation of the mm-wave and 14--150\,keV emission for 31 AGNs with high mm-wave surface brightnesses, exceeding an Eddington limit for the SF. A regression line is indicated by the black dashed line. 
    }
    \label{fig:sfrs} 
\end{figure}

\subsection{High Mm-wave Surface Brightness Objects}

In this and next subsections, we assume that among the SF-related components, the synchrotron and free-free components dominate the mm-wave emission. As a result, 
the observed mm-wave luminosities can be once converted to SFRs via Equation~\ref{eqn:sfr2lmm}. If an obtained SFR is larger than an appropriate value, we can infer the additional component from an AGN and by selecting such objects, we can assess correlations of possibly AGN-related mm-wave emission with X-ray emission. 
Under the above strategy, we first consider AGNs whose SFR surface densities ($\Sigma_{\rm SFR}$) based on mm-wave luminosities exceed an Eddington limit of the SF, above which its radiation-driven outflow blows out the surrounding gas, perhaps suppressing the SF. According to theoretical considerations \citep[e.g.,][]{Elm99,Tho05,You08} and observational results \cite[e.g.,][]{Soi00,Ima11highres}, the limit is expected to be $\sim 10^{13}$ $L_\odot/$kpc$^{2}$, corresponding to $\Sigma_{\rm SFR} \sim 2000\,M_\odot$ yr$^{-1}$ kpc$^{-2}$ for the FIR-to-SFR conversion factor of \cite{Ken98}. This conversion factor should be reasonable for such active SF regions, given that these regions would have a large amount of dust and emit IR photons by absorbing almost all of the UV/optical photons from SF. 
To find objects that exceed the limit and thus should have an important fraction of the AGN contribution, we estimate SFR per kpc$^2$ via Equation \ref{eqn:sfr2lmm} by ascribing the observed mm-wave luminosity solely to the SF. Here, the emitting area is set to the elliptical area of the ALMA beam. 
Consequently, 31 objects with mm-wave-based surface densities above the Eddington limit are identified. 
Although one might consider adopting the SFRs from the SED analysis of \cite{Ich19} and scaling them to beam sizes, these estimates would have large uncertainty, as described in Section~\ref{sec:sf} and Appendix~\ref{sec_app:sed}.
Figure~\ref{fig:sfrs} shows a correlation of the mm-wave and 14--150\,keV luminosities for the 31 AGNs. The correlation is significant and strong, as suggested from $P_{\rm cor} = 4.3\times10^{-7}$ and $\rho_{\rm P} = 0.77$. 
This result supports that AGN-related mm-wave emission contributes to forming the correlation between the mm-wave and X-ray emission. 

\begin{figure}
    \centering 
    \includegraphics[width=8.cm]{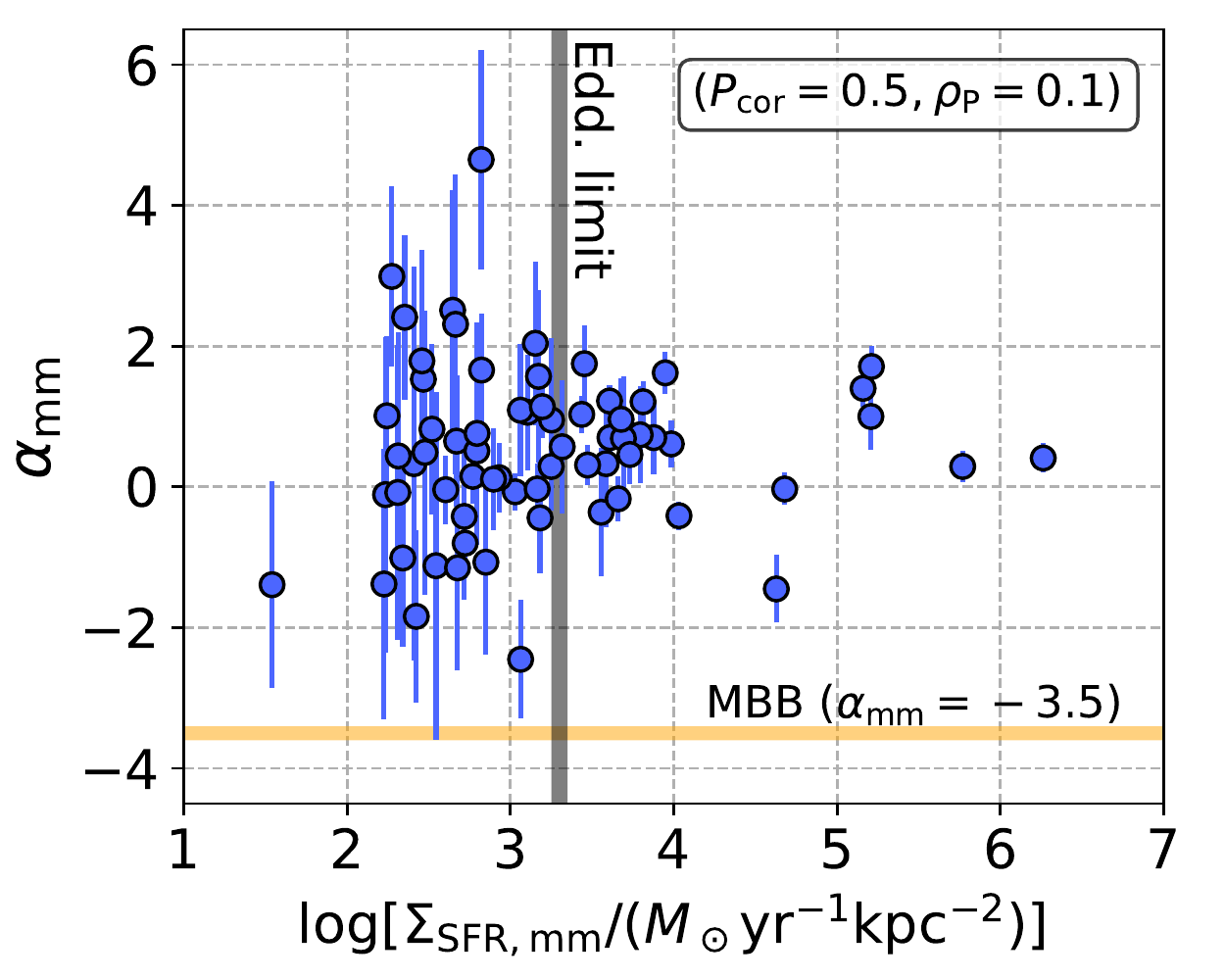}
    \caption{
    Spectral index versus SFR surface density, estimated by converting the observed mm-wave luminosity to the SFR. 
    An Eddington limit of 2000 $M_\odot$ yr$^{-1}$ kpc$^{-2}$ is indicated by the gray line, and the spectral index of $-3.5$, expected for thermal emission, is denoted by the yellow line. No correlation is found. 
    }
    \label{fig:sur_vs_index}
\end{figure}

Supplementarily, we examine a relation between $\alpha_{\rm mm}$ and $\Sigma_{\rm SFR,mm}$ to investigate whether the high surface densities are not due to the emission of heated dust. 
As shown in Figure~\ref{fig:sur_vs_index}, no negative trend is found which would have supported an increase in the contribution of heated dust emission. 
Recently, \cite{Per21} found for nearby ($z < 0.165$) ultraluminous infrared galaxies 
that their observed $\sim$ 220 GHz fluxes and the extrapolated ones from IR gray body components agree within a factor of two, suggesting significant thermal emission in the mm-wave band. This result is apparently discrepant with ours but would be just because our targets are less luminous (i.e., $\lesssim 10^{12}$ solar luminosities).

\subsection{AGN with Luminous Mm-wave Emission in Comparison with SFR}

\begin{figure}
    \centering
    \includegraphics[width=7.cm]{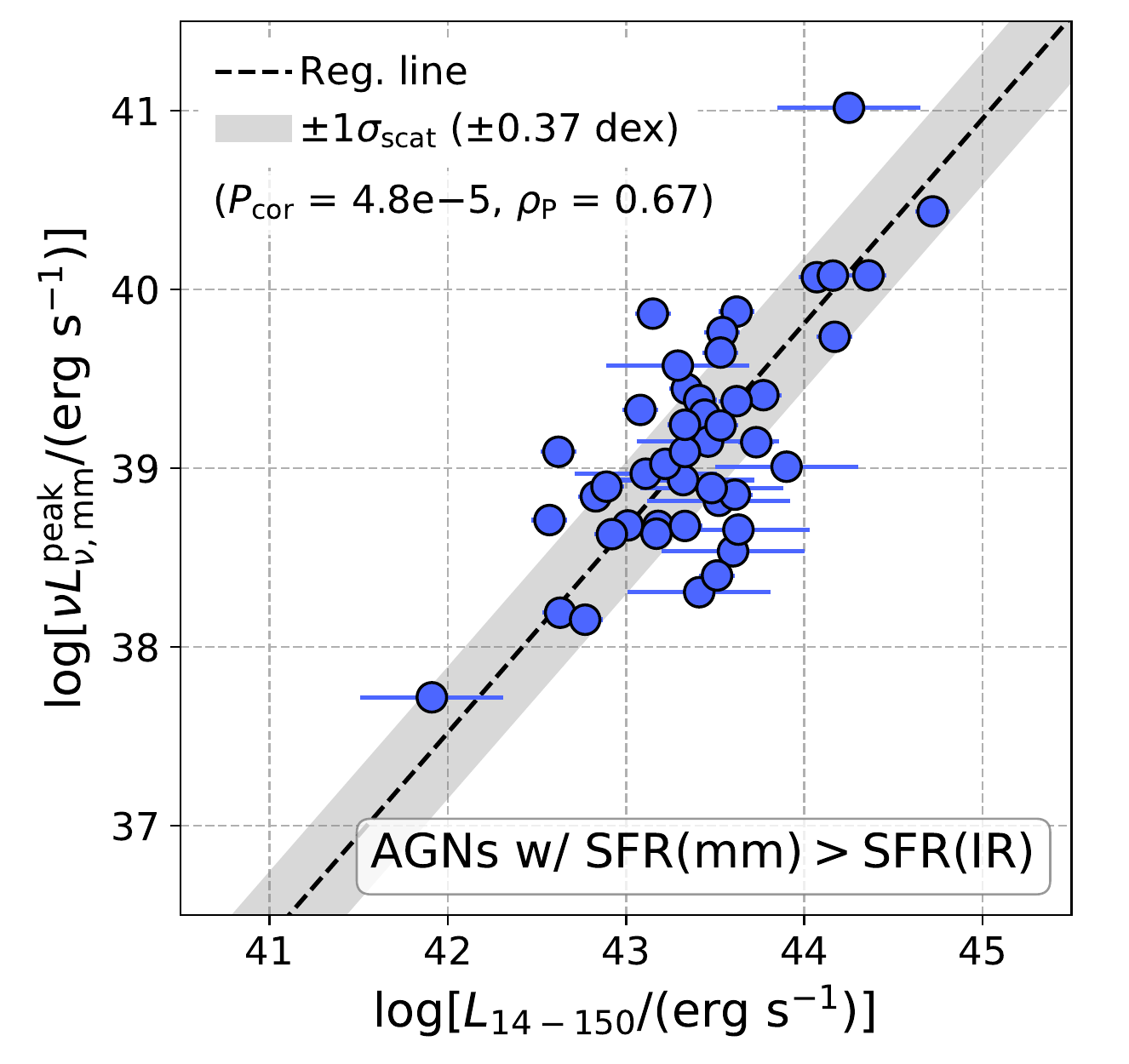}
    \caption{
    Luminosity correlation of the mm-wave and 14--150\,keV emission for AGNs with mm-wave-based SFRs being higher than those expected from the IR decomposition analysis.
    A regression line is indicated by the black dashed line. 
    }
    \label{fig:cor_for_high_sfrmm}
\end{figure}

We lastly test a correlation for a subsample of AGNs, selected by considering the SFR derived based on the observed mm-wave emission (Equation~\ref{eqn:sfr2lmm}) and that expected from the IR decomposition analysis \citep{Shi17,Ich19}. We find  mm-wave-based SFRs higher than the IR-based ones for $\sim 50$\% of our objects whose mm-wave emission thus could have a non-negligible AGN contribution. We emphasize that the identification is conservative given that the SFRs from the IR data were measured at resolutions of $> $ 6\arcsec\ larger than those of the ALMA data ($<$ 0\farcs6). 
For the subsample, a significant and moderately strong luminosity correlation is found with $P_{\rm cor} = 5\times10^{-5}$ and $\rho_{\rm P} = 0.67$ (Figure~\ref{fig:cor_for_high_sfrmm}). This result is consistent with the conclusion that has been drawn in this section. 

As a summary of the three subsections, in both assumptions (the dominant SF component is the thermal emission from dust and is the synchrotron plus free-free emission), we have found the significant correlations of mm-wave emission likely from an AGN and X-ray emission. 
Also, the correlation strengths, close to that derived for 
the entire sample, have been confirmed. 
This result could indicate that the correlation for the entire sample is also due to the AGN-related mm-wave emission. 

\section{Dust-extinction Free AGN Luminosity Measurement Using Mm-wave Emission}\label{sec:reg_sum}

Our results in Section~\ref{sec:cor4sigAGN} have suggested that 
the mm-wave emission from an AGN could form the correlation with AGN X-ray emission. 
Therefore, the nuclear mm-wave luminosity may be used as 
a proxy for the AGN luminosity.
A remarkable advantage of mm-wave emission is its high penetrating power, up to $N_{\rm H} \sim 10^{26}$ cm$^{-2}$ \citep{Hil83}. 
In Table~\ref{tab:cors4use}, we provide the relations (i.e., regression lines) of the AGN luminosities ($L_{14-150}$, $L_{2-10}$, $\lambda L^{\rm AGN}_{\lambda, 12\,\mu{\rm m}}$, and $L_{\rm bol}$,) with the mm-wave luminosity determined for our entire sample and also those for the clean sample of AGNs with high spectral indices ($\alpha_{\rm mm} - \alpha^{\rm e}_{\rm mm} > 0.3$; see Section~\ref{sec:cor4sigAGN}). 
Although the scatters for the whole and clean samples are almost the same of $\approx$ 0.3\,dex, the relations for the clean sample would be preferred for use in estimating the AGN luminosities. 

\begin{deluxetable}{cccccccc}
\tabletypesize{\footnotesize}
\tablecolumns{8}
\tablewidth{0pt}
\tablecaption{
Regression fits for estimating AGN luminosities from mm-wave luminosities\label{tab:cors4use}}
\tablehead{
\colhead{(1)} & 
 \colhead{(2)} & 
 \colhead{(3)} &  
 \colhead{(4)} \\
 \colhead{$\log Y$} & 
 \colhead{$\alpha$} & 
 \colhead{$\beta$} &  
 \colhead{$\sigma_{\rm scat}$} 
 } 
\startdata
\multicolumn{4}{c}{all AGNs} \\ \hline
 $\log[L_{\rm 14-150}/({\rm erg}\,{\rm s}^{-1})]$ & 
 $0.83$ & $10.79$ & $0.30$ \\ 
 $\log[L_{\rm 2-10}/({\rm erg}\,{\rm s}^{-1})]$ & 
 $0.93$ & $6.83$ & $0.45$ \\ 
 $\log[\lambda L^{\rm AGN}_{\lambda, \rm 12\mu m}/({\rm erg}\,{\rm s}^{-1})]$ & 
 $1.19$ & $- 3.36$ & $0.71$ \\ 
 $\log[L_{\rm bol}/({\rm erg}\,{\rm s}^{-1})]$ & 
 $1.16$ &  $-0.87$ & $0.51$ \\ \hline 
 \multicolumn{4}{c}{AGNs with $\alpha_{\rm mm}-\alpha^{\rm e}_{\rm mm} > 0.3$ } \\ \hline 
 $\log[L_{\rm 14-150}/({\rm erg}\,{\rm s}^{-1})]$ & 
 $0.99$ & $4.56$ & $0.35$ \\ 
 $\log[L_{\rm 2-10}/({\rm erg}\,{\rm s}^{-1})]$ & 
 $1.20$ & $-3.95$ & $0.47$ \\ 
 $\log[\lambda L^{\rm AGN}_{\lambda,\rm 12\mu m}/({\rm erg}\,{\rm s}^{-1})]$ & 
 $1.06$ & $1.58$ & $0.56$ \\ 
 $\log[L_{\rm bol}/({\rm erg}\,{\rm s}^{-1})]$ & 
 $1.60$ & $-18.63$ & $0.88$ \\ 
\enddata
\tablecomments{
(1,2,3) Parameters of a regression line represented as $\log Y = \alpha \times \log X + \beta$  where $X = \nu L^{\rm peak}_{\nu,{\rm mm}}$. 
(4) Intrinsic scatter. 
}
\end{deluxetable}

As a point to be noticed, the intrinsic scatters found for the correlations with the 14--150\,keV luminosity ($\approx$ 0.3\,dex) are comparable to those of relations between X-ray luminosity (e.g., 2--10\,keV and 14--195\,keV) and MIR luminosity (e.g., 9\,$\mu$m, 12\,$\mu$m, and 22\,$\mu$m) obtained in past studies of nearby AGNs \citep[$\sim$ 0.2--0.5\,dex; e.g., ][]{Gan09,Asm15,Ich12,Ich17}. Thus, the mm-wave relations have approximately the same reliability as the MIR ones. 
Note that if necessary, a 14--150\,keV luminosity can be converted to a 2--10\,keV one based on $L_{\rm 2-10}/L_{\rm 14-150} = 0.55$ where cut-off power-law emission with a typical photon index of 1.8 and a cut-off energy of 200\,keV is assumed \citep[e.g.,][]{Kaw16c,ric17c,Tor18,Bal20}.

Although we have presented only the correlations with the X-ray, MIR, and bolometric luminosities, it is also important to perform correlation analyses for the indicators of AGN luminosities in different wavelengths. Here, we focus on [O {\sc{iii}}]$\lambda$5007, 
[Si {\sc{vi}}]$\lambda$1.96, and [Si {\sc{x}}]$\lambda$1.43 lines \citep{Oh22,den22}. 
With the bootstrap method, we find significant luminosity correlations for the three lines with scatters of $\sim$ 0.8--0.9\,dex but find no significant correlations for their fluxes (Table~\ref{tab_app:corr}). 
The insignificant flux correlation for the [O {\sc{iii}}] line, despite the large sample size of 90, would indicate that the Malmquist Bias produces 
the luminosity correlation. 
We also comment that the [O {\sc{iii}}] fluxes were measured with a range of apertures \citep[1\arcsec--1\farcs6;][]{Oh22} and may be contaminated by a  heterogeneous amount of host-galaxy light. In fact, optical emission line diagnostics by \cite{Oh22} with the [N\,{\sc ii}]/H$\alpha$ and [O\,{\sc iii}]/H$\beta$ found that some BAT-selected AGNs are  located in non-Seyfert regions.
Thus, by extracting AGN-dominant [O {\sc{iii}}] emission, one might find a tighter luminosity correlation and also a significant flux correlation. 
As for the silicate lines, a further study with a larger sample of AGNs with significant line detections is desired to conclude whether they correlate with the mm-wave emission, given that we only use 17 AGNs and the Silicate lines are not detected for roughly half of them. 

\section{Physical Origin of AGN Mm-wave Emission}\label{sec:agn_mec}

\begin{figure*}
    \centering 
    \includegraphics[width=5.8cm]{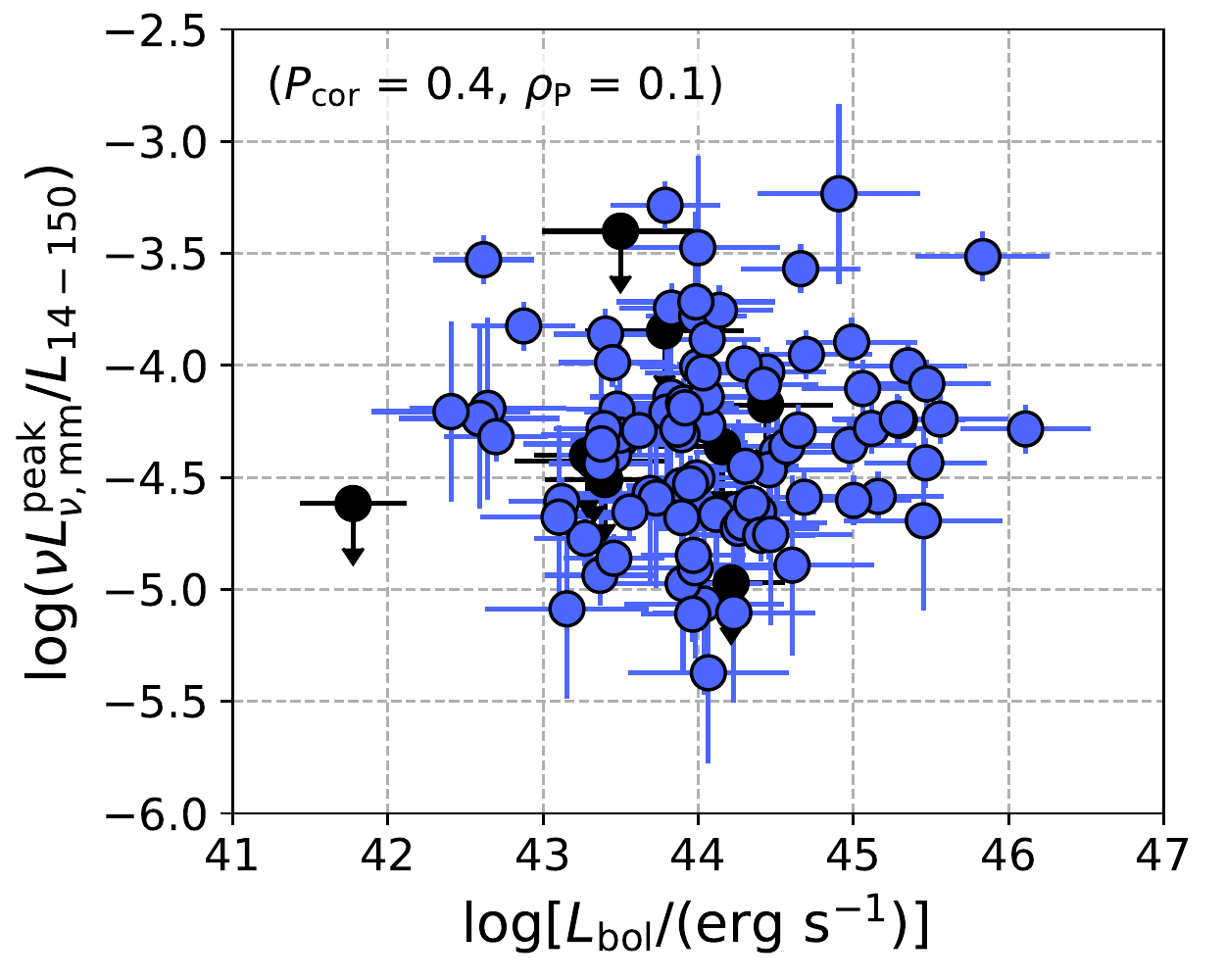} \hspace{-.cm}
    \includegraphics[width=5.8cm]{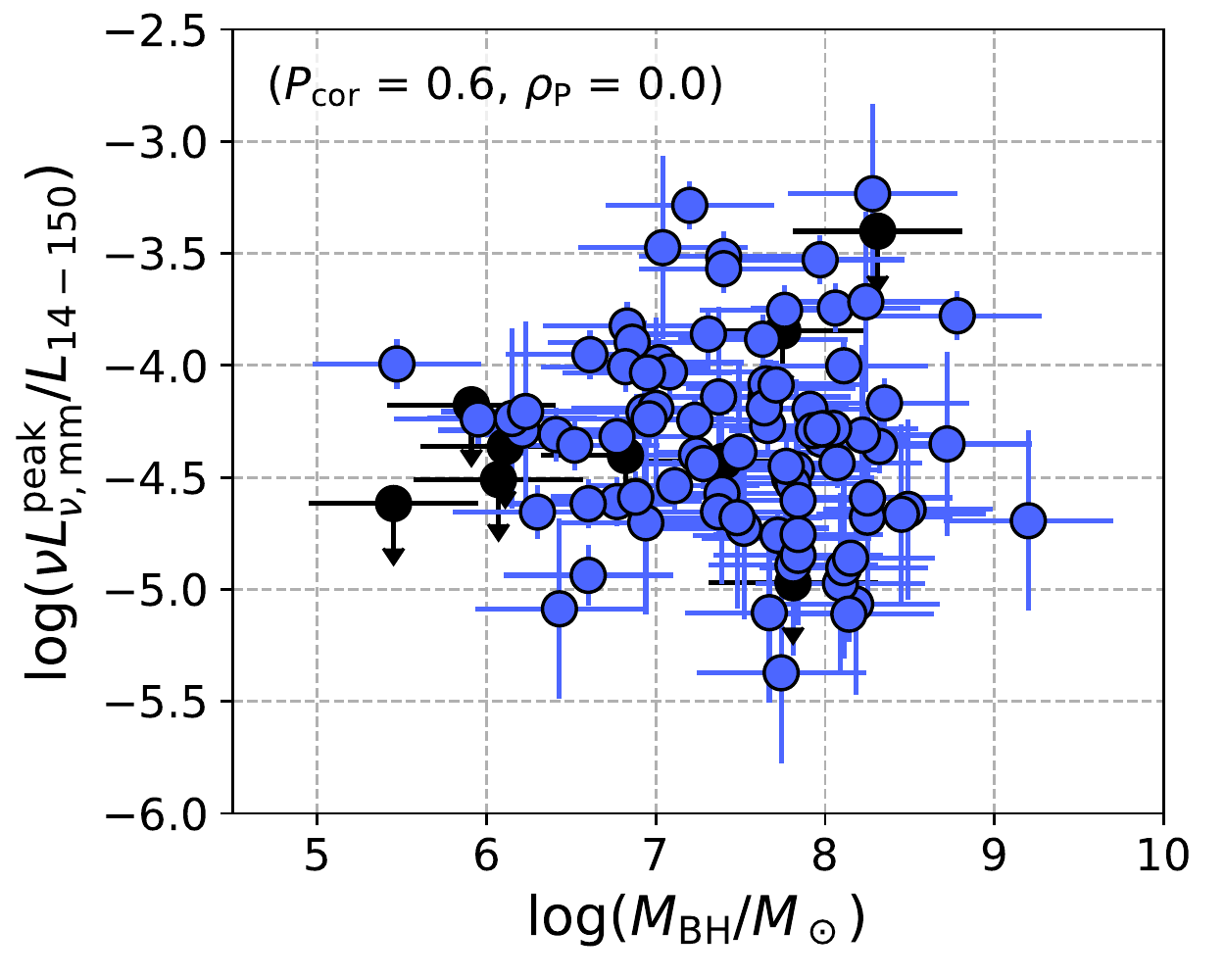}  \hspace{-.cm}
    \includegraphics[width=5.8cm]{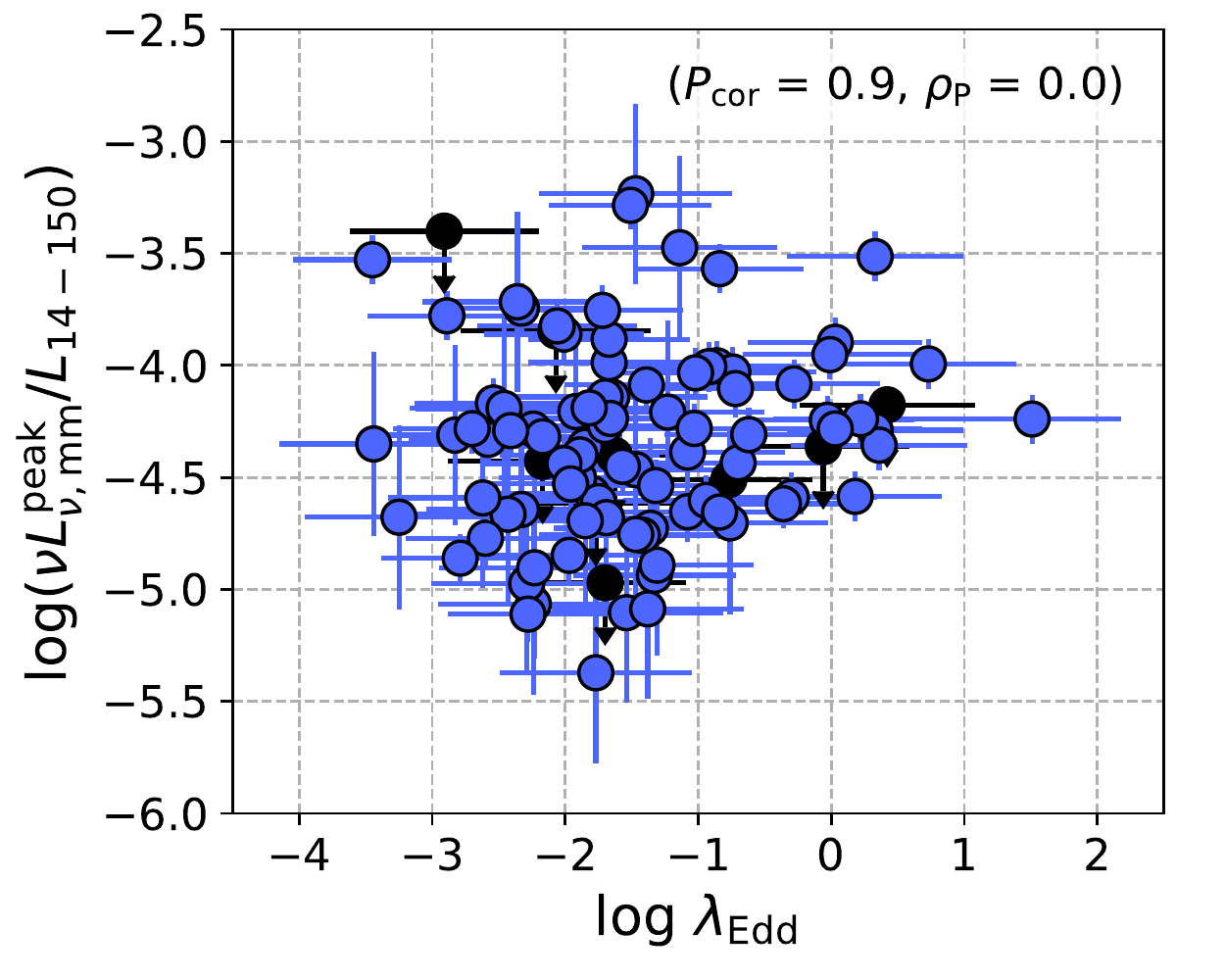}   
    \caption{
    Scatter plots of the $\nu L^{\rm peak}_{\nu,{\rm mm}}/L_{\rm 14-150}$ ratio versus the bolometric luminosity, the black hole mass, and the Eddington ratio. 
    AGNs with upper limits are shown as black circles. 
    No correlations are found. 
    }
    \label{fig:ratio_vs_agnpar}
\end{figure*}

We aim to identify the AGN mechanism responsible for the observed correlation between the mm-wave and 14--150\,keV luminosities.
Before proceeding to detailed scenarios, it is important to confirm whether the $\nu L^{\rm peak}_{\nu,{\rm mm}}/L_{\rm 14-150}$ ratio depends on the fundamental AGN parameters of $L_{\rm bol}$, $M_{\rm BH}$, and $\lambda_{\rm Edd}$, which could provide clues about the origin of the nuclear mm-wave emission.
Figure~\ref{fig:ratio_vs_agnpar} shows scatter plots of the $\nu L^{\rm peak}_{\nu,{\rm mm}}/L_{\rm 14-150}$ ratio for the three parameters. We find $p$-values for the three parameters to be 
0.4, 0.6, and 0.9, respectively, suggesting that 
the ratio is not strongly affected by the AGN parameters within the investigated ranges.

In the following, we discuss four AGN mechanisms: (1) thermal emission from dust heated by an AGN, 
(2) synchrotron emission originating around an X-ray corona, 
(3) outflow-driven emission, and 
(4) jet emission \citep[e.g.,][]{Mul13,Zak14,Beh15,Beh18,Ino18}.

\subsection{Emission from Dust Heated by an AGN}\label{sec:agn_dust}

We discuss whether the thermal emission from AGN-heated dust can account for a significant fraction of the observed mm-wave emission. The mm-wave flux of the dust component ($S^{\rm AGN(ext)}_{\nu,\rm mm}$) can be estimated by extrapolating the model for an AGN dusty torus fitted to an IR SED in \cite{Ich19}. 
Here, $\beta_{\rm BB}$ = 1.5 for $S_\nu \propto \nu^{\beta_{\rm BB}} F_{\rm BB}$ is adopted for the extrapolation. 
The index value was used in \cite{Ich19} \citep[see also][]{Mul11} and is supported by other studies \cite[e.g.,][]{Xu20}.  
For example, the SEDs in Figure~\ref{fig:sed} show a trend that the observed mm-wave emission is stronger than expected from the AGN-heated dust emission. To confirm whether this is generally seen or not, the flux ratio between the observed mm-wave emission and the thermal AGN emission is calculated, and 
the result is summarized as a histogram in the bottom panel of Figure~\ref{fig:ratios}. 
The histogram has a peak around 
$\log[S^{\rm peak}_{\nu,\rm mm}/S^{\rm AGN(ext)}_{\nu,\rm mm}]$ 
= 1--1.5 and indicates that the mm-wave emission is generally much stronger than the dust emission for our AGNs. 
This is also supported by the result that the observed spectral slopes ($\alpha^{\rm ave}_{\rm mm} = 0.5\pm1.2$)  are inconsistent with that expected for the dust emission (see the top panel of Figure~\ref{fig:freq1}). 
Thus, the AGN dust emission does not seem to be a dominant mm-wave source. 


\subsection{Relativistic Particles around an X-ray Corona}\label{sec:x-ray_ori}

The tight correlations we have found for the mm-wave and X-ray luminosities (14--150\,keV and 2--10\,keV) suggest that 
these emission may be energetically coupled, and perhaps the mm-wave emission could originate around and/or from where the X-ray corona forms.
According to a theoretical discussion of \cite{Lao08}, 
mm-wave emission can be produced by relativistic particles moving along magnetic field lines (i.e., synchrotron radiation), and 
observed emission of $\nu L^{\rm peak}_{\nu,{\rm mm}} \sim 10^{39}$ erg s$^{-1}$ can be reproduced only by considering a region on a scale of $10^{-4}$--$10^{-3}$ pc. This scale is consistent with the observed sizes of the X-ray coronae of SMBHs \cite[e.g.,][]{Mor08,Mor12}. 
Also, \cite{Ino14} similarly predicted the spectra of synchrotron radiation from relativistic electrons, and later in \cite{Ino18}, they showed, using ALMA data, that the synchrotron peak due to synchrotron self-absorption appears in the mm-wave band for nearby AGNs. 

\begin{figure} 
    \centering 
    \includegraphics[width=7.3cm]{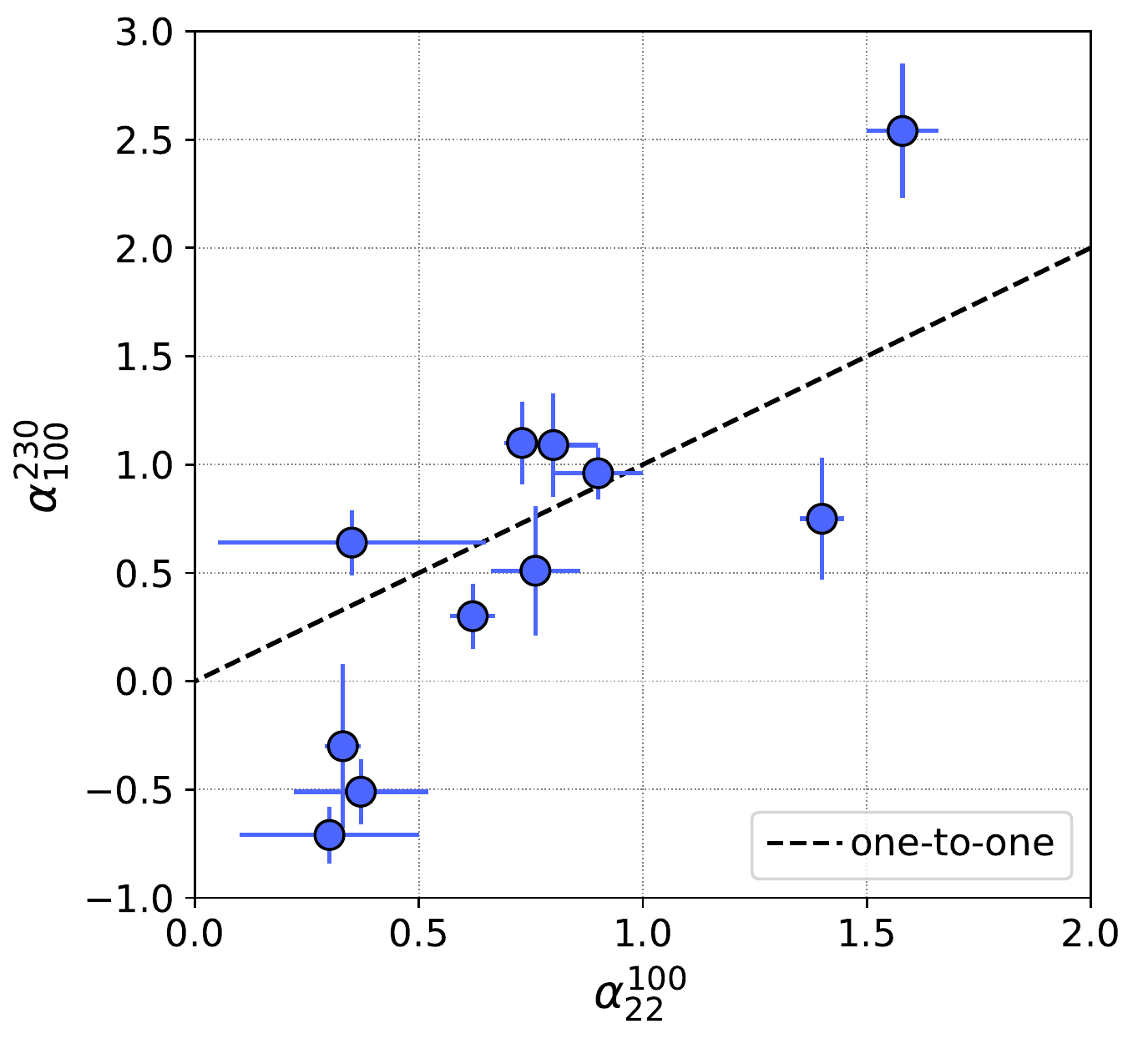}
    \vspace{-.5cm} 
    \caption{
    Scatter plot of the index derived from $\sim$ 230\,GHz and 100\,GHz data versus that from 100\,GHz and 22\,GHz data. 
    The dashed line indicates the one-to-one relationship. 
    }
    \label{fig:indices2}
\end{figure}

A supporting result for the presence of the synchrotron absorption peak can be obtained by comparing $\alpha^{230}_{100}$ (Section~\ref{sec:alma_data}) with an index between 22\,GHz and 100\,GHz ($\alpha^{100}_{\rm 22}$). 
Here, we use 100\,GHz peak fluxes from \cite{Beh18} at resolutions of $\sim$ 1\arcsec--2\arcsec\
and 22\,GHz fluxes measured within 1\arcsec\ aperture from \cite{smi20}. 
Figure~\ref{fig:indices2} plots $\alpha^{230}_{100}$ versus $\alpha^{100}_{22}$ for AGNs for which both values are calculated, and shows that three objects have $\alpha^{230}_{100} < \alpha^{100}_{22}$ and $\alpha^{230}_{100} < 0$, which are expected if there is a strong self-absorbed synchrotron component, as found for some AGNs \citep{Ino18,Ino20}. 

We note that while the Lorentz factor of an X-ray corona is $\approx 1$, considering that a typical range of electron temperature is $\sim$ 50\,keV \citep[e.g.,][]{ric17c,Tor18,Bal20}, the mm-wave synchrotron emission would be emitted from electrons with higher Lorentz factors \citep[$\gg$ 1; e.g.,][]{Ino18}. Therefore, an energy downgrade from $\gamma > 1$ to $\gamma \sim 1$ is needed if an electron emits X-ray and synchrotron emission \citep[see detailed discussion in][]{Lao08}.

\cite{Beh18} examined the relation between 100\,GHz and X-ray emission for 
26 BAT-selected AGNs using CARMA \cite[see also][]{Beh15,Pan19}, but did not find a significant trend. Nevertheless, we have succeeded in finding mm-wave correlations.
Our success is partly due to our sample size being more than $\sim$ three times larger than the previous sample.
Additionally, the sub-arcsec resolutions of our data, more than a few times better than in the previous work ($\sim$ 1\arcsec--2\arcsec), should help us to find the significant correlations by reducing the contamination from host-galaxy emission. 
Lastly, we comment that our choice of the 200--300\,GHz band could be a better option than the previously used lower frequencies ($\sim$ 100\,GHz).  
According to the discussion based on AGN SEDs \citep{Beh15,Ino18,Ino20}, 
the mm-wave excess, expected to be related to the X-ray emission, might typically become more prominent at higher frequencies.
For example, \cite{Ino20} reported that the nuclear 100\,GHz emission of NGC~1068 is dominated by free-free emission \citep{Gal04}.
If this is true at 100\,GHz for a non-negligible fraction of AGNs, it may be difficult to find a tight correlation between the nuclear X-ray and mm-wave emission. 

We caution here that it is unclear whether the correlation we find can hold for lower-luminosity or lower-Eddington-ratio AGNs (e.g.,
$L_{\rm 2-10} < 10^{42}$ erg s$^{-1}$ or $\lambda_{\rm Edd} < 10^{-3}$), not well sampled by our work. 
In fact, \cite{Beh15} found that while AGNs with $L_{\rm 2-10} \gtrsim 10^{42}$ erg s$^{-1}$ roughly follow a relation of $\nu L^{\rm peak}_{\nu,{\rm mm}}/L_{\rm 2-10} = 10^{-4}$, AGNs with $L_{\rm 2-10} < 10^{42}$ erg s$^{-1}$ and $\lambda_{\rm Edd} < 10^{-3}$, taken from \cite{Doi11}, tend to show relatively stronger mm-wave emission. 
Thus, AGNs with low accretion rates may have different nuclear structures \citep[e.g., ][]{Doi05,Ho08}. According to the model of the hot accretion flow of \cite{Yua14}, expected to apply to low-Eddington-ratio AGNs, the X-ray luminosity produced by Compton scattering decreases more rapidly than the mm-wave luminosity, due to less frequent Compton scattering \cite[see also][]{Mah97}. 
This prediction is qualitatively consistent with the finding of \cite{Beh15}. 
However, as the mm-wave fluxes of the low-luminosity AGNs were measured at coarser resolutions of $\sim$ 7\arcsec\ in \cite{Doi11}, observations of low-activity AGNs at high spatial resolutions are crucial to understand this better. 

In the following subsections, we check three suggestions relating X-ray and mm-wave emission made by \cite{Shi17}, \cite{Che20}, and \cite{Pes21}.  

\subsubsection{FIR Excess and Mm-wave Emission}

\cite{Shi16} studied FIR (70--500\,$\mu$m) emission of BAT-selected nearby AGNs ($z < 0.05$) and found 500\,$\mu$m ($\sim$ 600\,GHz) emission that exceeds a modified black-body component fitted to photometry data at shorter wavelengths (160\,$\mu$m, 250\,$\mu$m, and 350\,$\mu$m). 
The excesses were quantified as $E500 = (F_{\rm obs}-F_{\rm MBB})/F_{\rm MBB}$, where $F_{\rm obs}$ and $F_{\rm MBB}$ are observed and modified-black-body-model fluxes at 500\,$\mu$m, respectively, and were found to increase with 14--195\,keV luminosity. Accordingly, it was speculated that the FIR emission is associated with the AGN activity. 
Such excesses were also found at 100\,GHz by \cite{Beh15}, and the authors  interpreted these excesses to originate around an X-ray corona based on the fact  that the X-ray-to-100\,GHz luminosity ratio is close to that found for stellar coronae. 
Considering the possible connection between the FIR and 100\,GHz excesses and the interpretation of \cite{Beh15}, \cite{Shi16} consequently suggested that the FIR excess emission could also be from a region around the X-ray corona. 

\begin{figure}
    \centering 
    \includegraphics[width=8.3cm]{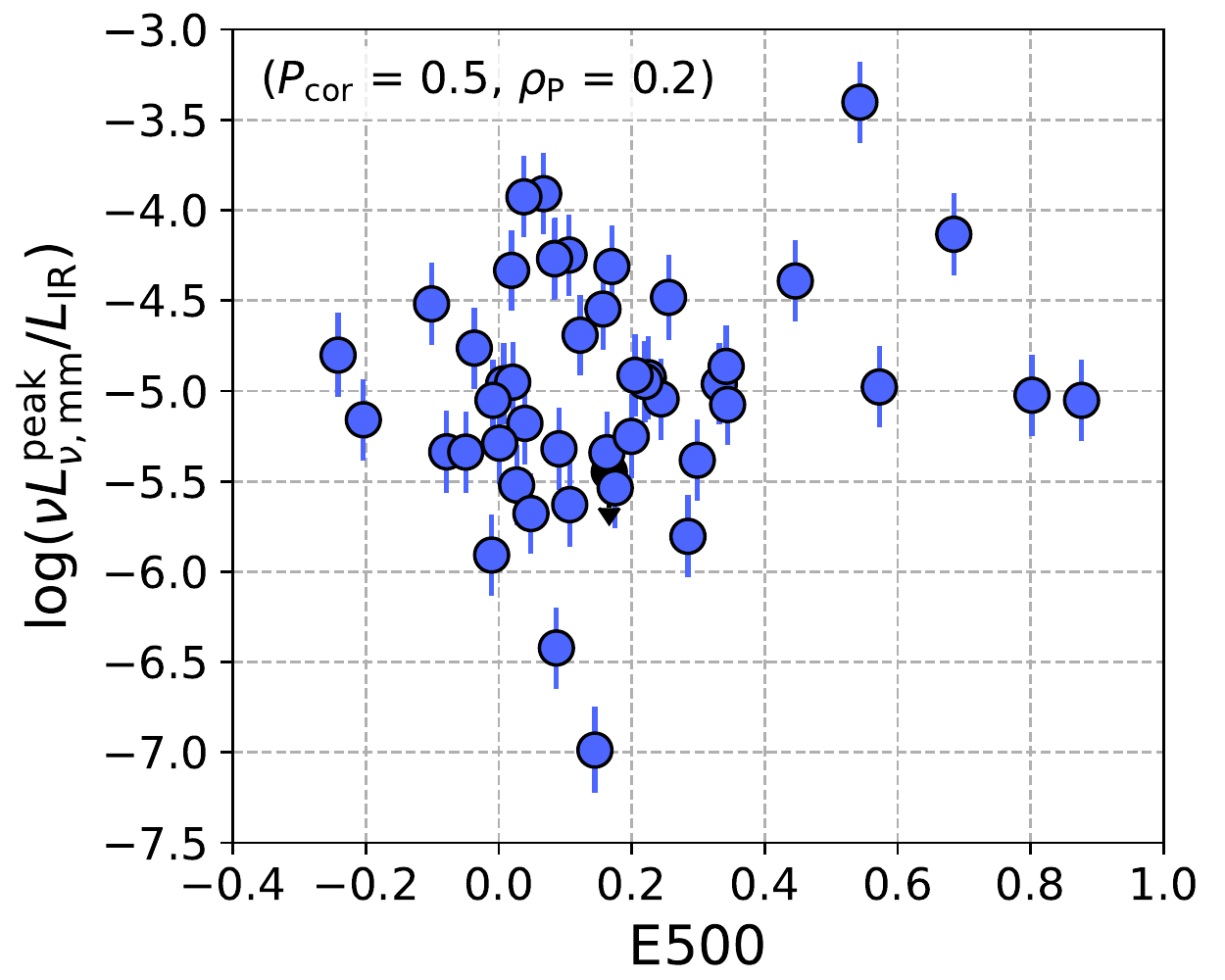}
    \caption{
    Scatter plot of the ratio between the mm-wave and host-galaxy IR luminosities versus the excess 500\,$\mu$m emission, whose values are taken from \cite{Shi17}. 
    AGNs with upper limits are shown as a black circle. 
    }
    \label{fig:e500}
\end{figure}

To test the 100\,GHz-to-FIR connection, a basis of the suggestion by \cite{Shi16}, we assess a relation between $\nu L^{\rm peak}_{\nu,{\rm mm}}/L_{\rm IR}$ and E500. 
Here, $L_{\rm IR}$ is IR (8--1000\,$\mu$m) luminosity of the host-galaxy SED model, and as the denominator of E500 (i.e., the modified black-body component) basically traces emission from the IR emission of a host galaxy, or SF regions \citep{Shi17}, we divide the mm-wave  luminosity by $L_{\rm IR}$ for a fair comparison with E500. 
A scatter plot of the two quantities is shown in Figure~\ref{fig:e500}, and no significant correlation is found for them ($P_{\rm cor} >$ 0.1). This result may be explained if extended emission missed by the high-resolution interferometer observation contributes to the FIR excess. This extended AGN emission could be related to a large-scale jet and/or  galaxy-scale AGN outflow.

\begin{figure}
    \centering
    \includegraphics[width=8cm]{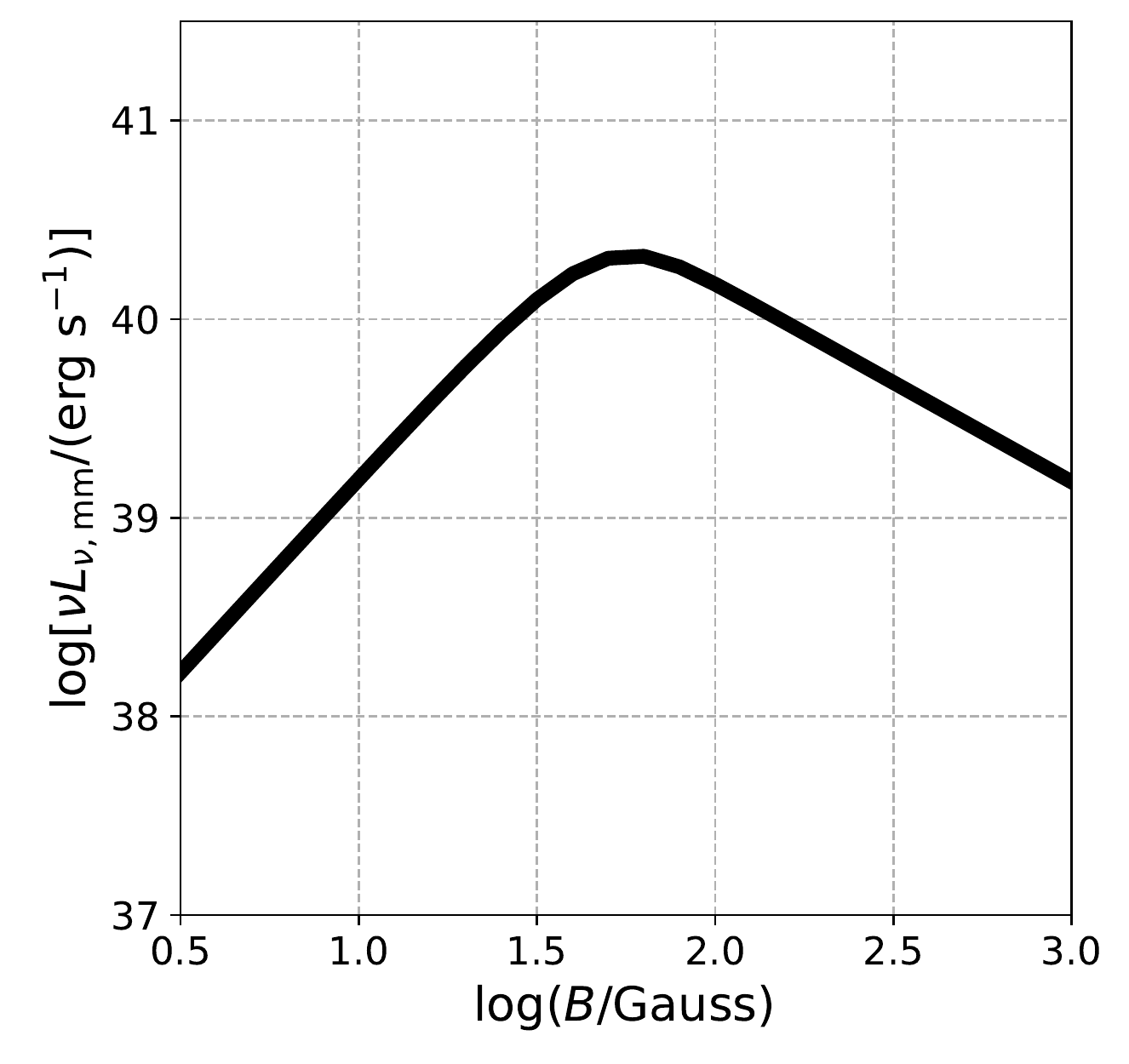}
    \hspace{1.2cm}
    \includegraphics[width=8cm]{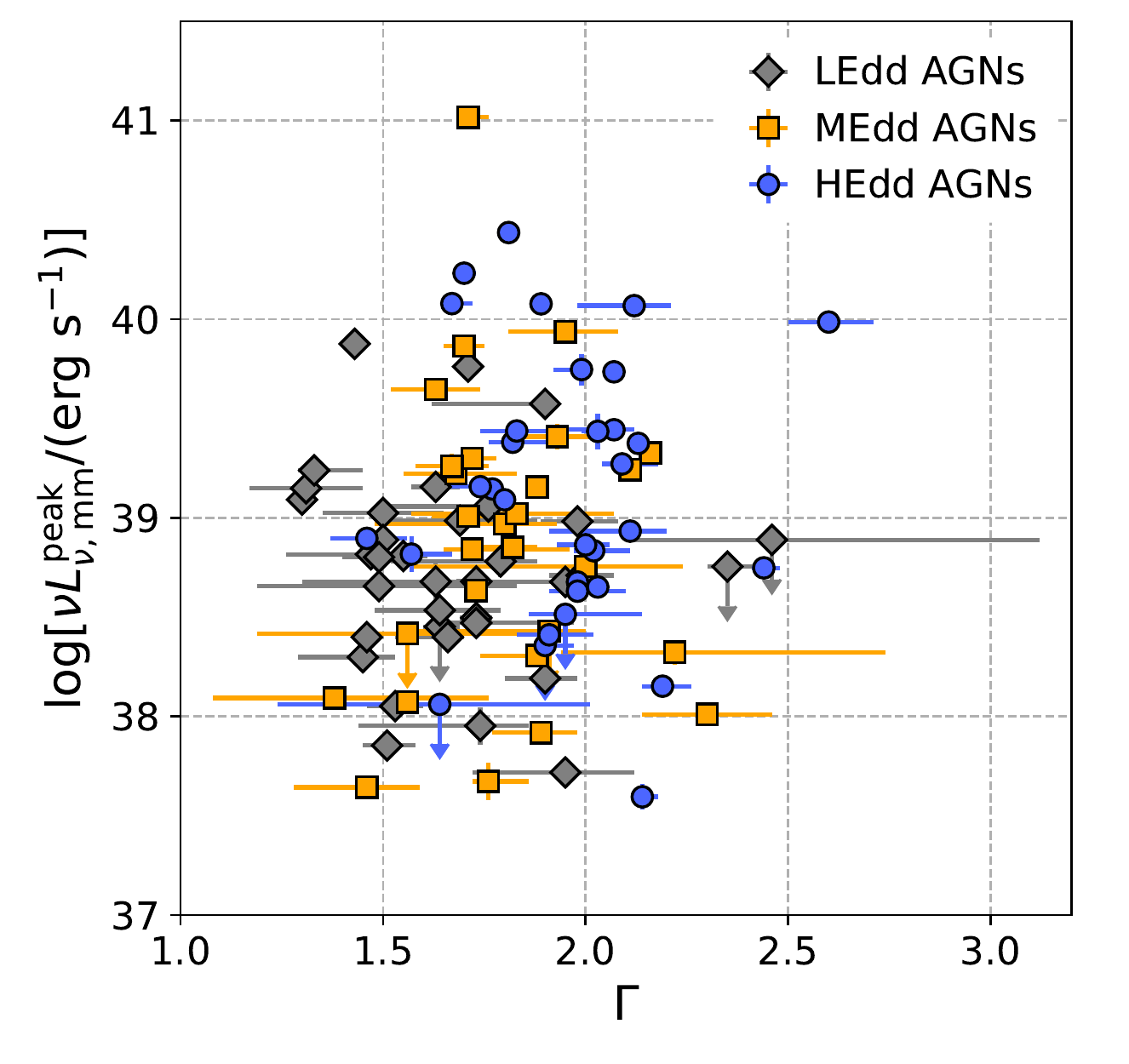} 
    \caption{
    Top: Theoretically expected mm-wave luminosity as a function of magnetic field strength. 
    Bottom: Scatter plot of the mm-wave luminosity versus the X-ray photon index for three subsamples divided by Eddington-ratio bins 
    of $\log \lambda_{\rm Edd} \geq -1.30$ (HEdd), $-1.30 > \log \lambda_{\rm Edd} \geq -1.85$ (MEdd) and $\log \lambda_{\rm Edd} < -1.85$ (LEdd). 
    No correlation is found for any of the subsamples. 
    }
    \label{fig:mm_vs_xgamma}
\end{figure}

\subsubsection{X-ray Magnetic-reconnection Model and Mm-wave Synchrotron Emission}

We examine a recent suggestion by \cite{Che20}, who created optical-to-X-ray spectral AGN models while considering magnetic reconnection as the heating source for an X-ray corona. 
According to \cite{Che20}, with increasing magnetic field strength ($B$), more energy can be transported into the X-ray corona. 
Thus, this predicts a harder spectrum, or a lower X-ray spectral index ($\Gamma$), with $B$.
The magnetic field is also a key parameter for synchrotron emission. By following \cite{Ino14}, the mm-wave luminosity can be calculated as a function of $B$, as shown in the top panel of Figure~\ref{fig:mm_vs_xgamma}. Here, the other parameters necessary to calculate the mm-wave luminosity are set to those obtained from a radio-to-mm-wave SED of IC 4329A \citep{Ino18}. 
If we consider a magnetic-field strength around 10\,G, as suggested for IC 4329A \citep{Ino18}, 
the mm-wave luminosity increases with $B$ in a range of $\sim$ 1--50\,G. 
Thus, these X-ray and mm-wave models \citep{Che20,Ino14} predict that the mm-wave emission becomes stronger with decreasing X-ray photon index. Motivated by this, we assess the correlation between the mm-wave luminosity and the X-ray photon index (Figure~\ref{fig:mm_vs_xgamma}). 
As the model of \cite{Che20} strongly depends on the Eddington ratio, to reduce the dependence on $\lambda_{\rm Edd}$, we divide the sample into three Eddington-ratio bins of $\log \lambda_{\rm Edd} \geq -1.30$, $-1.30 > \log \lambda_{\rm Edd} \geq -1.85$, and $\log \lambda_{\rm Edd} < -1.85$ that we label as HEdd, MEdd, and LEdd, respectively.
The boundaries are determined so that the HEdd, MEdd, and LEdd subsamples have even sizes of 33, 30, and 35, respectively. 
For any of the three bins, no significant correlation is found ($P_{\rm cor} > 0.1$). 
This result may suggest that the actual values of $B$ cover a wider range beyond 1--50\,G, and therefore no correlations are found. Otherwise, both or one of the X-ray and mm-wave models need to be revised. 

\subsubsection{Comparison with Hot Accretion Model}

\begin{figure}
    \centering
    \includegraphics[width=8.3cm]{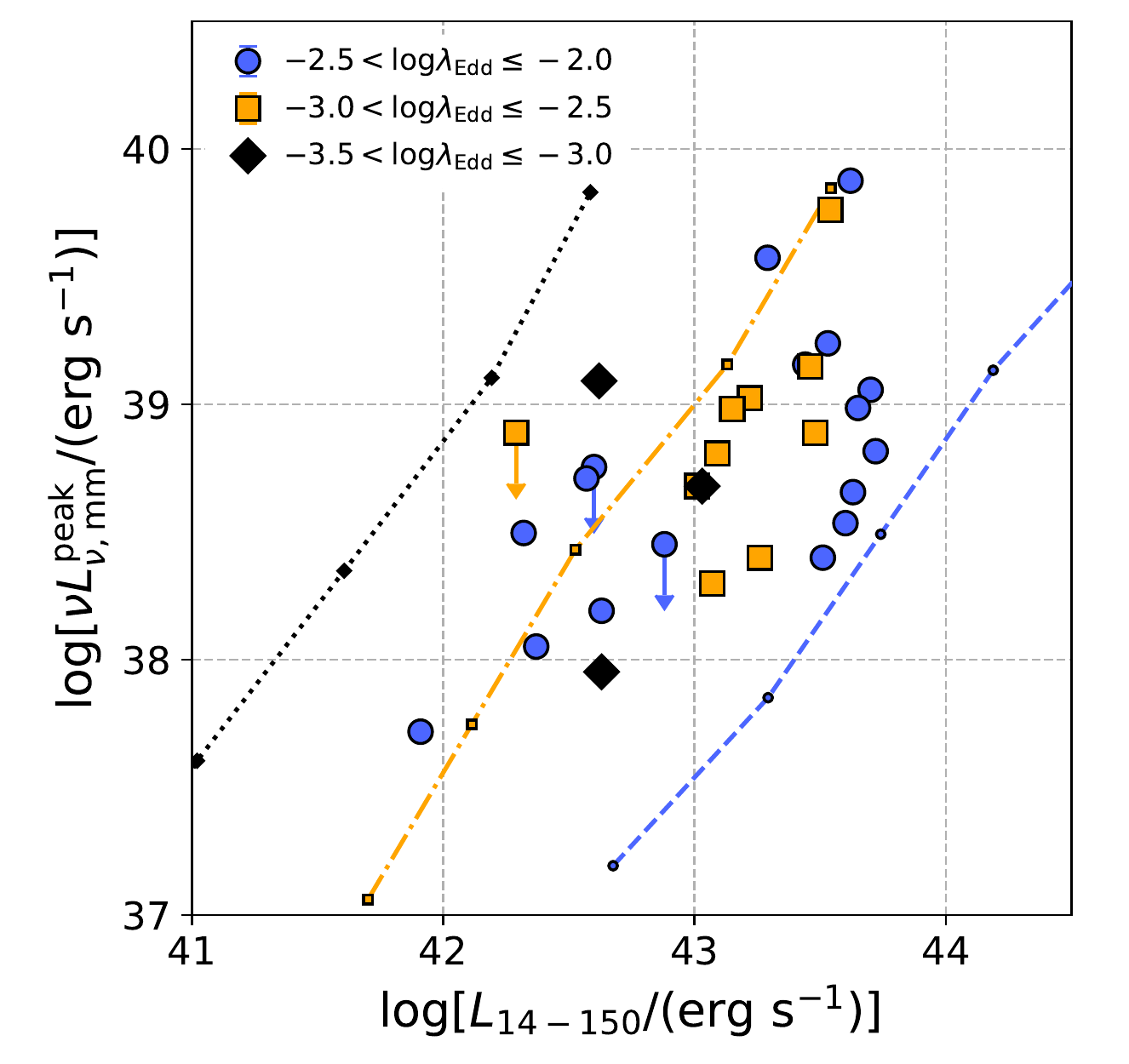}
    \caption{
    Mm-wave and X-ray luminosities of the sources with $\log \lambda_{\rm Edd}< -2$ and the predictions of the ADAF model of \cite{Pes21}. 
    Black points connected by a dotted line represent the cases with $\log \lambda_{\rm Edd} = -3.25$ for different black hole masses of $\log (M_{\rm BH}/M_\odot) = 6.5, 7.5, 8.5$, and 9.5 (from lower to higher luminosities). 
    In the same way, the orange and blue points accompanied by lines indicate the cases of $\log \lambda_{\rm Edd} = -2.75$ and $-2.25$, respectively. 
    }
    \label{fig:adaf}
\end{figure}

Our sample includes AGNs with low Eddington ratios of $\log \lambda_{\rm Edd} < -2$, and a hot accretion flow is expected to form around their SMBHs.  
Such an accretion flow is known to have a broadband spectrum covering the X-ray and mm-wave bands, and we examine whether a model of the advection-dominated accretion flow (ADAF) can reproduce the observed X-ray and mm-wave luminosities of the low-Eddington-ratio sources. 
We use the model formulated by \cite{Pes21}\footnote{A Python code that calculates ADAF spectra is available in  https://github.com/dpesce/LLAGNSED}, who followed the formalism described in \cite{Mah97}. The model has 
$\lambda_{\rm Edd}$ and $M_{\rm BH}$ as free parameters, and is valid in the range $\log \lambda_{\rm Edd} < -2$. The spectrum consists of three components: synchrotron emission, inverse Compton scattered emission, and free-free emission. Seed photons for the Compton scattering are provided by the synchrotron process. 
Mm-wave and 14--150\,keV luminosities are then calculated based on the model over 
$\log (M_{\rm BH}/M_\odot) = 6.5$--9.5 and $\log \lambda_{\rm Edd} = -3.25 \sim -2.25$, covering most of the masses and Eddington ratios of the sources with $\log \lambda_{\rm Edd} < -2$. 
Figure~\ref{fig:adaf} compares the predictions of the model and the observed data. 
To clarify the dependence on the Eddington ratio, we divide the sample into three bins. 
The ADAF model shows its great potential to reproduce the luminosities in both bands, but 
it appears that the dependence of the model on the Eddington ratio cannot be clearly seen in the observed data. This result could infer that
the accretion flow may not depend as strongly on the Eddington ratio as expected from the model.

\subsection{Outflow-driven Shock}\label{sec:of_ori}

\begin{figure}
    \centering
    \includegraphics[width=8.3cm]{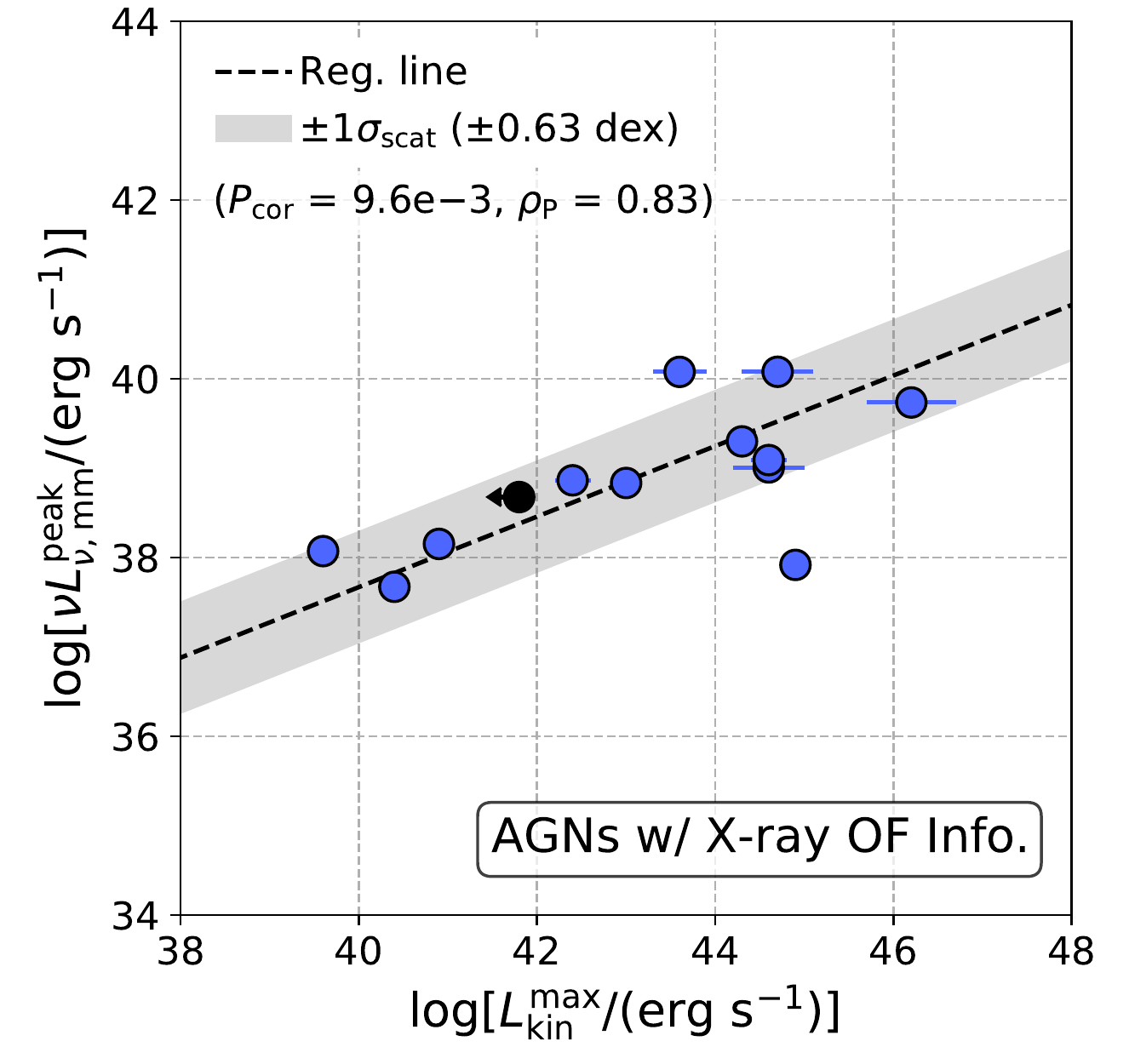}
    \includegraphics[width=8.3cm]{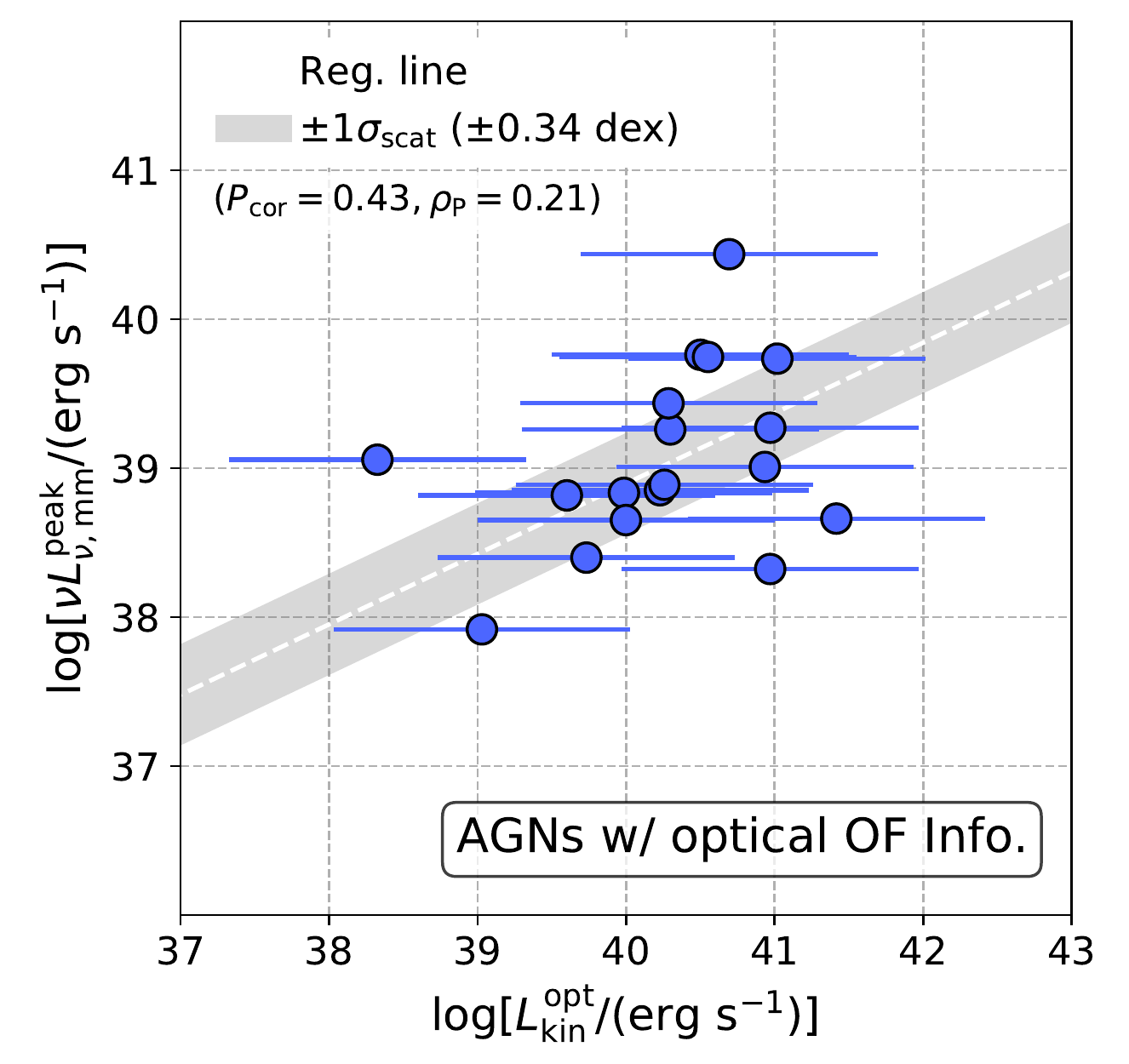}
    \caption{
    Top: Scatter plot for the mm-wave emission and the maximum energy carried by the X-ray outflow for the 14 AGNs for which the maximum energies were estimated in either \cite{Tom12} or \cite{Gof15}. 
    An AGN with an upper limit is shown as a black circle, and the AGN NGC 4395 is located outside in the lower left direction.
    A regression line is indicated by the black dashed line. 
    Bottom: Scatter plot for the mm-wave emission and the energy carried by the outflow traced with [O {\sc{iii}}] emission for 18 AGNs. The outflow data are from \cite{roj20}. 
    No significant correlation is found, and due to this, the regression line is drawn in white.
    }
    \label{fig:outflow}
\end{figure}

\begin{figure}
    \centering
    \includegraphics[width=8.cm]{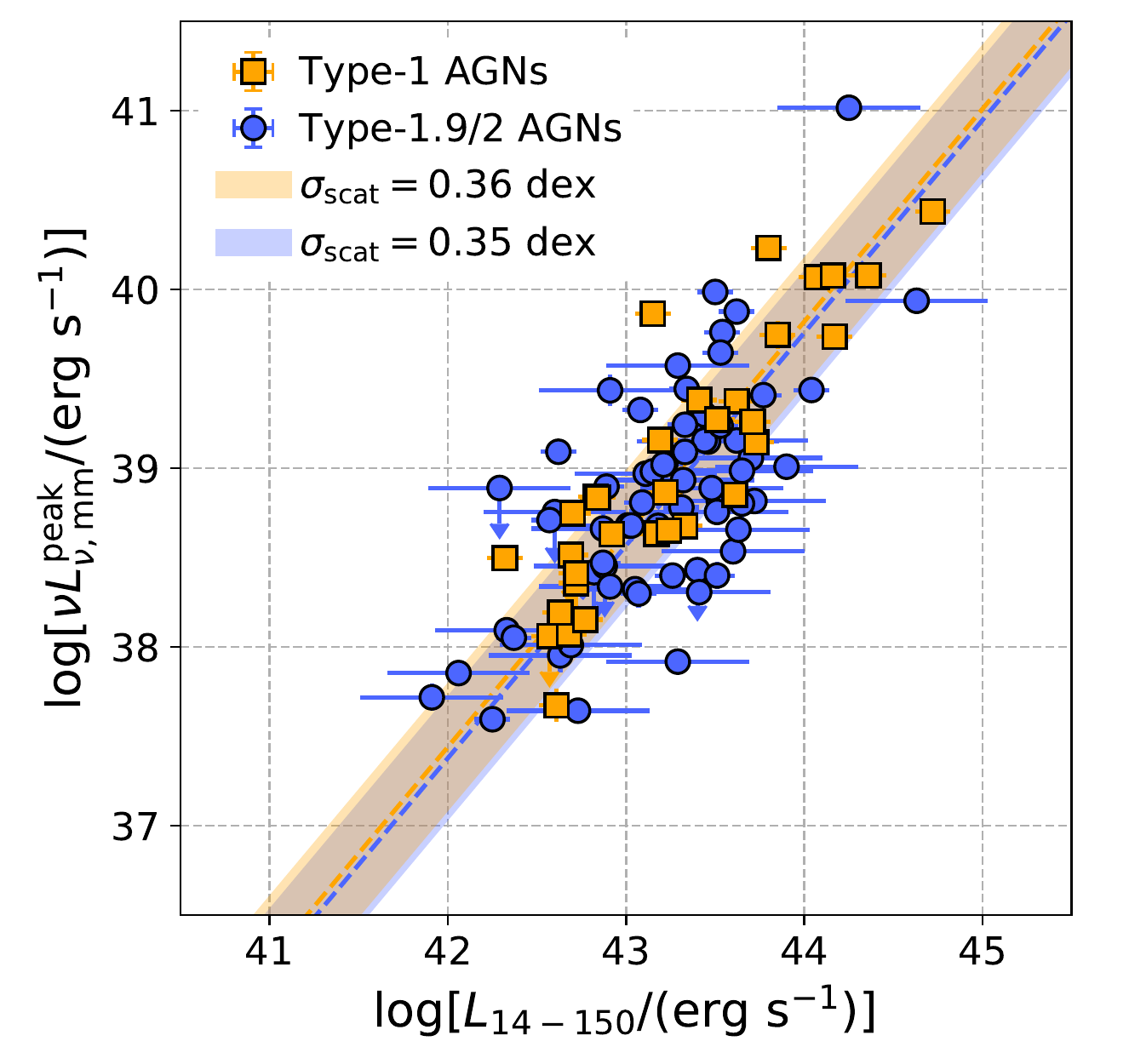} \hspace{.2cm}
    \caption{
    Correlations between the mm-wave and 14--150\,keV luminosities for type-1 AGN and type-2 AGN subsamples, indicated in orange and blue. 
    Orange and blue dashed lines indicate regression lines obtained for the type-1 and type-2 subsamples, respectively.  
    We mention the reason why a smaller 
    intrinsic scatter ($\sigma_{\rm scat}$)
    is obtained for type-2 AGNs in spite of the apparently larger scatter of their data point. The intrinsic scatter is derived by subtracting the scatter due to uncertainties in data points, and as a large fraction of type-2 AGNs have large uncertainty in their luminosities as shown, a much smaller intrinsic scatter than the apparent one can be obtained. 
    } 
    \label{fig:types}
\end{figure}

The collision between an AGN outflow and the surrounding gas may cause a shock in which electrons are accelerated, producing synchrotron emission \citep{Jia10,Hwa18}. Quantitatively, \cite{Nim15} suggested
that the ratio between synchrotron emission and AGN bolometric luminosity may be $10^{-5} \sim 10^{-6}$ if 0.5--5\% of the bolometric luminosity is converted into the kinetic energy of the outflow, and then $\sim$ 1\% of the outflow energy is used to produce relativistic particles that radiate synchrotron emission. 
The predicted ratio of \cite{Nim15} is consistent with our finding ($\nu L^{\rm peak}_{\nu,{\rm mm}}/L_{\rm bol} \sim 10^{-5}$; see Figure~\ref{fig:var_lum2}).

In what follows, we discuss the outflow scenario in three ways. First, we focus on AGNs for which outflows were observed in the X-ray band and examine the relation between the energy carried by the outflow and the mm-wave luminosity. 
Second, we perform the same test by focusing on outflows traced by optical [O {\sc{iii}}] emission. 
Finally, we discuss the relation between the mm-wave emission and the Eddington ratio on the basis of a larger sample. This approach is motivated by past works \citep[e.g.,][]{Fab09,Ric17nat}, indicating that, as the Eddington ratio increases, outflows become more common. Thus, if outflows produce the mm-wave emission, it is expected that the nuclear mm-wave luminosity is correlated with the Eddington ratio.

We collect information on X-ray outflows by referring to \cite{Tom12} and \cite{Gof15} \citep[see also][]{Tom10,Tom11,Gof13}. Using XMM-Newton archival data, \cite{Tom12} detected X-ray outflows through Fe absorption lines in 19 of 42 AGNs at $z < 0.1$. In \cite{Gof15}, 51 AGNs were investigated in a similar way using Suzaku archival data, and outflows 
were found in 20 of the 51 AGNs. 
In almost the same way, \cite{Tom12} and \cite{Gof15} derived the maximum and minimum values of the energy carried by the outflow. Here, we focus only on the maximum values ($L^{\rm max}_{\rm kin}$) since the minimum value is based on the strong assumption that the outflow always reaches the escape velocity. On the other hand, the maximum value was simply estimated by determining the spatial scale of an outflow from an ionization parameter. 
Cross-matching with their samples, we obtain X-ray-outflow information for 14 AGNs in our sample. 

The top panel of Figure~\ref{fig:outflow} shows a correlation between $\nu L^{\rm peak}_{\nu,{\rm mm}}$ and $L^{\rm max}_{\rm kin}$. With the bootstrap method, we confirm that it is significant with $P_{\rm cor} < 0.01$. A correlation in the flux space is, however, found to be insignificant with $P_{\rm cor} \approx 5\times10^{-2}$, favoring a larger sample to confirm the relation in the flux space. 
Interestingly, the Pearson's correlation coefficient of $\rho_{\rm P} = 0.83$ found for $\nu L^{\rm peak}_{\nu,{\rm mm}}$ vs. $L^{\rm max}_{\rm kin}$ is higher than that found for $\nu L^{\rm peak}_{\nu,{\rm mm}}$ vs. $L_{\rm 14-150}$ ($\rho_{\rm P} = 0.74$), while the difference is insignificant (i.e., $P_{\rm rz} > 0.1$). 
This is consistent with the scenario in which the mm-wave emission is driven by the AGN outflow. However, this may not be surprising, given that $L^{\rm max}_{\rm kin}$ is proportional to ionizing luminosity, or UV-to-X-ray luminosity. 

Furthermore, we discuss the outflow scenario focusing on ionized gas outflows traced by optical emission [O {\sc iii}]. 
We refer to a study of \cite{roj20}, who searched for [O {\sc iii}]$\lambda$5007 outflow signatures in 547 BAT-selected nearby ($z \lesssim 0.25$) AGNs and then found signatures for 178 AGNs. 
Although their single-slit spectroscopic data did not constrain spatial information directly, the spatial scales of the outflows 
were estimated to be $\sim$ 300\,pc--3\,kpc by using a relation of the
size of an outflow and [O {\sc iii}] luminosity (see more details in their paper). 
After cross-matching their sample with ours, we find that our sample includes 18 AGNs with kinetic energies ($L^{\rm opt}_{\rm kin}$) derived from [O {\sc iii}] outflows. 
We assign 1\,dex as the errors in $L^{\rm opt}_{\rm kin}$ considering that density, a poorly constrained parameter in the derivation of the energies, can range from $10^{2.5}$ to $10^{4.5}$ cm$^{-3}$ \citep{roj20}. 
The bottom panel of Figure~\ref{fig:outflow} shows a scatter plot of $\nu L^{\rm peak}_{\nu, {\rm mm}}$ versus $L^{\rm opt}_{\rm kin}$, and no significant correlation is found with $P_{\rm cor} = 0.2$.
This result is natural given the different scales probed by the mm-wave and [O {\sc iii}] emission. Spatially well-resolved outflow data are needed for further investigation.

Lastly, we discuss an outflow model where a higher Eddington ratio is preferred to launch an outflow \citep[e.g.,][]{Fab09,Ric17nat}. In addition, we consider that among AGNs having similar Eddington ratios, more luminous objects may carry more energy in the form of an outflow, as proposed observationally and theoretically \citep[e.g.,][]{Gof15,fio17,Nom17}. 
Under these ideas, it is predicted that the $\nu L^{\rm peak}_{\nu, {\rm mm}}/L_{\rm 14-150}$ ratio increases with the Eddington ratio. 
However, as shown in Figure~\ref{fig:ratio_vs_agnpar}, such a correlation is not found. 
This result could disfavor the outflow model as the ``general" mechanism for the mm-wave emission. However, given the previous result focusing on the X-ray outflows, there may be objects in which the outflow contributes to the mm-wave emission significantly.


\subsection{Unresolved Jet on a Scale $\lesssim$ 200 pc}\label{sec:jet_ori}

We discuss the possibility that the mm-wave emission is associated with a jet, unresolved even at resolutions of $\sim$ 1--200\,pc. 
Here, we only test a very simple well-collimated 
jet model where its apparent luminosity changes solely according to the inclination angle, although, in fact, the jet may bend, and its luminosity may be related to other various factors \cite[e.g., spin, accretion rate, black hole mass, and Eddington ratio;][]{Ho08,Van13,Bal18}. 
As a proxy of the inclination angle, we use the Seyfert type (i.e., type-1 and type-2) by assuming the inclination-dependent unified AGN model where the jet is aligned to the polar axis of a putative accretion disk \citep[e.g.,][]{Urr95}. Here, we define the angle so that it is 0$^\circ$ at the polar axis of the disk. 
Thus, if the jet is responsible for the mm-wave emission, by assuming that X-ray emission is isotropic, stronger mm-wave emission would be expected for type-1 AGNs, or AGNs with lower inclination angles. 

Figure~\ref{fig:types} shows the correlations 
between $\nu L^{\rm peak}_{\nu,{\rm mm}}$ and $L_{\rm 14-150}$ for type-1 and type-2 AGNs. We rely on the Seyfert-type classification by \cite{Kos22_catalog}, who considered three Seyfert types of type-1, type-1.9, and type-2. In our study, the intermediate type-1.9 objects are added to the type-2 sample. 
In addition to the regression lines constrained by leaving both the slope and intercept free to vary, we also estimate the intercepts of regression lines by fixing their slopes at 1.19, the average of the independently derived slopes. We then find a higher intercept for the type-1 AGN sample by 0.06\,dex than the type-2 AGN sample, but this is only $\approx$ 1-sigma. 
Supplementarily, we take into account the possibility that the Seyfert type cannot be used as a proxy of the inclination angle below an activity level because the broad line region intrinsically may disappear (i.e., true type-2 AGNs; e.g., \citealt{Mar12}).
For a sample of type-2 AGNs, \cite{Mar12} investigated the dependence of the presence of polarized broad line emission on AGN activity and suggested that polarized broad lines were found particularly for objects with $\log L_{\rm bol} > 43.90$ and $\log \lambda_{\rm Edd} > -1.9$. Following this result, we exclude objects with $\log \lambda_{\rm Edd} < -2$ and $\log (L_{\rm bol}/{\rm erg s^{-1}}) < 44$ and find that the same conclusion drawn above (no significant difference in the intercept) is obtained.

Furthermore, we perform the same analysis for subsamples created by dividing the whole sample into low-$N_{\rm H}$ AGNs and high-$N_{\rm H}$ AGNs. This analysis is motivated by the suggestion that the column density may also be used as an indicator of the inclination angle \citep[i.e., a higher column density for a higher angle; ][]{Fis14,Gup21}.
We here adopt $\log(N_{\rm H}/{\rm cm}^{-2}) = 22.5$ as the boundary so that the low-$N_{\rm H}$ and high-$N_{\rm H}$ 
subsamples have almost the same sizes of 48 and 50. 
In the same way as previously adopted, 
we find a higher intercept for the low-$N_{\rm H}$ subsample by $\approx$ 0.16\,dex, but the difference is insignificant, consistent with the result obtained by using the Seyfert type. 

In summary, we have not obtained evidence supporting the simple well-collimated jet model. More research is needed that compares more detailed modelings of the jet to understand the jet contribution better. 

\subsection{Short Summary of Discussion}

Throughout Section~\ref{sec:agn_mec}, we have discussed the four AGN-related mechanisms. The AGN-related dust emission would be unlikely, as the extrapolated luminosities from the IR AGN models are generally lower than the observed ones, and the observed mm-wave indices are generally inconsistent with that expected for the dust emission. 
As for the scenario that the mm-wave emission originates around where the X-ray corona forms, the tight correlation of the mm-wave and X-ray luminosities (i.e., 14--150\,keV and 2--10\,keV), inferring an energetic link of these emission, would be a supporting result. 
Also, although the sample size is small, some objects show $\alpha^{230}_{100} < \alpha^{100}_{22}$. Excesses inferred from the indices are consistent with the presence of self-absorbed synchrotron emission in the mm-wave band, suggested to originate from a compact region of a $\sim$ 40--50 Schwarzschild radius in some AGNs \citep{Ino18}. 
Regarding the outflow-driven scenario, we find the significant correlation between the mm-wave luminosity and the energy carried by outflows traced with Fe-K absorption lines. However, an increase in mm-wave luminosity with Eddington ratio which is expected from an Eddington-ratio dependent outflow model is not found. 
Thus, the outflow is possibly not a general mechanism but may be important in some objects. 
Lastly, we find that a simple jet model where its luminosity changes solely according to the inclination angle is not favored. To better understand the contribution of a jet,  further investigations that consider dependencies of the jet luminosity 
on other parameters (e.g., spin, accretion rate, black hole mass, and Eddington ratio) are crucial.

\section{Potential of Mm-wave Emission for AGN Studies}\label{sec:future}
 
Before summarizing this paper, we demonstrate that mm-wave observations 
have a great potential in searching for obscured AGNs. 
As summarized in Section~\ref{sec:reg_sum}, the mm-wave luminosity correlates with the X-ray luminosity with $\sigma_{\rm scat} \approx $ 0.3--0.4\,dex, thus serving as a good measure of the AGN luminosity. 
As mm-wave emission can be almost dust-extinction free up to $N_{\rm H} \sim 10^{26}$ cm$^{-2}$, the mm-wave observation can detect even heavily obscured systems (e.g., $N_{\rm H} > 10^{24}$\,cm$^{-1}$) without the severe effect of extinction. There should have been many such systems in the distant universe at $z \sim$ 1--3, where intense galaxy and black hole growth occurred \citep[e.g.,][]{Bou11,Bur13}, and are important for understanding the growth of galaxies and black holes. 


We calculate the detection limits of ALMA, ngVLA, Chandra, and Athena as a function of $z$. 
ngVLA is a next-generation observatory planned to achieve high sensitivity to emission in the band of $\approx$ 1--100\,GHz and thus will be able to be used to detect rest-frame 230\,GHz emission from objects at redshifts greater than $\approx$ 1. 
Chandra has detected very distant AGNs at $z \sim$ 6--8 (e.g., \citealt{Vit19}) and thus is selected for comparison as a representative X-ray observatory. In addition, Athena to be launched in the early 2030s is also considered as one of the biggest X-ray observatories planned. 

We consider two different exposures of 1 hr and 100 hrs. 
The latter assumes surveys in the large program category of ALMA ($>$ 50 hrs). 
The limits of ALMA in Band 6, Band 5, and Band 4, are calculated based on an exposure calculator\footnote{ https://almascience.eso.org/proposing/sensitivity-calculator}.
Here, the declination angle needed for determining airmass is set to that of the Hubble Ultra Deep Field. 
The field was observed in the ALMA large program ASPECS \cite[e.g.,][]{Dec19} and is one of the most extensively studied extra-galactic fields. Thus, this would be investigated with ngVLA by spending a large amount of time \citep{Dec18}.
We note that Band 3 is available and Band 1 and Band 2 will be available in the future to observe 230\,GHz emission at $z > 1$, but their best angular resolutions ($>$ 0\farcs042) are insufficient to achieve physical resolutions of $\lesssim$ 200\,pc at redshifts $>$ 1. Therefore, they are not considered. 
The limits of ngVLA are calculated based on information provided on the official website\footnote{https://ngvla.nrao.edu/page/performance}. Among the different resolution options listed, we consider 10 mas, required to achieve physical resolutions of $< 200$\,pc at redshifts greater than 1. 
For ALMA and ngVLA, we consider 5$\sigma$ detections. 

To derive the limits of Chandra and Athena, we focus on their detectors of the ACIS and WFI, respectively, and consider that at least $\approx$ 100 counts need to be obtained to infer X-ray luminosity by constraining basic X-ray parameters \cite[i.e., column density and photon index; e.g.,][]{Luo17}. 
For calculation, we simulate X-ray spectra for an obscured cutoff power-law model with $\Gamma = 1.8$, $E_{\rm cut} = 200$\,keV, and $N_{\rm H} = 10^{24}$ cm$^{-2}$ while considering the latest response files\footnote{https://cxc.harvard.edu/caldb/prop\_plan/imaging/ for Chandra and https://www.the-athena-x-ray-observatory.eu/resources/simulation-tools.html for Athena}. In particular, for Chandra, the background is ignored since a 0.5--6\,keV count rate required to obtain $\approx$ 100 counts in 100 hrs is $3\times10^{-4}$ counts s$^{-1}$, and this is higher than a typical background count rate of $\sim 10^{-6}$ counts s$^{-1}$ for a point source \citep{Luo17}. 
For the Athena/WFI, a response file averaged over the wide field of view (FoV) of 40\arcmin$\times$40\arcmin\ without any filters is adopted.

\begin{figure*}
    \centering
    \includegraphics[width=17cm]{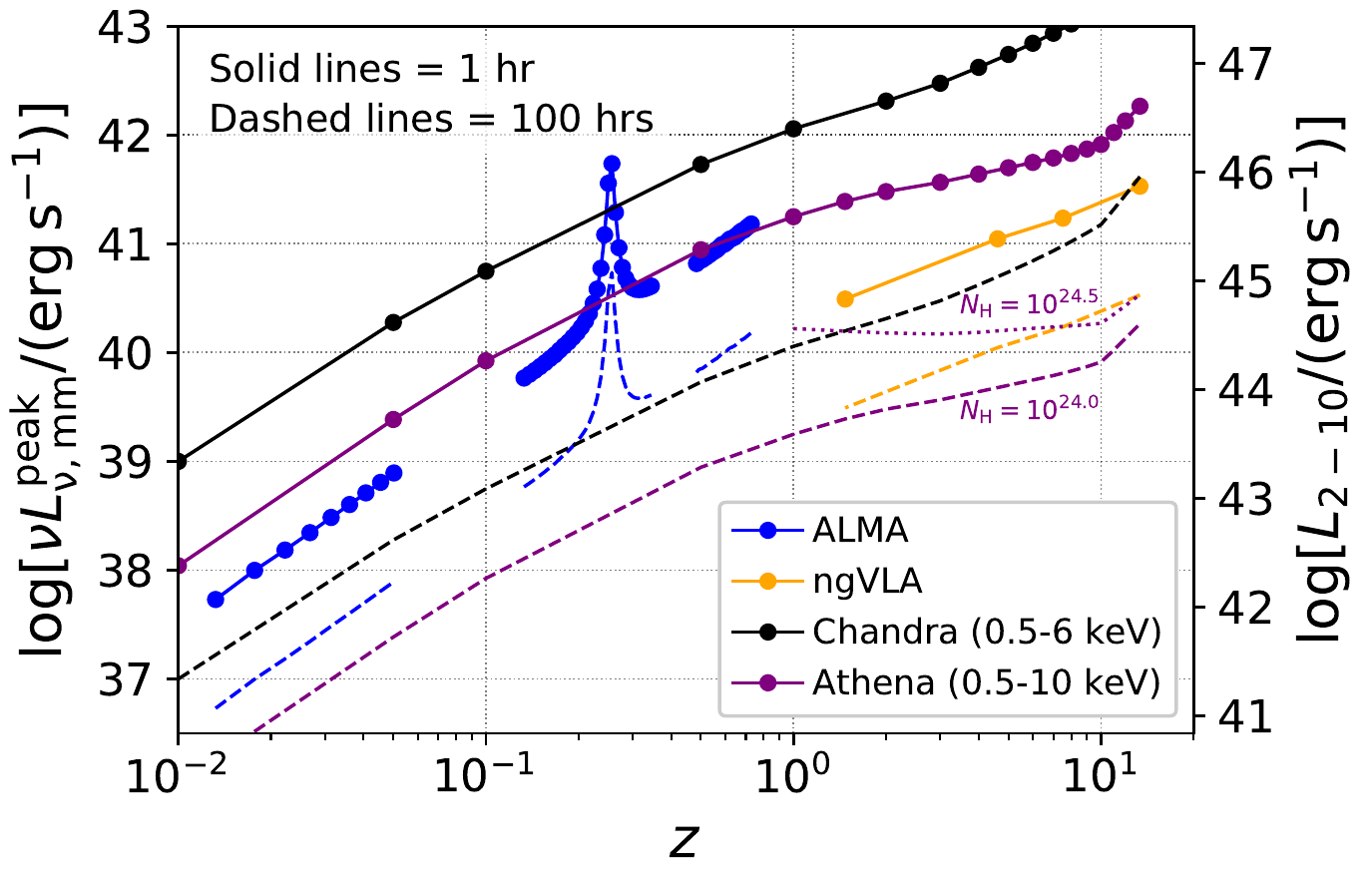}
    \caption{
    Detection limits of ALMA, ngVLA, Chandra (0.5--6\,keV), and Athena (0.5--10\,keV). The ACIS and WFI are considered as the detectors of Chandra and Athena, respectively. 
    The right X-ray-luminosity axis is adjusted to the left one by considering $\nu L^{\rm peak}_{\nu,{\rm mm}}/L_{\rm 14-150} \sim 10^{-4.6}$, 
    which was found for AGNs with $\alpha_{\rm mm} - \alpha^{\rm e}_{\rm mm} > 0.3$, and 
    $L_{\rm 2-10}/L_{\rm 14-150} = 0.55$, where a typical cutoff power-law model is assumed (i.e., $\Gamma = 1.8$ and $E_{\rm cut} = 200$\,keV).
    Sensitivities for the exposure of 1 hr are calculated at redshifts indicated as circles, and solid lines are shown to tie the data points.
    Likewise, for the exposure of 100 hrs, dashed lines are drawn to indicate its limits, calculated at the same redshifts. 
    For the ALMA and ngVLA observations, we consider 5$\sigma$ detections. 
    A poor sensitivity of ALMA at $z \sim $ 0.2--0.3, appearing as a flare, is because of severe atmospheric absorption. 
    The limits of Chandra (black) and Athena (purple) are calculated so that $\approx$ 100 counts can be obtained to infer X-ray luminosities. 
    For the calculation, we simulated X-ray spectra of an obscured cutoff power-law model with $\Gamma = 1.8$, $E_{\rm cut} = 200$\,keV, and $N_{\rm H} = 10^{24}$ cm$^{-2}$. 
    In addition, the case of $N_{\rm H} = 10^{24.5}$ cm$^{-2}$ for Athena is plotted at $z > 1$ by a dotted line. 
    }
    \label{fig:sens}
\end{figure*}

Figure~\ref{fig:sens} plots the estimated limits of the four observatories of ALMA, ngVLA, Chandra, and Athena. The figure shows that for the 1 hr exposure, ALMA and ngVLA would be able to reach fainter sources than the X-ray observatories. In the case of the 100 hrs, particularly at $z < 1$, Athena will detect fainter sources than ALMA and Chandra, and if we only consider ALMA and Chandra in operation, they can go down to a comparable level of luminosity, except for the low redshifts ($z < 0.05$) where ALMA can reach fainter objects by $\sim$ 0.4\,dex. 
At $z \gtrsim 1$, Athena will be slightly superior to ngVLA, and the notable ability of the Athena/WFI to achieve 5\arcsec\ resolution (half energy width) over its wide FoV (40\arcmin$\times$40\arcmin, or $\sim$ 0.4 deg$^2$) will detect more objects. In fact, in the band of 20\,GHz, low enough to detect rest-frame 230\,GHz emission at $z \sim 10$, ngVLA will have only a small FoV with FWHM $\approx$ 2\farcm1, or $\approx$ 7$\times 10^{-4}$ deg$^2$, which is $\approx$ 1/640 of the WFI. At lower redshifts, the areal ratio becomes larger as the beam size of ngVLA is smaller at a higher observing frequency. 
However, as plotted by simulating a different X-ray spectral model with $N_{\rm H} = 24.5$ cm$^{-2}$, the detection limit of the Athena/WFI is degraded, and ngVLA will be superior to the Athena/WFI. 

In summary, ALMA and ngVLA would be better options if the objective is to detect obscured systems in a relatively short time ($\sim$ 1 hr). If a much longer time ($\sim$ 100 hrs) is available, there is not much difference between ALMA and Chandra and also between ngVLA and Athena. However, the Athena/WFI is better in searching for objects due to its large FoV than ngVLA, but, in contrast, ngVLA will play an important role in finding heavily obscured systems that the Athena/WFI may miss. 



\section{Summary}\label{sec:sum}

To investigate the origin of nuclear continuum emission in the mm-wave band of AGNs, we have systematically analyzed Band-6 (211--275\,GHz) ALMA data of BAT-selected 98 AGNs at $z < 0.05$ (Table~\ref{tab_app:sample}). Almost all data were obtained at high resolutions better than $<$ 0\farcs6, corresponding to physical scales of $\lesssim$ 100--200 pc (Figure~\ref{fig:dist_vs_res}). 
The results obtained from the data and our arguments are as follows. 

\begin{itemize}
    \item We find significant correlations of the peak mm-wave luminosity with AGN luminosities (i.e., 2--10\,keV, 14--150\,keV, 12\,$\mu$m, and bolometric luminosities) in Section~\ref{sec:ll_cor} (Figures~\ref{fig:var_lum1} and \ref{fig:var_lum2}). The correlation found for the 14--150\,keV luminosity has the smallest intrinsic scatter of $\sigma_{\rm scat} = 0.36$\,dex. 
    \item We find that the ratio of the mm-wave and 14--150\,keV luminosities does not change much within the achieved resolution range of $\sim$ 1--200\,pc (Figure~\ref{fig:ratio_vs_res}). This can be interpreted as the ubiquitous presence of a compact mm-wave component on a scale of $< 10$\,pc related to the AGN X-ray emission. 
    \item To discuss the contribution of SF to the observed mm-wave flux, we have compared the fluxes expected from three possible SF components (i.e., free-free emission, synchrotron emission, and heated-dust emission) in Section~\ref{sec:sf_cont}. Among these components, it is found that the SF-related dust emission would be the strongest. 
    \item We have studied whether the SF-related dust emission contributes to the observed mm-wave emission based on the spectral index. 
    The comparison of the observed spectral indices and that expected for thermal dust emission (i.e., $\alpha_{\rm mm} = -3.5$; Section~\ref{sec:cor4sigAGN} and Figure~\ref{fig:freq1}) suggests that a significant contribution of dust emission would be unlikely. 
    \item We have restricted a sample to AGNs for which the SF contribution is likely to be small, by selecting them on the basis of the spectral index. For this sample, we find a similar correlation between $\nu L^{\rm peak}_{\nu,{\rm mm}}$ and $L_{\rm 14-150}$ to that obtained for the entire sample (Section~\ref{sec:cor4sigAGN}).
    This result suggests that the observed mm-wave emission is generally correlated with the AGN activity traced by the X-ray emission. 
    \item We have tabulated the relations of the mm-wave luminosity with AGN luminosities in Table~\ref{tab:cors4use}. Because the SF contribution would not be so strong, the relations may serve as good measures of the AGN luminosity, almost free from dust extinction. 
    The tightness of the relations for the 14--150\,keV luminosity suggests that they have approximately the same reliability as MIR relations with X-ray luminosity. 
    \item Among four AGN mechanisms that may be the origin of the mm-wave emission, the dust emission would not account for a large fraction of the observed mm-wave emission. This is suggested by the fact that the dust emission predicted from the AGN torus model in the IR band is generally weaker than observed (Section~\ref{sec:agn_dust}). 
    Also, this is supported by the observed spectral indices higher than expected from the dust emission. 
    \item The tight correlations between the mm-wave and X-ray luminosities (14--150\,keV and 2--10\,keV) perhaps suggest that the mm-wave emission originates around and/or from where an X-ray corona forms (Section~\ref{sec:x-ray_ori}), as inferred by past theoretical and observational studies. 
    Although this investigation is based on a small sample, we find that three objects show spectral indices, consistent with the presence of significant self-absorbed synchrotron emission from a compact region ($\lesssim 10^{-3}$\,pc; Figure~\ref{fig:indices2}). 
    \item Alternatively, relativistic particles created by outflow-driven shocks may produce synchrotron emission and contribute to the mm-wave emission. This is supported by the significant correlation between the mm-wave luminosity and the energy carried by an X-ray outflow and its high correlation strength (Figure~\ref{fig:outflow}). 
    However, we cannot find an increase in mm-wave luminosity with the Eddington ratio, which would have supported a radiation-driven AGN outflow model. Thus, the outflow may contribute to the mm-wave emission for some objects but is possibly not a common process. 
    \item As the fourth scenario, we also have discussed the possibility that the mm-wave emission is produced by an unresolved jet (i.e., relativistically beamed emission). 
    We have considered a very simple well-collimated jet model whose luminosity depends solely on the inclination angle. 
    By assessing the dependence of $\nu L^{\rm peak}_{\nu,{\rm mm}}/L_{\rm 14-150}$ on two different proxies for the inclination angle (i.e., Seyfert type and line-of-sight absorbing column density), we find that there is no significant increase in the ratio with decreasing inclination angle. This is inconsistent with the simple jet scenario. 
    \item Lastly, motivated by the tight correlation between the mm-wave and 14--150\,keV luminosity, we have demonstrated the potential of mm-wave emission observations for detecting AGNs by focusing on obscured AGNs with $N_{\rm H} \sim 10^{24}$ cm$^{-2}$ (Figure~\ref{fig:sens}). 
    By comparing the detection limits of ALMA, ngVLA, Chandra, and Athena, we find that in a relatively short exposure (e.g., $\sim$ 1 hr), ALMA and ngVLA will detect fainter objects than the X-ray observatories. If a much longer time ($\sim$ 100 hrs) is available, there is not much difference between ALMA and Chandra and also between ngVLA and Athena. However, regarding the next-generation telescopes of ngVLA and Athena,  there are two important points. The Athena with the WFI will provide a larger sample at a greater survey speed because of its large FoV. Instead, ngVLA will be able to find more heavily obscured systems down to lower luminosity. 
\end{itemize}

\clearpage 

\begin{acknowledgments} 
We thank the reviewer for the useful comments, which helped us improve the quality of the manuscript. 
We acknowledge support from FONDECYT Postdoctral Fellowships 3200470 (T.K.), 3210157 (A.R.) and 3220516 (M.J.T.), FONDECYT Iniciacion grant 11190831 (C.R.), FONDECYT Regular 1200495 and 1190818 (F.E.B.), ANID BASAL project FB210003 (C.R., F.E.B) and Millennium Science Initiative Program  – ICN12\_009 (F.E.B). T.K., T.I. and M.I. are supported by JSPS KAKENHI grant numbers JP20K14529, JP20K14531, and JP21K03632, respectively. 
K.O. acknowledges support from the National Research Foundation of Korea (NRF-2020R1C1C1005462).
M.B. acknowledges support from the YCAA Prize Postdoctoral Fellowship.
The scientific results reported in this article are based on data obtained from the Chandra Data Archive. This research has made use of software provided by the Chandra X-ray Center (CXC) in the application packages CIAO. 
This paper makes use of the following ALMA data: 
ADS/JAO.ALMA\#2012.1.00139.S, 
\#2012.1.00474.S, 
\#2013.1.00525.S, 
\#2013.1.00623.S, 
\#2013.1.01161.S, 
\#2015.1.00086.S, 
\#2015.1.00370.S, 
\#2015.1.00597.S, 
\#2015.1.00872.S, 
\#2015.1.00925.S,  
\#2016.1.00254.S, 
\#2016.1.00839.S, 
\#2016.1.01140.S, 
\#2016.1.01279.S, 
\#2016.1.01385.S, 
\#2016.1.01553.S, 
\#2016.2.00046.S,  
\#2016.2.00055.S,
\#2017.1.00236.S, 
\#2017.1.00255.S, 
\#2017.1.00395.S, 
\#2017.1.00598.S, 
\#2017.1.00886.L, 
\#2017.1.00904.S, 
\#2017.1.01158.S,  
\#2017.1.01439.S, 
\#2018.1.00006.S, 
\#2018.1.00037.S, 
\#2018.1.00211.S, 
\#2018.1.00248.S, 
\#2018.1.00538.S, 
\#2018.1.00576.S, 
\#2018.1.00581.S, 
\#2018.1.00657.S, 
\#2018.1.00978.S, 
\#2018.1.00986.S,
\#2018.1.01321.S,  
\#2019.1.00363.S,  
\#2019.1.00763.L,
\#2019.1.01229.S,
\#2019.1.01742.S, and 
\#2019.2.00129.S. 
ALMA is a partnership of ESO (representing its member states), NSF (USA) and NINS (Japan), together with NRC (Canada), MOST and ASIAA (Taiwan), and KASI (Republic of Korea), in cooperation with the Republic of Chile. The Joint ALMA Observatory is operated by ESO, AUI/NRAO and NAOJ.
Data analysis was in part carried out on the Multi-wavelength Data Analysis System operated by the Astronomy Data Center (ADC), National Astronomical Observatory of Japan.
\end{acknowledgments}

\clearpage


\appendix 
\setcounter{table}{0}
\setcounter{section}{0}
\renewcommand{\thetable}{\Alph{table}}

\section{Our Sample List and Correlation Results}

We provide a complete list of our 98 AGNs in Table~\ref{tab_app:sample}. In addition, 
in Table~\ref{tab_app:corr}, we tabulate the values of statistical parameters obtained in the correlation analyses.

\begin{longtable*}[c]{cccc}
\caption{Sample List\label{tab_app:sample}}\\
\multicolumn{4}{c}{\parbox{10.8cm}{
Notes. 
(1) Number in this paper and in Paper~II.
(2) Index adopted in the Swift/BAT 70-month catalog \citep{Bau13}. 
(3,4) Swift and counterpart names, taken from \cite{ric17c}.
}}\vspace{0.2cm}\\
\hline
(1) & (2) & (3) & (4) \\ 
Num. & BAT Index & Swift Name & Counterpart Name \\ 
\hline \endfirsthead \hline
(1) & (2) & (3) & (4) \\ 
Num. & BAT Index & Swift Name & Counterpart Name \\ 
\hline \endhead \hline \endfoot \hline\hline \endlastfoot
01   &    28  &           J0042.9$-$2332  &                    NGC 235A   \\       
02   &    31  &           J0042.9$-$1135  &           MCG $-$2$-$2$-$95   \\       
03   &    58  &           J0111.4$-$3808  &                     NGC 424   \\       
04   &    72  &           J0123.8$-$3504  &                    NGC 526A   \\       
05   &    84  &           J0134.1$-$3625  &                     NGC 612   \\       
06   &    102  &           J0201.0$-$0648  &                     NGC 788   \\       
07   &    112  &         J0209.5$-$1010D1  &                     NGC 833   \\       
08   &    112  &         J0209.5$-$1010D2  &                     NGC 835   \\       
09   &    131  &           J0231.6$-$3645  &                     IC 1816   \\       
10   &    134  &           J0234.6$-$0848  &                     NGC 985   \\       
11   &    144  &           J0242.6$+$0000  &                    NGC 1068   \\       
12   &    153  &           J0251.6$-$1639  &                    NGC 1125   \\       
13   &    156  &           J0252.7$-$0822  &           MCG $-$2$-$8$-$14   \\       
14   &    159  &           J0256.4$-$3212  &                 ESO 417$-$6   \\       
15   &    163  &           J0304.1$-$0108  &                    NGC 1194   \\       
16   &    182  &           J0331.4$-$0510  &           MCG $-$1$-$9$-$45   \\       
17   &    184  &           J0333.6$-$3607  &                    NGC 1365   \\       
18   &    216  &           J0420.0$-$5457  &                    NGC 1566   \\       
19   &    237  &           J0444.1$+$2813  &                  LEDA 86269   \\       
20   &    242  &           J0451.4$-$0346  &          MCG $-$1$-$13$-$25   \\       
21   &    252  &           J0502.1$+$0332  &                  LEDA 75258   \\       
22   &    260  &           J0508.1$+$1727  &   2MASX J05081967$+$1721483   \\       
23   &    261  &           J0510.7$+$1629  &           IRAS 05078$+$1626   \\       
24   &    266  &           J0516.2$-$0009  &                      Ark120   \\       
25   &    269  &           J0501.9$-$3239  &                ESO 362$-$18   \\       
26   &    272  &           J0521.0$-$2522  &           IRAS 05189$-$2524   \\       
27   &    301  &           J0543.9$-$2749  &                ESO 424$-$12   \\       
28   &    308  &           J0552.2$-$0727  &                    NGC 2110   \\       
29   &    313  &           J0557.9$-$3822  &                H 0557$-$385   \\       
30   &    319  &           J0601.9$-$8636  &                   ESO 5$-$4   \\       
31   &    330  &           J0623.8$-$3215  &                 ESO 426$-$2   \\       
32   &    404  &           J0804.2$+$0507  &                    Mrk 1210   \\       
33   &    416  &           J0823.4$-$0457  &                 Fairall 272   \\       
34   &    423  &           J0838.4$-$3557  &                Fairall 1146   \\       
35   &    453  &           J0920.8$-$0805  &          MCG $-$1$-$24$-$12   \\       
36   &    460  &           J0926.2$+$1244  &                     Mrk 705   \\       
37   &    465  &           J0934.7$-$2156  &                ESO 565$-$19   \\       
38   &    471  &           J0945.6$-$1420  &                    NGC 2992   \\       
39   &    472  &           J0947.6$-$3057  &          MCG $-$5$-$23$-$16   \\       
40   &    475  &           J0951.9$-$0649  &                    NGC 3035   \\       
41   &    480  &           J0959.5$-$2248  &                    NGC 3081   \\       
42   &    486  &           J1005.9$-$2305  &                ESO 499$-$41   \\       
43   &    497  &           J1023.5$+$1952  &                    NGC 3227   \\       
44   &    502  &           J1031.7$-$3451  &                    NGC 3281   \\       
45   &    518  &           J1048.4$-$2511  &                    NGC 3393   \\       
46   &    520  &          J1051.2$-$1704A  &                    NGC 3431   \\       
47   &    558  &           J1139.0$-$3743  &                    NGC 3783   \\       
48   &    576  &           J1152.1$-$1122  &               PG 1149$-$110   \\       
49   &    583  &           J1201.2$-$0341  &                    Mrk 1310   \\       
50   &    586  &           J1204.5$+$2019  &                    NGC 4074   \\       
51   &    607  &           J1217.3$+$0714  &                    NGC 4235   \\       
52   &    608  &           J1218.5$+$2952  &                    NGC 4253   \\       
53   &    615  &           J1225.8$+$1240  &                    NGC 4388   \\       
54   &    616  &           J1202.5$+$3332  &                    NGC 4395   \\       
55   &    626  &           J1235.6$-$3954  &                    NGC 4507   \\       
56   &    631  &           J1239.6$-$0519  &                    NGC 4593   \\       
57   &    641  &           J1252.3$-$1323  &                    NGC 4748   \\       
58   &    653  &           J1304.3$-$0532  &                    NGC 4941   \\       
59   &    655  &           J1305.4$-$4928  &                    NGC 4945   \\       
60   &    657  &          J1306.4$-$4025A  &                ESO 323$-$77   \\       
61   &    676  &           J1332.0$-$7754  &                  ESO 21$-$4   \\       
62   &    677  &           J1333.5$-$3401  &                ESO 383$-$18   \\       
63   &    678  &         J1334.8$-$2328D2  &                  LEDA 47848   \\       
64   &    679  &           J1336.0$+$0304  &                    NGC 5231   \\       
65   &    680  &           J1335.8$-$3416  &          MCG $-$6$-$30$-$15   \\       
66   &    694  &           J1349.3$-$3018  &                    IC 4329A   \\       
67   &    696  &           J1351.5$-$1814  &                     CTS 103   \\       
68   &    711  &           J1412.9$-$6522  &             Circinus Galaxy   \\       
69   &    712  &           J1413.2$-$0312  &                    NGC 5506   \\       
70   &    717  &           J1417.9$+$2507  &                    NGC 5548   \\       
71   &    719  &           J1419.0$-$2639  &                ESO 511$-$30   \\       
72   &    731  &           J1432.8$-$4412  &                    NGC 5643   \\       
73   &    733  &           J1433.9$+$0528  &                    NGC 5674   \\       
74   &    739  &           J1442.5$-$1715  &                    NGC 5728   \\       
75   &    751  &           J1457.8$-$4308  &                    IC 4518A   \\       
76   &    772  &           J1533.2$-$0836  &           MCG $-$1$-$40$-$1   \\       
77   &    783  &           J1548.5$-$1344  &                    NGC 5995   \\       
78   &    795  &           J1613.2$-$6043  &                LEDA 2793282   \\       
79   &    823  &           J1635.0$-$5804  &                ESO 137$-$34   \\       
80   &    841  &           J1652.9$+$0223  &                    NGC 6240   \\       
81   &    875  &           J1717.1$-$6249  &                    NGC 6300   \\       
82   &    896  &           J1737.5$-$2908  &     1RXS J173728.0$-$290759   \\       
83   &    970  &           J1824.3$-$5624  &                     IC 4709   \\       
84   &    986  &           J1836.9$-$5924  &                  Fairall 49   \\       
85   &    1042  &           J1937.5$-$0613  &                  LEDA 90334   \\       
86   &    1064  &           J2009.0$-$6103  &                    NGC 6860   \\       
87   &    1077  &         J2028.5$+$2543D1  &                    NGC 6921   \\       
88   &    1090  &           J2044.2$-$1045  &                     Mrk 509   \\       
89   &    1092  &           J2052.0$-$5704  &                     IC 5063   \\       
90   &    1127  &           J2148.3$-$3454  &                    NGC 7130   \\       
91   &    1133  &           J2200.9$+$1032  &                     Mrk 520   \\       
92   &    1135  &           J2201.9$-$3152  &                    NGC 7172   \\       
93   &    1157  &           J2235.9$-$2602  &                    NGC 7314   \\       
94   &    1161  &           J2236.7$-$1233  &                     Mrk 915   \\       
95   &    1182  &           J2303.3$+$0852  &                    NGC 7469   \\       
96   &    1183  &           J2304.8$-$0843  &                     Mrk 926   \\       
97   &    1188  &           J2318.4$-$4223  &                    NGC 7582   \\       
98   &    1198  &           J2328.9$+$0328  &                    NGC 7682
\end{longtable*}

\floattable
\renewcommand{\arraystretch}{1.1}
\begin{deluxetable}{ccccccccccc}
\tabletypesize{\scriptsize}
\thispagestyle{empty}
  \tablecaption{Values of Statistical Parameters}\label{tab_app:corr}
  \tablehead{
    (1) & (2) & (3) & (4) & (5) & (6) & (7) & (8) & (9)  \\ 
      $X$ & $Y$ & $\alpha$ & $\beta$ & $P_{\rm cor}$ & $\rho_{\rm P}$ & $\sigma_{\rm scat}$ &  Sample & Size
     }
    \startdata
\thispagestyle{empty} 
 $L_{\rm 14-150}$  & $\nu L^{\rm peak}_{\nu,\rm mm} $ & 
 $1.19^{+0.08}_{-0.05}$ & $-12.74^{+2.28}_{-3.26}$ & $4.8^{+293.2}_{-4.8}\times10^{-15}$ & $0.74\pm0.03$ & 0.36 & full & 98 \\
 $F_{\rm 14-150}$  & $\nu F^{\rm peak}_{\nu,\rm mm} $ & 
 $1.12^{+0.09}_{-0.08}$ & $-3.11^{+0.97}_{-0.81}$ & $3.3^{+77.2}_{-3.2}\times10^{-08}$ & $0.56\pm0.04$ & 0.38 & full & 98 \\ \hline
 $L_{\rm 2-10}$  & $\nu L^{\rm peak}_{\nu,\rm mm} $ &
 $1.08^{+0.06}_{-0.05}$ & $-7.41^{+2.25}_{-2.71}$ & $6.2^{+135.1}_{-6.0}\times10^{-13}$ & $0.67\pm0.03$ & 0.48 & full & 98 \\
 $F_{\rm 2-10}$  & $\nu F^{\rm peak}_{\nu,\rm mm} $ & 
 $0.97^{+0.08}_{-0.06}$ & $-4.32^{+0.83}_{-0.64}$ & $2.1^{+22.4}_{-1.9}\times10^{-07}$ & $0.48^{+0.04}_{-0.05}$ & 0.48 & full & 98 \\ \hline
 $\lambda L^{\rm AGN}_{\lambda, \rm 12\,\mu m}$  & $\nu L^{\rm peak}_{\nu,\rm mm} $ & 
 $0.84\pm0.04$ & $2.83^{+1.90}_{-1.80}$ & $8.6^{+96.2}_{-8.1}\times10^{-10}$ & $0.64\pm0.03$ & 0.59 & AGNs w/ $\lambda L^{\rm AGN}_{\lambda, \rm 12\,\mu m}$ & 85 \\
 $\lambda F^{\rm AGN}_{\lambda, \rm 12\,\mu m}$  & $\nu F^{\rm peak}_{\nu,\rm mm} $ & 
 $0.77^{+0.06}_{-0.05}$ & $-6.59^{+0.58}_{-0.49}$ & $8.8^{+38.2}_{-7.4}\times10^{-06}$ & $0.41\pm0.05$ & 0.57 & AGNs w/ $\lambda L^{\rm AGN}_{\lambda, \rm 12\,\mu m}$ & 85 \\ \hline
 $L_{\rm bol}$  & $\nu L^{\rm peak}_{\nu,\rm mm} $ & 
 $0.86^{+0.06}_{-0.05}$ & $0.74^{+2.07}_{-2.46}$ & $7.3^{+163.4}_{-7.2}\times10^{-09}$ & $0.60\pm0.04$ & 0.44 & AGNs w/ $M_{\rm BH}$ & 98 \\
 $F_{\rm bol}$  & $\nu F^{\rm peak}_{\nu,\rm mm} $ & 
 $0.77^{+0.06}_{-0.05}$ & $-7.41^{+0.58}_{-0.49}$ & $5.5^{+58.6}_{-5.2}\times10^{-05}$ & $0.39\pm0.05$ & 0.43 & AGNs w/ $M_{\rm BH}$ & 98 \\ \hline 
 $L_{\rm 14-150}$  & $\nu L^{\rm peak}_{\nu,\rm mm} $ & 
 $1.13^{+0.16}_{-0.11}$ & $-9.94^{+4.84}_{-7.00}$ & $4.3^{+14.3}_{-4.0}\times10^{-03}$ & $0.56^{+0.07}_{-0.09}$ & 0.33 & AGNs for the bias study & 25 \\ \hline
 $L_{\rm 14-150}$  & $\nu L^{\rm peak}_{\nu,\rm mm} $ & 
 $1.25^{+0.10}_{-0.08}$ & $-15.18^{+3.59}_{-4.32}$ & $2.3^{+43.3}_{-2.1}\times10^{-08}$ & $0.83^{+0.03}_{-0.04}$ & 0.38 & $\theta^{\rm ave}_{\rm beam} < 50$\,pc & 31 \\
 $F_{\rm 14-150}$  & $\nu F^{\rm peak}_{\nu,\rm mm} $ & 
 $1.12^{+0.14}_{-0.10}$ & $-3.07^{+1.40}_{-1.06}$ & $4.0^{+14.7}_{-3.1}\times10^{-03}$ & $0.56^{+0.06}_{-0.08}$ & 0.44 & $\theta^{\rm ave}_{\rm beam} < 50$\,pc & 31 \\
 $L_{\rm 14-150}$  & $\nu L^{\rm peak}_{\nu,\rm mm} $ & 
 $1.17$ & $-11.71^{+0.05}_{-0.06}$ & $2.7^{+58.0}_{-2.5}\times10^{-08}$ & $0.83^{+0.03}_{-0.04}$ & 0.39 & $\theta^{\rm ave}_{\rm beam} < 50$\,pc & 31 \\
 $L_{\rm 14-150}$  & $\nu L^{\rm peak}_{\nu,\rm mm} $ & 
 $1.10^{+0.14}_{-0.11}$ & $-8.62^{+4.78}_{-6.26}$ & $3.6^{+35.1}_{-3.2}\times10^{-05}$ & $0.73^{+0.06}_{-0.08}$ & 0.32 & $\theta^{\rm ave}_{\rm beam} > 100$\,pc & 27 \\
 $F_{\rm 14-150}$  & $\nu F^{\rm peak}_{\nu,\rm mm} $ & 
 $1.06^{+0.22}_{-0.17}$ & $-3.73^{+2.29}_{-1.76}$ & $1.5^{+5.6}_{-1.3}\times10^{-02}$ & $0.43^{+0.11}_{-0.13}$ & 0.32 & $\theta^{\rm ave}_{\rm beam} > 100$\,pc & 27 \\
 $L_{\rm 14-150}$  & $\nu L^{\rm peak}_{\nu,\rm mm} $ & 
 $1.17$ & $-11.72\pm0.06$ & $3.6^{+28.6}_{-3.3}\times10^{-05}$ & $0.73^{+0.06}_{-0.08}$ & 0.32 & $\theta^{\rm ave}_{\rm beam} > 100$\,pc & 27 \\
 \hline 
 $\nu L^{\rm peak}_{\nu,\rm mm}/L_{\rm 14-150}$  & $\theta^{\rm ave}_{\rm beam}$ & 
 $0.00065\pm0.00026$ & $-4.44\pm0.04$ & $1.9^{+2.1}_{-1.2}\times10^{-01}$ & $0.09\pm0.04$ & 0.39 & full & 98 \\ 
 %
 %
 %
 $\nu L^{\rm peak}_{\nu,\rm mm}/L_{\rm 14-150}$  & $\theta^{\rm ave}_{\rm beam}$ & 
 $0.00027^{+0.00086}_{-0.00081}$ & $-4.39^{+0.07}_{-0.08}$ & $5.9^{+2.8}_{-2.7}\times10^{-01}$ & $0.02\pm0.07$ & 0.40 & AGNs obs. high sens. & 49 \\ \hline 
 $L_{\rm 14-150}$  & $\nu L^{\rm peak}_{\nu,\rm mm} $ & 
 $1.21\pm0.11$ & $-13.31^{+4.68}_{-4.83}$ & $2.6^{+10.0}_{-2.3}\times10^{-03}$ & $0.61^{+0.08}_{-0.10}$ & 0.35  & AGNs w/ C or CB & 26 \\
 $L_{\rm 14-150}$  & $\nu L^{\rm peak}_{\nu,\rm mm} $ & 
 $0.96\pm0.06$ & $-2.29\pm2.72$ & $2.1^{+26.8}_{-2.0}\times10^{-05}$ & $0.74^{+0.06}_{-0.07}$ & 0.24 & AGNs w/ E or E plus some of the others & 28 \\
 $L_{\rm 14-150}$  & $\nu L^{\rm peak}_{\nu,\rm mm} $ & 
 $1.05^{+0.13}_{-0.11}$ & $-6.60^{+4.68}_{-5.47}$ & $7.3^{+50.1}_{-6.7}\times10^{-04}$ & $0.53^{+0.08}_{-0.10}$ & 0.27 & Unresolved Comp. & 38 \\ \hline  
 $\nu L^{\rm peak}_{\nu,\rm mm}/L_{\rm 14-150}$  & $R_{14-150}$ &
 $0.83^{+0.07}_{-0.06}$ & $-0.01^{+0.34}_{-0.31}$ & $1.6^{+15.1}_{-1.5}\times10^{-04}$ & $0.41\pm0.07$ & 0.37 & AGNs w/ $L_{\rm R}$ & 82 \\ \hline  
 $L_{\rm 14-150}$  & $\nu L^{\rm peak}_{\nu,\rm mm} $ & 
 $1.20^{+0.13}_{-0.10}$ & $-12.60^{+4.44}_{-5.55}$ & $3.0^{+17.2}_{-2.8}\times10^{-04}$ & $0.64^{+0.07}_{-0.08}$ & 0.30 & RL AGNs & 34 \\  
 $F_{\rm 14-150}$  & $\nu F^{\rm peak}_{\nu,\rm mm} $ & 
 $1.15^{+0.11}_{-0.10}$ & $-2.65^{+1.16}_{-1.02}$ & $2.0^{+6.9}_{-1.6}\times10^{-02}$ & $0.48^{+0.09}_{-0.10}$ & 0.33 & RL AGNs & 34 \\
 $L_{\rm 14-150}$  & $\nu L^{\rm peak}_{\nu,\rm mm} $ & 
 $1.19$ & $-12.38^{+0.05}_{-0.06}$ & $2.5^{+17.9}_{-2.3}\times10^{-04}$ & $0.64^{+0.07}_{-0.08}$ & 0.30 & RL AGNs & 34 \\
 $L_{\rm 14-150}$  & $\nu L^{\rm peak}_{\nu,\rm mm} $ & 
 $1.19^{+0.08}_{-0.06}$ & $-12.57^{+2.59}_{-3.30}$ & $3.4^{+47.6}_{-3.2}\times10^{-07}$ & $0.87\pm0.03$ & 0.23 & RQ AGNs & 35 \\ 
 $F_{\rm 14-150}$  & $\nu F^{\rm peak}_{\nu,\rm mm} $ & 
 $1.17^{+0.17}_{-0.13}$ & $-2.73^{+1.80}_{-1.30}$ & $8.6^{+61.2}_{-8.0}\times10^{-04}$ & $0.55^{+0.08}_{-0.09}$ & 0.24 & RQ AGNs & 35 \\ 
 $L_{\rm 14-150}$  & $\nu L^{\rm peak}_{\nu,\rm mm} $ & 
 $1.19$ & $-12.78\pm0.05$ & $2.9^{+32.3}_{-2.8}\times10^{-07}$ & $0.87^{+0.02}_{-0.03}$ & 0.23 & RQ AGNs & 35 \\ 
 \hline 
 $L_{\rm 14-150}$  & $\nu L^{\rm peak}_{\nu,\rm mm} $ & 
 $1.01^{+0.09}_{-0.08}$ & $-4.53^{+3.68}_{-3.87}$ & $3.0^{+19.0}_{-2.8}\times10^{-04}$ & $0.76^{+0.05}_{-0.06}$ & 0.35 & AGNs w/ $\alpha_{\rm mm} - \alpha^{\rm e}_{\rm mm}> 0.3$ & 19 \\ 
 $F_{\rm 14-150}$  & $\nu F^{\rm peak}_{\nu,\rm mm} $ & 
 $0.97^{+0.13}_{-0.12}$ & $-4.42^{+1.43}_{-1.29}$ & $1.4^{+5.4}_{-1.2}\times10^{-02}$ & $0.58^{+0.09}_{-0.11}$ & 0.31 & AGNs w/ $\alpha_{\rm mm} - \alpha^{\rm e}_{\rm mm}> 0.3$ & 19 \\ \hline
 $L_{\rm 14-150}$  & $\nu L^{\rm peak}_{\nu,\rm mm} $ & 
 $1.21\pm0.10$ & $-13.45^{+4.26}_{-4.40}$ & $4.3^{+45.0}_{-3.8}\times10^{-07}$ & $0.77^{+0.04}_{-0.05}$ & 0.40 & AGNs w/ high $\Sigma_{\rm mm}$ & 31 \\ 
 $F_{\rm 14-150}$  & $\nu F^{\rm peak}_{\nu,\rm mm} $ & 
 $1.06^{+0.11}_{-0.10}$ & $-3.58^{+1.04}_{-1.06}$ & $1.9^{+5.3}_{-1.5}\times10^{-02}$ & $0.51^{+0.07}_{-0.08}$ & 0.42 & AGNs w/ high $\Sigma_{\rm mm}$ & 31 \\ \hline 
 %
 %
 $L_{\rm 14-150}$  & $\nu L^{\rm peak}_{\nu,\rm mm} $ & 
 $1.14^{+0.08}_{-0.07}$ & $-10.57^{+3.08}_{-3.28}$ & $4.8^{+33.9}_{-4.4}\times10^{-05}$ & $0.67^{+0.05}_{-0.06}$ & 0.37 & AGNs w/ SFR(mm) $>$ SFR(IR) & 46 \\ 
 $F_{\rm 14-150}$  & $\nu F^{\rm peak}_{\nu,\rm mm} $ & 
 $1.14\pm0.09$ & $-2.82^{+0.97}_{-0.89}$ & $3.0^{+23.3}_{-2.7}\times10^{-04}$ & $0.52^{+0.07}_{-0.08}$ & 0.35 & AGNs w/ SFR(mm) $>$ SFR(IR) & 46 \\ \hline 
 $\nu L^{\rm peak}_{\nu,\rm mm} $ & $L_{\rm 14-150}$   & 
 $0.83\pm0.04$ & $10.79^{+1.62}_{-1.66}$ & $4.5^{+238.6}_{-4.4}\times10^{-15}$ & $0.74\pm0.03$ & 0.30 & full & 98 \\
 $\nu L^{\rm peak}_{\nu,\rm mm} $ & $L_{\rm 2-10}$  & 
 $0.93^{+0.04}_{-0.05}$ & $6.83^{+1.82}_{-1.68}$ & $4.6^{+115.4}_{-4.5}\times10^{-13}$ & $0.68\pm0.03$ & 0.45 & full & 98 \\  
 $\nu L^{\rm peak}_{\nu,\rm mm} $ & $\lambda L^{\rm AGN}_{\lambda, \rm 12\,\mu m}$ & 
 $1.19\pm0.06$ & $-3.36^{+2.34}_{-2.47}$ & $8.3^{+91.4}_{-7.9}\times10^{-10}$ & $0.64\pm0.03$ & 0.71 & AGNs w/ $\lambda L^{\rm AGN}_{\lambda, \rm 12\,\mu m}$ & 85 \\ 
 $\nu L^{\rm peak}_{\nu,\rm mm} $ & $L_{\rm bol}$  & 
 $1.16\pm0.07$ & $-0.87^{+2.69}_{-2.59}$ & $6.7^{+138.1}_{-6.6}\times10^{-09}$ & $0.60\pm0.04$ & 0.51 & AGNs w/ $M_{\rm BH}$ & 98 \\ \hline 
 $\nu L^{\rm peak}_{\nu,\rm mm} $ & $L_{\rm 14-150}$ & 
 $0.99^{+0.09}_{-0.08}$ & $4.56^{+3.03}_{-3.33}$ & $2.7^{+17.1}_{-2.5}\times10^{-04}$ & $0.76^{+0.05}_{-0.06}$ & 0.35 & AGNs w/ $\alpha_{\rm mm} - \alpha^{\rm e}_{\rm mm}> 0.3$ & 19 \\
 $\nu L^{\rm peak}_{\nu,\rm mm} $ & $L_{\rm 2-10}$  & 
 $1.20\pm0.09$ & $-3.95^{+3.53}_{-3.51}$ & $3.5^{+22.8}_{-3.1}\times10^{-04}$ & $0.75^{+0.04}_{-0.05}$ & 0.47 & AGNs w/ $\alpha_{\rm mm} - \alpha^{\rm e}_{\rm mm}> 0.3$ & 19 \\
 $\nu L^{\rm peak}_{\nu,\rm mm} $ & $\lambda L^{\rm AGN}_{\lambda, \rm 12\,\mu m}$   & 
 $1.06\pm0.09$ & $1.58^{+3.36}_{-3.66}$ & $5.0^{+10.4}_{-3.6}\times10^{-02}$ & $0.51^{+0.07}_{-0.10}$ & 0.56  & AGNs w/ $\alpha_{\rm mm} - \alpha^{\rm e}_{\rm mm}> 0.3$ & 18 \\
 $\nu L^{\rm peak}_{\nu,\rm mm} $ & $L_{\rm bol}$  & 
 $1.60^{+0.14}_{-0.15}$ & $-18.63^{+5.74}_{-5.56}$ & $7.4^{+9.4}_{-4.9}\times10^{-02}$ & $0.48^{+0.08}_{-0.07}$ & 0.88 & AGNs w/ $\alpha_{\rm mm} - \alpha^{\rm e}_{\rm mm}> 0.3$ & 19 \\ \hline 
 %
 $\nu L^{\rm peak}_{\nu,\rm mm} $ & $L_{\rm [O\,III]}$  & 
 $1.08^{+0.03}_{-0.05}$ & $-1.18^{+1.84}_{-1.32}$ & $3.3^{+3.6}_{-1.8}\times10^{-08}$ & $0.55\pm0.01$ & 0.78 & AGNs w/ [O III] fluxes & 90 \\ 
 $\nu F^{\rm peak}_{\nu,\rm mm} $ &  $F_{\rm [O\,III]}$ & 
 $1.21^{+0.06}_{-0.07}$ & $4.94^{+0.80}_{-1.08}$ & $1.5^{+0.9}_{-0.6}\times10^{-02}$ & $0.27\pm0.02$ & 0.81 & AGNs w/ [O III] fluxes & 90 \\ 
 $\nu L^{\rm peak}_{\nu,\rm mm} $ & $L_{\rm [Si\,VI]}$  & 
 $1.15^{+0.14}_{-0.09}$ & $-5.99^{+3.71}_{-5.34}$ & $2.0^{+7.3}_{-1.6}\times10^{-03}$ & $0.59^{+0.05}_{-0.07}$ & 0.86 & AGNs w/ [Si VI] fluxes & 17 \\ 
 $\nu F^{\rm peak}_{\nu,\rm mm} $ &  $F_{\rm [Si\,VI]}$ & 
 $0.86^{+0.19}_{-0.14}$ & $-2.08^{+2.69}_{-1.97}$ & $2.7^{+1.3}_{-1.1}\times10^{-01}$ & $0.30^{+0.09}_{-0.11}$ & 0.76 & AGNs w/ [Si VI] fluxes & 17 \\ 
 $\nu L^{\rm peak}_{\nu,\rm mm} $ & $L_{\rm [Si\,X]}$  & 
 $1.18^{+0.15}_{-0.11}$ & $-7.38^{+4.17}_{-5.81}$ & $8.2^{+44.0}_{-7.3}\times10^{-04}$ & $0.65^{+0.05}_{-0.08}$ & 0.82 & AGNs w/ [Si X] fluxes & 17 \\ 
 $\nu F^{\rm peak}_{\nu,\rm mm} $ &  $F_{\rm [Si\,X]}$ & 
 $0.76^{+0.16}_{-0.13}$ & $-4.01^{+2.32}_{-1.88}$ & $2.7^{+2.1}_{-1.4}\times10^{-01}$ & $0.31^{+0.11}_{-0.13}$ & 0.66 & AGNs w/ [Si X] fluxes & 17 \\  \hline 
 E500 & $\nu L^{\rm peak}_{\nu,\rm mm}/L_{\rm IR} $  & 
 $3.00\pm0.16$ & $-5.51\pm0.04$ & $5.0^{+2.8}_{-2.3}\times10^{-01}$ & $0.15\pm0.05$ & 0.85 & AGNs w/ E500 & 47 \\ \hline 
 $\Gamma$ & $\nu L^{\rm peak}_{\nu,\rm mm} $   & 
 $-2.58^{+5.62}_{-0.67}$ & $44.17^{+1.31}_{-11.00}$ & $5.7^{+2.8}_{-2.8}\times10^{-01}$ & $-0.00^{+0.10}_{-0.11}$ & 0.92 & HEdd AGNs & 33 \\ 
 $\Gamma$ & $\nu L^{\rm peak}_{\nu,\rm mm} $   & 
 $-3.34^{+6.35}_{-0.82}$ & $44.82^{+1.46}_{-11.53}$ & $7.4^{+1.8}_{-2.5}\times10^{-01}$ & $-0.07\pm0.09$ & 0.99 & MEdd AGNs  & 30 \\ 
 $\Gamma$ & $\nu L^{\rm peak}_{\nu,\rm mm} $   & 
 $-1.71^{+0.27}_{-0.29}$ & $41.58^{+0.46}_{-0.43}$ & $2.0^{+2.3}_{-1.2}\times10^{-01}$ & $-0.19^{+0.09}_{-0.10}$ & 0.53 & LEdd AGNs  & 35 \\ \hline 
 \enddata
 \thispagestyle{empty}
\tablenotetext{}{
Notes.--- (1,2,3,4) Investigated parameters, slope, and interception defined as $\log Y = \alpha \times \log X + \beta$. 
Particularly for $\theta^{\rm ave}_{\rm beam}$, $R_{\rm 14-150}$, E500, and $\Gamma$, we do not take their logarithms. 
(5,6) $p$-value and the Pearson's correlation coefficient. 
(7) Intrinsic scatter. 
(8,9) Sample used for the investigation and its size. 
}
\end{deluxetable}
\setlength{\topmargin}{0in}

\addtocounter{table}{-1}

\floattable
\renewcommand{\arraystretch}{1.1}
\begin{deluxetable}{ccccccccccc}
\tabletypesize{\scriptsize}
\thispagestyle{empty}
  \tablecaption{Values of Statistical Parameters}\label{tab_app:corr}
  \tablehead{
    (1) & (2) & (3) & (4) & (5) & (6) & (7) & (8) & (9)  \\ 
      $X$ & $Y$ & $\alpha$ & $\beta$ & $P_{\rm cor}$ & $\rho_{\rm P}$ & $\sigma_{\rm scat}$ &  Sample & Size
     }
    \startdata
    \thispagestyle{empty} 
 $L^{\rm max}_{\rm kin}$  & $\nu L^{\rm peak}_{\nu,\rm mm} $ & 
  $0.39\pm0.03$ & $21.88\pm1.23$ & $9.6^{+8.2}_{-5.5}\times10^{-03}$ & $0.83\pm0.02$ & 0.63 & AGNs w/ X-ray OF info. & 14 \\ 
 $F^{\rm max}_{\rm kin}$  & $\nu F^{\rm peak}_{\nu,\rm mm} $ &
 $0.22^{+0.04}_{-0.02}$ & $-12.19^{+0.37}_{-0.16}$ & $4.5^{+4.0}_{-1.9}\times10^{-02}$ & $0.63^{+0.06}_{-0.05}$ & 0.43  & AGNs w/ X-ray OF info. & 14  \\ \hline 
 $L^{\rm opt}_{\rm kin}$  & $\nu L^{\rm peak}_{\nu,\rm mm} $ & 
 $0.47^{+0.09}_{-0.87}$ & $19.99^{+35.17}_{-3.80}$ & $4.3^{+3.8}_{-3.0}\times10^{-01}$ & $0.21^{+0.19}_{-0.22}$ & 0.34 & AGNs w/ [O III] OF info. & 18  \\ 
 $F^{\rm opt}_{\rm kin}$  & $\nu F^{\rm peak}_{\nu,\rm mm} $ & 
 $0.41^{+0.07}_{-0.06}$ & $-9.12^{+0.94}_{-0.76}$ & $1.2^{+2.8}_{-1.0}\times10^{-01}$ & $0.41^{+0.15}_{-0.16}$ & 0.18 & AGNs w/ [O III] OF info. & 18  \\ \hline   
 $L_{\rm 14-150}$  & $\nu L^{\rm peak}_{\nu,\rm mm} $ & 
 $1.22^{+0.10}_{-0.08}$ & $-13.98^{+3.27}_{-4.44}$ & $1.4^{+35.1}_{-1.4}\times10^{-09}$ & $0.84^{+0.03}_{-0.04}$ & 0.30 & HEdd AGNs & 33 \\ 
 $F_{\rm 14-150}$  & $\nu F^{\rm peak}_{\nu,\rm mm} $ & 
 $0.86^{+0.08}_{-0.06}$ & $-5.59^{+0.79}_{-0.67}$ & $8.1^{+16.3}_{-5.5}\times10^{-03}$ & $0.58^{+0.06}_{-0.07}$ & 0.30 & HEdd AGNs & 33 \\ 
 $L_{\rm 14-150}$  & $\nu L^{\rm peak}_{\nu,\rm mm} $ & 
 $1.15$ & $-10.76\pm0.04$ & $1.3^{+36.7}_{-1.3}\times10^{-09}$ & $0.84^{+0.03}_{-0.04}$ & 0.30 & HEdd AGNs & 33 \\ 
 $L_{\rm 14-150}$  & $\nu L^{\rm peak}_{\nu,\rm mm} $ & 
 $1.32^{+0.12}_{-0.10}$ & $-18.40^{+4.45}_{-5.12}$ & $1.1^{+7.6}_{-1.0}\times10^{-04}$ & $0.77\pm0.05$ & 0.38 & MEdd AGNs & 30 \\ 
 $F_{\rm 14-150}$  & $\nu F^{\rm peak}_{\nu,\rm mm} $ & 
 $0.94^{+0.11}_{-0.09}$ & $-5.03^{+1.15}_{-0.96}$ & $2.4^{+8.1}_{-1.9}\times10^{-02}$ & $0.50^{+0.09}_{-0.10}$ & 0.38 & MEdd AGNs & 30 \\ 
 $L_{\rm 14-150}$  & $\nu L^{\rm peak}_{\nu,\rm mm} $ & 
 $1.15$ & $-10.96\pm0.06$ & $1.1^{+8.4}_{-1.0}\times10^{-04}$ & $0.77\pm0.05$ & 0.41 & MEdd AGNs & 30 \\ 
 $L_{\rm 14-150}$  & $\nu L^{\rm peak}_{\nu,\rm mm} $ & 
 $0.91^{+0.11}_{-0.10}$ & $-0.48^{+4.18}_{-4.77}$ & $3.4^{+17.8}_{-2.9}\times10^{-04}$ & $0.58^{+0.06}_{-0.07}$ & 0.31 & LEdd AGNs & 35 \\ 
 $F_{\rm 14-150}$  & $\nu F^{\rm peak}_{\nu,\rm mm} $ & 
 $1.11^{+0.12}_{-0.11}$ & $-3.12^{+1.25}_{-1.12}$ & $7.4^{+24.6}_{-6.0}\times10^{-03}$ & $0.53\pm0.08$ & 0.34 & LEdd AGNs & 35 \\ 
 $L_{\rm 14-150}$  & $\nu L^{\rm peak}_{\nu,\rm mm} $ & 
 $1.15$ & $-10.84\pm0.06$ & $3.5^{+21.8}_{-3.1}\times10^{-04}$ & $0.58^{+0.07}_{-0.08}$ & 0.33 & LEdd AGNs & 35 \\ \hline 
 $L_{\rm 14-150}$  & $\nu L^{\rm peak}_{\nu,\rm mm} $ & 
 $1.27^{+0.09}_{-0.06}$ & $-15.94^{+2.63}_{-3.76}$ & $1.8^{+13.1}_{-1.6}\times10^{-11}$ & $0.90^{+0.01}_{-0.02}$ & 0.36 & type-1 AGNs & 33 \\ 
 $F_{\rm 14-150}$  & $\nu F^{\rm peak}_{\nu,\rm mm} $ & 
 $1.31^{+0.20}_{-0.12}$ & $-1.06^{+2.12}_{-1.28}$ & $3.1^{+6.3}_{-2.1}\times10^{-03}$ & $0.49\pm0.05$ & 0.44 & type-1 AGNs & 33 \\ 
 $L_{\rm 14-150}$  & $\nu L^{\rm peak}_{\nu,\rm mm} $ & 
 $1.19$ & $-12.54\pm0.03$ & $2.0^{+13.9}_{-1.8}\times10^{-11}$ & $0.90^{+0.01}_{-0.02}$ & 0.36 & type-1 AGNs & 33 \\ 
 $L_{\rm 14-150}$  & $\nu L^{\rm peak}_{\nu,\rm mm} $ & 
 $1.12^{+0.10}_{-0.08}$ & $-9.55^{+3.56}_{-4.20}$ & $1.3^{+24.2}_{-1.3}\times10^{-06}$ & $0.60^{+0.05}_{-0.06}$ & 0.35 & type-2 AGNs & 65 \\ 
 $F_{\rm 14-150}$  & $\nu F^{\rm peak}_{\nu,\rm mm} $ & 
 $1.07^{+0.09}_{-0.08}$ & $-3.62^{+0.95}_{-0.80}$ & $3.2^{+38.8}_{-3.0}\times10^{-06}$ & $0.58^{+0.05}_{-0.06}$ & 0.35  & type-2 AGNs & 65 \\ 
 $L_{\rm 14-150}$  & $\nu L^{\rm peak}_{\nu,\rm mm} $ & 
 $1.19$ & $-12.60\pm0.05$ & $1.3^{+24.4}_{-1.3}\times10^{-06}$ & $0.59^{+0.05}_{-0.06}$ & 0.35 & type-2 AGNs & 65 \\ \hline 
 $L_{\rm 14-150}$  & $\nu L^{\rm peak}_{\nu,\rm mm} $ & 
 $1.24^{+0.08}_{-0.05}$ & $-14.64^{+2.34}_{-3.69}$ & $8.6^{+73.6}_{-7.6}\times10^{-14}$ & $0.88^{+0.01}_{-0.02}$ & 0.37 & low-NH AGNs & 48 \\ 
 $F_{\rm 14-150}$  & $\nu F^{\rm peak}_{\nu,\rm mm} $ & 
 $1.28^{+0.13}_{-0.09}$ & $-1.32^{+1.37}_{-0.99}$ & $4.1^{+13.8}_{-3.2}\times10^{-05}$ & $0.56\pm0.04$ & 0.41 & low-NH AGNs & 48 \\ 
 $L_{\rm 14-150}$  & $\nu L^{\rm peak}_{\nu,\rm mm} $ & 
 $1.18$ & $-12.07^{+0.02}_{-0.03}$ & $1.0^{+6.8}_{-0.9}\times10^{-13}$ & $0.87^{+0.01}_{-0.02}$ & 0.36 & low-NH AGNs & 48 \\
 $L_{\rm 14-150}$  & $\nu L^{\rm peak}_{\nu,\rm mm} $ & 
 $1.12^{+0.11}_{-0.10}$ & $-9.80^{+4.30}_{-4.71}$ & $3.4^{+34.9}_{-3.2}\times10^{-05}$ & $0.58^{+0.06}_{-0.07}$ & 0.33 & high-NH AGNs & 50 \\ 
 $F_{\rm 14-150}$  & $\nu F^{\rm peak}_{\nu,\rm mm} $ & 
 $1.08^{+0.11}_{-0.09}$ & $-3.58^{+1.19}_{-0.92}$ & $7.5^{+70.7}_{-7.0}\times10^{-05}$ & $0.57^{+0.06}_{-0.07}$ & 0.33 & high-NH AGNs & 50 \\ 
 $L_{\rm 14-150}$  & $\nu L^{\rm peak}_{\nu,\rm mm} $ & 
 $1.18$ & $-12.23\pm0.06$ & $3.0^{+37.4}_{-2.9}\times10^{-05}$ & $0.58^{+0.06}_{-0.07}$ & 0.33 & high-NH AGNs & 50 \\
 \hline
 \enddata
 \thispagestyle{empty}
\tablenotetext{}{
Notes.--- (1,2,3,4) Investigated parameters, slope, and interception defined as $\log Y = \alpha \times \log X + \beta$. Particularly for $\theta^{\rm ave}_{\rm beam}$, $R_{\rm 14-150}$, E500, and $\Gamma$, we do not take their logarithms. 
(5,6) $p$-value and the Pearson's correlation coefficient. 
(7) Intrinsic scatter. 
(8,9) Sample used for the investigation and its size. 
}
\end{deluxetable}
\setlength{\topmargin}{0in}

\clearpage

\section{
    Correlations of Mm-wave and AGN Fluxes 
}\label{sec_app:flux}

\begin{figure*}[!h]
    \centering
    \includegraphics[width=8.5cm]{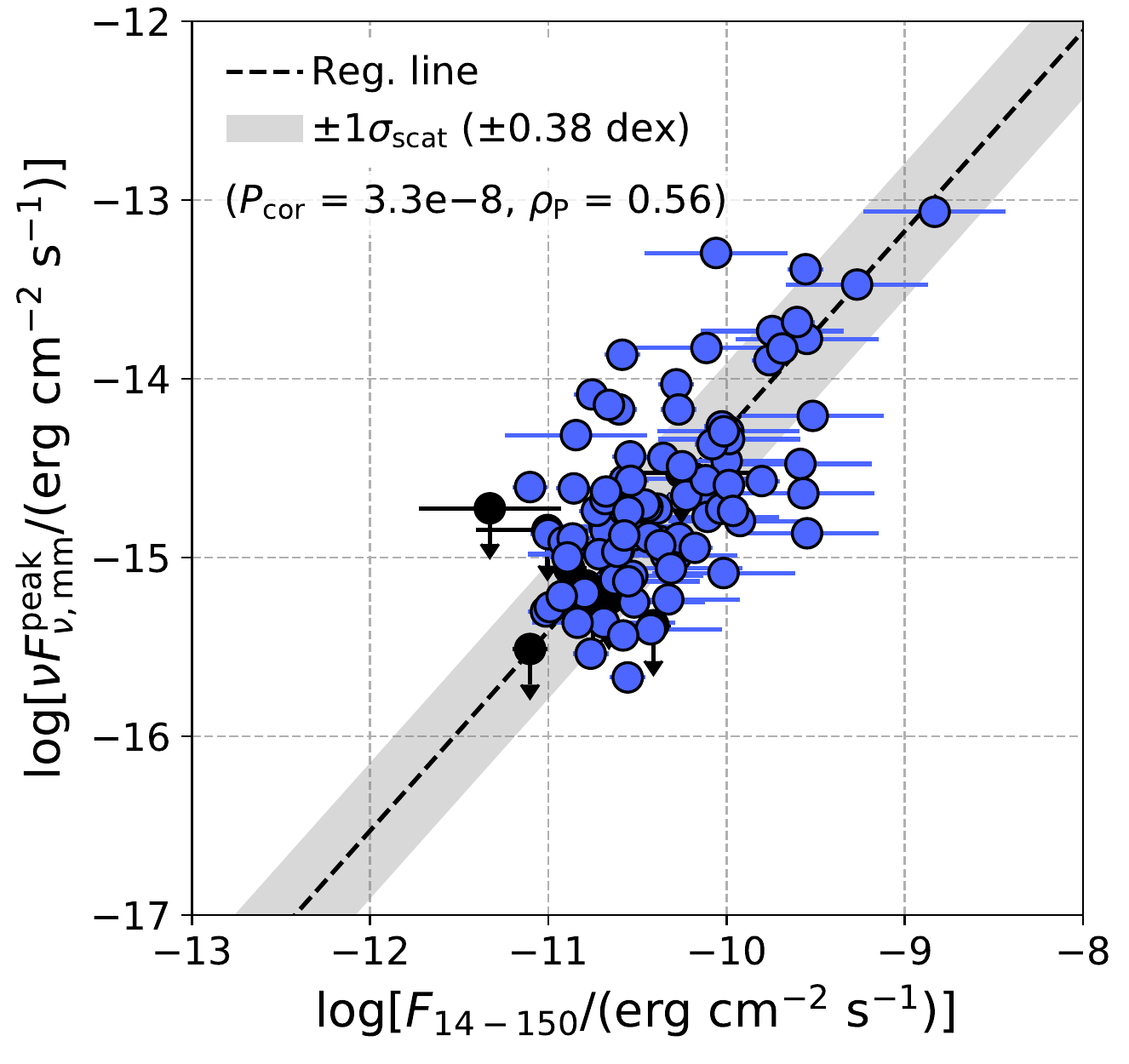}
    \includegraphics[width=8.5cm]{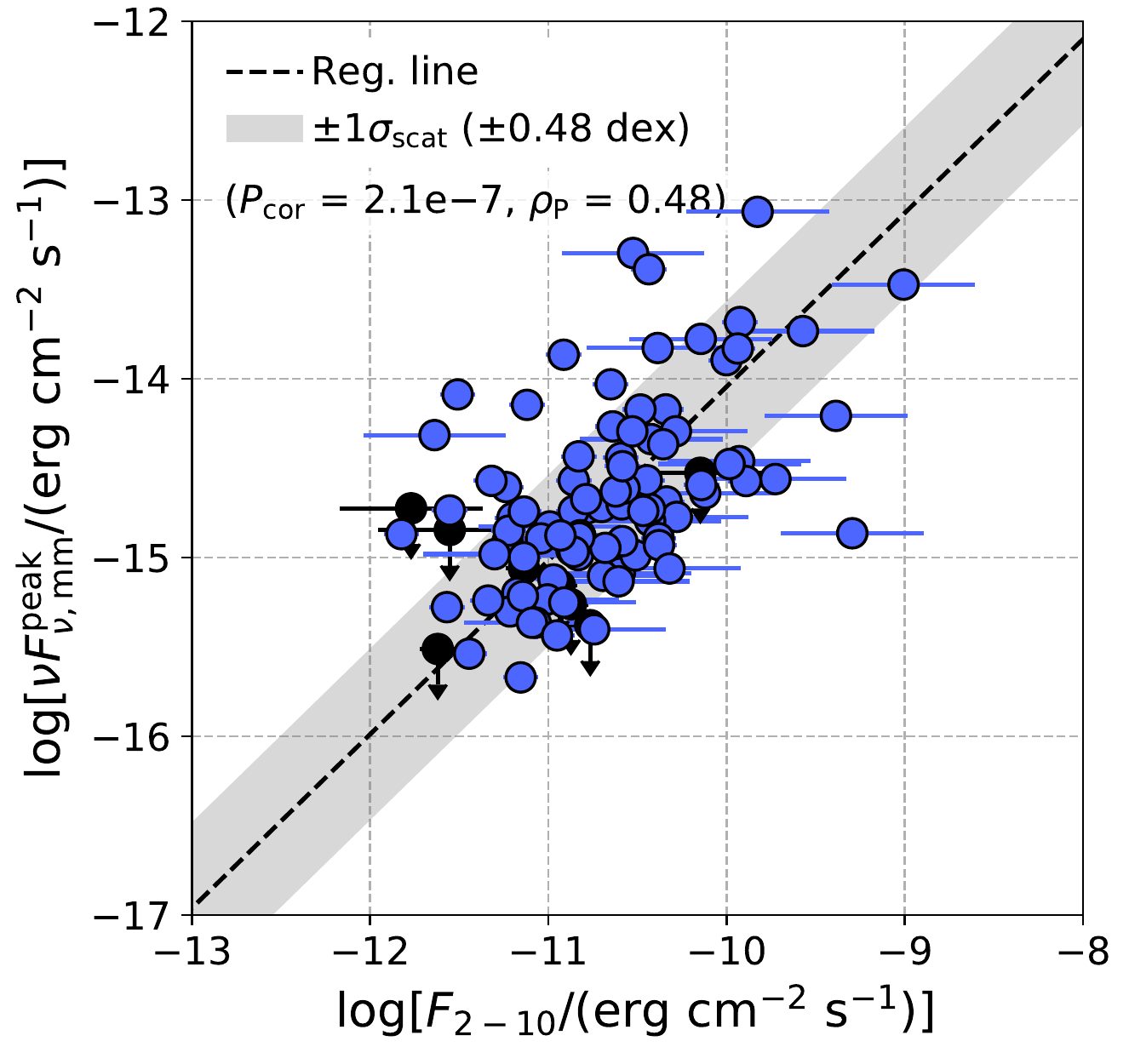} 
    \includegraphics[width=8.5cm]{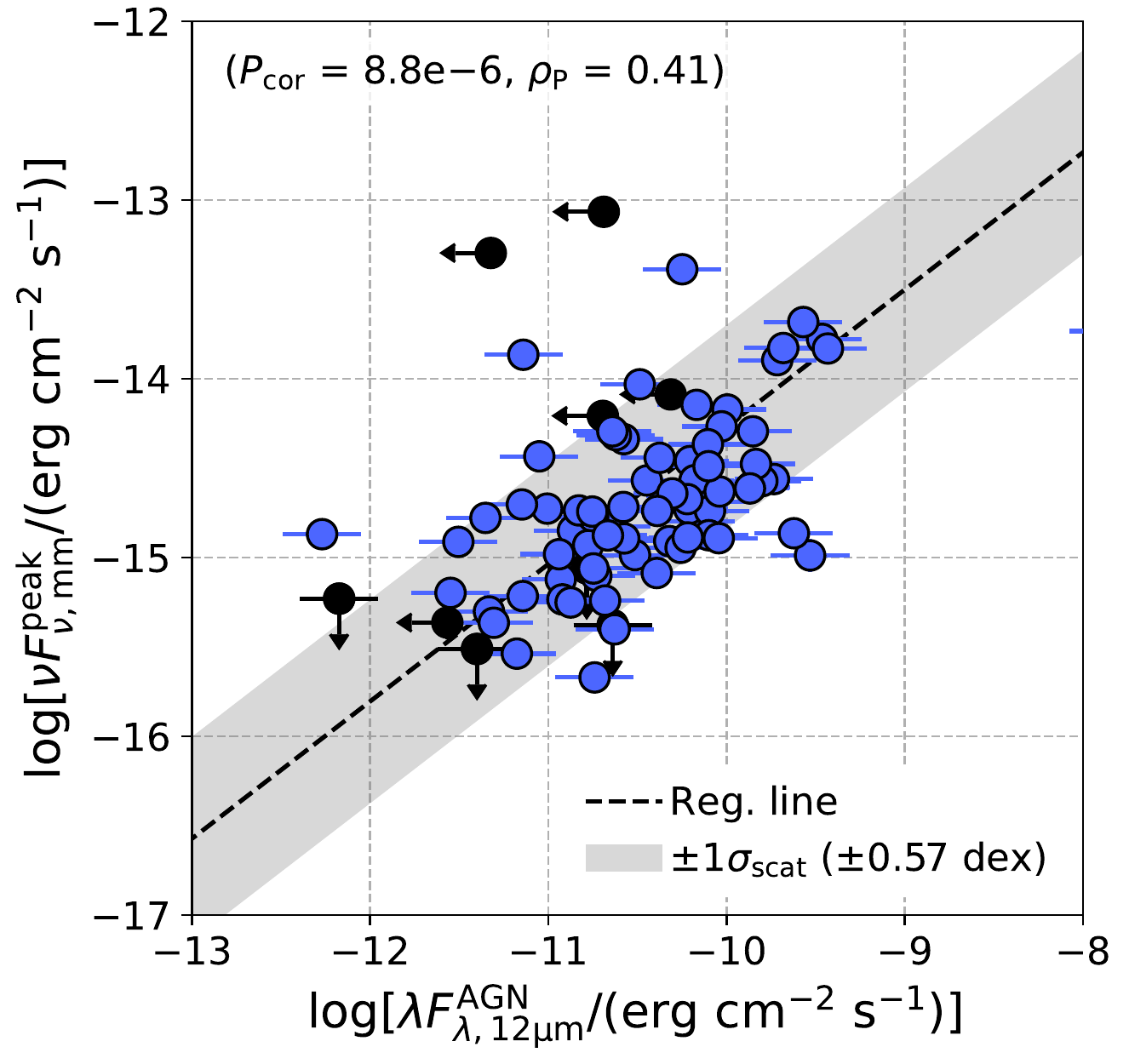}
    \includegraphics[width=8.5cm]{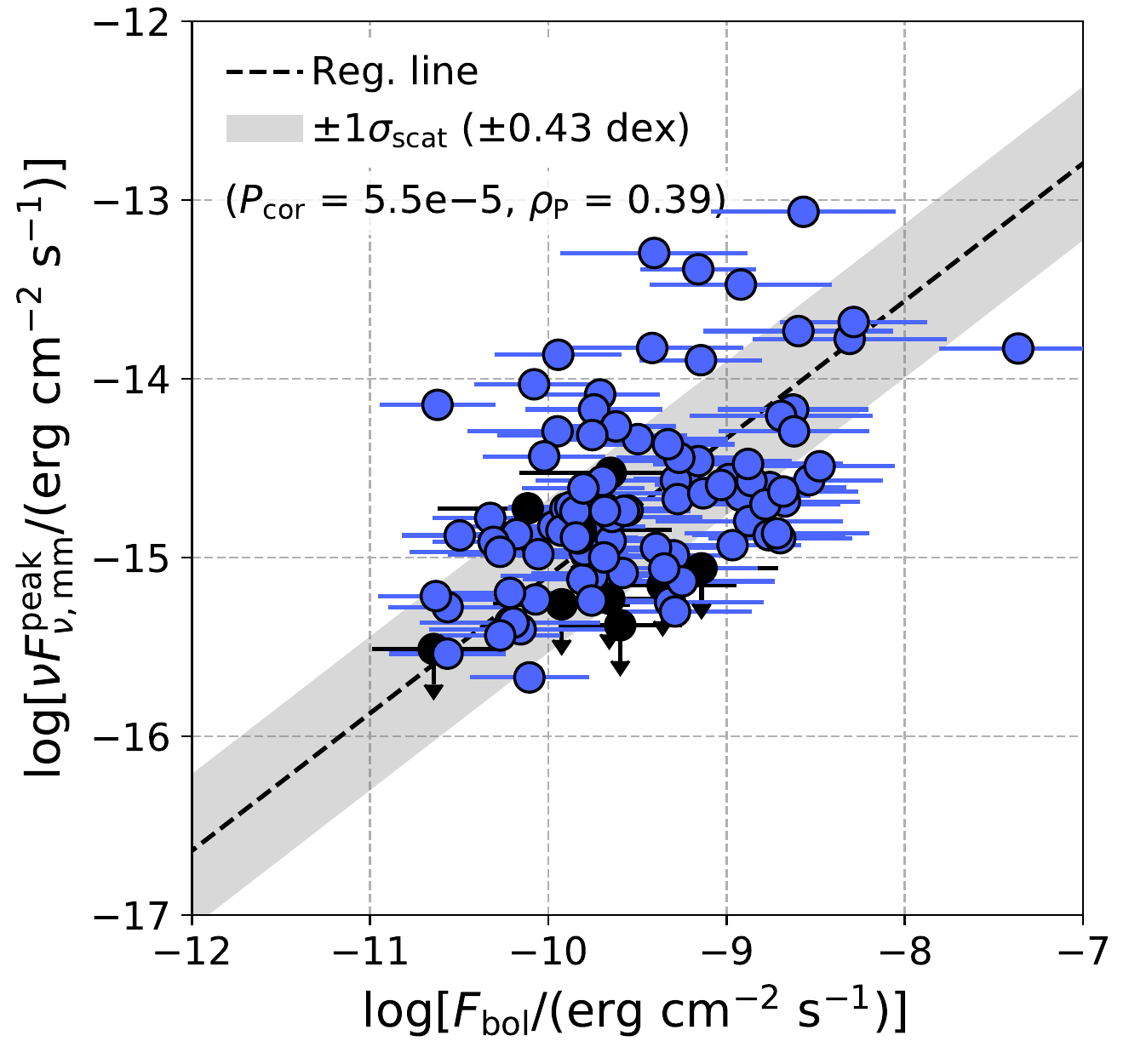}
    \caption{
    Mm-wave flux versus 14--150\,keV, 2--10\,keV, 12\,$\mu$m, and bolometric fluxes. 
    The 12\,$\mu$m fluxes take into account only an AGN component determined by the SED analysis of \cite{Ich19}.
    AGNs with upper limits are shown as black circles. 
    The black dashed lines indicate the best-fit linear regression lines, and the gray regions denote the $\pm1\sigma_{\rm scat}$ ranges.
    } 
    \label{fig_app:flux}
\end{figure*}

We show correlations of the mm-wave and AGN fluxes in Figure~\ref{fig_app:flux}


\clearpage 

\section{
    Observed mm-wave Emission Versus Expected One from Host galaxy
}\label{sec_app:sed}

In this section, we discuss the contribution of host-galaxy emission in the mm-wave band by comparing the observed mm-wave flux with that expected from the host-galaxy SED model determined in the IR band \cite[Section~\ref{sec:anc}; ][]{Ich19}. 
This discussion was introduced in Section~\ref{sec:sf}, but was omitted there.

\begin{figure}
    \centering
    \includegraphics[width=8.3cm]{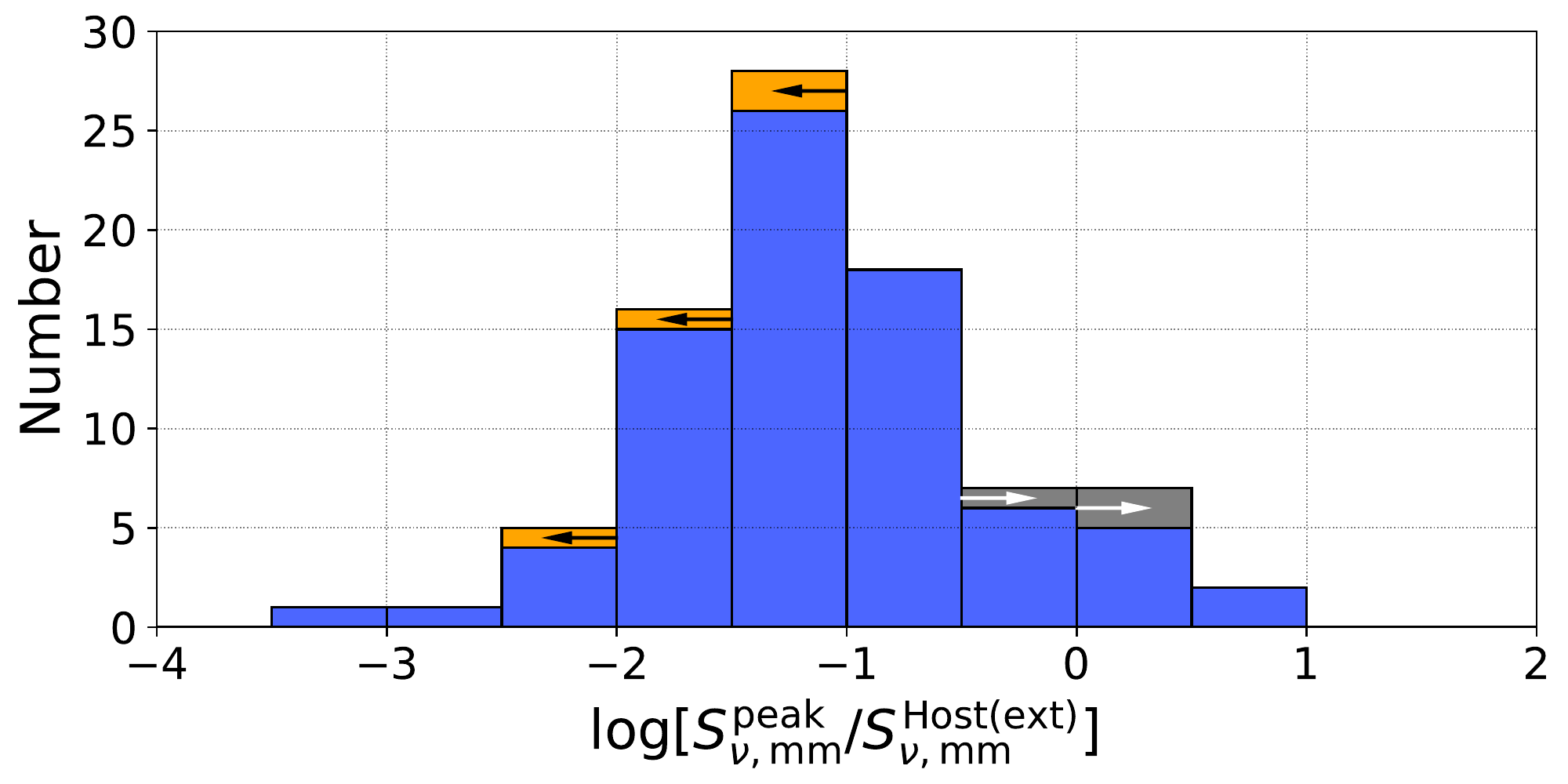}
    \caption{
    Histogram of the logarithmic ratio between the 
    peak mm-wave flux density and 
    the one predicted by extrapolating 
    the host-galaxy SED model. 
    Sources with upper and lower limits are indicated in orange and gray, respectively.
    }
    \label{fig_app:ratio}
\end{figure}

As a starting point for the discussion, we calculate the ratio between the observed mm-wave flux density and the extrapolated one from the IR host-galaxy SED model \citep{Ich19}. 
As the calculated ratios are summarized as a histogram in Figure~\ref{fig_app:ratio}, we find that the extrapolated flux typically exceeds the observed one by one order of magnitude. This indicates that the host-galaxy contribution is overestimated and some factors need to be considered to reconcile the discrepancy.

A crucial factor is the difference in the scales probed by the mm-wave observations and the IR models. Representatively, we consider their probed scales to be $\approx$ 0\farcs6 and 6\arcsec, respectively. 
The former corresponds to the angular resolutions achieved for most of our AGNs (Figure~\ref{fig:dist_vs_res}), and the latter is based on a result of \cite{Mus14}, who found that in almost all of their nearby AGNs ($z <$ 0.05) selected by BAT, a significant fraction of their FIR emission ($>$ 50\%) is concentrated within a 6\arcsec\ aperture. 
To infer the host-galaxy flux on the scale of the ALMA beam from the IR SED model, we refer to a result by \cite{Jen17}. The authors estimated the surface brightness radial profile of PAH emission for nearby AGNs in the central regions within $\sim$ 10--1000 pc, matching the scales which we focus on. 
Given that the PAH emission traces the SF region and the IR part of the host-galaxy emission is due to SF, the obtained profile can be used for inference. 
We note that PAH molecules may be destroyed in regions close to the AGN, and therefore may miss the SF regions around it. Thus, the PAH emission would indicate the lower contribution limit from the SF emission. 
Finally, by adopting a typical radial profile of PAH emission ($r^{-1.1}$) reported by \cite{Jen17}, the ratio of SF luminosities between the scales of 0\farcs6\ and 6\arcsec\ is calculated to be $\sim$ 0.1. 
By considering this factor, $S^{\rm peak}_{\nu,\rm mm}/S^{\rm Host(ext)}_{\nu, \rm mm,corr} $, corrected for the difference in the spatial scales, is $ \sim 0.1 \times 1/0.1 = 1$. 
This suggests that the galaxy component may be significant, but the observed spectral indices are inconsistent with that expectation (Figure~\ref{fig:freq1} and Section~\ref{sec:posindex}), suggesting that other factors need to be considered additionally.
 
Two ideas could mitigate the remaining discrepancy. First, we may underestimate the spatial scale of the IR emission, although we have adopted 6\arcsec\ following the result of \cite{Mus14}. 
The angular size of 6\arcsec\ was derived for FIR emission at 70\,$\mu$m, but this might not apply to emission at longer wavelengths.
A supportive result was reported by \cite{Shi16}. 
Using PACS and SPIRE data of BAT-selected nearby AGNs ($z < 0.05$), they found that a correlation between 70\,$\mu$m and 500\,$\mu$m is weak and suggested that the emission in these bands may not be closely related to each other. Therefore, if 
the actual size at longer wavelengths $> 500$\,$\mu$m is larger than we have assumed, the discrepancy can be reconciled somewhat. 
To constrain the size at such wavelengths, high-resolution FIR studies, for example, using ALMA, are crucial.

Second, the high-resolution ALMA interferometer observations may miss a fraction of emission by resolving out extended emission originating from SF. 
In fact, for a non-negligible fraction of our targets ($\sim$ 40\%), the maximum recoverable scales, adopted in the ALMA observatory team as a criterion of measuring 10\% of the total flux density of a uniform disk, are less than 3\arcsec.  Although the remaining objects were observed on larger scales up to $\sim$ 6\arcsec\ with a few exceptions with scales of 8\arcsec--9\arcsec, to constrain how much emission can be resolved out, a more detailed analysis (e.g., simulation of observations) is needed, and we may also need additional data obtained with larger beam sizes, recovering resolved-out emission.






\begin{figure*}
    \centering
    \includegraphics[width=5.5cm]{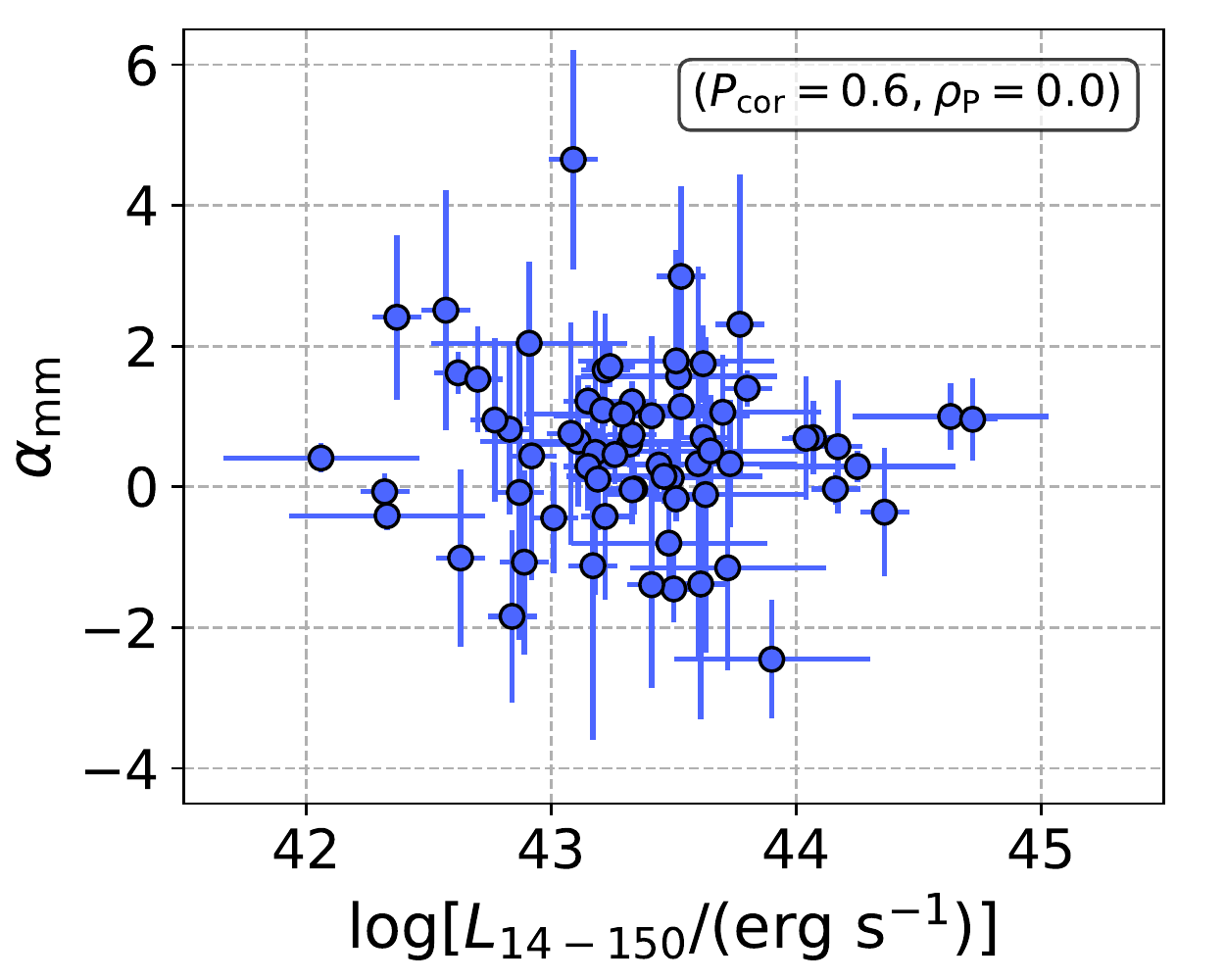}
    \includegraphics[width=5.5cm]{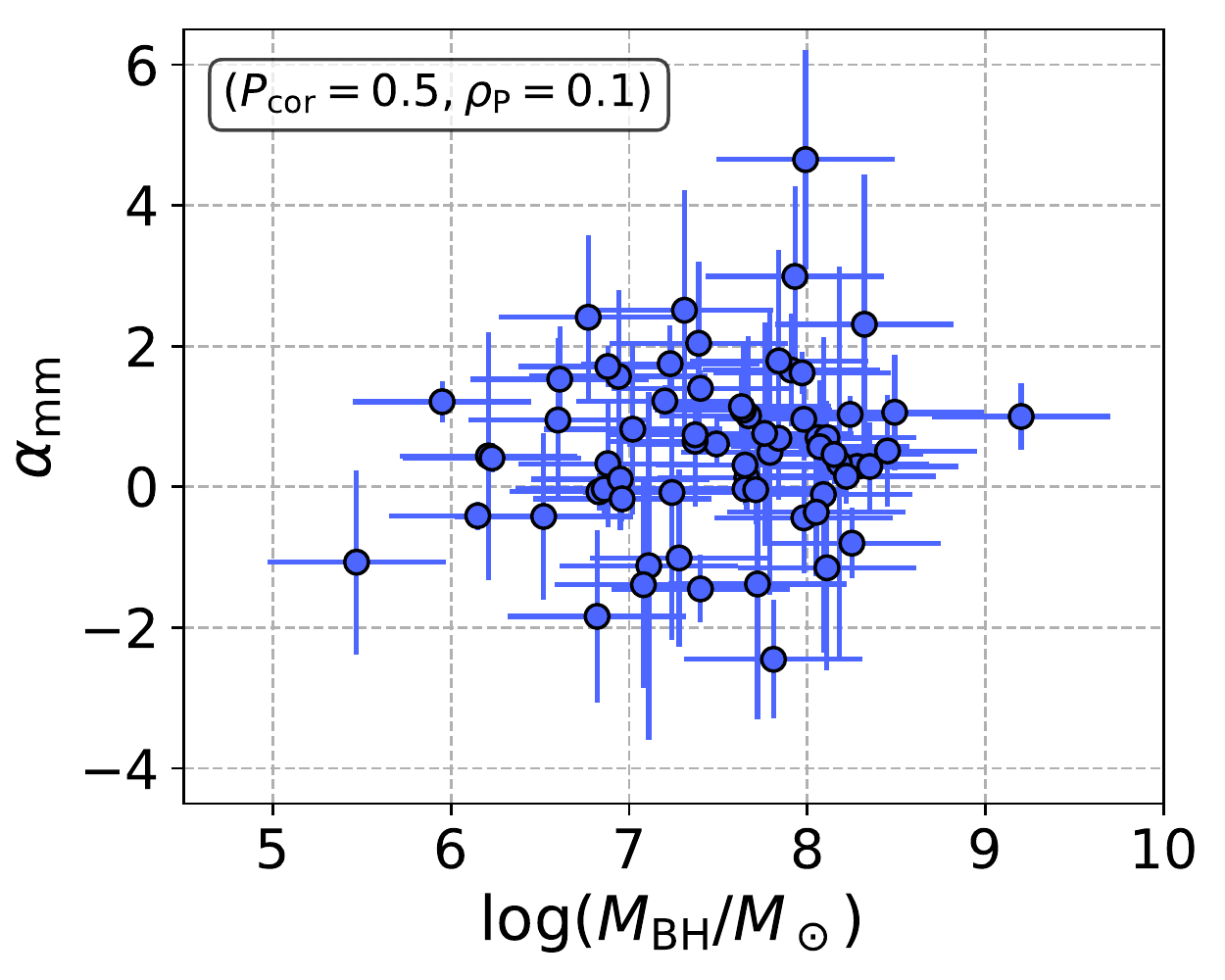}
    \includegraphics[width=5.5cm]{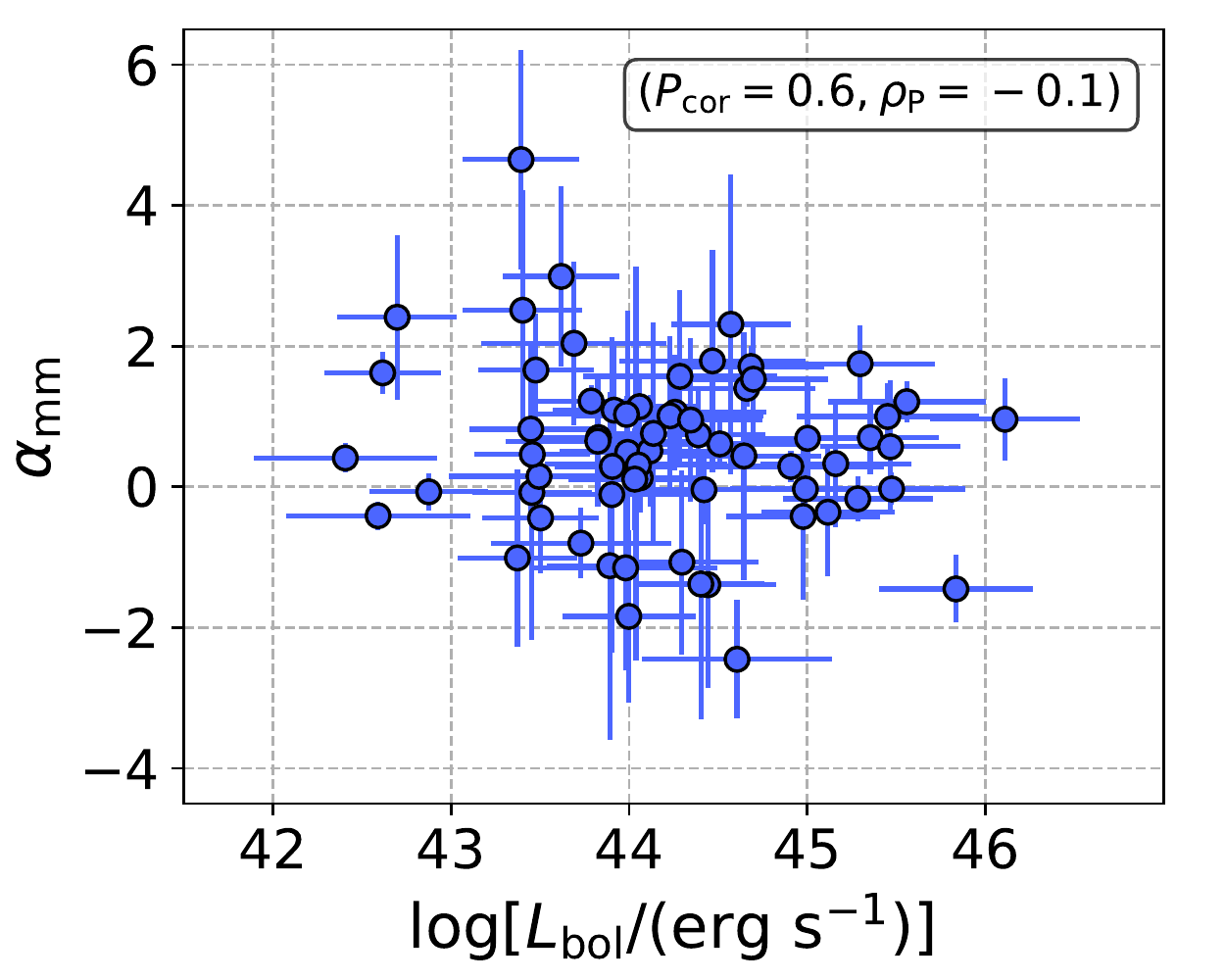}    
    \includegraphics[width=5.5cm]{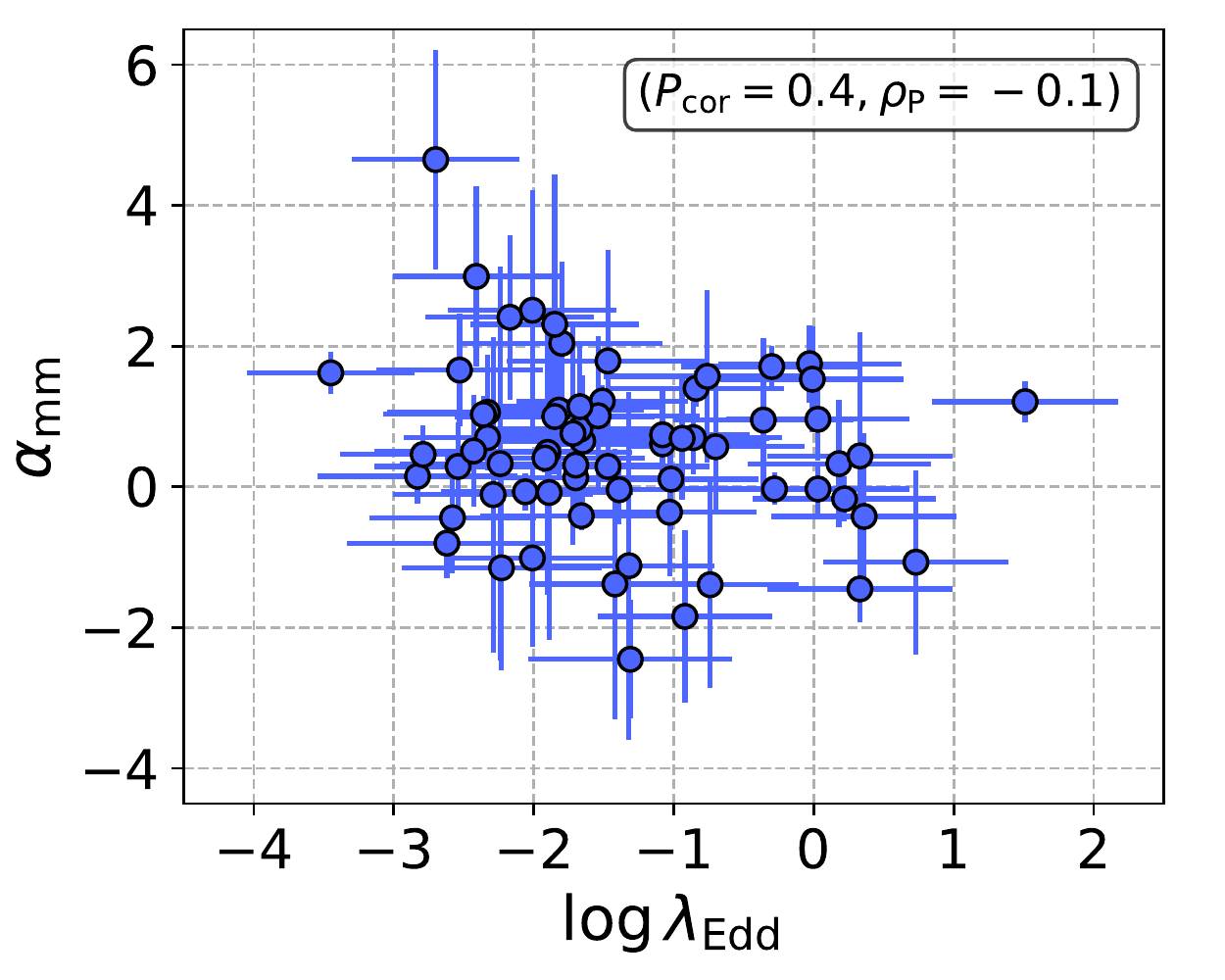}
    \includegraphics[width=5.5cm]{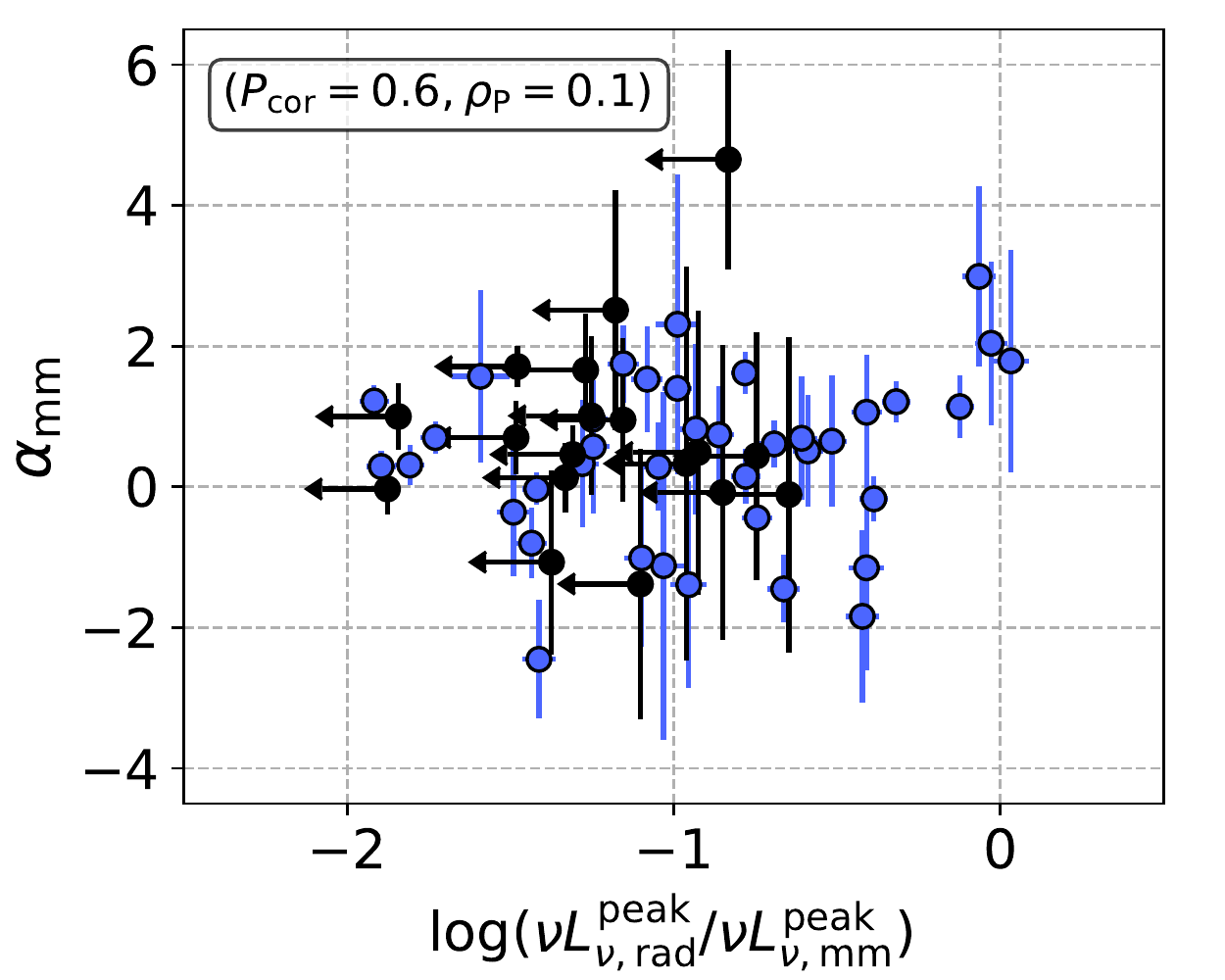}
    \includegraphics[width=5.5cm]{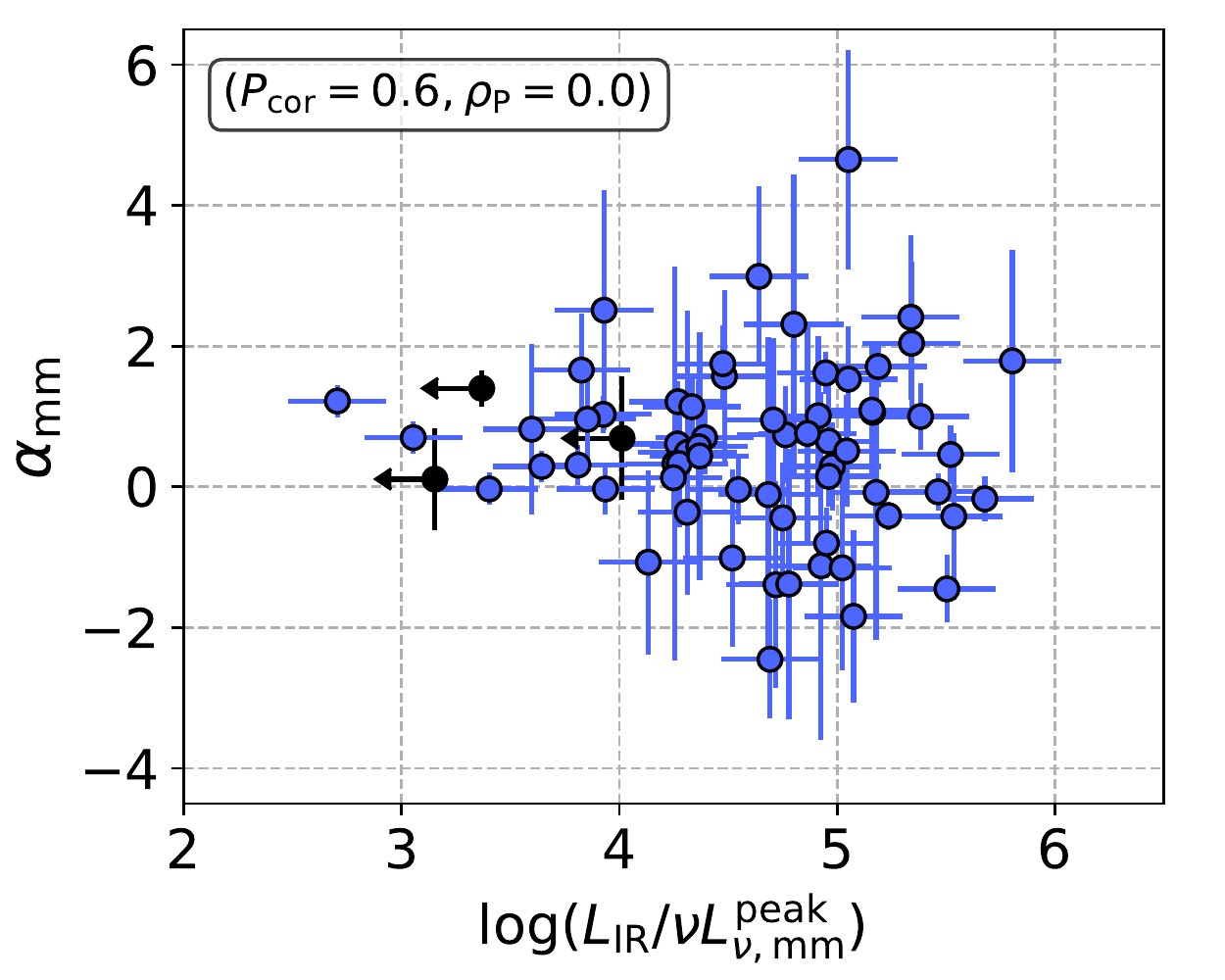}        
    \caption{
    Scatter plots of the spectral index for the 14--150\,keV luminosity,
    the black hole mass, the bolometric luminosity, the Eddington ratio, 
    the radio-to-mm-wave luminosity ratio, and 
    the IR(host-galaxy)-to-mm-wave luminosity ratio. 
    In each figure, a $p$-value ($P_{\rm cor}$) and a Pearson's correlation coefficient ($\rho_{\rm P}$) are indicated within a box. No correlations are found. 
    }
    \label{fig_app:indices}
\end{figure*}

\section{Spectral Index versus Other Physical Parameters}\label{sec_app:dep_inedx}

We briefly summarize some results obtained by evaluating the relations of the spectral index with some AGN parameters of $L_{\rm 14-150}$, $M_{\rm BH}$, $L_{\rm bol}$, and $\lambda_{\rm Edd}$.
Furthermore, we examine dependencies on the relative strengths 
of $\nu L_{\nu,{\rm rad}}/\nu L^{\rm peak}_{\nu,{\rm mm}}$ and 
$L_{\rm IR}/\nu L^{\rm peak}_{\nu,{\rm mm}}$, which are adopted as 
proxies of the contributions of cm-wave and SF spectral components to the mm-wave emission, respectively. 
Figure~\ref{fig_app:indices} shows all scatter plots. 
No correlation is found in any of the combinations. 
However, even if there is an intrinsic correlation, its confirmation 
may be hampered by the large uncertainties in the spectral indices. 

\section{Constraining an Unresolved Mm-wave Component by Visibility Fitting}\label{sec_app:fit}

\begin{figure}
    \centering
    \includegraphics[width=7.5cm]{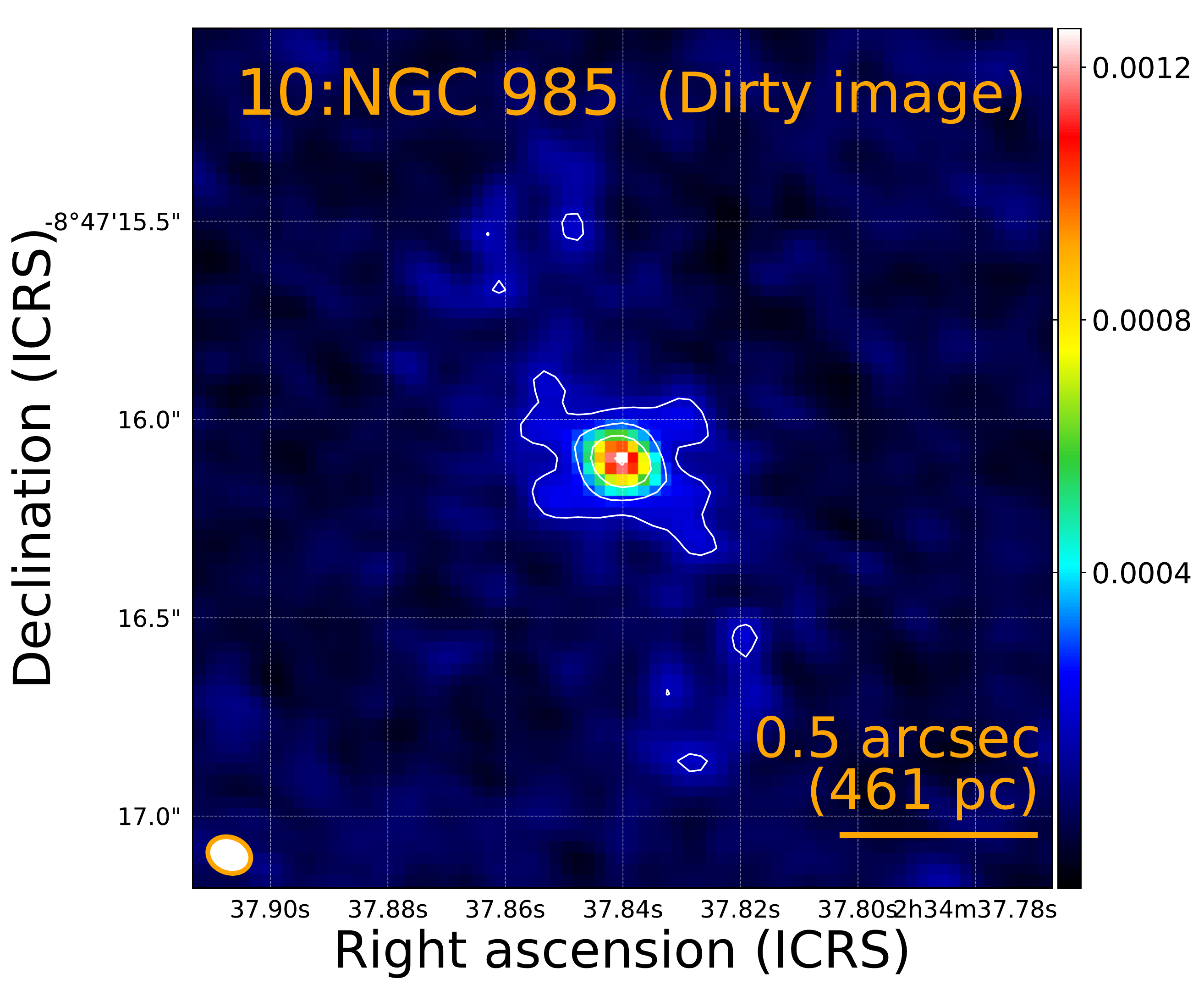}
    \includegraphics[width=7.5cm]{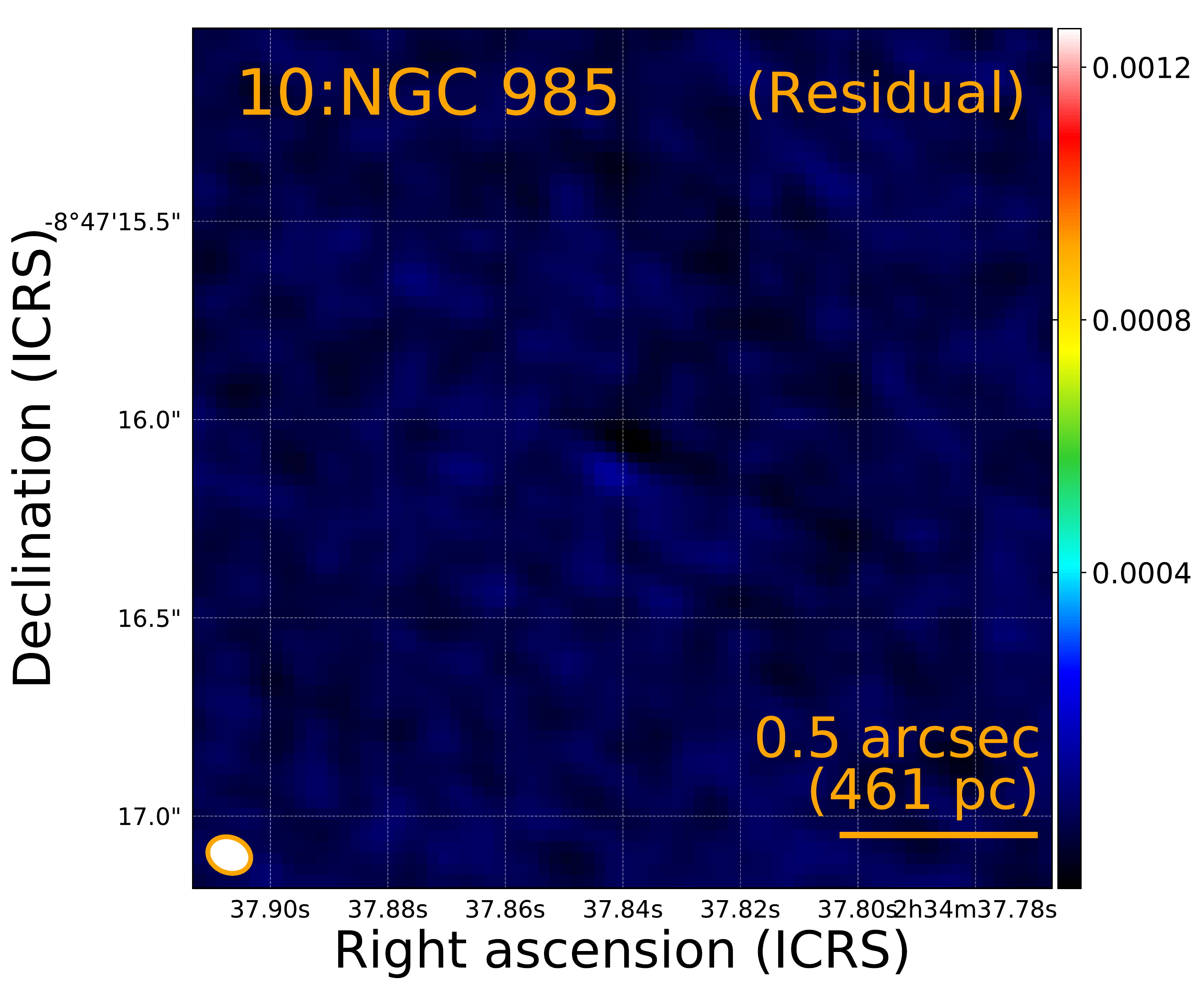}
    \caption{
    From top to bottom, a dirty image of NGC 985 and a residual image obtained by removing fitted un-resolved and extended components, which are modeled by Gaussian functions. We note that the clear extended morphologies seen in the dirty image do not necessarily indicate the presence of such emission in a reconstructed image produced by the cleaning task. 
    }
    \label{fig_app:visfit}
\end{figure}

To extract only the contribution of an unresolved component from the observed mm-wave emission and discuss its correlation in Section~\ref{sec:mor}, we fitted the observed visibility data of 38 AGNs using \textsc{UVMULTIFIT} \citep{Mar14}. 
As noted in the subsection, the AGNs were selected by considering their simple emission morphologies (i.e., pure C objects). Thus, the simple Gaussian function should be sufficient; indeed, we considered only two Gaussian functions to reproduce the observed data. One is for an unresolved component, and the other is for resolvable extended emission.
An example of our fit result is shown in Figure~\ref{fig_app:visfit}.  The residual image (bottom panel) indicates that our fit reproduces signals well in the dirty image (top panel). 


\clearpage

\bibliography{ref.bib}


\end{document}